\documentclass[11pt]{article}
\usepackage{latexsym}
\usepackage{amsmath}
\usepackage{multibox}
\usepackage{verbatim}
\usepackage{mathrsfs}

 \csname
@addtoreset\endcsname{equation}{section}

\def\prd{\pr \cdot}

\def\gz0{\gamma^{0}}

\def\nn{\nonumber}
\newcommand{\eq}[1]{(\ref{#1})}

\def\ket#1{|#1\rangle}
\def\scs#1{\section{\sc #1}}
\def\scss#1{\subsection{\sc #1}}
\def\scsss#1{\subsubsection{\sc #1}}

\def\g{\gamma}

\def\d{\delta}
\def\D{\Delta}
\def\e{\epsilon}

\def\z{\zeta}
\def\h{\eta}
\def\th{\theta}
\def\Th{\Theta}

\def\k{\kappa}
\def\l{\lambda}
\def\L{\Lambda}
\def\m{\mu}
\def\n{\nu}
\def\x{\xi}
\def\X{\Xi}

\def\P{\Pi}

\def\r{\rho}

\def\s{\sigma}

\def\t{\tau}

\def\vf{\varphi}
\def\c{\chi}

\def\o{\omega}
\def\O{\Omega}

\def\cB{{\cal B}}

\def\cE{{\cal E}}

\def\cJ{{\cal J}}
\def\cK{{\cal K}}
\def\cL{{\cal L}}

\def\cO{{\cal O}}

\def\cS{{\cal S}}

\def\cW{{\cal W}}

\def\cY{{\cal Y}}
\def\cZ{{\cal Z}}

\def\be{\begin{equation}}
\def\ee{\end{equation}}
\def\bea{\begin{eqnarray}}
\def\eea{\end{eqnarray}}
\def\ba{\begin{array}}
\def\ea{\end{array}}
\def\bec{\begin{center}}
\def\ec{\end{center}}
\def\ba{\begin{align}}
\def\ena{\end{align}}

\def\hpe{\pe \hat{\phantom{\! \! \pe}}}
\def\pe{\prime}
\def\12{\frac{1}{2}}
\def\fr{\frac}
\def\pr{\partial}
\def\prd{\partial \cdot}
\def\bra{\langle \,}
\def\ket{\, \rangle}
\def\comma{\,,\,}
\def\eq{\equiv}

\def\ra{\rightarrow}

\def\dsl{\not {\! \pr}}
\def\psl{\not {\hspace{-0.09cm} p}}
\def\dsll{\not {\! \pr}}

\def\psisl{\not {\!\! \psi}}
\def\cWsl{\not {\!\!\! \cal W}}

\def\esl{\not {\! \epsilon}}

\def\ssl{\not {\! \cal S}}

\def\xisl{\not {\! \xi}}
\def\xibsl{\not {\! \bar{\xi}}}

\def\xibsl{\not {\! \bar{\xi}}}
\def\ssl{\not {\! \cal S}}

\def\bep{\bar{\e}}

\begin{document}

\begin{flushright}
{\today}
\end{flushright}

\vspace{25pt}

\begin{center}


{\Large\sc Unconstrained Higher Spins of Mixed Symmetry \vskip 12pt II. Fermi Fields}\\


\vspace{25pt}
{\sc A.~Campoleoni${}^{\; a,\,}$\footnote{Address after October 1, 2009: Max-Planck-Institut f\"ur Gravitationsphysik, Albert-Einstein-Institut,
Am M\"uhlenberg 1, DE-14476 Golm, Germany.}, D.~Francia${}^{\; b}$, J.~Mourad$^{\; b}$ and A.~Sagnotti$^{\; a}$}\\[15pt]

{${}^a$\sl\small
Scuola Normale Superiore and INFN\\
Piazza dei Cavalieri, 7\\I-56126 Pisa \ ITALY \\
e-mail: {\small \it a.campoleoni@sns.it,
sagnotti@sns.it}}\vspace{10pt}

{${}^b$\sl\small AstroParticule et Cosmologie (APC) \footnote{Unit\'e mixte de Recherche du CNRS (UMR 7164.)}\\
Universit\'e Paris VII - Campus Paris Rive Gauche \\
10, rue Alice Domon et Leonie Duquet \\ F-75205 Paris Cedex 13\
FRANCE
\\ e-mail:
{\small \it francia@apc.univ-paris7.fr, mourad@apc.univ-paris7.fr}}\vspace{10pt}

\vspace{35pt} {\sc\large Abstract}\end{center}

{This paper is a sequel of arXiv:0810.4350 [hep-th], and is also devoted
to the local ``metric-like'' unconstrained Lagrangians and field
equations for higher-spin fields of mixed symmetry in flat space.
Here we complete the previous constrained on-shell formulation of
Labastida for Fermi fields, deriving the corresponding constrained Lagrangians
both via the Bianchi identities and via the requirement of
self-adjointness. We also describe two types of unconstrained
Lagrangian formulations: a ``minimal'' one, containing higher
derivatives of the compensator fields, and another non-minimal one,
containing only one-derivative terms. We identify classes of these
systems that are invariant under Weyl-like symmetry transformations.}

\setcounter{page}{1}

\pagebreak

{\linespread{0.75}\tableofcontents}

\newpage


\scs{Introduction}\label{sec:intro}


This paper is devoted to fermionic higher-spin fields \cite{solvay}
of mixed symmetry in flat space, and thus complements the
corresponding treatment for bosonic fields presented in
\cite{bose_mixed}. The general framework and the motivations for the
present analysis were already illustrated in detail in
\cite{bose_mixed} and will not be repeated here, where we shall
confine ourselves to some additional considerations.

Actions for generic massless symmetric spinor-tensors were first
constructed in the late seventies by Fang and Fronsdal
\cite{fangfronsdal} starting from the massive Singh-Hagen
\cite{singhagen} construction. They were thus able to display the
gauge symmetry underlying these massless Lagrangians, while also
inheriting from the massive theory some surprising algebraic
constraints on gauge parameters and gauge fields: in their
setup the former are in fact to be $\g$-traceless, while the
latter are to be triply $\g$-traceless. Other independent
justifications for the constraints were provided in
\cite{curt0, dewfr} shortly thereafter.

In the eighties, crucial inputs originating from the development of
String Field Theory \cite{sftheory} shifted the attention to tensors
of mixed symmetry. Important progress was then made by a number of
authors \cite{mixed,siegel} (with some relevant work actually dating
back to previous years \cite{curt}), and most notably Labastida
\cite{firstlaba, labastida} constructed bosonic Lagrangians
extending what Fronsdal had attained for symmetric tensors in
\cite{fronsdal}. Those constrained Lagrangians were recently taken
as a starting point for a theory of
unconstrained mixed-symmetry bosons presented in \cite{bose_mixed}. Indeed, as
shown in a number of previous works
\cite{bpt, fsoldnl, fsoldl1, fsold2, mixednl, fms, buchnew, dario07}, the need for
algebraic constraints on gauge fields and parameters can be
naturally foregone, either via the addition of non-local terms or
via the introduction of auxiliary fields. While this could well
help to simplify the structure of the interactions, one clear
motivation to delve more deeply into these matters is provided by
String Field Theory \cite{sftheory}, whose equations of
motion, in their simplest form and in the tensionless limit,
describe reducible chains of massless higher-spin modes, without
any algebraic constraints on gauge fields and gauge parameters
\cite{triplets}. In addition, in the free limit a geometrical formulation of interacting
higher-spin gauge fields along the lines of Einstein's
gravity\footnote{Both the present work and its companion paper
\cite{bose_mixed} proceed along the lines of the metric formalism
for gravity. An important alternative is the higher-spin counterpart
of the frame-like formalism \cite{lopvas,extraframe}, that lies at
the heart of the Vasiliev setting, where the interactions are
driven by an infinite-dimensional extension of the tangent-space
Lorentz algebra \cite{vasold,vasnew}.} would be expected to
rely somehow on the linearized higher-spin curvatures introduced by de Wit
and Freedman for the symmetric case \cite{dewfr}, or on their
generalizations for mixed-symmetry fields discussed in
\cite{mixednl}. Curvature tensors are defined in
terms of unconstrained gauge fields, while their properties rely on
gauge transformations involving unconstrained gauge parameters, which
calls for a formulation of the free dynamics that is capable of
avoiding from the outset the algebraic restrictions
of the Fang-Fronsdal-Labastida construction.

As we anticipated, both the Fronsdal constraints and their
generalizations can be bypassed in two quite distinct, albeit
related, ways. The first rests directly on the higher-spin
curvatures, and was indeed shown to be possible in
\cite{fsoldnl,fms} for symmetric tensors, at the price of allowing
\emph{non-local} contributions to the Lagrangians. Investigations in
the same spirit were then performed for bosons of mixed symmetry in
\cite{mixednl}, and more recently for massive symmetric Bose and
Fermi fields in \cite{dario07}, leading in particular to a
formulation of massive higher spins in which, in sharp contrast with
the original Singh-Hagen setting \cite{singhagen}, \emph{no}
auxiliary fields are needed. The second way rests on the
introduction of additional fields in more conventional local
Lagrangian formulations \cite{bpt,fsold2,fms}. In its ``minimal
form'' \cite{fsoldl1, fsold2, fms}, Fronsdal's constraints can then be
foregone at the price of introducing only two additional fields and
some higher-derivative terms that are anyway pure gauge. In
particular, unconstrained Lagrangians for symmetric Fermi fields
of spin $s+\frac{1}{2}$ were obtained~\footnote{Previous
investigations in this direction \cite{bpt}, based on BRST
techniques \cite{BRST}, led to constructions involving ${\cal
O}(s)$ different fields to describe unconstrained spin-$s$ modes.
The reduction of the corresponding equations to the minimal ones
involving the compensators was first presented in \cite{fsoldl1}.
See also \cite{schwinger} for an early discussion of the spin $3$ compensator.
More recently, an
alternative formulation without higher-derivative terms was obtained
in \cite{buchnew} for symmetric bosons and fermions by suitably
restricting the ``triplets'' \cite{triplets} of String Field Theory.
See also \cite{sorokvas} for a discussion of the triplets in the
frame-like formalism.} in \cite{fsold2}. Together with the gauge
potential $\psi_{\,\mu_1 \ldots\, \mu_s}$, they involve an additional
spin-$(s-2)$ compensator $\xi_{\,\mu_1 \ldots\, \mu_{s-2}}$,
that first emerges for spin $s = \frac{5}{2}$, and a spin-$(s-3)$
Lagrange multiplier, that first emerges for spin $s =
\frac{7}{2}$.

Even at the constrained level, the theory of mixed-symmetry
fermionic fields $\psi_{\m_1 \ldots\, \m_{s_1}\!,\, \n_1 \ldots\, \n_{s_2}, \ldots}$,
bearing several ``families'' of fully symmetric indices, was left somewhat incomplete in the metric-like
approach, since Labastida \cite{labferm} did not arrive at
Lagrangians generalizing the results of Fang and Fronsdal~\footnote{Actually, fermionic Lagrangians
were first built in \cite{siegel}, within a BRST-like formalism
along the lines of the first attempts to construct free covariant
string equations, in an approach that is rather remote from this
line of developments.}. He
nonetheless proposed free fermionic equations of motion and
identified the proper generalizations of the constraints on parameters and gauge
fields, that in our notation read
\be \g_{\,(\,i} \, \e_{\,j\,)} \ = \ 0 \ , \qquad T_{(\,ij}\,\g_{\, k\,)}\,
\psi \ = \ 0 \ . \label{laba2} \ee
Here $\g_{\,i}$ denotes a $\g$-trace in the $i$-th index family, $T_{ij}$
denotes a trace involving a pair of space-time indices belonging,
respectively, to the $i$-th and $j$-th families, and the
parentheses indicate, here and elsewhere in this paper, complete symmetrization of the enclosed indices,
with the minimal number of terms needed to this end and
with \emph{unit} overall normalization.
Hence, quite differently from what the symmetric case of
\cite{fangfronsdal} could naively suggest, but in strict analogy
with the bosonic construction of \cite{firstlaba, labastida},
\emph{not all} $\g$-traces of the gauge parameters and \emph{not
all} triple $\g$-traces of the gauge fields are forced to vanish in Labastida's formulation.

In this paper we present three main groups of results :
\begin{itemize}
\item we extend to the Lagrangian level the field equations
proposed by Labastida for Fermi fields of mixed symmetry in $D$ dimensions (see eqs.~\eqref{lagferconstr}
and \eqref{solgenf}).
In Section \ref{sec:dofs} we also show that their propagating degrees
of freedom fill representations of the little group $O(D-2)$;

\item we propose a number of alternative formulations in which all constraints
are removed at the price of introducing ``minimal''
sets of additional auxiliary fields, in the spirit of what was already done for bosons in \cite{bose_mixed} (see for instance eqs.~\eqref{laggenfunc} and \eqref{basiclagfer});

\item we show that the resulting Lagrangians can display Weyl-like symmetries for particular models in special space-time dimensions. In particular, in Section \ref{sec:reduction2f} we provide a full classification of Weyl-invariant Lagrangians for two-family Fermi fields, while in Sections \ref{sec:reductionf} and \ref{sec:irr_example} we set up the corresponding analysis for general $N$-family fields and display some interesting classes of examples.
\end{itemize}

For pedagogical reasons, in  Section \ref{sec:2fermi} we open our discussion constructing
Lagrangians and field equations for the simplest generalization of the fully symmetric
case, two-family fermionic gauge fields of the type $\psi_{\,\mu_1 \ldots\, \mu_{s_1},\, \nu_1
\ldots\, \nu_{s_2}}$, and  study their on-shell reduction to the constrained Labastida form
of \cite{labferm}. As in \cite{bose_mixed}, we discuss in detail the emergence of
Weyl-like symmetries and their role in the reduction procedure of the field equations to
the non-Lagrangian Labastida form. We conclude Section \ref{sec:2fermi} with the explicit
example of reducible rank-$(3,2)$ spinor-tensor fields, that are relatively simple but
suffice to illustrate a number of key points of our construction~\footnote{In contrast with the notation used for instance in \cite{fms}, in this paper the rank refers only to the vector indices carried by spinor-tensor fields.}. In Section
\ref{sec:generalf} we extend the discussion to multi-family fermionic fields, identifying via the Bianchi identities their unconstrained Lagrangians. When restricted
to the case of constrained fields, our results determine the off-shell Lagrangian
extension of the field equations of \cite{labferm}. However, we also
describe how, in the constrained case, requiring that the kinetic operator be self-adjoint, along the lines of what was previously done for bosons in \cite{labastida},
one can recover the same result. Moreover, we show that the non-Lagrangian Labastida equation of \cite{labferm} propagates in general the correct degrees of freedom and comment on the degenerate low-dimensional cases in which the Lagrangian equations do not reduce directly to it and Weyl-like symmetries emerge.

The key ingredients of our construction are \emph{reducible} $gl (D)$
spinor-tensors $\psi_{\,\mu_1 \ldots\, \mu_{s_1}
,\,\nu_1 \ldots\, \nu_{s_2},\,\ldots}$,
possessing several ``families'' of \emph{symmetric} index sets, in a
$D$-dimensional space time. Spinor-tensor fields of this kind
describe a portion of the massive fermionic
excitations of the five ten-dimensional superstrings, of their type-0
counterparts \cite{type0} and of various brane systems \cite{dms}.
However, in Section \ref{sec:irreducible} we
also describe how to adapt the results of this paper to
spinor-tensor fields transforming in \emph{irreducible} representations
of the Lorentz group, so as to clarify how the theory develops along
lines that are more closely related to what is usually done for low
spins. Both the Fang-Fronsdal-Labastida kinetic operators and the corresponding
Lagrangians take the same form for reducible and for Young-projected irreducible fields, but when considering the latter the theory involves fewer independent gauge parameters and fewer independent auxiliary fields, in a way that we spell out in detail. We also display an instructive off-shell counting argument that illustrates for irreducible two-family fields how the Labastida constraints of \cite{labferm} result in the propagation of irreducible representations of the Poincar\'e group. In Section \ref{sec:irr_example} we show that combining the conditions for irreducibility with the constraints selecting cases with Weyl-like symmetries allows one to find explicit solutions of the latter for an interesting class of mixed-symmetry fields.

In Section \ref{sec:multiforms} we adapt our formalism, following largely previous results of \cite{mixednl}, to
another relevant class of fields, multi-form spinor-tensors possessing several
``families'' of \emph{antisymmetric} index sets in a $D$-dimensional space
time, while in Section \ref{sec:lowerder} we discuss how the minimal
higher-derivative unconstrained Lagrangians of Sections \ref{sec:2fermi} and \ref{sec:generalf}
can be reduced systematically to others with a single derivative,
generalizing the results of \cite{dario07, bose_mixed}. Section
\ref{sec:conclusions} contains our conclusions, and the paper closes
with a number of appendices where our notation is carefully spelled
out and tedious intermediate derivations are elaborated upon.

Both here and in \cite{bose_mixed} our
notation is a natural extension of the index-free description used
in previous work on symmetric (spinor-) tensors  by some of us
\cite{fsoldnl,fsoldl1, fsold2, fms, dario07}. It involves explicit ``family'' indices,
that are needed to identify particular subsets of the actual tensor indices to be
treated differently by the various operators entering field
equations and Lagrangians. For instance, while the basic fermionic
field can still be simply denoted by $\psi$, as in the fully
symmetric case, its gauge transformation,
\be \delta \, \psi \ = \ \partial^{\,i} \, \epsilon_{\,i} \ , \ee
now involves explicit family indices. These identify
the various gauge parameters and the various
sets of tensor indices taking part in the gradient,
which are to be fully symmetrized (or antisymmetrized) \emph{only}
with others of the same family. At the same time, the gauge parameters
$\epsilon_{\,i}$ also carry one lower family index, and bear correspondingly
one less space-time index of the $i$-th family than the gauge
field $\psi$.

From the methodological viewpoint, our approach differs somewhat
from the original paper \cite{labastida}, since we are guided
throughout by the Bianchi identities and their $\g$-traces. This
choice was instrumental for deriving, in a relatively handy fashion, complete
local Lagrangians for Fermi fields of mixed symmetry and provides
a neat rationale for these rather complicated constructions.

This paper is admittedly rather long. However, Section \ref{sec:2fermi} can already provide a fair view of our methods and of the subtleties to be met in the general case, both in its early general portion and in the explicit examples of Section \ref{sec:examples2f}. On the other hand, Section \ref{sec:generalf} contains our main results (eqs.~\eqref{lagferconstr} and \eqref{solgenf}), but the derivation of the Lagrangians is somewhat involved and may be skipped in a first reading. Sections \ref{sec:irreducible} and \ref{sec:multiforms}, devoted to irreducible fields and to multi-forms, should be also accessible to a large extent after reading Section \ref{sec:2fermi}. Finally, a glance at the first part of Section \ref{sec:lowerder} can provide a fair view of the key ideas behind the formalism that we are proposing to reduce the unconstrained models to more conventional one-derivative Lagrangians.

\vskip 24pt


\scs{Two-family fermionic fields}\label{sec:2fermi}


The ``metric-like'' theory of fermionic fields of mixed symmetry is less developed
than its bosonic counterpart, since Labastida \cite{labferm} did identify both their field equations and the algebraic constraints on gauge fields and parameters but
did not arrive at the corresponding Lagrangians. One of the aims of
the present paper is therefore to build constrained and unconstrained metric-like Lagrangian formulations for Fermi fields of mixed symmetry in flat space.
We begin, as we did for Bose fields in \cite{bose_mixed}, with a discussion of the
relatively handy two-family fields, in this case spinor-tensors of the type $\psi_{\,\mu_1\ldots\, \mu_{s_1} , \, \nu_1 \ldots\, \nu_{s_2}}$, that can already exhibit the key features of
our construction.

\vskip 24pt


\scss{The Lagrangians}\label{sec:lagrangian2f}


In the following all vector and spinor indices will be left implicit, as in \cite{fsoldnl,fsoldl1, fsold2, fms, dario07, bose_mixed}, so that a generic spinor-tensor of the type $\psi_{\,\mu_1\ldots\, \mu_{s_1} , \, \nu_1 \ldots\, \nu_{s_2}}$ will be simply denoted by $\psi$. However, operators like gradients, divergences and ($\gamma$-)traces will bear ``family'' indices specifying the sets of space-time indices they are acting upon. In particular, divergences and gradients will be denoted exactly as in \cite{bose_mixed}, so that, for instance,
\begin{eqnarray}
\pr_{\,1} \, \psi & \equiv & \pr_{\,\l} \, \psi^{\,\l}{}_{\m_1 \ldots\, \m_{s_1-1} , \, \n_1 \ldots\, \n_{s_2}} \, , \nn \\
\pr^{\,1} \, \psi & \equiv & \pr_{\,(\,\m_1} \, \psi_{\,\m_2 \ldots\,
\m_{s_1+1} \,)\, , \, \n_1 \ldots\, \n_{s_2}} \, . \label{firstnot}
\end{eqnarray}
As in \cite{bose_mixed} and as we shall explain in detail in Appendix \ref{app:MIX}, an upper family index, as the one borne by the gradient above, will signal the addition of a space-time index to the corresponding group or family, while a lower family index, as the one borne by the divergence above, will signal its removal. In a similar spirit, even when dealing with Fermi fields it is convenient to introduce traces and metric tensors, carrying
respectively a pair of lower or upper family indices, so that for instance~\footnote{The reader should notice that a somewhat unconventional but convenient factor, $\12$, enters the definition of $\h^{12}$, as in \cite{bose_mixed}.}
\begin{eqnarray}
T_{12} \, \psi & \equiv & \psi^{\,\l}{}_{\m_1 \ldots\, \m_{s_1-1} , \, \l \, \n_1 \ldots\, \n_{s_2-1}} \, , \nn \\
\h^{12} \, \psi & \equiv & \12 \, \sum_{n\,=\,1}^{s_2+1} \, \h_{\,\m_n\,(\, \n_1 \, |} \, \psi_{\, \ldots\, \m_{r\,\neq\,n}\, \ldots\, , \, | \, \n_2 \ldots\, \n_{s_2+1}\, )} \, . \label{secondnot}
\end{eqnarray}
We should stress again that in this paper groups of (space-time or family) indices surrounded by parentheses are meant to be symmetrized via the minimal number of terms needed to this end, and to be normalized with unit overall coefficient, rather than with unit strength.
Aside from these building blocks, fermionic Lagrangians involve two new types of operators, both related to the space-time $\gamma$-matrix and bearing, respectively, one lower and one upper family index. The former,
$\gamma_{\,i}$, removes one Lorentz index belonging to the $i$-th family via a
$\gamma$-trace, while the latter, $\gamma^{\,i}$, adds an ordinary
$\gamma$-matrix bearing a Lorentz index of the $i$-th family, to be
properly symmetrized with all its peers. Thus, for
instance, in the two-family case one is also led to consider expressions of the type
\begin{eqnarray}
\g_{\,1}\, \psi & \equiv & \g_{\,\l} \, \psi^{\,\l}{}_{\m_1 \ldots\, \m_{s_1-1} , \, \n_1 \ldots\, \n_{s_2}} \, , \nn \\
\g^{\,1} \, \psi & \equiv & \g_{\,(\,\m_1} \, \psi_{\,\m_2 \ldots\,
\m_{s_1+1} \,)\, , \, \n_1 \ldots\, \n_{s_2}} \, . \label{gammanot}
\end{eqnarray}
Together with these family $\g$-matrices, we shall also need to
consider their antisymmetric combinations, that in the two-family
case are only $\g^{\,ij}$ and $\g_{\,ij}$. The former adds two
Lorentz indices borne by a conventional antisymmetric combination of
two $\g$'s, say $\gamma_{\m\n}$, while the latter signals an
antisymmetric combination of two $\g$-traces. In general, it is also
convenient to introduce the compact notations
\be
\dsl \, = \, \g_{\,\m} \, \pr^{\,\m} \ , \qquad\qquad \psisl_{\,i} \, = \, \g_{\,i} \, \psi \ ,
\ee
where $\gamma_\mu$ clearly denotes a conventional space-time
$\gamma$-matrix.

In complete analogy with the bosonic case, the starting point to
build the Lagrangians is the Fang-Fronsdal-Labastida kinetic tensor
\cite{labferm}
\be
\cS \, = \, i \left(\, \dsl \, \psi \, - \, \pr^{\,i} \! \psisl_i \,\right) \,
,
\label{fermikin}
\ee
supplemented with the \emph{unconstrained} gauge transformation of
the two-family spinor-tensor $\psi$,
\be
\d \, \psi \, = \, \pr^{\,i} \, \e_{\,i} \, . \label{gaugef}
\ee
As in the bosonic case, this gauge symmetry is reducible, and the
corresponding gauge-for-gauge transformations read
\be
\d \, \e_{\, i} \, = \, \pr^{\,j} \, \e_{\,[\, ij\, ]} \,
,\label{gaugexgaugef}
\ee
where the parameter $\e_{\,[\,ij\,]}$ is antisymmetric in its family
indices. At two families the chain stops here, while in the general
multi-family setting one must clearly face longer chains of
gauge-for-gauge transformations.

The gauge variation of (\ref{fermikin}),
\be \label{fermvar}
\d \, \cS \, = \, - \, \frac{i}{2} \ \pr^{\,i} \pr^{\,j} \, \g_{\,(\,i}\,\e_{\,j\,)} \,
,
\ee
is proportional to the Labastida constraints on the gauge parameters,
\be
\g_{\,(\,i}\,\e_{\,j\,)} \, = \, 0 \, , \label{labacfermg}
\ee
and can be disposed of introducing compensator fields $\xi_{\,ij}$ such that
\be
\d \, \x_{\,ij} \, = \, \12 \ \g_{\,(\,i} \, \e_{\,j\,)} \, . \label{fermicompens}
\ee
As in the bosonic case, however, these compensators are nicely covariant in family notation but can not be
independent in general, simply because the Labastida constraints are not independent. Indeed, gauge invariant combinations of the $\g$-traces of the $\xi_{\,ij}$ exist~\footnote{The example presented to this effect in \cite{bose_mixed} is actually incorrect, since the
combination in eq.~(2.11) of that paper \emph{is not} gauge invariant. A poignant example, however, is the more complicated expression in eq.~(2.49) of \cite{bose_mixed}, that on the contrary \emph{is} gauge invariant and is available in general from two families onward. A corresponding example for Fermi fields will be presented in eq.~\eqref{gauge_inv}. In the bosonic case there is another example that can be analyzed
by inspection, a combination of triple traces of the $\alpha_{\,ijk}$ compensators carrying a $\{5,3,1\}$ Young projection in the family indices. Such an expression is gauge invariant simply because one cannot recover its structure starting from four traces of the gauge parameters $\L_i$, since multiple $T_{ij}$ lead to Young diagrams with even numbers of boxes in their rows.}, so that they are to be regarded as $\gamma$-traces of
independent compensators $\Psi_i$ such that
\be
\xi_{\,ij}\,(\Psi) =
\frac{1}{2} \ \gamma_{\,(\,i} \,
\Psi_{j\,)}\, , \label{xipsi}
\ee
with
\be \label{deltapsi}
\delta\, \Psi_{\,i} \, = \, \e_{\,i} \, .
\ee
Now, in complete analogy with the bosonic construction of \cite{bose_mixed},
or with the one-family case of
\cite{fsold2,fms}, one can move directly to the unconstrained theory,
defining the gauge invariant tensor
\be \label{W}
\cW \, = \, \cS \, + \, i \ \pr^{\,i} \pr^{\,j} \, \x_{\,ij}\,(\Psi) \, ,
\ee
that in the following will be the basic kinetic tensor for the theory. All our results can be easily adapted to the constrained setting via a partial gauge fixing, and
for brevity in the
following we shall often write $\xi_{\,ij}$ to indicate these combinations, without displaying explicitly their dependence on the $\Psi_i$.
Notice that eq.~\eqref{W} involves \emph{two} derivatives of the compensator field.
In Section \ref{sec:lowerder} we shall see how more conventional forms of the kinetic operator
involving a single derivative obtain via the introduction of a few additional types of
fields.

Furthermore, as was the case for bosons, the composite compensator $\xi_{\,ij}(\Psi)$ of
eq.~(\ref{xipsi}) would emerge after performing in the
Fang-Fronsdal-Labastida tensor $\cS$ the Stueckelberg-like shift
\be
\psi \ \to \ \psi \,-\, \partial^{\,i}\, \Psi_{\,i} \label{stueckf}
\ee
to generate $\cW$, an approach we shall elaborate upon at the end of this section \footnote{M.A. Vasiliev
stressed to us the role of this shift in the symmetric case of
\cite{fsold2}. In the following we shall see how to formulate the
procedure in various ways for mixed-symmetry spinor-tensors.}.
Actually, this would only build the combination (\ref{xipsi}),
as in eq.~\eqref{W}, a circumstance that reflects the presence
of the constrained Labastida gauge symmetry \eqref{fermvar}.
Notice that under the gauge-for-gauge transformations
\eqref{gaugexgaugef} the $\Psi_i$ fields shift like gauge
parameters would, and thus like gauge fields. This is
as it should be: if they were inert like the higher-spin field
$\psi$, part of the gauge transformations would be ineffective on
them, in contradiction with the manifest possibility of removing the
$\Psi_i$ by simply undoing the Stueckelberg shift.

The Bianchi identities for $\cW$ are
\be \label{bianchi_W} \pr_{\,i}\, \cW \, - \, \12 \dsl \ \cWsl_i \, - \, \12 \ \pr^{\,j} \, T_{ij} \, \cW \, - \, \frac{1}{6} \ \pr^{\,j} \,
\g_{\,ij} \, \cW \, = \, \12 \ \pr^{\,j} \pr^{\,k} \, \cZ_{\,ijk} \, , \ee
where the gauge-invariant spinor-tensor constraints
\be \label{z}
\cZ_{\,ijk} = \, \frac{i}{3} \left\{ \, T_{(\,ij} \psisl_{k\,)}  - \, 2 \, \pr_{\,(\,i} \, \xi_{jk\,)}\, -  \dsl \, \g_{\,(\,i} \, \xi_{jk\,)}  - \, \pr^{\,l} \left(\, 2 \, T_{(\,ij} \, \xi_{k\,)\,l}  - \, T_{l\,(\,i} \, \xi_{jk\,)}  - \, \g_{\,l\,(\,i} \, \xi_{jk\,)} \,\right) \right\}
\ee
are totally
symmetric in their family indices, and when expressed in
terms of the independent compensators of eq.~\eqref{xipsi}
take the far simpler form
\be
\cZ_{\,ijk} \, = \, \frac{i}{3} \ T_{(\,ij}\, \gamma_{\, k\,
)} \left( \, \psi \, -\, \pr^{\,l} \, \Psi_{\, l} \, \right) .
\label{constrf_r}
\ee
After removing the compensators, eq.~\eqref{constrf_r}
identifies precisely the Labastida constraints
on the gauge field,
\be
T_{(\,ij} \psisl_{k\,)} \, = \, 0 \, , \label{labacf}
\ee
that can thus be read directly from the Bianchi identities for the
Fang-Fronsdal-Labastida tensor $\cS$ of \cite{labferm}.

The constraints \eqref{labacf} on the gauge fields are \emph{not} independent in general, just like the constraints \eqref{labacfermg} on the gauge parameters, and indeed, once the $\x_{\,ij}$ are expressed in terms of the independent compensators $\Psi_{i}$, the $\g$-traces of the $\cZ_{\,ijk}$ spinor-tensors satisfy algebraic relations, so that for instance
\be \label{comb}
\left(\, \g_{\,m\,(\,i}\, \cZ_{\,j\,)\,kl} \, + \, \g_{\,m\,(\,k}\, \cZ_{\,l\,)\,ij} \,\right) \, - \, 15 \ Y_{\{4,1\}} \, \left(\, T_{ij}\, \cZ_{\,klm} \, + \, T_{kl}\, \cZ_{\,ijm} \,\right) \, = \, 0 \, ,
\ee
where $Y_{\{4,1\}}$ denotes the Young projector onto the irreducible $\{4,1\}$ representation of the permutation group acting on the family indices~\footnote{The Young projector acts only on the last two terms in eq.~\eqref{comb} since, when they are expressed in terms of the fields via eq.~\eqref{constrf_r}, the first two terms automatically result in a $\{4,1\}$-projected combination. Our conventions for the permutation group are spelled out at the end of Appendix \ref{app:MIX}.}. The lack of independence of the Labastida constraints for the gauge parameters \eqref{labacfermg} and of their counterparts \eqref{labacf} for the gauge fields
can be traced to a common origin. Indeed, starting from eq.~\eqref{z} the part of the combination \eqref{comb} involving the $\x_{\,ij}$ is not forced to vanish a priori. Rather, in complete analogy with the bosonic case of \cite{bose_mixed} it reduces to
\be \label{gauge_inv}
\begin{split}
& \left(\, \g_{\,m\,(\,i}\, \cZ_{\,j\,)\,kl} \, + \, \g_{\,m\,(\,k}\, \cZ_{\,l\,)\,ij} \,\right) \, - \, 15 \ Y_{\{4,1\}} \, \left(\, T_{ij}\, \cZ_{\,klm} \, + \, T_{kl}\, \cZ_{\,ijm} \,\right) \, = \\
& = \, \frac{i}{3} \ \pr^{\,n} \left\{\, 8 \, T_{(\,ij}\,T_{kl\,)}\, \x_{\,mn} \, - 4 \, T_{mn}\, T_{(\,ij}\, \x_{\,kl\,)} \, + \, 4\, T_{m\,(\,i\,|}\, T_{n\,|\,j}\, \x_{\,kl\,)} \, - \, 2\, T_{(\,m\,|\,(\,i}\,T_{jk}\, \x_{\,l\,)\,|\,n\,)} \,\right\} \\
& + \, \frac{i}{3} \ \pr^{\,n} \left\{\, 2\, T_{(\,i\,|\,(\,m}\, \g_{\,n\,)\,|\,j}\, \x_{\,kl\,)} \, - \, 2 \, \g_{\,(\,m\,|\,(\,i}\, T_{jk}\, \x_{\,l\,)\,|\,n\,)} \,\right\} \, ,
\end{split}
\ee
where the left over combination of ($\g$-)traces of the $\x_{\,ij}$ is gauge invariant and is available in general from two families onward. The terms entering eq.~\eqref{gauge_inv} involve four $\g$-traces acting on the $\x_{\,ij}$, and in the two-family case it is not possible to obtain gauge invariant combinations containing smaller numbers of $\g$-traces \footnote{In order for the most general linear combination of three $\g$-traces
of the compensators to be gauge invariant several cancelations must
occur simultaneously. The resulting system apparently admits
a number of solutions, but a closer analysis shows that they all involve combinations of
$\g$-traces
of the compensators that are antisymmetric in three indices, and thus vanish identically
where only two index families are present.}. Therefore, the linear dependence of the Labastida constraints does not manifest itself in simple mixed-symmetry models with small numbers of space-time indices.

The form of the Bianchi identities \eqref{bianchi_W}
now suggests to start from the trial Lagrangian
\be \label{trial_ferm}
\cL_0 \, = \, \12 \, \bra \bar{\psi} \comma \cW \, - \, \12 \
\g^{\,i} \cWsl_i \, - \, \12 \ \h^{\,ij} \, T_{ij} \, \cW \, - \,
\frac{1}{12} \ \g^{\,ij} \, \g_{\,ij} \, \cW \ket \, + \, \textrm{h.c.} \ ,
\ee
where, as in \cite{bose_mixed}, we have introduced a convenient scalar product
that is defined in Appendix \ref{app:MIX}. The gauge variation of $\cL_0$ is then
\be
\begin{split}
\d \, \cL_0 \,  = \, & - \, \12 \, \bra \bep_{\,i} \comma \pr_{\,i}\, \cW
\, - \, \12 \dsl \, \cWsl_i \, - \, \12 \ \pr^{\,j} \, T_{ij} \, \cW \, -
\, \frac{1}{6} \ \pr^{\,j} \, \g_{\,ij} \, \cW \ket \\& + \, \frac{1}{4} \,
\bra \bep_{\,i} \comma \g^{\,j} \, \pr_{\,i}
\, \cWsl_j \, - \, \frac{1}{3} \ \g^{\,j} \! \dsl \, \g_{\,ij} \, \cW
\ket \, + \, \frac{1}{4} \ \bra \bep_{\,i} \comma \h^{\,jk} \,
\pr_{\,i} \, T_{jk} \, \cW \ket \\& + \, \frac{1}{24} \, \bra
\bep_{\,i} \comma \g^{\,jk} \, \pr_{\,i} \, \g_{\,jk} \, \cW \ket \, +
\, \textrm{h.c.} \label{deltal0}
\end{split}
\ee
and, as in the bosonic case, the key ingredients to build the complete
unconstrained Lagrangian are Young projections for the family
indices to distinguish in $\d \cL_0$ contributions related to
the gauge transformation (\ref{fermicompens}) of the composite compensators
$\xi_{\,ij}$ from others to be dealt with via suitable $\g$-traces of
the Bianchi identities. In the fermionic case, however, this identification is
not at all straightforward, and is actually fraught with new subtleties,
that we shall try to exhibit following a sequence of steps. The Bianchi identities \eqref{bianchi_W} can relate directly the first term of eq.~\eqref{deltal0}
to the $\cZ_{\,ijk}$ constraint tensors, while the other terms must be split
aforehand into different components that transform irreducibly under permutations of the
family indices. One is thus led to
\be \label{var1}
\begin{split}
\d \, \cL_0 \, = \, & - \, \frac{1}{12} \, \bra \pr_{\,(\,i} \, \pr_{\,j} \, \bep_{\,k\,)} \comma \cZ_{\,ijk} \ket \\
& + \, \frac{1}{16} \, \bra \bep_{\,(\, i} \, \g_{\,j\,)} \comma
\pr_{\,(\,i} \, \cWsl_{j\,)} \ket \, + \, \frac{1}{16}
\, \bra \bep_{\,[\,i} \, \g_{\,j\,]} \comma \pr_{\,[\,i} \cWsl_{j\,]} \, - \, \frac{2}{3} \dsl \, \g_{\,ij} \, \cW \ket \\
& + \, \frac{1}{72} \, \bra T_{(\,ij}\,\bep_{\,k\,)} \comma
\pr_{\,(\,i} \, T_{jk\,)} \, \cW \ket \, + \, \frac{1}{24} \, \bra
Y_{\{2,1\}} \, T_{jk}\,\bep_{\,i} \comma \left(\, 2 \, \pr_{\,i}
\, T_{jk} \, - \, \pr_{\,(\,j} \, T_{k\,)\,i} \,\right) \cW \ket \\&
+ \frac{1}{72} \, \bra Y_{\{2,1\}} \, \bep_{\,i} \, \g_{\,jk}
\comma \left(\, 2 \, \pr_{\,i} \, \g_{\,jk} \, - \,
\pr_{\,[\,j}\,\g_{\,k\,]\,i} \,\right) \cW \ket \, + \, \textrm{h.c.}
\ ,
\end{split}
\ee
after discarding the $\{1,1,1\}$ projection, that a priori
$\bep_{\,i} \, \g_{\,jk}$ would admit but is clearly not available with
only two families. Aside from ordinary parentheses, that as we stated above
denote symmetrizations, here we are also introducing square brackets. As is commonly the case,
these denote antisymmetrizations, here built via the minimum possible number of terms and
defined with unit overall coefficient.

The second and fourth terms of eq.~\eqref{var1} are both
symmetrically projected and involve combinations of the type
$\bep_{\,(\, i} \,
\g_{\,j\,)}$. Taking into account that, with the ``mostly plus''
convention for the metric tensor adopted in this paper,
\be
\delta \, \bar{\x}_{\,ij} \, = \, - \, \12 \ \bep_{\,(\, i} \, \g_{\,j\,)} \, ,
\ee
they can be removed introducing
\be
\begin{split}
\cL_1 \, = \, \frac{1}{4} \, \bra \bar{\xi}_{\,ij} \comma \12 \ \pr_{\,(\,i}  \cWsl_{j\,)}
\, + \, \frac{1}{6} \ \g^{\,k} \, \pr_{\,(\,i} \, T_{jk\,)} \, \cW \ket \, + \, \textrm{h.c.} \
,
\end{split}
\ee
that collects all compensator terms that are already visible in the
symmetric case of \cite{fsold2,fms}. The total variation then becomes
\be \label{var12}
\begin{split}
\d \left( \, \cL_0 + \cL_1 \right) \, = \, & - \, \frac{1}{12} \, \bra \pr_{\,(\,i} \, \pr_{\,j} \, \bep_{\,k\,)} \comma \cZ_{\,ijk} \ket \, + \, \frac{1}{8} \, \bra \bep_{\,i} \, \g_{\,j} \comma \pr_{\,[\,i} \cWsl_{j\,]} \, - \, \frac{2}{3} \dsl \, \g_{\,ij} \, \cW \ket  \\
& + \, \frac{1}{24} \, \bra Y_{\{2,1\}} \, T_{jk}\,\bep_{\,i} \comma \left(\, 2 \, \pr_{\,i} \, T_{jk} \, - \, \pr_{\,(\,j} \, T_{k\,)\,i} \,\right) \cW \ket \\
& + \frac{1}{72} \, \bra Y_{\{2,1\}} \, \bep_{\,i} \, \g_{\,jk}
\comma \left(\, 2 \, \pr_{\,i} \, \g_{\,jk} \, - \,
\pr_{\,[\,j}\,\g_{\,k\,]\,i} \,\right) \cW \ket \, + \,
\textrm{h.c.} \ .
\end{split}
\ee
Aside from the first contribution, that contains the $\cZ_{\,ijk}$ and, as we shall see, can be compensated by Lagrange multipliers $\l_{\,ijk}$, this remainder can be dealt with resorting to some consequences of the Bianchi identities. For the second term, the relevant one is the
$\g$-trace of \eqref{bianchi_W},
\be
\begin{split} \label{bianchigamma}
& \pr_{\,[\,i} \cWsl_{j\,]} \, - \, \frac{2}{3} \dsl \, \g_{\,ij} \,
\cW \, + \, \frac{1}{6} \
\pr^{\,k} \, \g_{\,ijk} \, \cW \, - \, \, \frac{1}{6} \ \pr^{\,k} \left(\, 3 \, T_{ik} \, \g_j \, +
\, T_{j\,[\,i} \, \g_{\,k\,]} \,\right) \cW \\
& = \, \pr^{\,k} \!\! \dsl \, \cZ_{\,ijk} \, + \, \12 \
\pr^{\,k}\pr^{\,l} \, \g_{\,j} \, \cZ_{\,ikl}\, ,
\end{split}
\ee
that actually admits two independent Young projections. The first projection is a
symmetric $\{2\}$, that reads
\be \label{bianchi2}
\pr^{\,k} \, T_{(\,ij} \, \cWsl_{k\,)} \, = \, - \, 6 \ \pr^{\,k} \!\! \dsl \, \cZ_{\,ijk} \, -
\, \frac{3}{2} \ \pr^{\,k}\pr^{\,l} \, \g_{\,(\,i\,} \cZ_{\,j\,)\,kl}\, ,
\ee
and in complete analogy with the bosonic construction of \cite{bose_mixed}
reflects an algebraic identity satisfied by $\cW$,
\be
T_{(\,ij}\,\cWsl_{k\,)} \, = \, 3 \ \Big\{\, - \, 2 \dsl \,
\cZ_{\,ijk} \, + \, \pr^{\,l}\,\g_{\,l}\,\cZ_{\,ijk}
\, - \, \12 \ \pr^{\,l} \, \g_{\,(\,i}\,\cZ_{\,jkl\,)}  \,\Big\} \,
. \label{triplegammaW}
\ee
Eq.~\eqref{bianchi2} contains indeed on the right-hand side the
gradient of this expression, as can be seen taking into account that
the $(k,l)$ symmetry induced by the additional gradient makes the
last two terms directly comparable. On the other hand, the second projection of
eq.~\eqref{bianchigamma} is an antisymmetric $\{1,1\}$, that reads
\be
\pr_{\,[\,i} \cWsl_{j\,]} \, - \, \frac{2}{3} \dsl \, \g_{\,ij} \, \cW \, = \, - \, \frac{1}{6} \ \pr^{\,k} \, \g_{\,ijk} \, \cW \, + \, \frac{1}{3} \ \pr^{\,k} \, T_{k\,[\,i} \, \g_{\,j\,]} \, \cW \, - \, \frac{1}{4} \ \pr^{\,k}\pr^{\,l} \, \g_{\,[\,i} \, \cZ_{\,j\,]\,kl} \label{bianchi11} \, ,
\ee
where the left-hand side is exactly as needed for eq.~(\ref{var12}).
It should be appreciated that for fully symmetric $\psi$ fields only
eq.~\eqref{bianchi2} would survive, so that the $\gamma$-trace of
the Bianchi identity would not contribute any useful information.

The remaining two terms in eq.~\eqref{var12} present some novel
features with respect to the bosonic case that are worthy of some
discussion. First of all, taking a further $\g$-trace of the Bianchi
identities gives rise to two distinct types of expressions, a trace
$T_{jk}$ and an antisymmetric $\g$-trace $\g_{jk}$. The first reads
\be \label{tbianchi2}
\begin{split}
&\12 \left(\, 2 \, \pr_{\,i} \, T_{jk} \, - \, \pr_{\,(\,j} \, T_{k\,)\,i} \,\right) \cW \, + \, \frac{1}{6} \ \pr_{\,(\,j} \, \g_{\,k\,)\,i} \, \cW \, -  \, \frac{1}{2} \dsl \ T_{jk} \, \cWsl_i \, - \, \12 \ \pr^{\,l} \, T_{jk}\,T_{il} \, \cW \\
& - \, \frac{1}{6} \
\pr^{\,l}\, T_{jk} \, \g_{\,il} \, \cW  \, = \, \pr^{\,l} \,
\pr_{\,(\,j}\, \cZ_{\,k\,)\,il} \, + \, \12 \ \pr^{\,l}\pr^{\,m} \,
T_{jk} \, \cZ_{\,ilm} \, ,
\end{split}
\ee
and admits a $\{3\}$ and a $\{2,1\}$ Young projection in family
indices, while the second reads
\be
\begin{split} \label{gbianchi2}
& \pr_{\,[\,i}\,\g_{\,jk\,]} \, \cW \, + \, \frac{1}{6} \ \pr_{\,[\,j}\,\g_{\,k\,]\,i} \, \cW \, + \, \12 \ \pr_{\,[\,j}\,T_{k\,]\,i} \, \cW \, - \, \frac{5}{6} \dsl \ \g_{\,ijk} \, \cW \, + \, \frac{1}{6} \dsl \ T_{i\,[\,j}\, \g_{\,k\,]} \, \cW \\
& - \, \frac{1}{6} \ \pr^{\,l} \, \g_{\,ijkl} \, \cW \, - \, \12 \ \pr^{\,l} \, T_{il}\, \g_{jk} \, \cW \, + \, \frac{1}{6} \ \pr^{\,l} \left(\, T_{i\,[\,j}\, \g_{\,k\,]\,l} \, - \, T_{l\,[\,j} \, \g_{\,k\,]\,i} \,\right) \cW \\
& + \, \frac{1}{6} \ \pr^{\,l} \, T_{i\,[\,j}\,T_{k\,]\,l} \, \cW \,
= \, - \  \pr^{\,l} \!\dsl \ \g_{\,[\,j} \, \cZ_{\,k\,]\,il} \, + \,
\pr^{\,l} \, \pr_{\,[\,j} \, \cZ_{\,k\,]\,il\,} \, + \, \12 \
\pr^{\,l}\pr^{\,m} \, \g_{jk} \, \cZ_{\,ilm} \, ,
\end{split}
\ee
and would admit in principle a $\{2,1\}$ and a $\{1,1,1\}$ Young projection in the
family indices. In analogy with the $\{2\}$ projection of
eq.~\eqref{bianchigamma}, the $\{3\}$ projection of
eq.~\eqref{tbianchi2} simply relates structures that are
proportional to the constraints $\cZ_{\,ijk}$, and furthermore the
$\{1,1,1\}$ projection of eq.~\eqref{gbianchi2} is clearly not
available in the two-family case. The remaining $\{2,1\}$ component
is common to both expressions, and will be the key
ingredient of the present construction. Indeed, letting
$\mathscr{B}_i$ denote for brevity the Bianchi identities
\eqref{bianchi_W}, the two resulting expressions read
\be \label{tbianchi2proj1}
\begin{split}
& Y_{\{2,1\}} \, T_{jk} \, \mathscr{B}_i \, : \quad \12 \left(\, 2 \, \pr_{\,i} \, T_{jk} \, - \, \pr_{\,(\,j} \, T_{k\,)\,i} \,\right) \cW \, + \, \frac{1}{6} \ \pr_{\,(\,j} \, \g_{\,k\,)\,i} \, \cW \\
& - \, \frac{1}{6} \dsl \, \left(\, 2 \, T_{jk} \, \g_{\,i} \, - \, T_{i\,(\,j}\, \g_{\,k\,)} \,\right) \cW \, - \, \frac{1}{6} \ \pr^{\,l} \left(\, 2 \, T_{jk}\,T_{il} \, - \, T_{i\,(\,j}\,T_{k\,)\,l} \,\right) \cW \\
& - \, \frac{1}{18} \ \pr^{\,l} \left(\, 2\, T_{jk}\, \g_{\,il} \, - \, T_{i\,(\,j}\, \g_{\,k\,)\,l} \,\right) \cW \, = \, - \, \frac{1}{3} \ \pr^{\,l} \left(\, 2\, \pr_{\,i} \, \cZ_{\,jkl} \, - \, \pr_{\,(\,j} \, \cZ_{\,k\,)\,il} \,\right) \\
& + \, \frac{1}{6} \ \pr^{\,l}\pr^{\,m} \left(\, 2\, T_{jk} \, \cZ_{\,ilm} \, - \, T_{i\,(\,j}\, \cZ_{\,k\,)\,lm} \,\right)
\end{split}
\ee
and
\begin{align}
& Y_{\{2,1\}} \, \g_{\,jk} \, \mathscr{B}_i \, : \quad - \, \frac{1}{18} \left(\, 2 \, \pr_{\,i}\, \g_{\,jk} \, - \, \pr_{\,[\,j}\,\g_{\,k\,]\,i} \,\right) \cW \, + \, \frac{1}{2} \ \pr_{\,[\,j}\, T_{k\,]\,i} \, \cW \, + \, \frac{1}{6} \dsl \ T_{i\,[\,j}\, \g_{\,k\,]} \, \cW \nn \\
& - \, \frac{1}{9} \ \pr^{\,l} \left(\, 2 \, T_{il}\, \g_{\,jk} \, - \, T_{l\,[\,j}\, \g_{\,k\,]\,i} \,\right) \cW \, + \, \frac{1}{6} \ \pr^{\,l} \, T_{i\,[\,j}\, \g_{\,k\,]\,l} \, \cW \, + \, \frac{1}{6} \ \pr^{\,l} \, T_{i\,[\,j}\, T_{\,k\,]\,l} \, \cW \nn \\
& = \, - \,  \pr^{\,l} \!\dsl \, \g_{\,[\,j} \, \cZ_{\,k\,]\,il} \, + \, \pr^{\,l} \, \pr_{\,[\,j} \, \cZ_{\,k\,]\,il\,} \, + \, \frac{1}{6} \ \pr^{\,l}\pr^{\,m} \left(\, 2\, \g_{\,jk} \, \cZ_{\,ilm} \, - \, \g_{\,i\,[\,j}\, \cZ_{\,k\,]\,lm} \,\right) \, . \label{gbianchi2proj1}
\end{align}

Clearly, \emph{neither} of these equations suffices to move past
eq.~\eqref{var12}, as was the case for \eqref{bianchi11} in the
previous step, or altogether for bosonic fields in \cite{bose_mixed}, since both the
third and fourth contributions in \eqref{var12} lack some of the
divergences present on the left-hand sides of
eqs.~\eqref{tbianchi2proj1} and \eqref{gbianchi2proj1}. In order to
proceed, one might therefore try to combine these two consequences
of the Bianchi identities, after making their manifest symmetries
compatible, since at this stage their differences reflect the nature
of the left entries of the corresponding scalar products in
\eqref{var12}. Let us therefore construct a consequence of
eq.~\eqref{gbianchi2proj1} that is manifestly symmetric in $(j,k)$
like eq.~\eqref{tbianchi2proj1}, and similarly a consequence of
eq.~\eqref{tbianchi2proj1} that is manifestly antisymmetric in
$(j,k)$ like eq.~\eqref{gbianchi2proj1}. The issue at stake is
whether suitable combinations of these two expressions could
reproduce the two sets of divergences in the last two terms of
eq.~\eqref{var12}. Unfortunately, however, the results of these
manipulations are
\be \label{gbianchi2proj2}
\begin{split}
& \g_{\,i\,(\,j} \, \mathscr{B}_{k\,)} \, : \quad \12 \left(\, 2 \, \pr_{\,i} \, T_{jk} \, - \, \pr_{\,(\,j} \, T_{k\,)\,i} \,\right) \cW \, + \, \frac{1}{6} \ \pr_{\,(\,j} \, \g_{\,k\,)\,i} \, \cW \\
& - \, \frac{1}{6} \dsl \, \left(\, 2 \, T_{jk} \, \g_{\,i} \, - \, T_{i\,(\,j}\, \g_{\,k\,)} \,\right) \cW \, - \, \frac{1}{6} \ \pr^{\,l} \left(\, 2 \, T_{jk}\,T_{il} \, - \, T_{i\,(\,j}\,T_{k\,)\,l} \,\right) \cW \\
& + \, \12 \ \pr^{\,l} \, T_{l\,(\,j}\,\g_{\,k\,)\,i} \, \cW \, - \, \frac{1}{6} \ \pr^{\,l} \left(\, 2 \, T_{jk} \, \g_{\,il} \, - \, T_{i\,(\,j}\, \g_{\,k\,)\,l} \, + \, T_{l\,(\,j}\, \g_{\,k\,)\,i} \,\right) \cW \\
& = - \, \pr^{\,l} \! \dsl \left(\, 2\, \g_{\,i} \, \cZ_{\,jkl}  -
\, \g_{\,(\,j} \, \cZ_{\,k\,)\,lm} \,\right) + \, \pr^{\,l} \left(\,
2\, \pr_{\,i} \, \cZ_{\, jkl}  - \, \pr_{\,(\,j}\,
\cZ_{\,k\,)\,il} \,\right) + \, \12 \ \pr^{\,l}\pr^{\,m} \,
\g_{\,i\,(\,j} \, \cZ_{\,k\,)\,lm}
\end{split}
\ee
and
\be\label{tbianchi2proj2}
\begin{split}
& T_{i\,[\,j}\,\mathscr{B}_{\,k\,]} \, : \quad - \, \frac{3}{2} \
\pr_{\,[\,j}\, T_{k\,]\,i} \, \cW \, + \, \frac{1}{6} \left(\, 2\,
\pr_{\,i}\, \g_{\,jk} \, +
\, \pr_{\,[\,j} \, \g_{\,k\,]\,i} \,\right) \cW \, - \, \12 \dsl \ T_{i\,[\,j} \cWsl_{k\,]} \\
& - \12 \ \pr^{\,l} \, T_{i\,[\,j}\,T_{\,k\,]\,l} \, \cW \, - \,
\frac{1}{6} \ \pr^{\,l} \, T_{i\,[\,j}\,\g_{\,k\,]\,l} \, \cW \, =
\, \pr^{\,l} \, \pr_{\,[\,j} \, \cZ_{\,k\,]\,il} \, + \, \12 \
\pr^{\,l}\pr^{\,m} \, T_{i\,[\,j} \, \cZ_{\,k\,]\,lm}\, ,
\end{split}
\ee
and contain divergence terms that are identical, respectively, to
those present in \eqref{tbianchi2proj1} and \eqref{gbianchi2proj1},
that have the same manifest symmetries. As a result, no progress can
be attained in this fashion.

The conclusion is that one can not forego the need to combine the
last two terms in eq.~\eqref{var12}. This is actually possible, if
perhaps surprising, since $T_{jk} \,
\bep_{\,i}$ and $\bep_{\,i} \, \g_{\,jk}$ are \emph{not} independent, and
can be related to one another up to some well-defined compensator
terms. Thus, for instance,
\be
Y_{\{2,1\}} \, \bep_{\,i} \, \g_{\,jk} \, = \, \frac{1}{3} \left(\, 2 \,
\bep_{\,i} \, \g_{\,jk} \, - \, \bep_{\,[\,j}\,\g_{\,k\,]\,i} \,\right) \, =
\, \frac{1}{3} \left(\, \bep_{\,(\,i}\,\g_{\,j\,)\,k} \, - \, \bep_{\,(\,i}\,\g_{\,k\,)\,j}
\,\right)\, ,
\ee
and each term in the last sum can be related to traces of the gauge parameters and to the compensators letting
\be
\bep_{\,(\,i}\,\g_{\,j\,)\,k} \, = \, \bep_{\,(\,i}\,\g_{\,j\,)} \, \g_{\,k} \, - \, \,T_{k\, (\, i}\, \bep_{\,j\, )} \, .
\ee
Similar considerations apply to the other combination present in
eq.~\eqref{var12}, so that one finally obtains
\bea\label{gammaandt}
Y_{\{2,1\}} \, T_{jk} \, \bep_{\,i} & = & \frac{2}{3} \left(\, 2
\,
\d \, \bar{\x}_{\,jk} \, \g_{\,i} \, - \, \d \, \bar{\x}_{\,i\,(\,j}\,
\g_{\,k\,)} \,\right) \, + \, \bep_{\,(\,j} \, \g_{\,k\,)\,i} \, , \label{ttogamma} \\
Y_{\{2,1\}} \, \bep_{\,i} \, \g_{jk} & = & - \, \frac{2}{3} \ \d
\, \bar{\x}_{\,i\,[\, j} \, \g_{\,k\,]} \, + \, \frac{1}{3} \
\, T_{i\, [\, j} \, \bep_{\,k\, ]} \, . \label{gammatot}
\eea

This circumstance clearly brings about some ambiguities, since one is
somehow working with a set of structures that are \emph{not}
linearly independent. The simplest way to proceed would be to choose
a basis, and in the next section we shall eventually do it. However,
it is quite instructive to first dwell upon these ambiguities,
introducing a new parameter, $\k$, to put some
emphasis on their role. Let us therefore cast a portion $\k$ of the
last two terms in \eqref{var12} in the $T_{jk}$ format, making use
of eq.~\eqref{gammatot} in the fourth term and
the remaining portion $(1-\k)$ in the $\g_{\,jk}$ format, making
use of eq.~\eqref{ttogamma} in the third term. Some of the resulting
contributions are then manifestly proportional to the variation of
the compensators, and can be canceled adding
\be
\cL_2 \, = \, \frac{1}{12} \, \bra \bar{\x}_{\,ij} \comma (\,\k-1\,) \ \g^{\,k} \left(\, 2 \, \pr_{\,k} \, T_{ij} \, - \, \pr_{\,(\,i} \, T_{j\,)\,k}  \,\right) \cW \, + \, \frac{\k}{3} \ \g^{\,k} \, \pr_{\,(\,i} \, \g_{\,j\,)\,k} \, \cW \ket \, + \, \textrm{h.c.} \ ,
\ee
so that
\be \label{var12k}
\begin{split}
\d \left( \, \cL_0 + \cL_1 + \cL_2 \right) \, & = \, - \, \frac{1}{12} \,
\bra \pr_{\,(\,i} \, \pr_{\,j} \, \bep_{\,k\,)} \comma \cZ_{\,ijk}
\ket \, + \, \frac{1}{8} \, \bra \bep_{\,i} \, \g_{\,j} \comma \pr_{\,[\,i}
\cWsl_{j\,]} \, - \, \frac{2}{3} \dsl \, \g_{\,ij} \, \cW \ket \\
& + \, \frac{\k}{24} \, \bra T_{jk} \, \bep_{\,i} \comma \left(\, 2 \, \pr_{\,i} \, T_{jk} \, - \, \pr_{\,(\,j} \, T_{k\,)\,i} \,\right) \cW \, + \, \frac{1}{3} \ \pr_{\,(\,j} \, \g_{\,k\,)\,i} \, \cW \ket \\
& - \, \frac{1-\k}{8} \, \bra \bep_{\,i} \, \g_{\,jk} \comma \pr_{\,[\,j}\, T_{k\,]\,i} \, \cW \, - \, \frac{1}{9} \left(\, 2\,
\pr_{\,i}\, \g_{\,jk} \, -
\, \pr_{\,[\,j} \, \g_{\,k\,]\,i} \,\right) \cW \ket \, + \, \textrm{h.c.} \ .
\end{split}
\ee

We have thus obtained two types of terms that can be taken care of
via the Bianchi identities, but before proceeding a further comment
is in order. The issue is that, for the special case of two-family
gauge fields, some of the $\g$-traces of $\cW$ that appear in
eqs.~\eqref{tbianchi2proj1},
\eqref{gbianchi2proj1}, \eqref{gbianchi2proj2} and \eqref{tbianchi2proj2} are
actually proportional to the constraints, although this is not
evident at first sight. Indeed, in these expressions the terms carrying four lower family
indices are a priori of the form
\begin{center}
\begin{picture}(240,30)(0,0)
\multiframe(0,20)(10.5,0){1}(10,10){}
\multiframe(10.5,20)(10.5,0){1}(10,10){}
\multiframe(0,9.5)(10.5,0){1}(10,10){}
\put(27,21){$\otimes$}
\multiframe(43,20)(10.5,0){1}(10,10){}
\put(63,20){$\eq$}
\multiframe(80,20)(10.5,0){1}(10,10){}
\multiframe(90.5,20)(10.5,0){1}(10,10){}
\multiframe(101,20)(10.5,0){1}(10,10){}
\multiframe(80,9.5)(10.5,0){1}(10,10){}
\put(120,20){$\oplus$}
\multiframe(140,20)(10.5,0){1}(10,10){}
\multiframe(150.5,20)(10.5,0){1}(10,10){}
\multiframe(140,9.5)(10.5,0){1}(10,10){}
\multiframe(150.5,9.5)(10.5,0){1}(10,10){}
\put(170,20){$\oplus$}
\multiframe(190,20)(10.5,0){1}(10,10){}
\multiframe(200.5,20)(10.5,0){1}(10,10){}
\multiframe(190,9.5)(10.5,0){1}(10,10){}
\multiframe(190,-0.8)(10.5,0){1}(10,10){}
\end{picture}
\end{center}
where the first Young diagram factor is dictated by the projection while
the additional box originates from the gradient. However, the subset
of terms containing both a trace and a two-index $\g$ actually admit
only the $\{3,1\}$ and $\{2,1,1\}$ projections, as one can see from
the composition rule
\begin{center}
\begin{picture}(195,30)(0,0)
\multiframe(0,20)(10.5,0){1}(10,10){}
\multiframe(0,9.5)(10.5,0){1}(10,10){}
\put(17,21){$\otimes$}
\multiframe(33,20)(10.5,0){1}(10,10){}
\multiframe(43,20)(10.5,0){1}(10,10){}
\put(63,20){$\eq$}
\multiframe(80,20)(10.5,0){1}(10,10){}
\multiframe(90.5,20)(10.5,0){1}(10,10){}
\multiframe(101,20)(10.5,0){1}(10,10){}
\multiframe(80,9.5)(10.5,0){1}(10,10){}
\put(120,20){$\oplus$}
\multiframe(140,20)(10.5,0){1}(10,10){}
\multiframe(150.5,20)(10.5,0){1}(10,10){}
\multiframe(140,9.5)(10.5,0){1}(10,10){}
\multiframe(140,-0.8)(10.5,0){1}(10,10){}
\end{picture}
\end{center}
\vspace{-5pt}
Here the symmetric boxes originate from the trace $T$ while the
antisymmetric ones originate from the two-index $\g$, and as we
stressed repeatedly with only two families three-row projections
like the $\{2,1,1\}$ are clearly impossible. As a result, one is
finally left only with the $\{3,1\}$ projection for these terms,
that as such is related to the $\cZ_{\,ijk}$. For example, a
contribution of this type is present in eq.~\eqref{tbianchi2proj1},
and as we just stressed coincides with its $\{3,1\}$ portion
\be \label{constr31}
\left(\, 2\, T_{jk}\, \g_{\,il} \, - \, T_{i\,(\,j}\, \g_{\,k\,)\,l} \,\right)
\cW \, \xrightarrow{\{3,1\}} \, \frac{1}{2} \ T_{(\,ij}\,\g_{\,k\,)\,l} \, \cW \, +
\, \frac{3}{4} \left(\, T_{(\,il}\,\g_{\,j\,)\,k} \, + \, T_{(\,il}\,\g_{\,k\,)\,j} \,\right)
\cW\, ,
\ee
that is proportional to the $\cZ_{\,ijk}$ on account of
eq.~\eqref{triplegammaW}, since
\be
T_{(\,ij}\,\g_{\,k\,)\,l}\, \cW \, = \, - \, \g_{\,l} \,
T_{(\,ij}\,\cWsl_{\,k\,)} \, + \, \frac{1}{4} \
\g_{\,(\,i}\,T_{jk}\,\cWsl_{\,l\,)} \, .
\ee
Actually, an alternative possibility exists to select the components not related to the constraints,
that is more suitable for an extension to generic fields with an arbitrary number of index
families. It amounts to working with linear combinations of
eqs.~\eqref{tbianchi2proj1} and \eqref{gbianchi2proj2}, or of
eqs.~\eqref{gbianchi2proj1} and
\eqref{tbianchi2proj2}, that do not contain the $\{3,1\}$ projection
of terms of this type. The relevant relations in this two-family
setting are
\be
\begin{split}
& \frac{9}{8} \ Y_{\{2,1\}} \left(\, T_{jk} \, \mathscr{B}_i \, - \, \frac{1}{9} \
\g_{\,i\,(\,j}\, \mathscr{B}_{\,k\,)} \,\right) \, : \quad \frac{1}{2}
\left(\, 2 \, \pr_{\,i}\, T_{jk} \, - \, \pr_{\,(\,j} \, T_{k\,)\,i} \,\right) \cW \, +
\, \frac{1}{6} \ \pr_{\,(\,j} \, \g_{\,k\,)\,i} \, \cW \\
& - \, \frac{1}{6} \dsl \, \left(\, 2 \, T_{jk} \, \g_{\,i} \, - \,
T_{i\,(\,j}\, \g_{\,k\,)} \,\right) \cW \, - \, \frac{1}{6} \
\pr^{\,l} \left(\, 2 \, T_{jk}\,T_{il} \, - \,
T_{i\,(\,j}\,T_{k\,)\,l} \,\right) \cW \\
& = \, \frac{1}{8} \ \pr^{\,l} \!\dsl \left(\, 2\, \g_{\,i}\, \cZ_{\,jkl} \, - \, \g_{\,(\,j}\, \cZ_{\,k\,)\,il} \,\right) \, - \, \12 \ \pr^{\,l} \left(\, 2\, \pr_{\,i} \, \cZ_{\,jkl} \, - \, \pr_{\,(\,j} \, \cZ_{\,k\,)\,il} \,\right) \\
& + \, \frac{3}{16} \ \pr^{\,l}\pr^{\,m} \left(\, 2\, T_{jk}\, \cZ_{\,ilm} \, - \, T_{i\,(\,j}\, \cZ_{\,k\,)\,lm}
\,\right) \, - \, \frac{1}{16} \ \pr^{\,l}\pr^{\,m} \, \g_{\,i\,(\,j}\,
\cZ_{\,k\,)\,lm}\, , \label{rightbianchi1}
\end{split}
\ee
together with their counterparts obtained by an antisymmetrization
in $(i,j)$ and a cyclic relabeling,
\be \label{rightbianchi2}
\begin{split}
& \frac{9}{8} \ Y_{\{2,1\}} \left(\, \frac{1}{3} \ \g_{\,jk}\, \mathscr{B}_i \, + \, T_{i\,[\,j}\, \mathscr{B}_{k\,]} \,\right) \, : \quad - \, \frac{3}{2} \
\pr_{\,[\,j}\, T_{k\,]\,i} \, \cW \, + \, \frac{1}{6} \left(\, 2\,
\pr_{\,i}\, \g_{\,jk} \, - \, \pr_{\,[\,j} \, \g_{\,k\,]\,i} \,\right) \cW \\
& - \, \12 \dsl \ T_{i\,[\,j} \cWsl_{k\,]} \, - \12 \ \pr^{\,l} \, T_{i\,[\,j}\,T_{\,k\,]\,l} \, \cW \, = \, - \, \frac{3}{8} \ \pr^{\,l} \!\dsl \ \g_{\,[\,j\,}\, \cZ_{\,k\,]\,il} \, + \, \frac{3}{2} \ \pr^{\,l} \, \pr_{\,[\,j} \, \cZ_{\,k\,]\,il} \\
& + \, \frac{9}{16} \
\pr^{\,l}\pr^{\,m} \, T_{i\,[\,j} \, \cZ_{\,k\,]\,lm} \, + \, \frac{1}{16} \ \pr^{\,l}\pr^{\,m} \left(\, 2\,
\g_{\,jk}\, \cZ_{\,ilm} \, - \, \g_{\,i\,[\,j}\, \cZ_{\,k\,]\,lm}
\,\right)\, .
\end{split}
\ee

These two consequences of the Bianchi identities now reduce the total
variation \eqref{var12k} to
\be\label{finaLagvar}
\begin{split}
& \d \left( \, \cL_0 + \cL_1 +\cL_2 \right) \, = \, - \, \frac{1}{4} \, \bra \d \bar{\l}_{\,ijk} \comma \cZ_{\,ijk} \ket \, + \, \frac{1}{24} \, \bra \bep_{\,i} \, \g_{\,j} \comma \pr^{\,k} \, T_{k\,[\,i}\, \g_{\,j\,]} \, \cW \ket \\
& + \, \frac{\k}{72} \, \bra T_{jk} \, \bep_{\,i} \comma \dsl \left(\, 2 \, T_{jk}\, \g_{\,i} \, - \, T_{i\,(\,j}\, \g_{\,k\,)} \,\right) \cW \, + \, \pr^{\,l} \left(\, 2\, T_{jk}\,T_{il}  \, - \, T_{i\,(\,j}\,T_{k\,)\,l} \,\right) \cW \ket \\
& + \, \frac{1-\k}{24} \, \bra \bep_{\,i} \, \g_{\,jk} \comma \dsl \ T_{i\,[\,j}\,\g_{\,k\,]} \, \cW \, + \, \pr^{\,l} \, T_{i\,[\,j}\, T_{k\,]\,l} \, \cW \ket \, + \, \textrm{h.c.} \ ,
\end{split}
\ee
where
\be \label{mult_ferm}
\begin{split}
\d \l_{\,ijk} \, & = \, \frac{1}{3} \ \pr_{\,(\,i\,}\pr_{\,j}\,\e_{\,k\,)} \, + \, \frac{1}{24} \ \g^{\,l} \left(\, \g_{\,(\,i}\,\pr_{\,j\,}\pr_{\,k\,)}\,\e_{\,l} \, - \, \g_{\,l} \, \pr_{\,(\,i\,}\pr_{\,j}\,\e_{\,k\,)} \,\right) \\
& + \, \frac{\k}{18} \ \pr^{\,l} \left(\, 2\, T_{(\,ij}\,\pr_{\,k\,)} \, \e_{\,l} \, - \, T_{l\,(\,i}\,\pr_{\,j}\,\e_{\,k\,)} \,\right) \, + \, \frac{1-\k}{6} \ \pr^{\,l} \, \g_{\,l\,(\,i}\,\pr_{\,j}\,\e_{\,k\,)} \\
& - \, \frac{\k}{72} \ \g^{\,l} \! \dsl \, \left(\, 2\, T_{(\,ij}\,\pr_{\,k\,)} \, \e_{\,l} \, - \, T_{l\,(\,i}\,\pr_{\,j}\,\e_{\,k\,)} \,\right) \, - \, \frac{1-\k}{24} \ \g^{\,l} \! \dsl \ \g_{\,l\,(\,i}\,\pr_{\,j}\,\e_{\,k\,)} \\
& - \frac{\k}{12} \ \h^{\,lm} \left(\, T_{lm} \, \pr_{\,(\,i\,}\pr_{\,j}\,\e_{\,k\,)} \, - \, T_{l\,(\,i}\,\pr_{\,j\,}\pr_{\,k\,)}\,\e_{\,m} \,\right) \, + \, \frac{1-\k}{4} \ \h^{\,lm} \, \g_{\,l\,(\,i}\,\pr_{\,j\,}\pr_{\,k\,)}\,\e_{\,m} \\
& + \, \frac{\k}{72} \ \g^{\,lm} \, T_{l\,(\,i}\,\pr_{\,j\,}\pr_{\,k\,)}\,\e_{\,m} \, - \, \frac{1-\k}{72} \ \g^{\,lm} \left(\, \g_{\,lm} \, \pr_{\,(\,i\,}\pr_{\,j}\,\e_{\,k\,)} \, + \, \g_{\,l\,(\,i}\,\pr_{\,j\,}\pr_{\,k\,)}\,\e_{\,m} \,\right) \, .
\end{split}
\ee
can be regarded as the gauge variation of a set of Lagrange multipliers $\l_{\,ijk}$. The rest of the variation \eqref{finaLagvar} can be partly canceled adding
\be \label{ell3}
\begin{split}
\cL_3 \, & = \, \12 \, \bra \bar{\psi} \comma \frac{1}{18} \ \h^{ij}\,\g^{\,k} \left(\, 2\, T_{ij}\,\g_{\,k} - T_{k\,(\,i}\, \g_{\,j\,)} \,\right) \cW
+ \frac{1}{36} \ \h^{ij}\,\h^{kl} \left(\, 2 \, T_{ij}\, T_{kl} - T_{i\,(\,k}\,T_{l\,)\,j} \,\right) \cW \ket \\
& + \, \textrm{h.c.} \ ,
\end{split}
\ee
but one is readily to face a novel feature of the fermionic
construction: an additional compensator term,
\be
\cL_4 \, = \, \frac{1}{72} \, \bra \bar{\x}_{\,ij} \comma \pr^{\,k} \, \left(\, 2 \, T_{ij}\,
\g_{\,k} \, - \, T_{k\,(\,i}\, \g_{\,j\,)} \,\right) \cW \ket \, + \,
\textrm{h.c.} \ ,
\ee
is needed to eliminate the symmetric portion of the gradient terms
produced when varying the first contribution in eq.~\eqref{ell3}. At
this point
\be \label{deltal04}
\begin{split}
& \d \left( \, \cL_0 + \ldots +\cL_4 \right) \, = \, - \, \frac{1}{4} \, \bra \d \bar{\l}_{\,ijk} \comma \cZ_{\,ijk} \ket \\
& - \, \frac{1-\k}{72} \, \bra T_{jk} \, \bep_{\,i} \comma \dsl \left(\, 2 \, T_{jk}\, \g_{\,i} \, - \, T_{i\,(\,j}\, \g_{\,k\,)} \,\right) \cW \, + \, \pr^{\,l} \left(\, 2\, T_{jk}\,T_{il}  \, - \, T_{i\,(\,j}\,T_{k\,)\,l} \,\right) \cW \ket \\
& + \, \frac{1-\k}{24} \, \bra \bep_{\,i} \, \g_{\,jk} \comma \dsl \ T_{i\,[\,j}\,\g_{\,k\,]} \, \cW \, + \, \pr^{\,l} \, T_{i\,[\,j}\, T_{k\,]\,l} \, \cW \ket \\
& - \, \frac{1}{72} \, \bra T_{jk} \, \bep_{\,i} \, \g_{\,l} \comma \pr_{\,i} \left(\, 2 \, T_{jk}\, \g_{\,l} \, - \, T_{l\,(\,j}\, \g_{\,k\,)} \,\right) \cW \ket \\
& - \, \frac{1}{288} \, \bra T_{jk} \, T_{lm} \, \bep_{\,i} \comma \pr_{\,i} \left(\, 2\, T_{jk}\,T_{lm} \, - \, T_{l\,(\,j}\, T_{k\,)\,m} \,\right) \cW \ket \, + \, \textrm{h.c.} \ ,
\end{split}
\ee
and one is thus confronted, again, with two apparently distinct types of
structures, those present in the two terms above and proportional to
$(1-\k)$. They can be combined making use again of
eqs.~\eqref{ttogamma} and \eqref{gammatot}, only to discover that
they fully cancel, up to some terms proportional to gauge variation
$\d\,\xi_{\,ij}$ of the compensators, that can be taken care of adding
\be
\begin{split}
\cL_5 \, & = \, \frac{1-\k}{36} \, \bra \bar{\x}_{\,ij} \comma \g^{k} \!\dsl \left(\, 2 \, T_{ij}\, \g_{\,k} \,-\, T_{k\,(\,i}\, \g_{\,j\,)} \,\right) \cW \, + \, \g^{\,k}\,\pr^{\,l} \left(\, 2\, T_{ij}\,T_{kl} \,-\, T_{k\,(\,i}\,T_{j\,)\,l} \,\right) \cW \ket \\
& +\, \textrm{h.c.} \ .
\end{split}
\ee
Finally, \emph{at two families}, the last two terms in
eq.~\eqref{deltal04} can be turned into a form that can be
manifestly canceled by the additional compensator terms in
\be
\begin{split}
\cL_6 \,  = \, & - \, \frac{1}{288} \, \bra \bar{\xi}_{\,ij} \comma \g^{\,kl} \left(\, 2 \, \g_{\,l} \, T_{(\,ij} \, \pr_{\,k\,)} \, - \, T_{l\,(\,i} \, \g_{\,j} \, \pr_{\,k\,)} \,\right) \cW \\
& + \, 2 \, \h^{\,kl} \left(\, 2 \, T_{ij} \, \g_{\,(\,k}\, \pr_{\,l\,)} \, + \, 2 \, T_{kl} \, \g_{\,(\,i} \, \pr_{\,j\,)} \, - \, \g_{\,(\,i} \, T_{j\,)(\,k} \, \pr_{\,l\,)} \, - \, \g_{\,(\,k} \, T_{l\,)(\,i} \, \pr_{\,j\,)} \,\right) \, \cW \ket \\
& - \, \frac{1}{288} \, \bra \bar{\xi}_{\,ij} \comma \h^{kl} \, \g^{\,m} \left(\, 2 \, T_{kl} \, T_{(\,ij} \, \pr_{\,m\,)} \, - \, T_{k\,(\,i\,|}
\, T_{l\,|\,j} \, \pr_{\,m\,)} \,\right) \cW \ket \, + \, \textrm{h.c.} \ .
\end{split} \ee
This is clearly a further manifestation of the absence of
Young tableaux with more than two rows. Its origin can be understood
observing that at two families the fourth term in
eq.~\eqref{deltal04} contains only the $\{3,1\}$ projection
\be
\begin{split}
& \bra T_{jk}\, \bep_{\,i}\, \g_{\,l} \comma \pr_{\,i} \left(\, 2\, T_{jk}\, \g_{\,l} \, - \, T_{l\,(\,j}\, \g_{\,k\,)} \,\right) \cW \ket \, \xrightarrow{\{3,1\}} \\
& \frac{1}{24} \, \bra 2\, T_{(\,ij}\, \bep_{\,k\,)}  \, \g_{\,l} \, - \, T_{l\,(\,i} \, \bep_{\,j} \, \g_{\,k\,)} \comma \left(\, 2 \, \g_{\,l} \, T_{(\,ij} \, \pr_{\,k\,)} - \, T_{l\,(\,i} \, \g_{\,j} \, \pr_{\,k\,)} \,\right) \cW \ket \, ,
\end{split}
\ee
and the $\{2,2\}$ projection
\be
\begin{split}
& \bra T_{jk}\, \bep_{\,i}\, \g_{\,l} \comma \pr_{\,i} \left(\, 2\, T_{jk}\, \g_{\,l} \, - \, T_{l\,(\,j}\, \g_{\,k\,)} \,\right) \cW \ket \, \xrightarrow{\{2,2\}} \\
& \frac{1}{8} \, \bra T_{jk}\, \bep_{\,(\,i} \, \g_{\,l\,)} \comma \left(\, 2 \, T_{jk} \, \g_{\,(\,i} \, \pr_{\,l\,)} \, + \, 2 \, T_{il} \,
\g_{\,(\,j} \, \pr_{\,k\,)} \, - \, \g_{\,(\,j} \, T_{k\,)(\,i} \, \pr_{\,l\,)} \, - \, \g_{\,(\,i} \, T_{l\,)(\,j} \, \pr_{\,k\,)} \,\right)
\cW \ket \, ,
\end{split}
\ee
both clearly related to the gauge variation of the
compensators. In a similar fashion, at two families the last term in
eq.~\eqref{deltal04} only admits  a $\{3,2\}$
projection, and can be directly related to the variation of the
compensators on account of
\be
\begin{split}
& \left(\, 2 \, T_{ij}\,T_{kl} \, - \, T_{i\,(\,k}\,T_{l\,)\,j} \,\right) \bar{\e}_{\,m} \xrightarrow{\{3,2\}} \\
& \frac{1}{4} \left\{\, 3 \, T_{ij} \, T_{(\,kl}\, \bar{\e}_{\,m\,)} \, + \, 3 \,
T_{kl} \, T_{(\,ij}\, \bar{\e}_{\,m\,)} \, - \, T_{(\,ij\,|} \left(\, 2\, T_{\,|\,kl\,)} \, \bar{\e}_{\,m} \, + \, T_{m\,|\,k}\, \bar{\e}_{\,l\,)} \,\right)  \,\right\} \label{32projection}
\, ,
\end{split}
\ee
that played a similar role also in the bosonic case of \cite{bose_mixed}.

As a result, for two-family fermionic fields we have obtained a
whole set of complete unconstrained Lagrangians labeled by the
parameter $\k$,
\be \label{lag_fer}
\cL_{fin} \, = \, \12 \, \bra \bar{\psi} \comma \cE \ket \, + \, \frac{1}{4} \, \bra \bar{\x}_{\,ij}
\comma \X_{\,ij}\,{(\,\k\,)} \ket + \, \frac{1}{4} \, \bra \bar{\l}_{\,ijk} \comma \cZ_{\,ijk} \ket \, + \, \textrm{h.c.} \ ,
\ee
with the $\k$-independent tensor
\be \label{lag_fer1}
\begin{split}
\cE \, & \equiv \, \cW \, - \, \12 \ \g^{\,i} \cWsl_i \, - \, \12 \ \h^{ij} \, T_{ij} \, \cW \, - \, \frac{1}{12} \ \g^{\,ij} \, \g_{\,ij} \, \cW \\
& + \frac{1}{18} \ \h^{ij}\,\g^{\,k} \left(\, 2\, T_{ij}\,\g_{\,k} - T_{k\,(\,i}\, \g_{\,j\,)} \,\right) \cW \, + \frac{1}{36} \ \h^{ij}\,\h^{kl} \left(\, 2 \, T_{ij}\, T_{kl} - T_{i\,(\,k}\,T_{\,l\,)\,j} \,\right) \cW \, ,
\end{split}
\ee
that generalizes the Rarita-Schwinger kinetic tensor, and
\be \label{lag_fer2}
\begin{split}
\X_{\,ij}\,{(\,\k\,)} \, & \equiv \, \12 \ \pr_{\,(\,i}  \cWsl_{j\,)}
\, + \, \frac{1}{6} \ \g^{\,k} \, \pr_{\,(\,i} \, T_{jk\,)} \, \cW + \, \frac{1}{18} \ \pr^{\,k} \, \left(\, 2 \, T_{ij}\, \g_{\,k} \, - \, T_{k\,(\,i}\, \g_{\,j\,)} \,\right) \cW \\
& - \, \frac{1-\k}{3} \ \g^{\,k} \, \left(\, 2 \, \pr_{\,k} \, T_{ij} \, - \, \pr_{\,(\,i} \, T_{j\,)\,k}  \,\right) \cW \, + \, \frac{\k}{9} \ \g^{\,k} \, \pr_{\,(\,i} \, \g_{\,j\,)\,k} \, \cW \\
& + \, \frac{1-\k}{9} \ \g^{k} \left\{ \dsl \left(\, 2 \, T_{ij}\, \g_{\,k} \,-\, T_{k\,(\,i}\, \g_{\,j\,)} \,\right) \cW \, + \, \pr^{\,l} \left(\, 2\, T_{ij}\,T_{kl} \,-\, T_{k\,(\,i}\,T_{j\,)\,l} \,\right) \cW \,\right\} \\
& - \, \frac{1}{36} \ \h^{\,kl} \left(\, 2 \, T_{ij} \, \g_{\,(\,k}\, \pr_{\,l\,)} \, + \, 2 \, T_{kl} \, \g_{\,(\,i} \, \pr_{\,j\,)} \, - \, \g_{\,(\,i} \, T_{j\,)(\,k} \, \pr_{\,l\,)} \, - \, \g_{\,(\,k} \, T_{l\,)(\,i} \, \pr_{\,j\,)} \,\right) \, \cW \\
& - \frac{1}{72} \ \g^{\,kl} \left(\, 2 \, \g_{\,l} \, T_{(\,ij} \, \pr_{\,k\,)} \, - \, T_{l\,(\,i} \, \g_{\,j} \, \pr_{\,k\,)} \,\right) \cW \\
& - \, \frac{1}{72} \ \h^{\,kl} \, \g^{\,m} \left(\, 2 \, T_{kl} \, T_{(\,ij} \, \pr_{m\,)} \, - \, T_{k\,(\,i\,|}
\, T_{l\,|\,j} \, \pr_{\,m\,)} \,\right) \cW \, .
\end{split}
\ee
At the same time, we have determined corresponding $\k$-dependent
gauge variations of the $\l_{\,ijk}$ Lagrange multipliers, that are
given in eq.~\eqref{mult_ferm}.

One can also explain the origin of these ambiguities noticing that,
in sharp contrast with the bosonic constructions of \cite{bose_mixed},
the non-trivial fermionic Bianchi identities that drive our construction carry Young
projections that are \emph{not} forbidden for the $\g$-traces of the
compensators. Via their traces or $\g$-traces, in fact, the
$\xi_{\,ij}$ give rise in general to Young diagrams with at least
two columns, that can also characterize the relevant
consequences of the Bianchi identities, like the $\{2,1\}$ projections
that we used repeatedly. On the other hand, for bosonic fields the
consequences of the Bianchi identities that drive the construction
are two-column projected, while compensator terms contain at least
three columns, so that the relevant consequences of the Bianchi identities and the constraints on the gauge parameters \eqref{labacfermg} correspond to orthogonal Young projections.
In the present example we can see explicitly this phenomenon
after separating out in the Lagrangian all terms proportional
to $(1-\k)$, so that
\be
\cL_{fin} \, = \,  \cL \, + \, (\,1-\k\,)\, \widetilde{\cL}\, ,
\ee
with
\be
\begin{split}
\cL \, & = \, \12 \, \bra \bar{\psi} \comma \cE \ket \, + \, \frac{1}{4} \, \bra \bar{\l}_{\,ijk} \comma \cZ_{\,ijk} \ket \\
& + \, \frac{1}{4} \, \bra \bar{\x}_{\,ij} \comma \12 \ \pr_{\,(\,i}  \cWsl_{j\,)}
\, + \, \frac{1}{6} \ \g^{\,k} \, \pr_{\,(\,i} \, T_{jk\,)} \, \cW + \, \frac{1}{9} \ \g^{\,k} \, \pr_{\,(\,i} \, \g_{\,j\,)\,k} \, \cW \\
& + \, \frac{1}{18} \ \pr^{\,k} \left(\, 2 \, T_{ij}\, \g_{\,k} \, - \, T_{k\,(\,i}\, \g_{\,j\,)} \,\right) \cW \, - \frac{1}{72} \ \g^{\,kl} \left(\, 2 \, \g_{\,l} \, T_{(\,ij} \, \pr_{\,k\,)} \, - \, T_{l\,(\,i} \, \g_{\,j} \, \pr_{\,k\,)} \,\right) \cW \\
& - \, \frac{1}{36} \ \h^{kl} \left(\, 2 \, T_{ij} \, \g_{\,(\,k}\, \pr_{\,l\,)} \, + \, 2 \, T_{kl} \, \g_{\,(\,i} \, \pr_{\,j\,)} \, - \, \g_{\,(\,i} \, T_{j\,)(\,k} \, \pr_{\,l\,)} \, - \, \g_{\,(\,k} \, T_{l\,)(\,i} \, \pr_{\,j\,)} \,\right) \, \cW \\
& - \, \frac{1}{72} \ \h^{kl} \, \g^{\,m} \left(\, 2 \, T_{kl} \, T_{(\,ij} \, \pr_{m\,)} \, - \, T_{k\,(\,i\,|}
\, T_{l\,|\,j} \, \pr_{\,m\,)} \,\right) \cW \ket \, + \, \textrm{h.c.} \label{bestl2fam}
\end{split}
\ee
and
\be
\begin{split}
\widetilde{\cL} \, = \, & - \, \frac{1}{12} \, \bra \bar{\x}_{\,ij} \, \g_{\,k} \comma \left(\, 2 \, \pr_{\,k} \, T_{ij} \, - \, \pr_{\,(\,i} \, T_{j\,)\,k}  \,\right) \cW \, + \, \frac{1}{3} \ \pr_{\,(\,i} \, \g_{\,j\,)\,k} \, \cW \\
& - \, \frac{1}{3} \dsl \left(\, 2 \, T_{ij}\, \g_{\,k} \,-\, T_{k\,(\,i}\, \g_{\,j\,)} \,\right) \cW \, - \, \frac{1}{3} \ \pr^{\,l} \left(\, 2\, T_{ij}\,T_{kl} \,-\, T_{k\,(\,i}\,T_{j\,)\,l} \,\right) \cW \ket \, + \, \textrm{h.c.} \ .
\end{split}
\ee
The key feature of this last set of terms is that they build
exactly the left-hand side of the Bianchi identities
\eqref{rightbianchi1}, so that they can be turned into
\be
\begin{split}
\widetilde{\cL} \, & = \, \frac{1}{36} \, \bra \bar{\cZ}_{\,ijk} \comma
\pr^{\,l} \left(\, 2\, \g_{\,l}\, \pr_{\,(\,i}\, \x_{\,jk\,)} \, - \, \g_{\,(\,i}\, \pr_{\,j}\, \x_{\,k\,)\,l} \,\right) \, - \, \frac{1}{4} \ \g^{\,l} \!\dsl \left(\, 2\, \g_{\,l}\, \pr_{\,(\,i}\, \x_{\,jk\,)} \, - \, \g_{\,(\,i}\, \pr_{\,j}\, \x_{\,k\,)\,l} \,\right) \\
& + \, \frac{3}{2} \ \h^{\,lm} \left(\, \g_{\,(\,i}\,\pr_{\,j\,}\pr_{\,k\,)} \, \x_{\,lm} \, - \, \g_{\,l}\, \pr_{\,(\,i\,}\pr_{\,j}\, \x_{\,k\,)\,m} \,\right) \, + \, \frac{1}{4} \ \g^{\,lm} \, \g_{\,l} \, \pr_{\,(\,i\,}\pr_{\,j}\, \x_{\,k\,)\,m} \ket \, + \, \textrm{h.c.} \ ,
\end{split}
\ee
which is a mere redefinition of the Lagrange multipliers
$\l_{\,ijk}$. A similar result, although involving a different
redefinition, would obtain collecting the terms proportional to $\k$. In conclusion, in the fermionic case there exist field redefinitions that modify at the same time compensator terms and gauge transformations of the Lagrange multipliers, without affecting the types of projections that enter those terms or the Rarita-Schwinger-like tensor of eq.~\eqref{lag_fer1}. In Section \ref{sec:lagrangianfunc} we shall present other classes of field redefinitions affecting the Rarita-Schwinger-like tensor, that lead to convenient alternative presentations of the unconstrained Lagrangians, while in eq.~\eqref{lag_fer} the special
choice $\k=1$ will lead to somewhat simpler expressions, and therefore
from now on we shall often stick to it.

In conclusion, our two-family Lagrangians are finally given in
eq.~\eqref{bestl2fam}, a result that can be also presented in the
more compact form
\be
\begin{split}
\cL \, & = \, \frac{1}{2} \, \bra \bar{\psi} \comma \cW \, - \, \12 \ \g^{\,i} \cWsl_i \, - \, \12 \ \h^{ij} \, T_{ij} \, \cW \, - \, \frac{1}{12} \ \g^{\,ij} \, \g_{\,ij} \, \cW \\
& + \, \frac{1}{6} \ \h^{ij}\,\g^{\,k} \, Y_{\{2,1\}} \, T_{ij}\,\g_{\,k} \, \cW \, + \,  \frac{1}{12} \ \h^{ij}\,\h^{kl} \, Y_{\{2,2\}} \, T_{ij}\, T_{kl} \, \cW \ket \\
& + \, \frac{1}{4} \, \bra \bar{\x}_{\,ij} \comma \12 \ \pr_{\,(\,i}  \cWsl_{j\,)}
\, + \, \frac{1}{6} \ \g^{\,k} \, \pr_{\,(\,i} \, T_{jk\,)} \, \cW + \, \frac{1}{9} \ \g^{\,k} \, \pr_{\,(\,i} \, \g_{\,j\,)\,k} \, \cW \\
& + \, \frac{1}{6} \ \pr^{\,k} \, Y_{\{2,1\}} \, T_{ij}\, \g_{\,k} \, \cW \, - \frac{1}{24} \ \g^{\,kl} \, \pr_{\,(\,i\,|}\, Y_{\{2,1\}} \, T_{|\,jk\,)}\, \g_{\,l} \, \cW \, - \, \frac{1}{6} \ \h^{kl} \, Y_{\{2,2\}} \, T_{ij} \, \g_{\,(\,k}\, \pr_{\,l\,)} \, \cW \\
& - \, \frac{1}{24} \ \h^{kl} \, \g^{\,m} \, \pr_{\,(\,i\,|} \, Y_{\{2,2\}} \, T_{|\,jm\,)} \, T_{kl}\, \cW \ket \, + \, \frac{1}{4} \, \bra \bar{\l}_{\,ijk} \comma \cZ_{\,ijk} \ket \, + \, \textrm{h.c.} \ , \label{lag}
\end{split}
\ee
where we have indicated concisely the Young projections that
give rise to the various structures in eq.~\eqref{bestl2fam}. The
corresponding gauge transformations for the Lagrange multipliers are
then
\be \label{lagmultk1}
\begin{split}
\d \l_{\,ijk} \, & = \, \frac{1}{3} \ \pr_{\,(\,i\,}\pr_{\,j}\,\e_{\,k\,)} \, + \, \frac{1}{24} \ \g^{\,l} \left(\, \g_{\,(\,i}\,\pr_{\,j\,}\pr_{\,k\,)}\,\e_{\,l} \, - \, \g_{\,l} \, \pr_{\,(\,i\,}\pr_{\,j}\,\e_{\,k\,)} \,\right) \\
& + \, \frac{1}{18} \ \pr^{\,l} \left(\, 2\, T_{(\,ij}\,\pr_{\,k\,)} \, \e_{\,l} \, - \, T_{l\,(\,i}\,\pr_{\,j}\,\e_{\,k\,)} \,\right) \, - \, \frac{1}{72} \ \g^{\,l} \! \dsl \, \left(\, 2\, T_{(\,ij}\,\pr_{\,k\,)} \, \e_{\,l} \, - \, T_{l\,(\,i}\,\pr_{\,j}\,\e_{\,k\,)} \,\right) \\
& - \frac{1}{12} \ \h^{\,lm} \left(\, T_{lm} \, \pr_{\,(\,i\,}\pr_{\,j}\,\e_{\,k\,)} \, - \, T_{l\,(\,i}\,\pr_{\,j\,}\pr_{\,k\,)}\,\e_{\,m} \,\right) \, + \, \frac{1}{72} \ \g^{\,lm} \, T_{l\,(\,i}\,\pr_{\,j\,}\pr_{\,k\,)}\,\e_{\,m} \, .
\end{split}
\ee%
In practice this type of choice favors, at every stage, structures
with higher numbers of $T_{ij}$ tensors, that enter the various
expressions via their most antisymmetric projections. These have a
relatively simple structure, and as we shall see shortly this choice
will prove particularly effective in leading to Lagrangians for
fields with an arbitrary number of index families.
In the one-family case, the $\kappa$-dependent Lagrangians of
eq.~(\ref{lag_fer}) clearly reduce to the result of \cite{fsold2,fms}.
In a notation that is essentially as in those papers but with the scalar product of Appendix \ref{app:MIX}, they would read
\be {\cal L} \, = \, \frac{1}{2} \, \bra \bar{\psi} \comma \cW \, - \, \12 \, \g \, \cWsl \, - \, \12 \, \h \, \cW^{\,\pe} \ket \, + \, \frac{1}{4} \, \bra \bar{\x} \comma \pr\, \cdot \,  \cWsl \ket \, - \, \frac{1}{8} \, \bra \xibsl \comma \pr \cdot \cW^{\,\pe} \ket \, + \, \frac{1}{4} \, \bra \bar{\l} \comma \cZ \ket
\, + \,\textrm{h.c.} \ .  \ee
In addition, the terms not involving the compensators or the
Lagrange multipliers, that are independent of $\k$, are relatively
simple and yet quite interesting, since they determine completely
the \emph{constrained} two-family fermionic Lagrangian leading to
the (non-Lagrangian) Labastida equation of \cite{labferm}:
\be \label{constrlag2f}
\begin{split}
\cL \, & = \, \12 \ \bra \bar{\psi} \comma \cS \, - \, \12 \ \g^{\,i}\! \ssl_i \, - \, \12 \ \h^{ij} \, T_{ij} \, \cS \, - \, \frac{1}{12} \ \g^{\,ij} \, \g_{\,ij} \, \cS \\
& + \frac{1}{18} \ \h^{ij}\,\g^{\,k} \left(\, 2\, T_{ij}\,\g_{\,k} - T_{k\,(\,i}\, \g_{\,j\,)} \,\right) \cS \, + \frac{1}{36} \ \h^{ij}\,\h^{kl} \left(\, 2 \, T_{ij}\, T_{kl} - T_{i\,(\,k}\,T_{\,l\,)\,j} \,\right) \cS \ket \,
+ \,
\textrm{h.c.} \ .
\end{split}
\ee

Let us conclude by noticing that, in analogy with the bosonic case of \cite{bose_mixed},
once the $\xi_{\,ij}$ are expressed in terms of the independent
$\Psi_i$ multipliers, eq.~\eqref{lag_fer} acquires a
further gauge symmetry, related to shifts of the Lagrange
multipliers of the type
\be
\delta \l_{\,ijk} \, = \, \eta^{\,lm}\, M_{\,ijk,\,lm} \, +\, \g^{\,lm}\, N_{\,ijk;\,lm} \, , \label{lambdalsym}
\ee
where $M_{\,ijk,\,lm}$ and $N_{\,ijk;\,lm}$  are $\{4,1\}$ projected
in their family indices and are related according to~\footnote{Here
and in the following a semi-colon separates symmetric and
antisymmetric subsets of \emph{family} indices, so that, for instance, $N_{\,ijk;\,lm}$ is symmetric under interchanges of $(i,j,k)$ and is antisymmetric under interchanges of $l$ and $m$.}
\be
N_{\,ij\,(\,k\,;\,l\,)\,m} \, + \, N_{\,kl\,(\,i\,;\,j\,)\,m} \, = \, - \, \12 \left(\, M_{\,ijm,\,kl} + \, M_{\,klm,\,ij} \,\right) \, .\label{MNrel}
\ee
In fact, making use of eq.~\eqref{constrf_r} one can see that the shift \eqref{lambdalsym}
gives rise to
\be
\begin{split}
\delta \, \cL \, & = \, \frac{i}{24} \ \bra \bar{M}_{\,ijk,\,mn} \comma T_{mn}\,
T_{(\, ij}\, \g_{\, k\,)} \left(\, \psi \, - \, \pr^{\,l}
\Psi_l \,\right) \ket \\
& - \, \frac{i}{12} \ \bra \bar{N}_{\,ijk;\,mn} \comma \left(\,
\g_{\,mn\,(\,i}\,T_{jk\,)} \, - \,
T_{(\,ij}\,T_{k\,)\,[\,m}\,\g_{\,n\,]} \,\right) \left(\,
\psi \, - \, \pr^{\,l}
\Psi_l \,\right) \ket \, + \, \textrm{h.c.} \, . \label{compmn}
\end{split}
\ee
Indeed, the first scalar product can give
contributions in the $\{5\}$, $\{4,1\}$ and $\{3,2\}$ representations. On the other
hand, the first term in the second scalar product (that actually
vanishes when only two families are present) can give contributions
in the $\{3,1,1\}$ and $\{2,1,1,1\}$, and finally the last can give
contributions in the $\{4,1\}$ and $\{3,1,1\}$. Therefore, if the
$M$ and $N$ parameters are $\{4,1\}$ projected and are related to
one another as in eq.~\eqref{MNrel}, an interesting symmetry of the
Lagrangian \eqref{lag_fer} emerges, that actually continues to hold in the
presence of an \emph{arbitrary} number of index families. This
symmetry, as we have already stressed in \cite{bose_mixed} when
discussing its bosonic counterpart, reflects the lack of
independence of the Labastida constraints on the gauge fields, that
we displayed in eq.~\eqref{comb}.
As we shall see in Section \ref{sec:reduction2f}, it manifests
itself rather clearly when one tries to reduce the Lagrangian field equations to
the Labastida form, since all contributions that can be shifted away by
\eqref{lambdalsym} are left undetermined. Furthermore, as in the
bosonic case, an important lesson to be drawn from this symmetry is
that only part of the gauge transformation \eqref{mult_ferm} of the
Lagrange multipliers $\l_{\,ijk}$ is effective when one works with
the independent $\Psi_{\,i}$ compensators.

One can also present the Lagrangians \eqref{lag_fer} in a
different form, that proves very convenient when deriving their
field equations. A similar rewriting was already described in
\cite{bose_mixed} for Bose fields. As in that case, it can be
built starting from the Lagrangians for fields $\psi$
\emph{not} subject to any constraints but realizing nonetheless only
the constrained gauge transformations of eqs.~\eqref{gaugef} and
\eqref{labacfermg},
\be \label{constrmod}
\begin{split}
\cL_C \, & = \, \12 \ \bra \bar{\psi} \comma \cS \, - \, \12 \ \g^{\,i} \! \ssl_i \, - \, \12 \ \h^{ij} \, T_{ij} \, \cS \, - \, \frac{1}{12} \ \g^{\,ij} \, \g_{\,ij} \, \cS \, + \frac{1}{6} \ \h^{ij}\,\g^{\,k} \, Y_{\{2,1\}}\, T_{ij}\,\g_{\,k} \, \cS \\
& + \frac{1}{12} \ \h^{ij}\,\h^{kl} \, Y_{\{2,2\}} \, T_{ij}\,
T_{kl} \, \cS \ket \, + \, \frac{i}{12} \ \bra \bar{\z}_{\,ijk}
\comma T_{(\,ij} \psisl_{k\,)} \ket \, + \, \textrm{h.c.} \ ,
\end{split}
\ee
obtained introducing the $\z_{\,ijk}$ Lagrange multipliers. These
enforce on-shell the Labastida constraints \eqref{labacf}, and as a
result are to transform as the $\l_{\,ijk}$ did in
eq.~\eqref{mult_ferm}. Any of the $\k$-dependent transformations \eqref{mult_ferm}
grants the gauge invariance of \eqref{constrmod}, on account of the
previous discussion, while the unconstrained Lagrangians of
eq.~\eqref{lag_fer} can be recovered performing the $\k$-dependent
Stueckelberg-like shifts
\begin{align}
& \psi \, \rightarrow \, \psi \, - \, \pr^{\,i}\, \Psi_{\,i} \, , \nn \\
& \z_{\,ijk} \, \rightarrow \, \l_{\,ijk} \, - \, \Delta_{\,ijk}(\Psi) \, , \label{firststueck}
\end{align}
where $\Delta_{\,ijk}(\Psi)$ is the combination that maps the
$\z_{\,ijk}$ into gauge invariant expressions. From the technical
viewpoint, the explicit rewriting of these unconstrained Lagrangians
\eqref{lag_fer} involves precisely the same steps of the previous
direct derivation, but this presentation makes it possible to relate
the equations of motion for $\bar{\psi}$, the $\bar{\Psi}_{\,i}$ and the
$\bar{\l}_{\,ijk}$ to the simpler ones of eq.~\eqref{constrmod}.

Finally, we would like to call to the reader's attention that, as in
the bosonic case, one could also work with independent $\xi_{\,ij}$
compensators, at the cost of adding to the Lagrangian \eqref{lag_fer} new
\emph{gauge invariant} multipliers $\r_{\,ij}$ according to
\be \label{lagextra2}
\cL_{7} \, = \, \bra \bar{\r}_{\,ij} \comma \frac{1}{2} \ \g_{\,(\,i}\Psi_{j\,)} \, - \, \x_{\,ij} \ket \, + \, \textrm{h.c.} \ .
\ee

\vskip 24pt


\scss{The field equations}\label{sec:motion2f}


In the previous section we described three different classes of Lagrangians for ``two-family'' Fermi fields
or, equivalently, for ``two-row'' reducible spinorial representations of the
Poincar\'e group. We first presented, in eq.~\eqref{lag_fer}, the
Lagrangians for unconstrained spinor-tensors $\psi$. We then
recovered from them, in eq.~\eqref{constrlag2f}, their counterparts for
fields that are subject to the triple $\g$-trace
constraints \eqref{labacf}. In this second presentation,
the Lagrangians are only invariant under gauge transformations whose
parameters are subject to the $\g$-trace constraints
\eqref{labacfermg}. Finally, in eq.~\eqref{constrmod} we
exhibited Lagrangians that involve unconstrained fields but are
still only invariant under constrained gauge transformations. Now we
would like to derive the field equations that follow from all these
Lagrangians. To this end, it is convenient to start from
eq.~\eqref{constrmod}.

The field equations for the Lagrange multipliers $\z_{\,ijk}$  read
\be \label{eql2f1}
E_{\,\bar{\z}} \, : \ \frac{i}{12} \ T_{(\,ij} \psisl_{k\,)} \, = \, 0\, .
\ee
On the other hand, the field equation for the gauge field $\bar{\psi}$
requires more work, since one must combine the terms displayed in
eq.~\eqref{constrmod} with their hermitian conjugates, and this step
requires an extensive use of some of the identities collected in
Appendix \ref{app:fermi}. The end result is
\be \label{epsi1}
\begin{split}
E_{\,\bar{\psi}} \, : \ & \cE  - \, \12 \ \h^{ij} \g^{\,k}
\, \cY_{\,ijk} \, - \, \frac{i}{12} \,
\Big\{ \, \h^{ij} \pr^{\,k} \, T_{(\,ij} \psisl_{k\,)} - \, \frac{1}{6} \ \h^{ij} \g^{\,k} \pr^{\,l} \left(\, \g_{\,j}\, T_{(\,ik} \psisl_{l\,)} - \, \g_{\,k}\, T_{(\,ij} \psisl_{l\,)} \,\right) \\
& + \, \frac{1}{8} \ \h^{ij}\,\h^{kl} \, Y_{\{2,2\}} \, \pr^{\,m} \left(\, T_{ij} \, T_{(\,kl} \psisl_{\,m\,)} \, + \, T_{kl} \, T_{(\,ij} \psisl_{\,m\,)} \,\right) \, \Big\} \, = \, 0 \, ,
\end{split}
\ee
where
\be \label{rarita}
\begin{split}
\cE \, & \equiv \, \cS \, - \, \12 \ \g^{\,i}\! \ssl_i \, - \, \12 \ \h^{ij} \, T_{ij} \, \cS \, - \, \frac{1}{12} \ \g^{\,ij} \, \g_{\,ij} \, \cS \\
& + \frac{1}{18} \ \h^{ij}\,\g^{\,k} \left(\, 2\, T_{ij}\,\g_{\,k} - T_{k\,(\,i}\, \g_{\,j\,)} \,\right) \cS \, +
\frac{1}{36} \ \h^{ij}\,\h^{kl} \left(\, 2 \, T_{ij}\, T_{kl} - T_{i\,(\,k}\,T_{\,l\,)\,j} \,\right) \cS
\end{split}
\ee
can be regarded as the fermionic analogue of the Einstein tensor of
\cite{labastida,bose_mixed}, and the
\be \label{y2f}
\begin{split}
\cY_{\,ijk} \, & \equiv \, i\, \Big\{\, \z_{\,ijk} - \, \frac{1}{6} \ T_{(\,ij}\,\pr_{\,k\,)} \, \psi \, + \, \frac{1}{72} \ \g^{\,l} \left(\, 2 \, \g_{\,l}\,T_{(\,ij}\,\pr_{\,k\,)} \, -
\, T_{l\,(\,i}\,\g_{\,j}\,\pr_{\,k\,)} \,\right) \psi \\
& + \, \frac{1}{72} \ \h^{lm} \left(\, 2 \,
T_{lm}\,T_{(\,ij}\,\pr_{\,k\,)} \, - \, T_{l\,(\,i\,|}\,
T_{m\,|\,j}\,\pr_{\,k\,)} \,\right) \psi \, \Big\}
\end{split}
\ee
are totally symmetric under interchanges of their family labels.
Eq.~\eqref{rarita} generalizes the Rarita-Schwinger spinor-tensor
to the case of mixed-symmetry higher-spin Fermi fields, and can be
also recovered gauging away the compensators in
eq.~\eqref{lag_fer1}. For the constrained theory with Lagrange multipliers of
eq.~\eqref{constrmod}, combining eqs.~\eqref{eql2f1} and
\eqref{epsi1} leads to
\be \label{eqconstr} E_{\,\bar{\psi}} \, : \ \cE \, - \, \12 \
\h^{ij} \g^{\,k} \, \cY_{\,ijk} \, = \, 0  \, .\ee

Let us add a few comments on the role of eq.~\eqref{eqconstr} in the fully constrained framework, where naively the spinor-tensors $\cY_{\,ijk}$ should not be present. In fact, when dealing with constrained fields one cannot conclude directly that
\be
\cE \, = \, 0 \, ,
\ee
since in general
\be
T_{(\,ij}\,\g_{\,k\,)} \, \cE \, \neq \, 0 \, ,
\ee
in apparent contrast with the properties of the gauge field $\psi$. However, a field equation satisfying the same constraints as the gauge field obtains after eliminating the $\cY_{\,ijk}$ tensors from eq.~\eqref{eqconstr}. The proper projection is thus encoded in the tensorial structure of the $\cY_{\,ijk}$, that hence can play a role even in the constrained theory. We shall provide an explicit example of this fact in eqs.~\eqref{constreq32} and \eqref{solthree32}, where we shall display various presentations of the equations of motion for a rank-$(3,2)$ field. Notice that the projected equations couple to external currents satisfying the Labastida constraints, thus extending the Fang-Fronsdal symmetric setting. However, field equations and currents that are jointly redefined to satisfy different constraints lead to the same physical current exchanges.

Finally, the equations of the unconstrained Lagrangians \eqref{lag_fer} follow from eqs.~\eqref{eql2f1}
and \eqref{epsi1}, taking into account that
\be
\begin{split}
\d \, \cL_C \, & = \, \bra \d \, \bar{\psi} \comma E_{\,\bar{\psi}} \ket \, + \, \bra \d \, \bar{\z}_{\,ijk} \comma (E_{\,\bar{\z}})_{\,ijk} \ket \, + \, \textrm{h.c.} \\
& = \, \bra \d \, \bar{\psi} \comma E_{\,\bar{\psi}} \ket \, + \, \bra \d \, \bar{\Psi}_{\,i} \comma \pr_{\,i}\, E_{\,\bar{\psi}} \, - \, \widetilde{\Delta}_{\,ijk}(E_{\,\bar{\z}}) \ket \, + \, \bra \d \, \bar{\l}_{\,ijk} \comma (E_{\,\bar{\z}})_{\,ijk}\ket \, + \, \textrm{h.c.} \, ,
\end{split}
\ee
where the $\widetilde{\Delta}_{\,ijk}$ are obtained integrating by parts the terms in the $\Delta_{\,ijk}$ of eq.~\eqref{firststueck}. In conclusion, one thus obtains
\begin{alignat}{2}
& E_{\,\bar{\psi}} \, : & & \ \cE \,  - \, \12 \ \h^{ij} \, \g^{\,k}
\, \cY_{\,ijk} \, - \, \frac{1}{4} \,
\Big\{ \, \h^{ij} \, \pr^{\,k} \, \cZ_{\,ijk} \, - \, \frac{1}{6} \ \h^{ij} \, \g^{\,k} \, \pr^{\,l} \,
\g_{\,[\,j} \, \cZ_{\,k\,]\,il} \nn \\
& & & \ + \, \frac{1}{12} \ \h^{ij}\,\h^{kl} \, \pr^{\,m} \left(\, T_{ij} \, \cZ_{\,klm} \, + \, T_{kl} \,
\cZ_{\,ijm} \, - \, T_{k\,(\,i} \, \cZ_{\,j\,)\,lm} \,\right) \, \Big\} \, = \, 0 \, , \label{epsi} \\[5pt]
& E_{\,\bar{\Psi}} \, : & & \ \pr_{\,i}\, E_{\,\bar{\psi}} \, - \, \pr^{\,j} \pr^{\,k} \, (E_{\,\bar{\l}})_{\,ijk} \, - \, \frac{1}{8} \ \g^{\,j} \, \g_{\,[\,i\,|} \, \pr^{\,k}\pr^{\,l}\, (E_{\,\bar{\l}})_{\,|\,j\,]\,kl} \nn \\
& & & \ - \, \frac{\k}{4} \ \h^{\,jk} \, T_{k\,[\,i\,|}\, \pr^{\,l}\pr^{\,m} \, (E_{\,\bar{\l}})_{\,|\,j\,]\,lm} \, - \, \frac{\k}{12} \ \h^{\,jk} \, \g_{\,ij} \, \pr^{\,l}\pr^{\,m} \, (E_{\,\bar{\l}})_{\,klm} \nn \\
& & & \ + \frac{1-\k}{8} \ \g^{\,jk} \, T_{ij} \, \pr^{\,l}\pr^{\,m} \, (E_{\,\bar{\l}})_{\,klm} \, - \, \frac{1-\k}{24} \ \g^{\,jk} \, \g_{\,k\,(\,i\,|} \, \pr^{\,l}\pr^{\,m}\, (E_{\,\bar{\l}})_{\,|\,j\,)\,lm} \, = \, 0 \, , \\[5pt]
& E_{\,\bar{\l}} \, : & & \ \frac{1}{4} \ \cZ_{\,ijk} \, = \, 0 \, , \label{eql2f}
\end{alignat}
where now $\cE$ denotes the Rarita-Schwinger-like tensor of eq.~\eqref{lag_fer1}, while the  $\cY_{\,ijk}$ tensors are built performing the Stueckelberg shifts \eqref{firststueck} in eq.~\eqref{y2f}, barring some subtleties that we shall discuss shortly. Notice that the equations of motion for the compensators $\bar{\Psi}_{\,i}$, here written for brevity in a form that is not ``normal ordered'', simply provide the conservation conditions for an external current coupled to $\psi$. The latter is in fact always conserved as a result of the unconstrained gauge symmetry, as in the symmetric construction of \cite{fsold2, fms}.

We would like to conclude this section with some comments on the $\l_{\,ijk}$ Lagrange multipliers and on the Stueckelberg shifts \eqref{firststueck} since, as we anticipated at the end of Section \ref{sec:lagrangian2f},
some parts of the naive gauge transformation \eqref{mult_ferm} of the $\l_{\,ijk}$ have no effect on the Lagrangian when the compensators are expressed in terms of the $\Psi_i$. These parts could well be discarded, and different choices give rise to different definitions of the Stueckelberg shifts for the $\z_{\,ijk}$ in eq.~\eqref{firststueck}. The reader familiar with the construction of \cite{bose_mixed} will probably notice a close analogy with the behavior of bosonic Lagrange multipliers. In fact, in \cite{bose_mixed} we identified redundant portions of their gauge transformations that made the $\cB_{\,ijkl}$ tensors, bosonic analogs of the $\cY_{\,ijk}$
tensors, not gauge invariant. This, however, did not affect the gauge invariance of
$\h^{ij}\,\h^{kl}\,\cB_{\,ijkl}$, the combination that enters the field equations of the bosonic gauge fields. In a similar fashion, the $\cY_{\,ijk}$ tensors that enter eq.~\eqref{epsi} are not invariant under
the gauge transformations of eqs.~\eqref{gaugef} and \eqref{fermicompens} when these are combined with the \emph{naive} gauge transformations of the $\l_{\,ijk}$ multipliers given in eq.~\eqref{mult_ferm}, since
\begin{align}
& \d \, \cY_{\,ijk} \, = \, - \, \frac{i}{12} \ \h^{\,lm} \, Y_{\{4,1\}} \, \Big\{\,
\k \left(\, T_{lm} \, \pr_{\,(\,i\,}\pr_{\,j}\, \e_{\,k\,)} \, - \,
T_{l\,(\,i}\,\pr_{\,j\,}\pr_{\,k\,)}\,\e_{\,m} \,\right) \, - \, 3\, (\,1-\k\,)\,
\g_{\,l\,(\,i}\,\pr_{\,j\,}\pr_{\,k\,)}\,
\e_{\,m} \, \Big\} \nn \\
& + \, \frac{i}{72} \ \g^{\,lm}\, Y_{\{4,1\}} \, \Big\{\, \k \
T_{l\,(\,i}\,\pr_{\,j\,}\pr_{\,k\,)}\, \e_{\,m} \, - \, (\,1-\k\,)\, \left(\, \g_{\,lm} \,
\pr_{\,(\,i\,}\pr_{\,j}\,\e_{\,k\,)} \, + \,
\g_{\,l\,(\,i}\,\pr_{\,j\,}\pr_{\,k\,)}\, \e_{\,m} \,\right) \, \Big\} \, , \label{gaugeY}
\end{align}
but proceeding as in \eqref{compmn} one can see that $\h^{ij}\,\g^{\,k}\,\cY_{\,ijk}$ is properly gauge invariant. For both
Bose and Fermi fields, dropping the redundant terms one can define gauge transformations
for the multipliers that make the $\cB_{\,ijkl}$ or the $\cY_{\,ijk}$ directly gauge invariant. Alternatively, if one were to work with the naive transformations of the multipliers, the resulting gauge variation \eqref{gaugeY} of the $\cY_{\,ijk}$ tensors could be compensated by a particular shift under the additional gauge symmetry of eq.~\eqref{lambdalsym}. This, as we have seen, gives rise to the transformations
\be \label{shiftY}
\delta\, \cY_{\,ijk} \, = \, \eta^{\,lm}\, M_{\,ijk,\,lm} \, +\, \g^{\,lm}\, N_{\,ijk;\,lm} \, ,
\ee
with two $\{4,1\}$ projected parameters, and the two contributions present in eq.~\eqref{gaugeY} also satisfy eq.~\eqref{MNrel}, that relates to one another the two parameters $M$ and $N$ of the additional shift symmetry. At any rate, the additional shift symmetry \eqref{shiftY} is present independently of the choice of gauge transformation for the multipliers, provided one works in terms of the $\Psi_i$.

These observations are particularly important when the unconstrained theory is built making use of Stueckelberg shifts starting from the Lagrangian \eqref{constrmod}: in this case the $\cY_{\,ijk}$ are gauge invariant by construction, but if one starts from the naive gauge transformations of the multipliers the result is
\emph{not} expressible in terms of the composite compensators $\xi_{\,ij}(\Psi)$. In order to illustrate this fact, it is convenient to split the expression obtained via the \emph{naive} Stueckelberg shift \eqref{firststueck} in eq.~\eqref{y2f} as
\be \label{Ydec}
\cY_{\,ijk} \, = \, \cY^{\,(1)}{}_{ijk} \, + \, \g^{\,l} \, \cY^{\,(2)}{}_{ijk\,,\,l} \, + \, \h^{\,lm} \, \cY^{\,(3)}{}_{\,ijk\,,\,lm} \, + \, \g^{\,lm} \, \cY^{\,(4)}{}_{\,ijk\,;\,lm} \, .
\ee
The first term contains the contributions that are already visible in the symmetric case~\footnote{Notice, however, that in the symmetric, or one-family, case the $\k$ dependence disappears.},
\be
\begin{split}
\cY^{\,(1)}{}_{ijk} \, & = \, \, i \, \bigg\{\, \l_{\,ijk} \, - \, \frac{1}{6} \ T_{(\,ij}\, \pr_{\,k\,)} \, \psi \, + \, \frac{1}{6} \ \Box \, \g_{\,(\,i}\, \x_{\,jk\,)} \, + \, \frac{4\,\k-3}{18} \ \pr^{\,l} \, \g_{\,l} \, \pr_{\,(\,i}\, \x_{\,jk\,)} \\
& - \, \frac{2\,\k-3}{18} \ \pr^{\,l} \, \pr_{\,(\,i}\, \g_{\,j}\, \x_{\,k\,)\,l} \,\bigg\} \, .
\end{split}
\ee
The second term adds some contributions that first emerge in the two-family case but are still insensitive to the redundant portion of the gauge transformations of the Lagrange multipliers, and at two families reads
\be
\begin{split}
\cY^{\,(2)}{}_{ijk\,,\,l} \, & = \, \frac{i}{72} \, \bigg\{\, \left(\, 2\, \g_{\,l}\, T_{(\,ij}\, \pr_{\,k\,)} \, - \, T_{l\,(\,i}\, \g_{\,j}\, \pr_{\,k\,)} \,\right) \, \psi \, - \, 2 \, \left(\, \pr_{\,(\,i\,} \pr_{\,j}\, \x_{\,k\,)\,l} \, - \, \pr_{\,l\,} \pr_{\,(\,i}\, \x_{\,jk\,)} \,\right) \\
& - \, 2 \ \Box \, \g_{\,l\,(\,i}\, \x_{\,jk\,)} \, + \, 2\, (\,1-\k\,) \dsl \left(\, 2\, \g_{\,l} \, \pr_{\,(\,i}\, \x_{\,jk\,)} \, - \, \pr_{\,(\,i} \, \g_{\,j} \, \x_{\,k\,)\,l} \,\right) \\
& - \, \12 \ \pr^{\,m} \left(\, 2\, T_{lm}\, \pr_{\,(\,i}\, \x_{\,jk\,)} \, + \, 2\, T_{(\,ij}\, \pr_{\,k\,)}\, \x_{\,lm} \, - \, T_{l\,(\,i}\, \pr_{\,j}\, \x_{\,k\,)\,m} \, - \, T_{m\,(\,i}\, \pr_{\,j}\, \x_{\,k\,)\,l} \,\right) \\
& - \, \frac{1}{4} \ \pr^{\,m} \left(\, 2\, \g_{\,lm}\, \pr_{\,(\,i}\, \x_{\,jk\,)} \, + 3\, \g_{\,l\,(\,i}\, \pr_{\,j}\, \x_{\,k\,)\,m} \, + \, \g_{\,m\,(\,i}\, \pr_{\,j}\, \x_{\,k\,)\,l} \,\right) \bigg\} \, .
\end{split}
\ee
On the other hand, the naive form of the third and the fourth terms in eq.~\eqref{Ydec} contains some contributions with bare $\Psi_{\,i}$ compensators, and reads
\begin{align}
& \cY^{\,(3)}{}_{\,ijk\,,\,lm} \, = \, \frac{i}{24} \, Y_{\{3,2\}} \bigg\{\, T_{lm}\,T_{(\,ij}\,\pr_{\,k\,)} \, \psi + \, 2\,(\,1-\k\,) \left(\, 2 \, \g_{\,(\,i}\,\pr_{\,j\,}\pr_{\,k\,)}\, \x_{\,lm} + \, \g_{\,(\,l}\,\pr_{\,m\,)\,}\pr_{\,(\,i}\, \x_{\,jk\,)} \,\right) \nn \\
& - \, \frac{1}{2}\ \Box \left(\, T_{lm} \, \g_{\,(\,i}\,\x_{\,jk\,)} \, + \, T_{(\,ij}\,\g_{\,k\,)} \,
\x_{\,lm} \, + \, \g_{\,(\,l\,|}\,T_{(\,ij}\,\x_{\,k\,)\,|\,m\,)} \, \right) \\
& - \, \frac{1}{4} \ \pr^{\,n} \left(\, \g_{\,n}\, T_{(\,ij}\,\pr_{\,k\,)}\,\x_{\,lm} \, + \, \g_{\,n}\, T_{lm} \,
\pr_{\,(\,i}\,\x_{\,jk\,)} \, + \, T_{lm}\, \pr_{\,(\,i}\,
\g_{\,j}\,\x_{\,k\,)\,n} \, + \, T_{(\,ij}\,\pr_{\,k\,)}\, \g_{\,(\,l}\, \x_{\,m\,)\,n} \,\right) \bigg\} \nn \\[5pt]
& + \, \frac{i}{24} \, Y_{\{4,1\}} \bigg\{\,
\k \left(\, 2\, T_{lm} \, \pr_{\,(\,i\,}\pr_{\,j\,} \Psi_{\,k\,)} - \,
T_{(\,l\,|\,(\,i}\,\pr_{\,j\,}\pr_{\,k\,)\,}\Psi_{\,|\,m\,)} \,\right) - \, 3\, (\,1-\k\,)\,
\g_{\,(\,l\,|\,(\,i}\,\pr_{\,j\,}\pr_{\,k\,)\,}
\Psi_{\,|\,m\,)} \, \bigg\} \nn
\end{align}
for the third term and
\be
\begin{split}
\cY^{\,(4)}{}_{\,ijk\,;\,lm} \, = \, & - \, \frac{i}{144} \ Y_{\{4,1\}} \, \Big\{\, \k \
T_{[\,l\,|\,(\,i}\,\pr_{\,j\,}\pr_{\,k\,)}\, \Psi_{\,|\,m\,]} \\
& - \, (\,1-\k\,)\, \left(\, 2\, \g_{\,lm} \,
\pr_{\,(\,i\,}\pr_{\,j}\,\Psi_{\,k\,)} \, + \,
\g_{\,[\,l\,|\,(\,i}\,\pr_{\,j\,}\pr_{\,k\,)}\, \Psi_{\,|\,m\,]} \,\right) \, \Big\}
\end{split}
\ee
for the last one. The terms containing bare $\Psi_{\,i}$ compensate precisely the gauge variation \eqref{gaugeY} of the previous, $\x_{\,ij}$ dependent, $\cY_{\,ijk}$ tensors.

\vskip 24pt


\scss{On-shell reduction to $\cS=0$}\label{sec:reduction2f}


We can now conclude our discussion of two-family fields by showing
how the three sets of Lagrangian field equations
described in the preceding section are equivalent
to the Labastida equation of motion~\footnote{In Section \ref{sec:dofs} we shall see that
eq.~\eqref{eqlaba} and the (constrained) gauge symmetry
\eqref{gaugef} suffice in general to turn the spinorial $gl (D)$
representations carried by $\psi$ fields into their $o (D-2)$ counterparts, as expected in view of the uniqueness of $\cS$.}
\be \cS \, = \, 0 \, . \label{eqlaba} \ee
In analogy with the case of Bose fields \cite{bose_mixed}, in the
following we shall discuss how this result can be attained either
directly or after fixing extra Weyl-like gauge
symmetries that emerge for the Lagrangians in sporadic
low-dimensional cases.

To begin with, we would like to recall that the
equations of motion for $\bar{\psi}$ that follow from the Lagrangians \eqref{constrlag2f} and \eqref{constrmod} can both be cast in the form \eqref{eqconstr}. In addition, combining eqs.~\eqref{epsi} and \eqref{eql2f} and gauging away the compensators, the field equation for $\bar{\psi}$ following from the Lagrangian \eqref{lag_fer} can be also reduced to eq.~\eqref{eqconstr}. Thus, for all three Lagrangian theories that we presented one is led to discuss the reduction of
\begin{align}
E_{\,\bar{\psi}} \, : \ \ & \cS \, - \, \12 \ \g^{\,i}\! \ssl_i \, -
\, \12 \ \h^{ij} \, T_{ij} \, \cS \, - \, \frac{1}{12} \ \g^{\,ij}
\, \g_{\,ij} \, \cS \, + \, \frac{1}{18} \ \h^{ij}\,\g^{\,k}
\left(\, 2\, T_{ij}\,\g_{\,k} - T_{k\,(\,i}\, \g_{\,j\,)} \,\right)
\cS \nn \\
& + \, \frac{1}{36} \ \h^{ij}\,\h^{kl} \left(\, 2 \, T_{ij}\, T_{kl}
- T_{i\,(\,k}\,T_{\,l\,)\,j} \,\right) \cS \, - \, \12 \ \h^{ij}
\g^{\,k} \, \cY_{\,ijk} \, = \, 0 \, , \label{eqconstr2} \\[5pt]
E_{\,\bar{\z}} \, : \ \ & \frac{i}{12} \ T_{(\,ij} \psisl_{k\,)} \, = \, 0\quad \Longrightarrow  \quad T_{(\,ij} \ssl_{k\,)} \, = \, 0 \, , \label{eqmultred}
\end{align}
to the Labastida form \eqref{eqlaba}, where of course in the constrained case the second condition holds even off shell.

To this end, one can conveniently adapt the procedure developed in \cite{bose_mixed} for $N$-family bosonic fields. The picture one has in mind is that whenever eq.~\eqref{eqconstr2} \emph{does not} reduce directly to the conditions
\begin{align}
& \cS \, = \, 0 \, , \nn \\
& \cY_{\,ijk} \, = \, 0 \, , \label{solhom}
\end{align}
additional symmetries are present. In fact, this is equivalent to stating that the homogeneous equation \eqref{eqconstr2} admits non-trivial solutions aside from eq.~\eqref{solhom}, and now we would like to classify and describe these ``pathological'' cases. Actually, we already came across a phenomenon of this kind in Section \ref{sec:motion2f}, where we saw that eq.~\eqref{eqconstr2} is left invariant by the shifts of the $\cY_{\,ijk}$ defined in eq.~\eqref{shiftY}. In this case the undetermined
quantities do not affect the portion of the field equation involving $\cS$, and are thus immaterial insofar as its reduction to the Labastida form is concerned. At the same time, they can be gauged away directly, making use of the shift symmetry of the Lagrange multipliers identified in eq.~\eqref{lambdalsym}.

We can now show that, in sporadic examples with low space-time dimensions, some $\g$-traces of the spinor-tensor $\cS$ are left undetermined, a phenomenon that is however accompanied by the emergence of new Weyl-like gauge symmetries for the field $\psi$. These are precisely as needed to gauge away
the undetermined quantities, thus making it possible to complete the reduction to the
Labastida form. Let us stress that this peculiar behavior is already visible for symmetric spinor-tensors, since the Rarita-Schwinger equation \eqref{eqconstr2} for a gravitino can be cast in the form
\be
\cS_{\,\m}\, - \, \12 \ \g_{\,\m}\, \g \cdot \cS \, = \, 0 \, . \label{2draritasc}
\ee
In two dimensions this formal expression only involves the $\g$-traceless part of $\cS$, and thus apparently one cannot use it to set to zero its $\g$-trace. However, precisely in two dimensions Weyl-like shifts of the form
\be
\d \, \psi_{\,\m} \, = \, \g_{\,\m} \, \O  \label{shiftgravitino}
\ee
become a symmetry and make it possible to gauge away the undetermined quantity.
Again in partial analogy with the case of two-dimensional gravity, this behavior is accompanied by the vanishing of the Rarita-Schwinger Lagrangian, and thus of eq.~\eqref{2draritasc} when it is expressed in
terms of $\psi$. This property is manifest in the usual presentation of the Lagrangian,
\be \label{gamma_rarita}
\cL \, = \, \frac{i}{2} \ \psi_{\,\m}\, \g^{\,\m\n\r}\, \pr_{\,\n}\, \psi_{\,\r} \, + \, \textrm{h.c.} \ ,
\ee
simply because the fully antisymmetric $\g^{\,\m\n\r}$ vanishes in two dimensions. In the following we shall see that the whole class of one-column fully-antisymmetric fields exhibits a similar behavior in low enough space-time dimensions, but in generic mixed-symmetry cases the situation is more complicated, and in particular there are Weyl-invariant theories whose Lagrangians \emph{do not} vanish identically.

Before starting our analysis of generic two-family fields, let us recall that the relevant $\cS$ spinor-tensors are to satisfy the Bianchi identities, that reduce to
\be \label{bianchired}
\mathscr{B}_i \, : \ \pr_{\,i}\, \cS \, - \, \12 \dsl \ssl_i \, - \, \12 \ \pr^{\,j} \, T_{ij} \, \cS \, - \, \frac{1}{6} \ \pr^{\,j} \,
\g_{\,ij} \, \cS \, = \, 0
\ee
after making use of the constraints \eqref{eqmultred}, and are in addition to preserve the constraints
\be \label{constrred}
T_{(\,ij} \ssl_{k\,)} \, = \, 0 \, .
\ee
As we have seen, these conditions are identically satisfied in the constrained case, while in the unconstrained case they hold after enforcing the field equations for the Lagrange multipliers. At any rate, out of the non-trivial solutions of eq.~\eqref{eqconstr2}, or equivalently out of its shift symmetries, one must thus consider only those satisfying both \eqref{bianchired} and \eqref{constrred}. This key observation greatly simplifies the search for the field equations that do not reduce directly to the Labastida form, since non-trivial shifts of the Bianchi identities provide a most convenient starting point to attack the problem. The Bianchi identities really lie at the heart of the Rarita-Schwinger-like tensors, and we shall see shortly that all their shift symmetries also leave the field equations invariant.

In principle, eq.~\eqref{eqconstr2} can admit shift symmetries of the form \footnote{Covariance justifies
this assertion, although we cannot exclude degenerate cases where further cancelations take place. An example of this type is indeed the highly degenerate case of the $\{2,2\}$ field in two dimensions, that however is also captured by our analysis.}
\be \label{shift1}
\d \, \cS \, = \, \g^{\,i}\, \O_{\,i} \, ,
\ee
and now we would like to classify those preserving the Bianchi identities \eqref{bianchired}. In general, under shifts of the form \eqref{shift1} they vary according to \footnote{With a slight abuse of notation, both
in this section and in Section \ref{sec:reductionf}, with $\mathscr{B}_i$ and ``Bianchi identities'' we shall often denote the left-hand sides of eqs.~\eqref{bianchired}.}
\be \label{shiftB1}
\begin{split}
\d \, \mathscr{B}_i \, = \, & - \, \12 \dsl \, \Big[\, (\,D-2\,)\, \O_{\,i} \, + \, 2 \, S^{\,j}{}_{\,i}\, \O_{\,j} \,\Big] \, - \, \frac{1}{6} \ \pr^{\,j}\,\g_{\,[\,i\,|} \Big[\, (\,D-2\,)\, \O_{\,|\,j\,]} \, + \, 2 \, S^{\,k}{}_{\,|\,j\,]}\, \O_{\,k} \,\Big] \\
& + \, \g^{\,j}\, \mathscr{B}_i\, (\,\O_{\,j}\,) \, ,
\end{split}
\ee
where $\mathscr{B}_i\,(\,\O_{\,j}\,)$ denotes the Bianchi identities for the parameters $\O_{\,j}$.
Thus, if the conditions
\be \label{shiftcond1}
(\,D-2\,)\, \O_{\,i} \, + \, 2 \, S^{\,j}{}_{\,i}\, \O_{\,j} \, = \, 0
\ee
are satisfied by non-trivial parameters $\O_{\,i}$ that are themselves subject to the Bianchi identities, the corresponding shift \eqref{shift1} preserves indeed eq.~\eqref{bianchired}. Before presenting the solutions of this eigenvalue problem, we would like to stress that eq.~\eqref{shiftcond1} is clearly a \emph{sufficient} condition to identify a symmetry of the Bianchi identities. In the general case, in fact, the contributions that appear in eq.~\eqref{shiftB1} could in principle balance each other. However, one can proceed to look for solutions in an iterative manner, analyzing directly more complicated shifts of the Bianchi identities that involve more than one ``naked'' $\g$-matrix.

At two families a more general shift involving pairs of $\g$-matrices is fully available, so that one can also consider
\be \label{shift2}
\d \, \cS \, = \, \h^{\,ij}\, \O_{\,ij} \, + \, \g^{\,ij}\, \widetilde{\O}_{\,ij} \, ,
\ee
where the two parameters are symmetric and antisymmetric, respectively, in their family indices. This new type of shift gives rise to a more complicated variation
of the Bianchi identities,
\be \label{shiftB2}
\begin{split}
\d \, \mathscr{B}_i \, = \, & - \, \12 \dsl \ \g^{\,j}\, \Big[\, \O_{\,ij} \, - \, 2 \, S^{\,k}{}_{\,(\,i}\, \widetilde{\O}_{\,j\,)\,k} \,\Big] \, - \dsl \ \g^{\,j}\, \Big[\, (\,D-3\,)\, \widetilde{\O}_{\,ij} \, - \, S^{\,k}{}_{\,[\,i}\, \widetilde{\O}_{\,j\,]\,k} \,\Big] \\
& - \, \12 \ \pr^{\,j}\, \Big[\, (\,D-2\,)\, \O_{\,ij}\, + \, S^{\,k}{}_{\,(\,i}\, \O_{\,j\,)\,k}\, + \, 2 \, S^{\,k}{}_{\,(\,i}\, \widetilde{\O}_{\,j\,)\,k} \,\Big] \\
& + \, \frac{1}{3} \ \pr^{\,j} \, \g^{\,k}{}_{\,[\,i\,|}\, \Big[\, (\,D-3\,)\, \widetilde{\O}_{\,|\,j\,]\,k} \, - \, 2 \, S^{\,l}{}_{\,|\,j\,]} \, \widetilde{\O}_{\,kl} \, + \, \12 \ \O_{\,|\,j\,]\,k} \,\Big] \\
& + \, \frac{D+2}{3} \ \pr^{\,j}\, \Big[\, (\,D-3\,)\, \widetilde{\O}_{\,ij} \, - \, S^{\,k}{}_{\,[\,i}\, \widetilde{\O}_{\,j\,]\,k} \,\Big] \, + \, \h^{\,jk} \, \mathscr{B}_i\, (\,\O_{\,jk}\,) \, + \, \g^{\,jk}\, \mathscr{B}_i\, (\,\widetilde{\O}_{\,jk}\,) \, ,
\end{split}
\ee
where we have conveniently introduced the new quantities
\be
\g^{\,i}{}_{\,j} \, = \, \g^{\,i}\,\g_{\,j} \, - \, S^{\,i}{}_{\,j} \, ,
\ee
that belong to a class of operators described in Appendix \ref{app:fermi}. One can thus identify some purely algebraic conditions granting that \eqref{shiftB2} vanish,
\begin{align}
& \O_{\,ij} \, = \, 2\, S^{\,k}{}_{\,(\,i}\, \widetilde{\O}_{\,j\,)\,k} \, , \nn \\[2pt]
& (\,D-3\,)\, \widetilde{\O}_{\,ij} \, - \, S^{\,k}{}_{\,[\,i}\, \widetilde{\O}_{\,j\,]\,k} \, = \, 0 \, . \label{shiftcond2}
\end{align}
These eliminate all terms involving $\not\!\! \pr$ from eq.~\eqref{shiftB2}. All the rest then evidently vanishes, provided the $\O_{\,ij}$ and $\widetilde{\O}_{\,ij}$ satisfy the Bianchi identities, with the single exception of the second line \footnote{Notice that these new parameters \emph{are not} special cases of the previous ones, as one might naively think, since they are selected by different conditions.}. However, this can be turned into
the second of eqs.~\eqref{shiftcond2} using the first and noticing that
\be \label{commS}
S^{\,k}{}_{(\,i}\, S^{\,l}{}_{j\,)}\, \widetilde{\O}_{\,kl} \, = \, \12 \, [\, S^{\,k}{}_{(\,i\,|} \comma S^{\,l}{}_{|\,j\,)} \,]\, \widetilde{\O}_{\,kl} \, = \, - \, S^{\,k}{}_{(\,i}\, \widetilde{\O}_{\,j\,)\,k} \, .
\ee

The present setting becomes degenerate at the next
step, since a fully antisymmetric $\g_{\,ijk}$ is clearly not available with only two families.
As a result, the next shift in the iteration takes the form
\be \label{shift3}
\d \, \cS \, = \, \h^{\,ij}\, \g^{\,k}\, \O_{\,ij\,;\,k} \, ,
\ee
and gives rise to the variation
\begin{align}
& \d \, \mathscr{B}_l \, = \,  - \, \12 \dsl \ \h^{\,ij} \Big[\, (\,D-2\,)\,\O_{\,ij\,;\,l} + \, \O_{\,l\,(\,i\,;\,j\,)} + \, 2\, S^{\,m}{}_l\, \O_{\,ij\,;\,m} \,\Big] - \, \frac{1}{4} \dsl \ \g^{\,ij}\, \O_{\,l\,[\,i\,;\,j\,]} \nn \\
& - \, \frac{1}{6}\ \h^{\,ij}\, \pr^{\,m}\, \g_{\,[\,l\,|} \Big[\, (\,D-2\,)\, \O_{\,ij\,;\,|\,m\,]} + \, \O_{\,|\,m\,]\,(\,i\,;\,j\,)} + \, 2\, S^{\,n}{}_{|\,m\,]}\, \O_{\,ij\,;\,n} \,\Big] - \, \frac{1}{12}\ \g^{\,ij}\,\pr^{\,m}\, \g_{\,[\,l}\, \O_{\,m\,]\,[\,i\,;\,j\,]} \nn \\
& - \, \12 \ \g^{\,k}\,\pr^{\,m} \Big[\, (\,D-2\,)\, \O_{\,lm\,;\,k} + \, \O_{\,k\,(\,l\,;\,m\,)} + \, S^{\,n}{}_{(\,l}\, \O_{\,m\,)\,n\,;\,k} - \, \frac{1}{6}\, S^{\,n}{}_{[\,l}\, \O_{\,m\,]\,(\,k\,;\,n\,)} \,\Big] \nn \\
& - \, \frac{1}{12} \ \g^{\,k}\,\pr^{\,m} \Big[\, 2\,(\,D-1\,)\, \O_{\,k\,[\,l\,;\,m\,]} - \, 3\, S^{\,n}{}_{[\,l}\, \O_{\,m\,]\,[\,k\,;\,n\,]} \,\Big] + \, \h^{\,ij}\, \g^{\,k}\, \mathscr{B}_l\,(\,\O_{\,ij\,;\,k}\,)  \label{shiftB3}
\end{align}
of the Bianchi identities. Again, the conditions that eliminate the terms with $\not \! \pr$,
\begin{align}
& (\,D-2\,)\, \O_{\,ij\,;\,k} \, + \, \O_{\,k\,(\,i\,;\,j\,)} \, + \, 2 \, S^{\,l}{}_k\, \O_{\,ij\,;\,l} \, = \, 0 \, , \nn \\
& \O_{\,k\,[\,i\,;\,j\,]} \, = \, 0 \, , \label{shiftcond3}
\end{align}
set to zero all other contributions if the parameters satisfy themselves the Bianchi identities. Indeed, the second condition in \eqref{shiftcond3} implies that all combinations of parameters giving rise to a $\{2,1\}$ Young projection in family indices
vanish, and consequently the first condition reduces to
\begin{align}
& 3\,D\, \widehat{\O}_{\,ijk} + 2\, S^{\,l}{}_{(\,i}\, \widehat{\O}_{\,jk\,)\,l} \, = \, 0 \, , \nn \\
& S^{\,l}{}_{\,[\,i}\, \widehat{\O}_{\,j\,]\,kl} \, = \, 0 \, , \label{shiftcond3bis}
\end{align}
where
\be \label{defomegahat}
\widehat{\O}_{\,ijk} \, = \, Y_{\{3\}}\, \O_{\,ij\,;\,k} \, = \, \frac{1}{3}\ \O_{\,(\,ij\,;\,k\,)} \, .
\ee
We shall see shortly that eqs.~\eqref{shiftcond3bis} admit solutions at two families, while in Section \ref{sec:reductionf} we shall also see that $\{2,1\}$-projected combinations of parameters become available in the presence of
more than two index families, when the second condition in eq.~\eqref{shiftcond3} acquires further contributions.

In principle one should move on and consider shifts of the form
\be \label{shift4}
\d \, \cS \, = \, \h^{ij}\, \h^{kl}\, \O_{\,ij\,,\,kl} \, + \, \h^{ij}\, \g^{\,kl}\, \widetilde{\O}_{\,ij\,;\,kl}\, ,
\ee
since only two types of parameters with four indices are available with only two families. The corresponding variation of the Bianchi identities reads
\begin{align}
& \d \, \mathscr{B}_l \, = \ \dsl \ \h^{ij}\, \g^{\,k} \Big[\, (\,D-3\,)\, \widetilde{\O}_{\,ij\,;\,kl} - \, \widetilde{\O}_{\,l\,(\,i\,;\,j\,)\,k} - \, \O_{\,ij\,,\,kl} + 2\, S^{\,m}{}_l\, \widetilde{\O}_{\,ij\,;\,km} \,\Big] \nn \\
& - \, \frac{1}{3}\ \pr^{\,m}\, \h^{ij}\, \g^{\,k}{}_{[\,l\,|} \Big[\, (\,D-3\,)\, \widetilde{\O}_{\,ij\,;\,k\,|\,m\,]} - \, \widetilde{\O}_{\,|\,m\,]\,(\,i\,;\,j\,)\,k} - \, \O_{\,ij\,,\,k\,|\,m\,]} + 2\, S^{\,n}{}_{|\,m\,]}\, \widetilde{\O}_{\,ij\,;\,kn} \,\Big] \nn \\
& - \, \h^{ij}\, \pr^{\,k} \Big[\, (\,D-2\,)\, \O_{\,ij\,,\,kl} + \, \O_{\,i\,(\,k\,,\,l\,)\,j} + \, S^{\,m}{}_{(\,k}\, \O_{\,l\,)\,m\,,\,ij} + \, S^{\,m}{}_{(\,k\,|}\, \widetilde{\O}_{\,ij\,;\,|\,l\,)\,m} \,\Big] \nn \\
& - \, \frac{1}{3}\, \h^{ij}\, \pr^{\,k} \Big[\, (\,D+2\,)\,\left(\, (\,D-3\,)\, \widetilde{\O}_{\,ij\,;\,kl} - \, S^{\,m}{}_{[\,k\,|}\, \widetilde{\O}_{\,ij\,;\,|\,l\,]\,m} \,\right) -\, 2\,(\,D-1\,)\, \widetilde{\O}_{\,i\,[\,k\,;\,l\,]\,j} \nn \\
& - \, S^{\,m}{}_{[\,k}\, \widetilde{\O}_{\,l\,]\,(\,i\,;\,j\,)\,m}\,\Big] - \, \12\ \g^{\,ij}\, \pr^{\,k} \Big[\, (\,D-2\,)\, \widetilde{\O}_{\,kl\,;\,ij} + \, 2\, \widetilde{\O}_{\,i\,(\,k\,;\,l\,)\,j} + \, S^{\,m}{}_{(\,k}\, \widetilde{\O}_{\,l\,)\,m\,;\,ij} \,\Big] \nn \\
& + \, \frac{1}{6}\ \g^{\,ij}\, \pr^{\,k} \Big[\, 2\,(\,D-2\,)\, \widetilde{\O}_{\,i\,[\,k\,;\,l\,]\,j} + \, \O_{\,i\,[\,k\,,\,l\,]\,j} + \, S^{\,m}{}_{[\,k}\, \widetilde{\O}_{\,l\,]\,m\,;\,ij} + \, 2\, S^{\,m}{}_{\,[\,k}\, \widetilde{\O}_{\,l\,]\,[\,i\,;\,j\,]\,m} \,\Big] \nn \\
& + \, \h^{ij}\,\h^{mn}\, \mathscr{B}_l\,(\,\O_{\,ij\,,\,mn}\,) + \, \h^{ij}\, \g^{\,mn}\, \mathscr{B}_l\,(\,\widetilde{\O}_{\,ij\,;\,mn}\,)  \, ,
\end{align}
so that now the conditions
\be \label{shiftcond4}
(\,D-3\,)\, \widetilde{\O}_{\,ij\,;\,kl} \, - \, \widetilde{\O}_{\,l\,(\,i\,;\,j\,)\,k} \, - \, \O_{\,ij\,,\,kl} \, + \, 2 \, S^{\,m}{}_l \, \widetilde{\O}_{\,ij\,;\,km} \, = \, 0 \,
\ee
eliminate the $\not\! \! \pr$ terms. However, at the end of the present section we shall prove that these conditions do not admit any solution for two-family fields, in contrast with the previous ones.

At two families the iteration stops here. Indeed, eqs.~\eqref{constrred} imply that all combinations of more than four $\g$-traces acting on $\cS$ vanish when only two index families are present, and therefore the two chains of shifts
\begin{align}
& \d \, \cS \, = \, \h^{i_1j_1} \ldots\, \h^{i_pj_p}\, \g^{\,k}\, \O_{\,i_1j_1,\, \ldots\,,\,i_pj_p;\,k} \, ,
\label{shift5} \\
& \d \, \cS \, = \, \h^{i_1j_1} \ldots\, \h^{i_pj_p}\, \O_{\,i_1j_1,\, \ldots\,,\,i_pj_p} \, +
\, \h^{i_1j_1} \ldots\, \h^{i_{p-1}j_{p-1}}\, \g^{\,kl}\, \widetilde{\O}_{\,i_1j_1,\, \ldots\,,\,i_{p-1}j_{p-1};\,kl} \, ,
\label{shift6}
\end{align}
are expected to yield no further information, a fact that we verified using similar techniques.

We have thus classified the conditions characterizing the relevant symmetries of the Bianchi identities. While we shall discuss in some detail the corresponding eigenvalue problems in the second part of this section, we can now elaborate further upon the conditions imposed by eq.~\eqref{constrred}, before showing that all the resulting transformations actually leave eq.~\eqref{eqconstr2} invariant when they are combined with proper shifts of the $\cY_{\,ijk}$. The symmetrized triple $\g$-traces of eq.~\eqref{shift1} read
\be \label{trace1}
T_{(\,ij}\, \g_{\,k\,)}\, \d \, \cS \, = \, T_{(\,ij\,|} \Big[\, (\,D-2\,)\, \O_{\,|\,k\,)} \, + \, 2 \, S^{\,l}{}_{\,|\,k\,)}\, \O_{\,l} \,\Big] \, - \, \g^{\,l}\, T_{(\,ij}\, \g_{\,k\,)}\, \O_{\,l} \, ,
\ee
and thus lead directly to Labastida-like constraints for the $\O_{\,i}$ satisfying the condition \eqref{shiftcond1}. In a similar fashion, the symmetrized triple $\g$-traces of eq.~\eqref{shift2} read
\be \label{trace2}
\begin{split}
T_{(\,ij}\, \g_{\,k\,)}\, \d \, \cS \, & = \, \g_{\,(\,i\,|} \Big[\, D\, \O_{\,|\,jk\,)}\, + \, S^{\,l}{}_{\,|\,j}\, \O_{\,k\,)\,l} \, - \, 2\, S^{\,l}{}_{\,|\,j}\, \widetilde{\O}_{\,k\,)\,l} \,\Big] \\
& + \, \g^{\,l}\, T_{(\,ij\,|} \Big[\, \O_{\,|\,k\,)\,l} \, - \, 2 \left(\, S^{\,m}{}_{\,|\,k\,)}\, \widetilde{\O}_{\,lm} \, + \, S^{\,m}{}_{\,l}\, \widetilde{\O}_{\,|\,k\,)\,m} \,\right) \,\Big] \\
& + \, 2\, \g^{\,l}\, T_{(\,ij\,|} \Big[\, (\,D-3\,)\, \widetilde{\O}_{\,|\,k\,)\,l} \, -  \left(\, S^{\,m}{}_{\,|\,k\,)}\, \widetilde{\O}_{\,lm} \, - \, S^{\,m}{}_{\,l}\, \widetilde{\O}_{\,|\,k\,)\,m} \,\right) \,\Big] \\
& + \h^{\,lm}\, T_{(\,ij}\, \g_{\,k\,)}\, \O_{\,lm} \, + \, \g^{\,lm}\, T_{(\,ij}\, \g_{\,k\,)}\, \widetilde{\O}_{\,lm} \, ,
\end{split}
\ee
and when eq.~\eqref{shiftcond2} holds they lead again to Labastida-like constraints for the $\O_{\,ij}$ and $\widetilde{\O}_{\,ij}$ parameters. As we shall see, however, this particular set of conditions does not play a role at two families, since all solutions of eq.~\eqref{shiftcond2} correspond to fields bearing only a few space-time indices, for which the Labastida constraints are trivial. On the other hand, for the shift \eqref{shift3} the conditions \eqref{constrred} read
\begin{align}
& T_{(\,ij}\, \g_{\,k\,)}\, \d \, \cS \, = \, \h^{\,lm}\, T_{(\,ij\,|} \Big[\, D\, \O_{\,lm\,;\,|\,k\,)} + \, 2\, S^{\,n}{}_{|\,k\,)}\, \O_{\,lm\,;\,n} \,\Big] + \, \12 \ \g^{\,lm} \, T_{(\,ij}\, \O_{\,k\,)\,[\,l\,;\,m\,]} \nn \\
& + \Big[\, D\,(\,D+2\,)\, \O_{\,(\,ij\,;\,k\,)} + \, 2\,D\, S^{\,l}{}_{(\,i}\, \O_{\,jk\,)\,;\,l} + \, (\,D+2\,)\, S^{\,l}{}_{(\,i\,|}\, \O_{\,l\,|\,j\,;\,k\,)} + 2\, S^{\,l}{}_{(\,i}\, S^{\,m}{}_{j}\, \O_{\,k\,)\,m\,;\,l} \,\Big] \nn \\
& - \, \g^{\,l}\, \g_{\,(\,i\,|} \Big[\, D\, \O_{\,|\,jk\,)\,;\,l} + \, S^{\,m}{}_{|\,j}\, \O_{\,k\,)\,m\,;\,l} \,\Big] \, - \, \h^{\,lm}\, \g^{\,n}\, T_{(\,ij\,} \g_{\,k\,)} \, \O_{\,lm\,;\,n} \, . \label{trace3}
\end{align}
They play a crucial role in the ensuing analysis, and reduce once more to Labastida-like constraints on these additional parameters when \eqref{shiftcond3} holds. This becomes manifest after eliminating the $\{2,1\}$ component in the family indices from the parameters, as demanded by eq.~\eqref{shiftcond3}, and making use of the rewriting in eq.~\eqref{shiftcond3bis}.

We can now close the circle, showing that the shift symmetries with parameters satisfying Bianchi identities, Labastida-like triple $\g$-trace constraints and the algebraic conditions of eqs.~\eqref{shiftcond1}, \eqref{shiftcond2} and \eqref{shiftcond3} leave the equation of motion \eqref{eqconstr2} invariant, when they are combined with proper $\d\,\cY_{\,ijk}$ shifts. Let us begin
by considering the shift \eqref{shift1}, that at two families affects the Rarita-Schwinger-like tensor of eq.~\eqref{rarita} as follows:
\be \label{shiftE1}
\begin{split}
\d \, \cE \, = \, & - \, \12 \ \g^{\,i} \Big[\, (\,D-2\,)\, \O_{\,i} + \, 2\, S^{\,j}{}_i\, \O_{\,j} \,\Big] \, - \, \frac{1}{12}\ \g^{\,ij}\, \g_{\,[\,i\,|} \Big[\, (\,D-2\,)\, \O_{\,|\,j\,]} + \, 2\, S^{\,k}{}_{|\,j\,]}\, \O_{\,k} \,\Big] \\
& + \, \frac{1}{6}\ \h^{\,ij}\, \g^{\,k}\, Y_{\{2,1\}}\, T_{ij} \Big[\, (\,D-2\,)\, \O_{\,k} + \, 2\, S^{\,l}{}_k \, \O_{\,l} \,\Big] \, - \, \frac{1}{6}\ \h^{ij}\, \g^{\,k}\, T_{(\,ij}\, \O_{\,k\,)} \\
& - \, \frac{1}{3}\ \h^{ij}\,\h^{kl}\, Y_{\{3,1\}} \left(\, Y_{\{2,1\}}\, T_{ij}\,\g_{\,k} \,\right) \O_{\,l} - \, \frac{1}{6}\ \h^{ij}\,\g^{\,kl}\, Y_{\{3,1\}} \left(\, Y_{\{2,1\}}\, T_{ij}\,\g_{\,k} \,\right) \O_{\,l} \\
& + \frac{1}{12}\ \h^{ij}\,\h^{kl}\,\g^{\,m}\, Y_{\{3,2\}} \left(\, Y_{\{2,2\}}\, T_{ij}\,T_{kl} \,\right) \O_{\,m} \, .
\end{split}
\ee
The first three groups of terms vanish whenever eq.~\eqref{shiftcond1} holds, while at two families the others carry Young projections with more than two columns. As a result, in the variation of \eqref{eqconstr2} they can be compensated by proper $\O$-dependent shifts of the $\cY_{\,ijk}$ tensors. In particular, the contraction of all indices with invariant tensors makes it possible to factor out from all surviving terms the symmetric combinations $\h^{(\,ij}\,\g^{\,k\,)}$ that accompany the $\cY_{\,ijk}$, so that one can indeed conclude that the equation of motion \eqref{eqconstr2} is left invariant by the transformations
\begin{align}
& \d \, \cS \, = \, \g^{\,i}\, \O_{\,i} \, , \nn \\
& \d \, \cY_{\,ijk} \, = \, - \, \frac{1}{3} \left\{\, T_{(\,ij}\, \O_{\,k\,)} - \, \frac{1}{12}\ \g^{\,l} \left(\, 2\, \g_{\,l}\, T_{(\,ij}\, \O_{\,k\,)} - \, T_{l\,(\,i}\, \g_{\,j}\, \O_{\,k\,)} \,\right) \right. \nn \\
& \phantom{\d \, \cY_{\,ijk} \, =}\, + \left. \, \frac{1}{12}\ \h^{\,lm} \left(\, 2\, T_{lm}\, T_{(\,ij}\, \O_{\,k\,)} - \, T_{l\,(\,i\,|}\, T_{m\,|\,j}\, \O_{\,k\,)} \,\right) \,\right\} \, ,
\end{align}
with the $\O_{\,i}$ satisfying the conditions \eqref{shiftcond1}. However, not all resulting solutions are relevant, since, as we repeatedly stressed, one is also to preserve eq.~\eqref{eqmultred}.

As we shall see, for the simple shifts \eqref{shift2} that solve eq.~\eqref{shiftcond2} the invariance of the field equation \eqref{eqconstr2} can be verified almost by inspection. However, in order to better understand the behavior of the equation for $\bar{\psi}$ under Weyl-like transformations it is interesting to look at the variation of the Rarita-Schwinger-like tensor induced by generic shifts of the type \eqref{shift2}:
\begin{align}
& \d \, \cE \, = \, - \, \12\ \h^{ij} \Big[\, D\, \O_{\,ij} + \, S^{\,k}{}_{(\,i}\, \O_{\,j\,)\,k} - \, 2\, S^{\,k}{}_{(\,i}\, \widetilde{\O}_{\,j\,)\,k} \,\Big] \nn \\
& + \, \frac{1}{6}\ \g^{\,ij} \Big[\, (\,D-4\,)\,(\,D-3\,)\, \widetilde{\O}_{\,ij} - \, (\,2\,D-7\,)\, S^{\,k}{}_{[\,i}\, \widetilde{\O}_{\,j\,]\,k}  - \, \12\, S^{\,k}{}_{[\,i}\, \O_{\,j\,]\,k} + \, 2\, S^{\,k}{}_{[\,i\,|}\,S^{\,l}{}_{|\,j\,]}\, \widetilde{\O}_{\,kl} \,\Big] \nn \\
& - \, \frac{1}{6}\ \h^{ij}\,\g^{\,k}\, Y_{\{2,1\}}\, \g_{\,(\,i\,|} \Big[\, D\, \O_{\,|\,j\,)\,k} + \left(\, S^{\,l}{}_{|\,j\,)}\, \O_{\,kl} + \, S^{\,l}{}_k\, \O_{\,|\,j\,)\,l} \,\right) - \, 2 \left(\, S^{\,l}{}_{|\,j\,)}\, \widetilde{\O}_{\,kl} + \, S^{\,l}{}_k\, \widetilde{\O}_{\,|\,j\,)\,l} \,\right) \,\Big] \nn \\
& + \, \frac{1}{6}\, \h^{ij}\,\h^{kl}\, Y_{\{2,2\}}\, T_{ij}\, \Big[\, D\, \O_{\,kl} + \, S^{\,m}{}_{(\,k}\, \O_{\,l\,)\,m} - \, 2\, S^{\,m}{}_{(\,k}\, \widetilde{\O}_{\,l\,)\,m} \,\Big] \nn \\ & - \, \frac{1}{6}\ \h^{ij}\,\g^{\,k}\, \g_{\,(\,i}\, \O_{\,jk\,)}
\, - \, \frac{1}{12}\ \h^{ij}\,\h^{kl}\, T_{(\,ij}\, \O_{\,kl\,)} - \, \12\ \h^{\,ij}\,\g^{\,kl}\, Y_{\{3,1\}}\, T_{ij}\, \widetilde{\O}_{\,kl} - \, \frac{1}{12}\ \h^{\,ij}\,\g^{\,kl}\, Y_{\{3,1\}}\, \g_{\,kl}\, \O_{\,ij} \nn \\
& + \frac{D}{9}\ \h^{\,ij}\,\g^{\,kl}\, Y_{\{3,1\}} \left(\, 2\, T_{ij}\, \widetilde{\O}_{\,kl} - T_{k\,(\,i}\, \O_{\,j\,)\,l} \,\right) - \, \frac{2}{3}\ \h^{\,ij}\,\g^{\,kl}\, Y_{\{3,1\}}\, \left(\, Y_{\{2,1\}}\, T_{ij}\, S^{\,m}{}_k \,\right) \widetilde{\O}_{\,lm} \nn \\
& + \, \frac{1}{9}\, \h^{ij}\,\h^{kl}\,\g^{\,m}\, Y_{\{3,2\}} \Big[\, 6 \left(\, Y_{\{2,1\}}\, T_{ij}\,\g_{\,k} \,\right) \widetilde{\O}_{\,lm} -  \left(\, 2\, T_{ij}\,\g_{\,(\,k}\, \widetilde{\O}_{\,l\,)\,m} - \g_{\,(\,i}\,T_{j\,)\,(\,k}\, \widetilde{\O}_{\,l\,)\,m} \,\right) \,\Big] \, .
\end{align}
The last three lines contain terms that can be compensated by shifts of the $\cY_{\,ijk}$, while the two-column projected terms are to vanish by themselves. Although in this case the conditions are more complicated than those emerging from the Bianchi identities, one can verify that all these terms vanish whenever eq.~\eqref{shiftcond2} holds. In particular, the first, third and fourth groups of terms involve only rewritings of the second line of eq.~\eqref{shiftB2}, while the second group can be disposed of expressing everything in terms of $\widetilde{\O}$. Although one cannot exclude that these complicated conditions admit other solutions aside from those of eq.~\eqref{shiftcond2}, as we pointed out at the beginning of this section we are only interested in ``physical'' solutions, that as such also preserve the Bianchi identities \eqref{bianchired} and the Labastida-like constraints \eqref{constrred}.

The last class of relevant shift transformations appears in eq.~\eqref{shift3}. We already pointed out that the conditions \eqref{shiftcond3} imply that the $\{2,1\}$ family-index component of the parameters vanishes, and this suffices to prove the invariance of the field equation \eqref{eqconstr2}. In fact, eq.~\eqref{shiftcond3bis} suffices to replace all terms containing the $S^{\,i}{}_j$ operators with others where only the $\O$ parameters appear. As a result, the variation of the Rarita-Schwinger-like tensor can be manifestly compensated by suitable $\d\, \cY_{\,ijk}$ transformations, since all combinations of $\g$-traces of the parameters only admit projections in their family indices with at least three columns. This concludes the proof that the relevant shift symmetries of the Bianchi identities give rise to shift symmetries of the field equation \eqref{eqconstr2}. Furthermore, the direct variation of the field equation shows that the Bianchi identities greatly simplify the analysis of the relevant conditions, that could prove a non-trivial technical task otherwise.

\vskip 24pt


\scsss{Weyl-like symmetries} \label{sec:weyl2}


In the first part of this section we identified the conditions that forbid the direct reduction of the field equations to the Labastida form, and in the following we would like to present a list of the types of two-family fields exhibiting this phenomenon. As a first step, however, we shall display Weyl-like transformations for the $\psi$ fields that can gauge away the undetermined quantities. Let us begin from the simplest class of fields with pathological equations of motion, those admitting the shift symmetries \eqref{shift1} with parameters satisfying eq.~\eqref{shiftcond1}. A Weyl-like shift of $\psi$ of the form
\be \label{shiftpsi1}
\d \, \psi \, = \, \g^{\,i}\, \Theta_{\,i}
\ee
gives rise to the corresponding variation of the Fang-Fronsdal-Labastida tensor,
\be \label{shiftS1}
\d \, \cS \, = \, - \, i \, \pr^{\,k} \left[\, (\,D-2\,)\, \Theta_{\,k} \, + \, 2 \, S^{\,l}{}_{\,k}\, \Theta_{\,l} \,\right] \, - \, \g^{\,k}\, \cS\,(\,\Theta_{\,k}\,) \, .
\ee
Eq.~\eqref{shiftcond1} is an algebraic condition that depends only on the space-time dimension $D$ and on the tensorial structure of the parameters involved, and whenever it admits non-trivial solutions eq.~\eqref{shiftS1} reduces to
\be
\d \, \cS \, = \, - \, \g^{\,k}\, \cS\,(\,\Theta_{\,k}\,) \, ,
\ee
while the variation of the Labastida constraints or, in the unconstrained case, the field equations \eqref{eqmultred} for the Lagrange multipliers, reduce to
\be \label{tripleOi}
T_{(\,ij}\,\g_{\,k\,)} \, \d \, \psi \, = \, \g^{\,l}\, T_{(\,ij}\,\g_{\,k\,)} \, \Theta_{\,l} \, .
\ee
Notice that the $\cS\,(\,\Theta_{\,i}\,)$ satisfy the conditions
\be
(\,D-2\,)\, \cS\, (\,\Theta_{\,i}\,) \, + \, 2 \, S^{\,j}{}_{\,i} \, \cS\, (\,\Theta_{\,j}\,) \, = \, 0 \, ,
\ee
since the $S^{\,i}{}_{j}$ commute with the Fang-Fronsdal-Labastida $\cS$ operators.
Furthermore, they satisfy the constraints \eqref{constrred} provided the shifts
\eqref{shiftpsi1} are supplemented by the conditions
\be
T_{(\,ij}\,\g_{\,k\,)} \, \Theta_{\,l} \, = \, 0 \, ,
\ee
that are needed to preserve the Labastida constraints \eqref{labacf} on $\psi$. These conditions also
imply that the $\cS\, (\,\Theta_{\,i}\,)$ satisfy the Bianchi identities, so that
one can use them to gauge away the corresponding undetermined quantities in
the field equations.

In a similar fashion a shift of the form
\be \label{shiftpsi2}
\d \, \psi \, = \, \h^{\,ij}\, \Theta_{\,ij} \, + \, \g^{\,ij}\, \widetilde{\Theta}_{\,ij}
\ee
gives rise to the following variation of the Labastida tensor:
\be \label{shiftS2}
\begin{split}
\d \, \cS \, = \, & - \, i \, \pr^{\,j}\, \g^{\,k} \, \left[\, \Theta_{\,jk} \, - \, 2\, S^{\,l}{}_{\,(\,j}\, \widetilde{\Theta}_{\,k\,)\,l}  \,\right] \, - \, 2\, i \, \pr^{\,j}\, \g^{\,k} \, \left[\, (\,D-3\,)\, \widetilde{\Theta}_{\,jk} \, - \, S^{\,l}{}_{\,[\,j}\, \widetilde{\Theta}_{\,k\,]\,l} \,\right] \\
& + \, \h^{\,ij} \, \cS\,(\,\Theta_{\,ij}\,) \, + \, \g^{\,ij}\, \cS\, (\, \widetilde{\Theta}_{\,ij} \,) \, .
\end{split}
\ee
Even in this case, whenever the Lorentz structure of the gauge field $\psi$ guarantees
that the conditions \eqref{shiftcond2} be satisfied, one can properly choose the
parameters $\Theta_{\,ij}$ and $\widetilde{\Theta}_{\,ij}$ so that
\begin{align}
& \d \, \cS \, = \, \h^{\,ij}\, \cS\,(\,\Theta_{\,ij}\,) \, + \, \g^{\,ij}\, \cS\,(\,\widetilde{\Theta}_{\,ij}\,) \, , \nn \\
& T_{(\,ij}\,\g_{\,k\,)}\, \d \, \psi \, = \, \h^{\,lm}\, T_{(\,ij}\,\g_{\,k\,)} \,
\Theta_{\,lm} \, + \, \g^{\,lm}\, T_{(\,ij}\,\g_{\,k\,)} \, \widetilde{\Theta}_{\,lm} \,
,\label{shiftequation2}
\end{align}
with the $\cS\,(\,\Theta_{\,ij}\,)$ and $\cS\,(\,\widetilde{\Theta}_{\,ij}\,)$ satisfying
the conditions of eq.~\eqref{shiftcond2}. As in the previous case, these quantities also
satisfy the Bianchi identities once one requires that $\d \, \psi$ preserve the Labastida constraints. Therefore, further undetermined quantities can
be shifted away, as in the previous example, by the Weyl-like gauge transformation
\eqref{shiftpsi2}, with parameters satisfying the conditions \eqref{shiftcond2} and the
corresponding Labastida-like triple $\g$-trace constraints.

The last class of shift symmetries that we identified involves three $\g$-matrices, and
one can dispose of it via the Weyl-like gauge transformations
\be \label{shiftpsi3}
\d \, \psi \, = \, \h^{ij}\, \g^{\,k}\, \Theta_{\,ij\,;\,k} \, ,
\ee
that affect the Fang-Fronsdal-Labastida tensor according to
\be \label{shiftS3}
\begin{split}
\d \, \cS \, = \, & - \, i \, \h^{ij}\, \pr^{\,k} \left[\, (\,D-2\,)\, \Theta_{\,ij\,;\,k} + \, \Theta_{\,k\,(\,i\,;\,j\,)} + 2\, S^{\,l}{}_k\, \Theta_{\,ij\,;\,l} \,\right] \, - \, \frac{i}{2}\ \g^{\,ij}\, \pr^{\,k}\, \Theta_{\,k\,[\,i\,;\,j\,]} \\
& - \, \h^{ij}\, \g^{\,k}\, \cS\,(\,\Theta_{\,ij\,;\,k}\,) \, .
\end{split}
\ee
Again, whenever the conditions \eqref{shiftcond3} hold, one can factor out the invariant
tensors, and a shift of $\cS$ emerges that is capable of removing the undetermined
quantities. Once more, the leftover Weyl parameters $\widehat{\Theta}_{\,lmn}$ are fully
symmetric in their family indices and must satisfy triple $\g$-trace constraints of the
form
\be
T_{(\,ij}\, \g_{\,k\,)} \, \widehat{\Theta}_{\,lmn} \, = \, 0 \, ,
\ee
in order to leave eqs.~\eqref{eqmultred} and the Bianchi identities invariant.

These observations actually suggest an alternative starting point to classify the fields
whose equations of motion do not reduce directly to the Labastida form \eqref{eqlaba}:
one can classify the Weyl-like shifts of $\psi$, that by construction also
leave the Bianchi identities invariant. We defer
further comments to Section \ref{sec:reductionf}, where we shall adopt this convenient
viewpoint to study the reduction procedure in the general $N$-family case.

We can now take a closer look at the previous conditions. As we anticipated, and as we
are going to see in detail shortly, while eqs.~\eqref{shiftcond1}, \eqref{shiftcond2} and
\eqref{shiftcond3} admit non-trivial solutions, eqs.~\eqref{shiftcond4} do not. In all
these cases one is facing eigenvalue problems for the $S^{\,i}{}_j$ operators that can
identify Weyl-like gauge transformations, but whose solutions are to be subjected by
further conditions arising from triple $\g$-trace constraints. At two families the
eigenvalue problems greatly simplify, since one can resort to the $gl(2)$ techniques
already described in \cite{bose_mixed} for Bose fields. In fact, turning to the
suggestive notation
\be
L_+ \, = \, S^{\,1}{}_2 \ , \quad L_- \, = \, S^{\,2}{}_1 \ , \quad
L_3 \, = \, \frac{1}{2} \left(\, S^{\,1}{}_1 \, - \, S^{\,2}{}_2\,
\right) \, ,\label{ang_mom}
\ee
one readily recovers the angular momentum algebra
\be
[\, L_3 \, , \, L_\pm \, ]\, = \, \pm \, L_\pm \ , \qquad [\,L_+ \,
, \, L_- \, ]\, = \, 2 \, L_3 \ , \label{ang_momalg}
\ee
while looking at their definition in eq.~\eqref{flip} one can recognize that
$S^{\,1}{}_1$ and $S^{\,2}{}_2$ act diagonally on generic spinor-tensors $\psi$ of rank
$(s_1,s_2)$, with
\be
S^{\,1}{}_1 \, \psi \, = \, s_1 \, \psi \, , \qquad S^{\,2}{}_2 \, \psi \, = \, s_2 \, \psi \, .
\ee

In order to proceed, it is thus convenient to split the problems into combinations of
simpler ones, expanding in bases on which the ``total angular momentum'' $\mathbf{L}^2$
acts diagonally. For instance, a gauge field $\psi$ of rank $(s_1,s_2)$ can be decomposed
according to
\be \label{decomposition}
\psi \, = \, \sum_{n\,=\,0}^{s_2} \, Y_{\{s_1+s_2-n,\,n\}} \, \psi \, \equiv\, \sum_{n\,=\,0}^{s_2} \, \psi^{\,\{s_1+s_2-n,\,n\}} \, ,
\ee
where the various components are characterized by the same $L_3$ eigenvalue, that is
fixed by the rank of $\psi$ to be
\be \label{m_fixed}
m \, = \, \frac{s_1 \, - \, s_2}{2} \ .
\ee
The terms of the sum \eqref{decomposition} corresponding to Young projections of the type $Y_{\{s_1+s_2-n,\,n\}}$ can be related to null eigenvectors of $L_{+}$ via the standard descent relations
\be
\psi^{\,\{s_1+s_2-n,\,n\}} \, = \, (L_{-})^{\,s_2-n}\, \widehat{\psi}^{\,\{s_1+s_2-n,\,n\}} \, , \qquad  L_{+}\, \widehat{\psi}^{\,\{s_1+s_2-n,\,n\}} \, = \, 0 \, ,
\ee
so that they are all eigenvectors of $\mathbf{L}^2$, but in general with different
eigenvalues. The $\widehat \psi$ lie at the tips of the chains and are clearly $gl(D)$
\emph{irreducible} in their vector indices on account of eq.~\eqref{flip}. In fact, they
are annihilated by $L_+$, so that a vanishing result obtains if one tries to extend the
symmetrization beyond the first family. On the other hand, the lower members of the
chains in the decomposition \eqref{decomposition} are built acting on the $\widehat \psi$
with powers of $L_-$, and as a result their symmetry is not manifest. However, they are
still eigenvectors of $Y_{\{s_1+s_2-n,\,n\}}$, that commutes with $L_-$ simply because
the permutation group acts irreducibly within individual $Y$ eigenspaces. All in all, the
operator
\be
L_- L_+ \, = \, \mathbf{L}^2 - L_3^2 - L_3 \label{totang}
\ee
is diagonal when acting on this basis, whose members $\psi^{\,\{s_1+s_2-n,\,n\}}$ are characterized by
the \emph{fixed} value \eqref{m_fixed} for $m$ and by a \emph{range} of values of $\ell$,
\be
\ell \, = \, \frac{s_1\,+\,s_2}{2} \, - \, n \, , \qquad\qquad \frac{s_1\,-\,s_2}{2} \, \le \, \ell \, \le \, \frac{s_1\,+\,s_2}{2} \ ,
\ee
so that
\be \label{numcomp}
L_- L_+ \, \psi^{\,\{s_1+s_2-n,\,n\}} \, = \, (\,n-s_1-1\,)\,(\,n-s_2\,)\, \psi^{\,\{s_1+s_2-n,\,n\}} \, .
\ee

In the rest of this section we shall make extensive use of these tools to analyze the algebraic conditions that we previously identified, and we shall conclude with some comments on the solutions and with the identification of a special subclass of fields whose Lagrangians vanish identically, as is the case for the Rarita-Schwinger field in two dimensions. The discussion of the eigenvalue problems for the $S^{\,i}{}_j$ operators is somewhat technical, and the reader may wish to skip the following pages and move directly to the tables that collect their outcome. Their two columns are labeled by $\ssl$ and $\cS^{\,\pe}$ and collect the types of fields whose equations of motion leave some combinations of single or double $\g$-traces of $\cS$ undetermined. In fact, the solutions of the eigenvalue problem \eqref{shiftcond3} only act on the symmetrized triple $\g$-traces of $\cS$, and therefore will be discarded.


\subsubsection*{\sc Analysis of eq.~\eqref{shiftcond1}}


Using the notation of eq.~\eqref{ang_mom} and taking into account the previous considerations, for a two-family field $\psi$ of rank $(s_1,s_2)$ the conditions \eqref{shiftcond1} become
\begin{align}
& (\,D+2\,s_1-4\,)\, \O_{\,1} \, + \, 2\, L_{-}\, \O_{\,2} \, = \, 0 \, , \nn \\
& (\,D+2\,s_2-4\,)\, \O_{\,2} \, + \, 2\, L_{+}\, \O_{\,1} \, = \, 0 \, . \label{syst1}
\end{align}
Barring a couple of degenerate cases to be treated separately, this system can be solved by direct substitution. Thus, expressing $\O_{\,2}$ in terms of $\O_{\,1}$ and decomposing it into irreducible $gl(D)$ components, using the notation of eq.~\eqref{decomposition} leads to
\be
(\,D+2\,n-4\,)\,(\,D+2\,s_1+2\,s_2-2\,n-4\,)\, (\O_{\,1})^{\,\{s_1+s_2-n-1,\,n\}} \, = \, 0 \, , \qquad 0 \leq n \leq s_2 \, .
\ee
The second factor vanishes only in the degenerate cases where the direct
substitution is not possible. These correspond to the two-dimensional gravitino, that we have already
discussed, and to another two-dimensional field, the $\{1,1\}$ irreducible spinor-tensor, for which
however $\cS$ vanishes identically when expressed in terms of $\psi$. On the other hand, the first factor vanishes in $D=2$ for $n=1$ and in
$D=4$ for $n=0$, but for all $s_1\geq2$ and $s_2\geq2$. These solutions are to be subjected
to the double trace conditions
\be \label{tripleomega1} T_{(\,ij}\, \g_{\,k\,)} \, \Omega_{\,l} \ = \ 0  \, , \ee
that, as we shall see shortly, eliminate some of them. The final results are collected in
the first column of Table~\ref{table2fam_rid}.

Each value of $n$ selects a particular irreducible component for the vector indices
carried by the parameters, and consequently not all irreducible components of the kinetic
spinor-tensor (or of the field, if one takes eq.~\eqref{shiftpsi1} as a starting point)
can be affected. In particular, for these two classes of solutions the resulting shifts
of $\cS$,
\be \label{shiftexample}
\d \, \cS \, = \, \g^{\,1}\, \O_{\,1} \, - \, \frac{2}{D + 2\,s_2 -4} \ \g^{\,2}\, L_+ \, \O_{\,1} \, ,
\ee
satisfy the conditions
\begin{alignat}{2}
& L_-L_+ \,\d\,\cS \, = \, (\,s_1-1\,)\,(\,s_2-2\,)\, \d\,\cS \, , \qquad\quad & & D=2\,,\ n=1\, , \nn \\
& L_-L_+ \,\d\,\cS \, = \, s_1\,(\,s_2-1\,)\, \d\,\cS \, , \qquad\quad & & D=4\,,\ n=0 \,
. \label{irrcond1}
\end{alignat}
Comparing these eigenvalues with eq.~\eqref{numcomp}, one can see that
eq.~\eqref{shiftexample} affects only the $\{s_1+s_2-2,2\}$ component of $\cS$ in $D=2$
and its $\{s_1+s_2-1,1\}$ component in $D=4$. However, the Weyl-like transformation
\eqref{shiftexample} corresponding to the first solution in \eqref{irrcond1} disappears
if the parameter is subject to the condition \eqref{tripleomega1}, but the corresponding
undetermined quantity in the field equation $E_{\,\bar{\psi}}$ disappears as well on
account of $E_{\,\bar{\zeta}}$. The $\{2,2\}$ field in two dimensions is a notable exception, since it is not subject to any Labastida constraints, but this case is also highly degenerate, since the corresponding Lagrangian vanishes identically even before expressing
$\cS$ in terms of $\psi$. All these results can be obtained analyzing explicitly in
components the various expressions, and the irreducible field components that are left undetermined are collected in the first column of Table~\ref{table2fam_irr}.


\subsubsection*{\sc Analysis of eq.~\eqref{shiftcond2}}


Moving on to eq.~\eqref{shiftcond2}, one should notice that at two families the
antisymmetric $\widetilde{\O}_{\,ij}$ reduce to the single
$\widetilde{\O}_{\,12}$. For a rank-$(s_1,s_2)$ field, the first set of conditions in
eq.~\eqref{shiftcond2} then relates $\widetilde{\O}_{\,12}$ to the symmetric parameters
according to
\begin{align}
\O_{\,11} & = \, 4 \, L_-\, \widetilde{\O}_{\,12} \, , \nn \\
\O_{\,12} & = \, -\,2\, (\,s_1-s_2\,)\, \widetilde{\O}_{\,12} \, , \nn \\
\O_{\,22} & = \, -\, 4\, L_+\, \widetilde{\O}_{\,12} \, , \label{syst2a}
\end{align}
while the second condition becomes
\be \label{syst2b}
(\,D+s_1+s_2-5\,)\, \widetilde{\O}_{\,12} \, = \, 0 \, .
\ee
Eq.~\eqref{syst2b} admits non-trivial solutions only for $s_1=2$ and $s_2=1$ in $D=2$ and
for $s_1=1$ and $s_2=1$ in $D=3$, and these are indeed the only cases when non-trivial
$\O_{\,ij}$ are admissible. In both cases only symmetric and hooked irreducible field
components are available, and the pathological behavior clearly concerns the hooked
components, since we know already that a gravitino in $D=2$ is the only degenerate
symmetric example. At any rate, it is also possible to confirm this expectation by a
direct calculation similar to those of eq.~\eqref{irrcond1}.
In this case eq.~\eqref{constrred} does not impose any further conditions, since these fields are too simple and are not subject to any Labastida constraints. However, one is to discard
the $\{2,1\}$ case, that is degenerate: in this case the Weyl-like transformation
\eqref{shiftpsi2} leaves $\cS$ invariant, so that the $\g$-traces that are left undetermined by the Lagrangian field equation vanish
identically. Again, this can be seen analyzing explicitly these expressions in
components. The final results are collected in the second columns of
Tables~\ref{table2fam_rid} and \ref{table2fam_irr}.


\subsubsection*{\sc Analysis of eq.~\eqref{shiftcond3}}


As we already noticed, the second condition in eq.~\eqref{shiftcond3} annihilates the
$\{2,1\}$ combinations of the shift parameters, so that one can conveniently rewrite the
relations directly in terms of the fully symmetric $\widehat{\O}_{\,ijk}$. This was done
in eq.~\eqref{shiftcond3bis}, whose first line contains a square system of equations for
the four $\widehat{\O}_{\,ijk}$ that are available in the two-family case:
\begin{align}
& (\,D+2\,s_1-6\,)\, \widehat{\O}_{\,111} + \, 2\, L_- \, \widehat{\O}_{\,112} \, = \, 0 \, , \nn \\
& 2\, L_+ \, \widehat{\O}_{\,111} + \, (\,3\,D+4\,s_1+2\,s_2-10\,)\, \widehat{\O}_{\,112} + \, 4\, L_- \, \widehat{\O}_{\,122} \, = \, 0 \, , \nn \\
& 4\, L_+ \, \widehat{\O}_{\,112} + \, (\,3\,D+2\,s_1+4\,s_2-10\,)\, \widehat{\O}_{\,122} + \, 2\, L_- \, \widehat{\O}_{\,222} \, = \, 0 \, , \nn \\
& 2\, L_+ \, \widehat{\O}_{\,122} + \, (\,D+2\,s_2-6\,)\, \widehat{\O}_{\,222} \, = \, 0 \, . \label{syst3a}
\end{align}
One can solve this system after multiplying the equations by suitable powers of $L_-$ in
order to convert them into others characterized by the same value of $m$. The operation
corresponds to performing some symmetrizations, in order that all
equations finally carry like sets of symmetric space-time indices. However, these sets can be
displaced in different positions on the same parameter, so that for instance for a
rank-$(4,2)$ field $\psi$ the two distinct combinations
\be
\widehat{\O}^{\,(112)}{}_{\m_1\m_2\,,\,\n}  \qquad {\rm and} \qquad \widehat{\O}^{\,(112)}{}_{\n\,(\,\m_1\,,\,\m_2\,)}
\ee
can appear. This problem can be solved projecting the result into irreducible $gl(D)$
components, since at two families the various combinations become proportional for any
given component. In our algebraic language, this means that the resulting expressions can
be manipulated in order to leave only the diagonal $L_-L_+$ operator acting on the
various components or on their $L_-$-shifted analogues. Suppressing for brevity the
superscript introduced in eq.~\eqref{numcomp} to identify the irreducible components, when all parameters admit the $\{s_1+s_2-n-3,n\}$ Young
projection the system \eqref{syst3a} reduces to the following chain of square systems of
equations labeled by $n$:
\begin{align}
& (\,D+2\,s_1-6\,)\, \widehat{\O}_{\,111} + \, 2\, L_- \, \widehat{\O}_{\,112} \, = \, 0 \, , \nn \\
& 2\, (\,n-s_1+2\,)\,(\,n-s_2\,)\, \widehat{\O}_{\,111} + \, (\,3\,D+4\,s_1+2\,s_2-10\,)\, L_-\,\widehat{\O}_{\,112} + \, 4\, L_-^2 \, \widehat{\O}_{\,122} \, = \, 0 \, , \nn \\
& 4\, (\,n-s_1+1\,)\,(\,n-s_2+1\,)\, L_- \, \widehat{\O}_{\,112} + \, (\,3\,D+2\,s_1+4\,s_2-10\,)\, L_-^2\, \widehat{\O}_{\,122} + \, 2\, L_-^3 \, \widehat{\O}_{\,222} \, = \, 0 \, , \nn \\
& 2\, (\,n-s_1\,)\,(\,n-s_2+2\,)\, L_-^2 \, \widehat{\O}_{\,122} + \, (\,D+2\,s_2-6\,)\, L_-^3\, \widehat{\O}_{\,222} \, = \, 0 \, . \label{syst3b}
\end{align}
Their determinants read
\be
\begin{split}
& \left(\,D+2\,n-2\,\right)\left(\,3\,D+4\,(s_1+s_2)-2\,n-18\,\right)\left(\,D+2\,(s_1+s_2)-2\,n-6\,\right) \\
& \times\left(\,3\,D+2\,(s_1+s_2)+2\,n-14\,\right) \, ,
\end{split}
\ee
and vanish if and only if $D=2$ and $n=0$, since $0\leq n\leq s_2$ and the whole system
is available only if $s_1\geq3$ and $s_2\geq3$, and in the remaining degenerate cases no new solutions emerge.
In $D=2$, the $n=0$ system has rank three and admits the solution
\begin{align}
& \widehat{\O}_{\,111} \, = \, - \, \frac{1}{(s_1-2)\,(s_1-1)\,s_1}\ L_-^3 \, (\widehat{\O}_{\,222})^{\,\{s_1+s_2-3,0\}} \, , \nn \\
& \widehat{\O}_{\,112} \, = \, \frac{1}{(s_1-1)\,s_1}\ L_-^2 \, (\widehat{\O}_{\,222})^{\,\{s_1+s_2-3,0\}} \, , \nn \\
& \widehat{\O}_{\,122} \, = \, - \, \frac{1}{s_1}\ L_- \, (\widehat{\O}_{\,222})^{\,\{s_1+s_2-3,0\}} \, , \label{shiftsol3}
\end{align}
where the fully symmetric $(\widehat{\O}_{\,222})^{\,\{s_1+s_2-3,0\}}$ is also the only
non-vanishing component of $\widehat{\O}_{\,222}$. Actually, this solution also satisfies
the second condition in eq.~\eqref{shiftcond3bis}, but disappears if one imposes the
triple $\g$-trace constraints
\be \label{tripleomega3} T_{(\, ij}\, \g_{\, k\,)} \, \widehat{\O}_{\,lmn}\,=\, 0 \, ,
\ee
so that it must be rejected.

More in detail, this can be seen identifying the single irreducible component of the
fields that is affected by these transformations, since only $n=0$ parameters are
involved, that are fully symmetric in their space-time indices. Substituting the solution
\eqref{shiftsol3} in eq.~\eqref{shift3} and applying $L_-L_+$ one finds
\be L_-L_+ \,\d\,\cS \, = \, (\,s_1-2\,)\,(\,s_2-3\,)\, \d\,\cS \, , \qquad\quad D=2\,,\
n=0\, , \ee
so that only the $\{s_1+s_2-3,3\}$-component of $\cS$ is affected. This type of terms can
now be analyzed in components, showing that indeed the remaining independent parameter
satisfying eq.~\eqref{tripleomega3} cannot lead to a non-trivial shift of the form
\eqref{shiftsol3}. This concludes the classification of the $gl(D)$ irreducible fields
whose equations of motion do not reduce to the Labastida form, and the results are
summarized in Table \ref{table2fam_irr}.
\begin{table}[htb]
\begin{center}
\begin{tabular}{||c||c|c||c|c||}
\cline{2-5}
\multicolumn{1}{c|| }{} & \multicolumn{2}{|c|| }{$\ssl$} & \multicolumn{2}{|c|| }{$\cS^{\, \pe}$} \\
\hline
\hline
$D$  & $s_1$ & $s_2$ & $s_1$ & $s_2$ \\
\hline
\hline
$2$  &  $ 1 $  &  $ 0 $ &  &   \\
\hline
$2$  &  $2$  &  $2$ &  &  \\
\hline
$3$  &  &  & 1 & 1 \\
\hline
$4$ &  $\geq 1$  &  $\geq 1$ &   & \\
\hline
\end{tabular}
\end{center}
\caption{$gl(D)$-\emph{reducible} fields with Weyl-like symmetries} \label{table2fam_rid}
\end{table}
%


\subsubsection*{\sc Analysis of eq.~\eqref{shiftcond4}}


We can finally show that eq.~\eqref{shiftcond4} admits no solutions at two families. Its
decomposition into irreducible Young components for the family indices is more intricate
that for the previous conditions, and in order to obtain the result it is convenient to
proceed via some alternative manipulations. For instance, antisymmetrizing
eq.~\eqref{shiftcond4} in the $(k,l)$ indices leads to
\be
2\,(\,D-3\,)\, \widetilde{\O}_{\,ij\,;\,kl} + \, \widetilde{\O}_{\,k\,(\,i\,;\,j\,)\,l} - \, \widetilde{\O}_{\,l\,(\,i\,;\,j\,)\,k} - \, 2\, S^{\,m}{}_{[\,k\,|}\, \widetilde{\O}_{\,ij\,;\,|\,l\,]\,m} \, = \, 0 \, ,
\ee
that at two families reduces to three simple conditions for the three available
$\widetilde{\O}_{\,ij\,;\,kl}$:
\begin{align}
& 2\,(\,D+s_1+s_2-6\,)\, \widetilde{\O}_{\,11\,;\,12} \, = \, 0 \, , \nn \\
& 2\,(\,D+s_1+s_2-6\,)\, \widetilde{\O}_{\,12\,;\,12} \, = \, 0 \, , \nn \\
& 2\,(\,D+s_1+s_2-6\,)\, \widetilde{\O}_{\,22\,;\,12} \, = \, 0 \, .
\end{align}
These immediately imply that no solutions exist for $D>2$, since the
$\widetilde{\O}_{\,ij\,,\,kl}$ parameters are only available if $s_1+s_2\geq4$ and the
$\O_{\,ij\,,\,kl}$ are directly related to them via
\be
\O_{\,ij\,;\,kl} \, = \, \frac{1}{2} \left(\, S^{\,m}{}_{(\,i\,|}\, \widetilde{\O}_{\,kl\,;\,|\,j\,)\,m} + \, S^{\,m}{}_{(\,k\,|}\, \widetilde{\O}_{\,ij\,;\,|\,l\,)\,m} \,\right) \, ,
\ee
that emerges symmetrizing eq.~\eqref{shiftcond4} in the $(k,l)$ indices and also
enforcing the symmetry under interchanges of the  $(i,j)$ and $(k,l)$ pairs. Still, in
principle a $\{3,1\}$ or a $\{2,2\}$ field in two dimensions could admit non-trivial
Weyl-like symmetries, but in these cases all parameters are scalars and the equations
obtained only symmetrizing \eqref{shiftcond4} in $(k,l)$ simply reduce to
\begin{align}
& \O_{\,11\,,\,12} \, = \, \widetilde{\O}_{\,11\,;\,12} \, , \nn \\
& \O_{\,11\,,\,12} \, = \, - \, \widetilde{\O}_{\,11\,;\,12} \, ,
\end{align}
for the $\{3,1\}$ case and similarly to
\begin{align}
& \O_{\,11\,,\,22} \, = \, \widetilde{\O}_{\,12\,;\,12} \, , \nn \\
& \O_{\,11\,,\,22} \, = \, - \, \widetilde{\O}_{\,12\,;\,12} \, ,
\end{align}
for the $\{2,2\}$ case. In conclusion, the conditions of eq.~\eqref{shiftcond4} do not admit any solution,
as we anticipated.

\begin{table}[htb]
\begin{center}
\begin{tabular}{||c||c|c||c|c||}
\cline{2-5}
\multicolumn{1}{c|| }{} & \multicolumn{2}{|c|| }{$\ssl$} & \multicolumn{2}{|c|| }{$\cS^{\, \pe}$} \\
\hline \hline
$D$  & $s_1$ & $s_2$ & $s_1$ & $s_2$ \\
\hline \hline
$2$  &  $ 1 $  &  $ 0 $ &  &   \\
\hline
$2$  &  $2$  &  $2$ &  &  \\
\hline
$3$  &  &  & 1 & 1 \\
\hline
$4$ &  $s$  &  $1$ &   & \\
\hline
\end{tabular}
\end{center}
\caption{$gl(D)$-\emph{irreducible} fields with Weyl-like symmetries}
\label{table2fam_irr}
\end{table}
%


\subsubsection{\sc Lagrangians that are total derivatives} \label{sec:topology}


In the preceding pages we have classified the fields whose equations of motion do not
directly reduce to the Labastida form \eqref{eqlaba}. We have seen that this does not
happen for $D>4$, and that new Weyl-like symmetries emerge that can gauge away the
undetermined quantities. We have also seen that some solutions of the eigenvalue
equations, that identify symmetries of eq.~\eqref{eqconstr2}, are not compatible with the
triple $\g$-trace conditions \eqref{eqmultred}, and therefore must be discarded
\footnote{For Bose fields the analogues of these considerations involve double trace
constraints on Weyl-like parameters, that were correctly identified in \cite{bose_mixed}.
However, we failed to notice that they also eliminate some of the solutions,
namely the reducible $D=2$ cases $(s_1 \geq 3,s_2 \geq 3)$ of Table 1 of \cite{bose_mixed}, or the
corresponding irreducible cases $\{s,3\}$ of Table 2 of \cite{bose_mixed}.}.

The paradigmatic example provided by the two-dimensional gravitino suggests that
those symmetries could well signal the emergence of Lagrangians that are at most total
derivatives, and indeed the cases listed in Table~\ref{table2fam_irr} correspond to
representations carrying no degrees of freedom~\footnote{If an $o(n)$ bosonic representation
disappears for a given $n$, the fermionic representation with the same tensorial
structure disappears as well. The $\g$-tracelessness condition for $o (n)$
spinor-tensors is in fact stronger than the tracelessness condition for bosonic
irreducible representations, so that the tensorial contribution to the dimension is
always smaller in the fermionic case. One can verify this also looking at
eqs.~\eqref{dim1} and \eqref{dim2} for these dimensions. On the other hand, a general
result of representation theory for $o (n)$ groups states that (see, for instance, the
first reference in \cite{branching}, \textsection $10$-$6$), if the total number of boxes
in the first two columns of a tableau exceeds $n$, the corresponding traceless tensor
vanishes.}.  On the other hand, the example of the two-dimensional gravitino and its role
in the dynamics of superstrings should make it clear that models of this type are nonetheless potentially quite rich.

The possibility that the pathological cases be characterized by
vanishing actions can be verified directly at the level of the field equations, that
should vanish identically for this type of models. We have seen how, after relating the
$\cY_{\,ijk}$ tensors to the kinetic spinor-tensor $\cS$, one ends up with a combination
of $\cS$ and its $\g$-traces. If $\cS$ does not vanish identically, this combination can only vanish if it coincides with one of its vanishing $o(D)$ components, and when this is the case the remaining $o(D)$ components can be clearly shifted at will and a Weyl-like symmetry
emerges. This is precisely what happens for the gravitino field in two dimensions, where
the Rarita-Schwinger tensor coincides with the vanishing $\g$-traceless projection of
$\cS$, and in principle it can also happen for other fields in low space-time dimensions.
It is thus interesting to try and understand whether all fields appearing in Table
\ref{table2fam_irr} behave exactly like a gravitino in $D=2$ or, consistently with the
expectation that one could build from the corresponding analysis performed in
\cite{bose_mixed} for bosons, whether some Weyl-invariant models of a different type
exist for mixed-symmetry fermions. First of all, one can observe that the example
provided by the gravitino is not so exotic, since a whole class of similar theories
actually exists. In fact, the rewriting \eqref{gamma_rarita} of the Rarita-Schwinger
Lagrangian suggests to consider the class of Lagrangians
\be \label{lag_anti}
\cL \, = \, i \, \frac{(-1)^{\,\frac{q(q-1)}{2}}}{2\,q\,!} \ \bar{\psi}_{\,\m_1 \ldots\, \m_q}\, \g^{\,\m_1 \ldots\, \m_q  \l \, \n_1 \ldots\, \n_q} \, \pr_{\,\l}\, \psi_{\,\n_1 \ldots\, \n_q}  + \, \textrm{h.c.}
\ee
for one-column (\emph{fully antisymmetric}) spinor-tensors \footnote{Similar types
of Lagrangians were recently discussed in \cite{zinoviev}, that appeared while this paper was being typed.}.
In Section \ref{sec:lagrangianf} we shall derive the Lagrangians for arbitrary
mixed-symmetry Fermi fields, and the reader can verify that the two presentations
actually coincide, since the fields in this class are particularly simple and are not
subject to any constraints. Moreover, in Section \ref{sec:2fermi} we already saw that
the structure of the Lagrangians is fixed by the requirement of gauge invariance, and
\eqref{lag_anti} is indeed manifestly gauge invariant, simply because the gradients
acting on the gauge parameters are antisymmetrized with the derivative that is already
present. Therefore, spinor-tensor fields carrying $q$ antisymmetric vector indices
define in general a class of interesting analogs of the two-dimensional Rarita-Schwinger
system for $q \leq\! D \!\leq 2\,q$. And indeed their only two-family representative, a
$\{1,1\}$ irreducible field, appears in Table \ref{table2fam_irr} for $D=3,4$, while we
left out the $D=2$ case, that is somewhat degenerate since the corresponding $\cS$ vanishes
identically. As we anticipated, another interesting example is provided by an irreducible $\{2,2\}$ field
in two dimensions, whose Rarita-Schwinger-like tensor $\cE$ vanishes identically even before expressing $\cS$ in terms of $\psi$.

Not all models listed in Table~\ref{table2fam_irr}, however, have Lagrangians that are
total derivatives. A rather simple counterexample is provided by a $\{2,1\}$-irreducible
$gl(D)$ spinor-tensor in $D=4$: one can verify that the Rarita-Schwinger-like tensor
\eqref{rarita} does not vanish
in this case. This suffices to reach our conclusion, since the $\{2,1\}$ fields are
intrinsically unconstrained, simply because the combinations of $\g$-traces that identify
the Labastida constraints are not available for them. Deferring to future work a more
detailed analysis, we can already conclude that for mixed-symmetry Fermi fields the
emergence of Weyl-like symmetries is not necessarily accompanied by vanishing actions,
in analogy with the behavior of mixed-symmetry Bose fields discussed in \cite{bose_mixed}.

\vskip 24pt


\scss{An example: reducible spinor-tensor fields of rank $(3,2)$}\label{sec:examples2f}


In the previous sections we presented Lagrangians and field equations
for general two-family spinor-tensor fields. Our concise notation made it
possible to discuss simultaneously results that apply to all fields of
this type, independently of their actual Lorentz labels,
but the presentation is possibly less palatable for the reader than more
conventional ones. We thus deem it appropriate, in analogy
with what we did for Bose fields in \cite{bose_mixed}, to dwell upon
the explicit space-time structure of a relatively simple and yet very
instructive example, a \emph{reducible} spinor-tensor of rank $(3,2)$,
whose explicit space-time form is $\psi_{\,\m_1\m_2\m_3\,,\,\n_1\n_2}$.

\vskip 24pt


\scsss{Lagrangian and field equations} \label{sec:lagexample}


As for the bosonic examples presented in \cite{bose_mixed}, in this
section we shall resort to a shorthand notation to denote traces so
that, for instance,
\bea
\h^{\,\m_1\m_2} \, \psi_{\,\m_1\m_2\m_3\,,\,\n_1\n_2} & \equiv & \psi^{\, \pe_\m}{}_{\m_3\,,\,\n_1\n_2} \, , \nonumber \\
\h^{\,\m_1\n_1} \, \psi_{\,\m_1\m_2\m_3\,,\,\n_1\n_2} & \equiv & \psi^{\, \hpe}{}_{\m_2\m_3\,,\,\n_2} \, .
\eea
The $(3,2)$ example has the virtue of being relatively simple and yet well structured, since the two distinct
gauge parameters that characterize generic two-family fields are already present in this case, so that the
explicit gauge variation of the field reads
\be
\d \, \psi \, = \, \pr^{\,1} \, \e_{\,1} \, + \, \pr^{\,2} \, \e_{\,2} \, \equiv \, \pr_{\,(\,\m_1} \, \e^{\,(1)}{}_{\m_2\m_3\,)\,,\,\n_1\n_2} \, + \, \pr_{\,(\,\n_1\,|\,} \e^{\,(2)}{}_{\m_1\m_2\m_3\,,\,|\,\n_2\,)} \, .
\ee
The two $\Psi_i$ compensators that accompany the two gauge parameters in the unconstrained theory are
\bea
\Psi_{\,1} & \equiv & \Psi^{\,(1)}{}_{\m_1\m_2\,,\,\n_1\n_2} \, , \nonumber \\
\Psi_{\,2} & \equiv & \Psi^{\,(2)}{}_{\m_1\m_2\m_3\,,\,\n} \, ,
\eea
but, as we have seen in general, they enter the construction only via
their symmetrized $\g$-traces. A generic two-family fermionic field
makes use of three different combinations of this sort, a fact that is
fully visible in the present example, where they take the form
\bea
\xi_{\,11} & = & \g_{\,1} \, \Psi_{\,1} \, \equiv \, \x^{\,(1)}{}_{\m\,,\,\n_1\n_2} \, , \nonumber \\
\xi_{\,12} & = & \12 \left(\, \g_{\,1}\, \Psi_{\,2} + \g_{\,2}\, \Psi_{\,1} \,\right) \equiv \, \x^{\,(2)}{}_{\m_1\m_2\,,\,\n} \, , \\
\xi_{\,22} & = & \g_{\,2} \, \Psi_{\,2} \, \equiv \, \x^{\,(3)}{}_{\m_1\m_2\m_3} \, . \nonumber
\eea
On the other hand, the four Labastida constraints that annihilate the
symmetrized triple $\g$-traces of generic two-family
spinor-tensors reduce in the present $(3,2)$ example to only three conditions:
\bea \label{labconstrf32}
& & T_{11}\, \g_{\,1}\, \psi \ \equiv \ \g^{\,\l} \, \psi^{\,\pe_\m}{}_{\l\,,\,\n_1\n_2} \, = \ 0\, , \nonumber \\
& & \left(\, T_{11}\, \g_{\,2} \, + \, 2\, T_{12}\, \g_{\,1} \,\right) \psi \ \equiv \ \g^{\,\l} \left(\, \psi^{\,\pe_\m}{}_{\m\,,\,\n\l} \, + \, 2\, \psi^{\,\hpe}{}_{\l\m\,,\,\n} \,\right) \, = \ 0 \, , \\
& & \left(\, T_{22}\, \g_{\,1} \, + \, 2\, T_{12}\, \g_{\,2} \,\right) \psi \ \equiv \ \g^{\,\l} \left(\, \psi^{\,\pe_\n}{}_{\l\m_1\m_2} \, + \, 2\, \psi^{\,\hpe}{}_{\m_1\m_2\,,\,\l} \,\right) \, = \ 0 \, . \nonumber
\eea
Notice that the Labastida constraints \eqref{labconstrf32} also
imply that the symmetrized double traces of the fermionic gauge
field vanish,
\bea
& & T_{11} \, T_{12} \, \psi \ \equiv \ \psi^{\,\pe_\m \hpe}{}_{\n} \, = \ 0 \, , \nonumber \\
& & \left(\, T_{11} \, T_{22} \, + \, 2\, T_{12}\, T_{12}
\,\right)\, \psi \ \equiv \ \psi^{\, \pe_\m \pe_\n}{}_{\m} \, + \, 2
\, \psi^{\, \hpe\,\hpe}{}_{\m} \, = \ 0 \, ,
\eea
and these conditions, in their turn, imply that all expressions
containing at least five $\g$-traces vanish. This is
counterpart, for two-family Fermi fields, of the vanishing of all triple traces
for two-family Bose fields, that was widely used in \cite{bose_mixed}.

As we saw in eq.~\eqref{triplegammaW}, the combinations of
$\g$-traces of the kinetic tensor $\cW$ that have the same structure
as the Labastida constraints are proportional to the $\cZ_{\,ijk}$
tensors. Two hooked triple $\g$-traces of $\cW$, however, are not
related to the constraints, and are the analogs in this setting of
the non-trivial double trace $a_{\m\n}$ present in the $(4,2)$ bosonic example
discussed in \cite{bose_mixed}. In the following these expressions
will be often denoted as follows:
\bea \label{hooktriplef}
\left(\, T_{11}\, \g_{\,2} \, - \, T_{12}\, \g_{\,1} \,\right) \cW & \equiv & \g^{\,\l} \left(\, \cW^{\,\pe_\m}{}_{\m\,,\,\n\l} \, - \, \cW^{\,\hpe}{}_{\l\m\,,\,\n} \,\right) \, \equiv \, v_{\,\m\,,\,\n} \, , \nonumber \\
\left(\, T_{22}\, \g_{\,1} \, - \, T_{12}\, \g_{\,2} \,\right) \cW & \equiv & \g^{\,\l} \left(\, \cW^{\,\pe_\n}{}_{\l\m_1\m_2} \, - \, \cW^{\,\hpe}{}_{\m_1\m_2\,,\,\l} \,\right) \, \equiv \, u_{\,\m_1\m_2} \, .\eea
In addition, the $\{2,2\}$-projected double trace of the gauge
field,
\be \label{windowf}
\left(\, T_{11}\, T_{22} \, - \, T_{12}\, T_{12} \,\right) \psi \, \equiv \, \psi^{\, \pe_\m \pe_\n}{}_{\m} \, - \, \psi^{\, \hpe\,\hpe}{}_{\m} \, \equiv \, \o_{\,\m} \, ,
\ee
is not related to the constraints, and can also contribute
significantly to the Lagrangian. Finally, the structure of the
Labastida constraints
\eqref{labconstrf32} reflects itself in the three Lagrange
multipliers,
\be
\l_{\,111} \, \equiv \, \l^{\,(1)}{}_{\n_1\n_2} \, , \qquad\quad
\l_{\,112} \, \equiv \, \l^{\,(2)}{}_{\m\,,\,\n} \, , \qquad\quad \l_{\,122} \, \equiv \, \l^{\,(3)}{}_{\m_1\m_2} \, ,
\ee
that complete the field content of this model.

In an explicit space-time notation, the unconstrained Lagrangian for $\psi_{\,\m_1\m_2\m_3\,,\,\n_1\n_2}$ reads
\be \label{lag32f}
\begin{split}
\cL \, & = \, \12 \, \bar{\psi}^{\,\m_1\m_2\m_3\,,\,\n_1\n_2} \, (\,\cE_{\,\psi}\,)_{\,\m_1\m_2\m_3\,,\,\n_1\n_2} \\
& + \, \frac{3}{2} \, \bar{\x}^{\,(1) \, \m\,,\,\n_1\n_2} \, \X^{\,(1)}{}_{\m\,,\,\n_1\n_2} \, + \, 3 \, \bar{\x}^{\,(2)\, \m_1\m_2\,,\,\n} \, \X^{\,(2)}{}_{\m_1\m_2\,,\,\n} \, + \, \12 \, \bar{\x}^{\, (3) \, \m_1\m_2\m_3} \, \X^{\,(3)}{}_{\m_1\m_2\m_3}  \\
& + \, \frac{3}{2} \, \bar{\l}^{\,(1)\, \n_1\n_2} \, \cZ^{\,(1)}{}_{\,\n_1\n_2} \, + \, 9 \, \bar{\l}^{\,(2)\, \m\,,\,\n} \, \cZ^{\,(2)}{}_{\m\,,\,\n} \, + \, \frac{9}{2} \, \bar{\l}^{\,(3)\, \m_1\m_2} \, \cZ^{\,(3)}{}_{\m_1\m_2} \, + \, \textrm{h.c.} \ ,
\end{split}
\ee
with the Rarita-Schwinger-like tensor
\begin{align}
\cE_{\,\psi} \, & = \, \cW_{\,\m_1\m_2\m_3\,,\,\n_1\n_2} \, - \, \12 \, \g_{\,(\,\m_1} \, \g^{\,\l}\, \cW_{\,\m_2\m_3\,)\,\l\,,\,\n_1\n_2} \, - \, \12 \, \g_{\,(\,\n_1|} \, \g^{\,\l}\, \cW_{\,\m_1\m_2\m_3\,,\,|\,\n_2)\,\l} \nn \\
& - \, \12 \, \h_{\,(\,\m_1\m_2} \, \cW^{\,\pe_\m}{}_{\m_3\,)\,,\,\n_1\n_2} \, - \, \12 \, \h_{\,(\,\n_1|\,(\,\m_1} \, \cW^{\,\hpe}{}_{\m_2\m_3\,)\,,\,|\,\n_2\,)} \, - \, \12 \, \h_{\,\n_1\n_2} \, \cW^{\,\pe_\n}{}_{\m_1\m_2\m_3} \label{einstein32f} \\
& + \, \frac{1}{6} \, \g_{\,(\,\n_1|\,(\,\m_1}\, \g^{\,\l\r} \, \cW_{\,\m_2\m_3\,)\,\l\,,\,|\,\n_2\,)\,\r} \, + \, \frac{1}{18} \left(\, 2\, \h_{\,(\,\m_1\m_2|} \,\g_{\,(\,\n_1|} - \, \h_{\,(\n_1|\,(\,\m_1}\g_{\,\m_2|} \,\right) v_{\,|\,\m_3\,)\,,\,|\,\n_2\,)} \nn \\
& + \, \frac{1}{18} \left(\, 2\, \h_{\,\n_1\n_2}\g_{\,(\,\m_1|} - \, \g_{\,(\,\n_1}\h_{\,\n_2\,)\,(\,\m_1\,|} \,\right) u_{\,|\,\m_2\m_3\,)} + \, \frac{1}{18} \left(\, 2\ \h_{\, \n_1 \n_2} \,
\h_{\,(\,\m_1\m_2\,|} - \, \h_{\, \n_1 \,(\, \m_1}
\h_{\,\m_2\,|\,\n_2} \,\right) \o_{\,|\,\m_3\,)} \, , \nn
\end{align}
where the symbols $u$, $v$ and $\o$ were defined in
eqs.~\eqref{hooktriplef} and \eqref{windowf} and where, as we saw in
Section \ref{sec:lagrangian2f}, the $\X^{\,(i)}$ tensors, that we do not display here for brevity,
are built
from suitable projections of the divergences of eq.~\eqref{einstein32f}.

The equations for the Lagrange multipliers are in this
case
\be \label{eqz32}
\cZ^{\,(1)}{}_{\n_1\n_2} \, = \, 0 \, , \qquad\quad \cZ^{\,(2)}{}_{\m\,,\,\n} \, = \, 0 \, , \qquad\quad \cZ^{\,(3)}{}_{\m_1\m_2} \, = \, 0 \, ,
\ee
and after enforcing them the equation for an unconstrained gauge field $\psi_{\,\m_1\m_2\m_3\,,\,\n_1\n_2}$ becomes
\be \label{eqpsi32}
\begin{split}
E_{\,\bar{\psi}} \, & : \ \cE_{\,\psi} \, - \, \12 \, \h_{\,(\,\m_1\m_2\,}\g_{\,\m_3\,)} \, \cY^{\,(1)}{}_{\n_1\n_2} \, - \, \12 \left(\, \h_{\,(\,\m_1\m_2|} \,\g_{\,(\,\n_1|} + \, \h_{\,(\n_1|\,(\,\m_1}\g_{\,\m_2|} \,\right) \cY^{\,(2)}{}_{|\,\m_3\,)\,,\,|\,\n_2\,)} \\
& - \, \12 \left(\, \h_{\,\n_1\n_2}\g_{\,(\,\m_1|} + \, \g_{\,(\,\n_1}\h_{\,\n_2\,)\,(\,\m_1\,|} \,\right) \cY^{\,(3)}{}_{|\,\m_2\m_3\,)} \, =\, 0 \, ,
\end{split}
\ee
where the $\cY^{\,(i)}$ spinor-tensors denote suitable combinations of the fields $\psi$, $\xi^{\,(i)}$ and $\l^{\,(i)}$, that we also do not display here for brevity.

As we saw in Section \ref{sec:motion2f}, the equation for a constrained reducible field
$\psi_{\,\m_1\m_2\m_3\,,\,\n_1\n_2}$ can also be cast in the form \eqref{eqpsi32}. However, as in the bosonic case, the field equation that originates from the constrained two-family Lagrangian
\eqref{constrlag2f} calls for some care if one wants to present it without introducing a set
of $\cY^{\,(i)}$ spinor-tensors. In fact, differently from the Fang-Fronsdal or
one-family case, the symmetrized triple $\g$-traces of the Rarita-Schwinger tensor
\eqref{einstein32f},
\bea
& & \g^{\,\l}\, \cE^{\,\pe_\m}_{\psi}{}_{\l\,,\,\n_1\n_1} \, = \, \frac{D+2}{9} \ v_{\,(\,\n_1\,,\,\n_2\,)} \, - \, \frac{4}{9} \ u_{\,\n_1\n_2} \, , \nonumber \\
& & \g^{\,\l} \left(\, \cE^{\,\pe_\m}_{\psi}{}_{\m\,,\,\n\l} \, + \, 2 \, \cE^{\,\hpe}_{\psi}{}_{\,\l\m\,,\,\n} \,\right) = \, - \, \frac{2}{9} \ v_{\,(\,\m\,,\,\n\,)} \, - \, \frac{D}{9} \ u_{\,\m\n} \, , \\
& & \g^{\,\l} \left(\, \cE^{\,\pe_\n}_{\psi}{}_{\,\l\m_1\m_2} \, + \, 2 \, \cE^{\,\hpe}_{\psi}{}_{\,\m_1\m_2\,,\,\l} \,\right) = \, - \, \frac{D+6}{9} \ v_{\,(\,\m_1\,,\,\m_2\,)} \, + \, \frac{2\,(D+2)}{9} \ u_{\,\m_1\m_2} \, , \nonumber
\eea
do not vanish identically. At any rate, combining eq.~\eqref{eqpsi32} with its symmetrized triple $\g$-traces one can eliminate the $\cY^{\,(i)}$ and, after gauging away the $\Psi_i$ compensators, one can finally obtain the constrained field equation
\be \label{constreq32}
\begin{split}
& E_{\,\bar{\psi}}^{\,constr} : \ \, \cE_{\,\psi} \,  + \, \frac{1}{18\,(3D+4)} \, \bigg\{\, \frac{6}{3D-2} \ \h_{\,(\m_1\m_2\,}\g_{\,\m_3)} \left[\, 4 \, u_{\,\n_1\n_2} \, - \, (3D+2) \, v_{\,(\n_1\,,\,\n_2)} \,\right] \\
& + \, \frac{2}{3D-2} \left(\, \h_{\,(\m_1\m_2|} \,\g_{\,(\n_1|} + \, \h_{\,(\n_1|\,(\m_1}\g_{\,\m_2|} \,\right) \left[\, 3\,D \, u_{\,|\m_3)\,|\n_2)} - \, 2 \, v_{\,|\m_3)\,,\,|\n_2)} - 2 \, v_{\,|\n_2)\,,\,|\m_3)} \,\right] \\
& - \, \frac{2}{3D-2} \left(\, \h_{\,\n_1\n_2}\g_{\,(\m_1|} + \, \g_{\,(\n_1}\h_{\n_2)\,(\m_1|} \,\right) \left[\, 2\, (3D+2) \, u_{\,|\m_2\m_3)} \, - \, 3\,(D+2) \, v_{\,
|\m_2\,,\,\m_3)} \,\right] \\
& + \, \frac{1}{D^2-2} \left(\, 2 \, \h_{\,\n_1\n_2}\, \h_{\,(\m_1\m_2\,}\o_{\,\m_3)} - \, \h_{\,\n_1(\m_1\,}\h_{\,\m_2|\,\n_1}\,\o_{\,|\m_3)} \,\right) \, \bigg\} = \, 0 \, .
\end{split}
\ee

This expression builds the symmetrized triply $\g$-traceless projection of $\cE_{\,\psi}$ that, as we shall see in a moment, couples to currents that satisfy the same set of constraints as the gauge field, precisely as was the case for the symmetric spinor-tensors of \cite{fms}. As we already pointed out in Section \ref{sec:motion2f}, this presentation mimics the Fang-Fronsdal setting, and is quite convenient if one wants to work only with the gauge field $\psi$, but the resulting equations depend explicitly both on the space-time dimension and on the Lorentz labels carried by the gauge fields.

\vskip 24pt


\scsss{On-shell reduction}\label{sec:reduction32}


As we saw in Section \ref{sec:reduction2f}, the reduction of the equation of motion
\eqref{eqpsi32} to the non-Lagrangian Labastida form
\be
\cS_{\,\m_1\m_2\m_3 \,,\, \n_1\n_2} \, = \, 0 \,
\ee
involves new subtleties that are foreign to the symmetric case of \cite{fsold2,fms}.
First of all, the $\cY^{\,(i)}$ tensors do not fully vanish on-shell, but their
undetermined components do not contribute to the equations of motion, as a consequence of
the shift symmetry \eqref{lambdalsym} of the fermionic Lagrangians. The other
difficulties can be traced to the anti-commutation relations \eqref{anticomgamma}, that
introduce the $S^{\,i}{}_{j}$ operators of Appendix \ref{app:MIX} in the $\g$-traces of
eq.~\eqref{epsi}, together with some additional technical complications related to the
rather involved form of the fermionic Rarita-Schwinger-like tensors. In order to better
illustrate the main steps of the reduction procedure, it is instructive to supplement the
general discussion of Section \ref{sec:reduction2f} with a more direct derivation of the
relevant results for the $(3,2)$-example, in an explicit space-time notation.

One can reduce eq.~\eqref{eqpsi32} to the Labastida form $\cS=0$
combining it with \emph{all} its admissible $\g$-traces. One can begin by taking five $\g$-traces
of the field equation \eqref{eqpsi32}, which yields
\begin{alignat}{2} \label{fivegamma32}
& \g \cdot E^{\,\pe_\m\hpe}_{\bar{\psi}} & & : \quad  \cY^{\,(1)\,\pe_\n} \, + \, \r_5  \, \cY^{\,(2)\, \hpe} \, - \, 3 \, \g^{\,\l\r} \, \cY^{\,(2)}{}_{\l\,,\,\r} \, + \, \r_2 \, \cY^{\,(3)\,\pe_\m} \, = \, 0 \, , \nonumber \\
& \g \cdot E^{\,\pe_\m\pe_\n}_{\bar{\psi}} & & : \quad \r_0\, \cY^{\,(1)\,\pe_\n} \, + \, 6\, \cY^{\,(2)\,\hpe} \, + \, \g^{\,\l\r} \, \cY^{\,(2)}{}_{\l\,,\,\r} \, + \, \r_2\, \cY^{\,(3)\,\pe_\m} \, = \, 0  \, ,  \\
& \g \cdot E^{\,\hpe\,\hpe}_{\bar{\psi}} & & : \quad \cY^{\,(1)\,\pe_\n} \, + \, 2\,\r_2\, \cY^{\,(2)\,\hpe} \, + \, 2\, \g^{\,\l\r}\, \cY^{\,(2)}{}_{\l\,,\,\r} \, + \, 6\, \cY^{\,(3)\,\pe_\m}\, = \, 0 \, , \nonumber
\end{alignat}
where the coefficients
\be
\label{rhok}
\rho_{\,k} \, \equiv \, D \, + \, k \, ,
\ee
depend on the space-time dimension $D$. Notice that we obtained in this fashion
\emph{three} relations, those in eq.~\eqref{fivegamma32}, that however involve the
\emph{four} available double $\g$-traces of the $\cY^{\,(i)}$ tensors. As a result, one
can at most link three of these double $\g$-traces to the remaining one, and here for
definiteness we make the choice of relating them to $\cY^{\,(1)}$, so that
\be \label{resfivegamma32}
\begin{split}
& \cY^{\,(2)\,\hpe} \, = \, \frac{1}{6} \ \cY^{\,(1)\,\pe_\n} \, , \\
& \g^{\,\l\r}\, \cY^{\,(2)}{}_{\l\,,\,\r} \, = \, - \, \frac{\rho_{-1}}{6} \ \cY^{\,(1)\,\pe_\n} \, , \\
& \cY^{\,(3)\,\pe_\m} \, = \, - \, \frac{2}{3} \ \cY^{\,(1)\,\pe_\n} \, .
\end{split}
\ee
Still, decomposing the $\cY^{\,(i)}$ tensors into their doubly
$\g$-traceless parts and corresponding remainders, according to
\begin{alignat}{2} \label{traceless32}
& \cY^{\,(1)}{}_{\n_1\n_2} & & = \, \cY^{\,(1)}_{\,T}{}_{\n_1\n_2} \, + \, \frac{1}{\r_0} \ \h_{\,\n_1\n_2} \, \cY^{\,(1)\,\pe_\n} \, , \nonumber \\
& \cY^{\,(2)}{}_{\m\,,\,\n} & & = \, \cY^{\,(2)}_{\,T}{}_{\m\,,\,\n} + \, \frac{1}{\r_0} \ \h_{\m\n} \, \cY^{\,(2)\,\hpe} \, - \, \frac{1}{\r_0\,\r_{-1}} \ \g_{\,\m\n} \, \g^{\,\l\r} \, \cY^{\,(2)}{}_{\l\,,\,\r} \, , \nonumber \\
& \cY^{\,(3)}{}_{\m_1\m_2} & & = \, \cY^{\,(3)}_{\,T}{}_{\m_1\m_2} \, + \, \frac{1}{\r_0} \ \h_{\,\m_1\m_2} \, \cY^{\,(3)\,\pe_\m} \, ,
\end{alignat}
\hspace{-5pt}and imposing the relations \eqref{resfivegamma32}, one can verify
that the remaining independent quantity disappears from \eqref{eqpsi32}.
As in the bosonic case of \cite{bose_mixed}, this fact simply reflects the shift symmetry of
eq.~\eqref{shiftY}, that for this example reduces to
\bea
\d \, \cY_{\,111} & = & \h^{\,22}\, M_{\,111,\,22} \, , \nonumber \\
\d \, \cY_{\,112} & = & 2\, \h^{\,12}\, M_{\,112,\,12} \, + \, 2\, \g^{\,12}\, N_{\,112;\,12} \, , \\
\d \, \cY_{\,122} & = & \h^{\,11}\, M_{\,122,\,11} \, . \nonumber
\eea
With only two index families, the parameter $N_{ijk;\,lm}$ admits only the $\{4,1\}$ component, and thus the projection does not impose any constraint on it, while $M_{ijk,\,lm}$ must
satisfy the conditions
\begin{alignat}{2}
& Y_{\{5\}} \, M_{\,ijk,\,lm} \, = \, 0 \quad \ & & \Rightarrow \quad \ M_{\,111,\,22} \, + \, 6 \, M_{\,112,\,12} \, + \, 3 \, M_{\,122,\,11} \, = \, 0 \, ,  \nonumber \\
& Y_{\{3,2\}} \, M_{\,ijk,\,lm} \, = \, 0 \quad \ & & \Rightarrow \quad \ M_{\,111,\,22} \, - \, 2 \, M_{\,112,\,12} \, + \, M_{\,122,\,11} \, = \, 0 \, .
\end{alignat}
However, $N_{\,ijk;\,lm}$ is linked to $M_{ijk,\,lm}$ via
eq.~\eqref{MNrel}, that in the $(3,2)$ case reduces to
\be
N_{\,112;\,12} \, = \, \frac{1}{12} \, M_{\,111,\,22} \, ,
\ee
so that the shift symmetry rests eventually on the single spinorial
parameter $M = M_{111,\,22}$, and in the space-time notation of this
section takes the explicit form
\begin{alignat}{2} \label{spaceshift32}
&\d \, \cY^{\,(1)}{}_{\n_1\n_2} & & = \, \h_{\,\n_1\n_2} \, M \, , \nonumber \\
&\d \, \cY^{\,(2)}{}_{\m\,,\,\n} & & = \, \frac{1}{6} \ \h_{\m\n} \, M \, + \, \frac{1}{6} \ \g_{\,\m\n} \, M \, , \\
&\d \, \cY^{\,(3)}{}_{\m_1\m_2} & & = \, - \, \frac{2}{3} \ \h_{\m_1\m_2} \, M \, . \nonumber
\end{alignat}
Substituting \eqref{resfivegamma32} in \eqref{traceless32}, the
possibility of gauging away the independent component left over by
the reduction procedure via \eqref{spaceshift32} becomes
evident.

In the generic two-family setting we inferred that, after gauging away the undetermined $\g$-traces of the $\cY_{\,ijk}$ tensors, the remaining quantities are set to zero by the homogeneous field equations. In the present explicit space-time approach this means, in particular, that after using the shift symmetry \eqref{shiftY} to reduce the number of independent $\g$-traces of the $\cY^{\,(i)}$ tensors, one is properly left with square systems of equations that are not singular provided $D \neq 2,4$ (see Section \ref{sec:weyl2}). In contrast with the symmetric case of \cite{fsold2,fms}, however, it is impossible to disentangle the equations that set to zero the $\g$-traces of $\cW$ from those that do the same for the $\cY^{\,(i)}$ tensors. This is due to the existence of non-vanishing symmetrized triple $\g$-traces of $\cE_{\,\psi}$. However, one can first express the $\cY^{\,(i)}$ in terms of the $\g$-traces of $\cW$, using the combinations of $\g$-traces that annihilate $\psi$ in the constrained theory, thus recovering an expression that has the same form as the constrained Lagrangian field equation \eqref{constreq32}. The $\g$-traces of $\cW$ can then be set to zero using the remaining $\g$-traces of the field equation.

Continuing with the various $\g$-traces of eq.~\eqref{eqpsi32}, one is to
consider $E_{\bar{\psi}}^{\,\pe_\m\hpe}{}_{\n}$, $(E_{\bar{\psi}}^{\,\pe_\m \pe_\n} + \, 2\,
E_{\bar{\psi}}^{\,\hpe\,\hpe})_{\,\m}$, $\g^{\,\l\r}\,
E_{\bar{\psi}}^{\,\pe_\m}{}_{\l\,,\,\r\n}$ and $\g^{\,\l\r}\,
E_{\bar{\psi}}^{\,\hpe}{}_{\,\l\m\,,\,\r}$, that can all be obtained combining $\g$-traces of the symmetrized triple $\g$-traces of $E_{\bar{\psi}}$ and read, respectively,
\be \label{foursym32}
\begin{bmatrix}
\r_2 & 3D+8 & 2 & 4 \\[5pt]
2 & 4 & 3D+10 & 3D+8 \\[5pt]
\r_{-1}\r_{2} & - \, \frac{D^2+3D+4}{2} & 2D+3 & - \r_0 \\[5pt]
\r_0 & -\,3\,\r_2 & (D+2)^{\,2} & - \,(D^2+2D-2)
\end{bmatrix} \cdot
\begin{bmatrix}
\g^{\,\l}\, \cY^{\,(1)}_{\,T}{}_{\l\m}  \\[5pt]
\g^{\,\l}\, \cY^{\,(2)}_{\,T}{}_{\l\,,\,\m}  \\[5pt]
\g^{\,\l}\, \cY^{\,(2)}_{\,T}{}_{\m\,,\,\l} \\[5pt]
\g^{\,\l}\, \cY^{\,(3)}_{\,T}{}_{\l\m}
\end{bmatrix} \, = \,
\begin{bmatrix}
0 \\[7pt] 0 \\[7pt] \frac{\r_{-2}}{9} \\[7pt] - \, \frac{\r_{-2}}{9}
\end{bmatrix} \, \o_{\,\m}\, ,
\ee
together with the additional $\{2,2\}$-projected trace
\be \label{window32}
(\,E^{\,\pe_\m \pe_\n}_{\bar{\psi}} - \, E^{\,\hpe\,\hpe}_{\bar{\psi}}\,)_{\,\m} : \quad \frac{\r_{-2}\, \r_1}{6} \ \o_{\,\m} + \, \g^{\,\l} \left\{\, \cY^{\,(1)}_{\,T}{}_{\l\m} - \, \cY^{\,(2)}_{\,T}{}_{\l\,,\,\m} - \cY^{\,(2)}_{\,T}{}_{\m\,,\,\l} + \, \cY^{\,(3)}_{\,T}{}_{\l\m} \,\right\} \, = \, 0 \, .
\ee
Notice that only the doubly $\g$-traceless tensors
$\cY^{\,(i)}_{\,T}$ appear in these equations, on account of the
previous discussion and that, as expected, the number of unknowns now matches the number of equations. Eqs.~\eqref{foursym32} then suffice to obtain
\begin{alignat}{2}
&\g^{\,\l}\, \cY^{\,(1)}_{\,T}{}_{\l\n} & & = \, \frac{\r_{-2}}{6\,(D^2-2)} \ \o_{\,\n}  \, , \nonumber \\
&\g^{\,\l}\, \cY^{\,(2)}_{\,T}{}_{\l\,,\,\n} & & = \, - \, \frac{\r_{-2}}{18\,(D^2-2)} \ \o_{\,\n} \, , \nonumber \\
&\g^{\,\l}\, \cY^{\,(2)}_{\,T}{}_{\m\,,\,\l} & & = \, - \, \frac{\r_{-2}}{18\,(D^2-2)} \ \o_{\,\m} \, , \nonumber \\
&\g^{\,\l}\, \cY^{\,(3)}_{\,T}{}_{\l\m} & & = \, \frac{\r_{-2}}{18\,(D^2-2)} \ \o_{\,\m} \, , \label{solfour32}
\end{alignat}
while eq.~\eqref{window32} will eventually set $\o$ to zero. Considering
the symmetrized triple $\g$-traces of the field equations and
substituting into them the results collected in eq.~\eqref{solfour32}
now gives
\be
\begin{split}
& \begin{bmatrix}
\frac{\r_0\r_2}{2} & 2\,\r_2 & 4 \\[5pt]
2\,\r_2 & \frac{3D^2+18D+40}{4} & 4\,\r_3 \\[5pt]
4 & 4\,\r_3 & \frac{\r_2\,(3D+8)}{2}
\end{bmatrix} \cdot
\begin{bmatrix}
\cY^{\,(1)}_{\,T}{}_{\m_1\m_2} \\[5pt] \cY^{\,(2)}_{\,T}{}_{(\m_1\,,\,\m_2)} \\[5pt] \cY^{\,(3)}_{\,T}{}_{\m_1\m_2}
\end{bmatrix} \\
& = \, \frac{1}{9}
\begin{bmatrix}
-\,4 \\[7pt] -\,\r_0 \\[7pt] 2\,\r_2
\end{bmatrix} u_{\,\m_1\m_2} \, + \, \frac{1}{9}
\begin{bmatrix}
\r_2 \\[7pt] 2\\[7pt]
- \, \r_6
\end{bmatrix} v_{\,(\m_1\,,\,\m_2)} \, - \, \frac{\r_{-2}}{36\,(D^2-2)}
\begin{bmatrix}
\r_0 \\[7pt] -\,\r_0 \\[7pt] \r_0
\end{bmatrix} \g_{\,(\m_1}\, \o_{\,\m_2)} \, ,
\end{split}
\ee
together with the antisymmetric component of the second equation,
that simply states that
\be
\g^{\,\l}\,(\,E_{\bar{\psi}}{}^{\,\pe_\m}{}_{[\m\,,\,\n]\,\l} +\, 2\, E_{\bar{\psi}}{}^{\,\hpe}{}_{\l\,[\m\,,\,\n]}\,) \, : \quad \cY^{\,(2)}_{\,T}{}_{[\m\,,\,\n]} \, = \, 0 \, .
\ee
All together, these equations finally give
\begin{alignat}{2} \label{solthree32}
& \cY^{\,(1)}_{\,T}{}_{\!\n_1\n_2} \!& & = \, - \, \frac{1}{6\,(3D+4)} \left\{ \frac{4}{3D-2} \left[\, 4 \, u_{\,\n_1\n_2} - \, (3D+2) \, v_{\,(\n_1\,,\,\n_2)} \,\right] +  \frac{\r_0}{D^2-2} \, \g_{\,(\n_1}\,\o_{\,\n_2)} \right\} , \nonumber \\
& \cY^{\,(2)}_{\,T}{}_{\!\m\,,\,\n} \!& & = \, - \, \frac{1}{18\,(3D+4)} \left\{ \frac{4}{3D-2} \left[\, 3\,\r_0 \, u_{\,\m\n} - \, 2 \, v_{\,(\m\,,\,\n)} \,\right] -  \frac{\r_0}{D^2-2} \, \g_{\,(\m}\,\o_{\,\n)} \right\} , \\
& \cY^{\,(3)}_{\,T}{}_{\!\m_1\m_2} \!& & = \, \frac{1}{18\,(3D+4)} \left\{ \frac{8}{3D-2} \left[\, (3D+2) \, u_{\,\m_1\m_2} - \, \frac{3}{2}\,\r_2 \, v_{\,(\m_1\,,\,\m_2)} \,\right] - \frac{\r_0}{D^2-2} \, \g_{\,(\m_1}\,\o_{\,\m_2)} \right\} . \nonumber
\end{alignat}

Substituting \eqref{solthree32} in the unconstrained equation of motion \eqref{eqpsi32},
one can recover \eqref{constreq32}. Let us stress again that after eliminating the
compensators $\Psi_i$ this turns into a convenient presentation of the Lagrangian
equations for the constrained theory. After these steps, the reduction to the Labastida
form would proceed as in the constrained theory, and in fact much as in the bosonic case
of \cite{bose_mixed}, barring some inevitable technical
complications that we shall not dwell upon here. Indeed, for the sake of brevity we shall
confine ourselves here with these cursory remarks, since the actual procedure is
straightforward if quite tedious. At any rate, in the next section we shall exhibit the
explicit form of the propagator for a $(3,2)$ field. This expression can be obtained with
similar techniques, and actually embodies in a clearcut fashion all relevant information
concerning the subtleties that can be met when performing the reduction in special (low
numbers of) dimensions where Weyl-like symmetries emerge.

\vskip 24pt


\scsss{Current exchanges}\label{sec:current}


Coupling a rank-$(3,2)$ gauge field
$\psi_{\,\m_1\m_2\m_3\,,\,\n_1\n_2}$ to an external current leads to the inhomogeneous equation
\be \label{eqpsi32J}
\begin{split}
E_{\,\bar{\psi}} \, & : \ \cE_{\,\psi} \, - \, \12 \, \h_{\,(\,\m_1\m_2\,}\g_{\,\m_3\,)} \, \cY^{\,(1)}{}_{\n_1\n_2} \, - \, \12 \left(\, \h_{\,(\,\m_1\m_2|} \,\g_{\,(\,\n_1|} + \, \h_{\,(\n_1|\,(\,\m_1}\g_{\,\m_2|} \,\right) \cY^{\,(2)}{}_{|\,\m_3\,)\,,\,|\,\n_2\,)} \\
& - \, \12 \left(\, \h_{\,\n_1\n_2}\g_{\,(\,\m_1|} + \, \g_{\,(\,\n_1}\h_{\,\n_2\,)\,(\,\m_1\,|} \,\right) \cY^{\,(3)}{}_{|\,\m_2\m_3\,)} \, = \, \cJ_{\,\m_1\m_2\m_3\,,\,\n_1\n_2} \, ,
\end{split}
\ee
that can be manipulated using the approach followed for the
homogeneous equation \eqref{eqpsi32}. In particular, the same steps of the preceding discussion can reduce
eq.~\eqref{eqpsi32J} to the form
\be
E_{\,\bar{\psi}}^{\,constr} \, = \, \cK_{\,\m_1\m_2\m_3\,,\,\n_1\n_2} \, ,
\ee
where $E_{\,\bar{\psi}}^{\,constr}$ was defined in
eq.~\eqref{constreq32} and $\cK_{\,\m_1\m_2\m_3\,,\,\n_1\n_2}$ is an
effective current whose symmetrized triple $\g$-traces vanish. We would like to stress again that this form is not unique, since one can also adsorb different parts of the $\cY^{\,(i)}$ in the definition of the effective current $\cK$, but it is anyway quite convenient, since it naturally extends the symmetric constrained setting, were the external current is subject to the same trace constraints as the gauge field.

One can compute further $\g$-traces of the equation of motion \eqref{eqpsi32J}, and combining these results one can recast it in the form
\be \label{eqdescr}
\cW_{\,\m_1\m_2\m_3\,,\,\n_1\n_2} \, = \, \cJ_{\,\m_1\m_2\m_3\,,\,\n_1\n_2} \, + \, \ldots\, ,
\ee
where the omitted terms are obtained expressing the $\g$-traces of $\cW$ and
$\cY^{\,(i)}$ in terms of $\cJ$. In order to give rise to the correct current exchanges, the combination on the right-hand side of eq.~\eqref{eqdescr} must define a combination that is \emph{$\g$-traceless in $D-2$ dimensions}, and in the following we shall see that this
is actually the case for the $(3,2)$-example. From the technical viewpoint, one can reach
the form \eqref{eqdescr} dealing separately with the various irreducible Young
projections of the $\g$-traces of $E_{\,\bar{\psi}}$, in order to extract square systems of equations
from the results. As we have already pointed out in Section \ref{sec:reduction2f}, this
decomposition is the space-time counterpart of the independent treatment of terms
corresponding to different values of $\ell$, that we used to identify the degenerate
cases where the gauge symmetry is enlarged to accommodate Weyl-like shifts.

As we repeatedly stressed, the $\g$-traces of the equation of motion contain the $S^{\,i}{}_{\,j}$ operators. These can displace indices from one set to another, and as a result the propagators of mixed-symmetry fields involve \emph{all} structures obtained displacing indices within the $\g$-traces of the current $\cJ$. In conclusion, the field equation \eqref{eqpsi32J} can be cast in the form
\begin{align}
&\cW_{\,\m_1\m_2\m_3 ,\,\n_1\n_2} \, = \, \cJ_{\,\m_1\m_2\m_3 ,\,\n_1\n_2} + \, \g_{\,(\,\m_1|} K_{11} \cdot \underline{\cJ}{}_{\,1}{}^a{}_{|\,\m_2\m_3)\,,\,\n_1\n_2} + \, \g_{\,(\,\n_1|} K_{12} \cdot \underline{\cJ}{}_{\,1}{}^b{}_{\m_1\m_2\m_3,\,|\,\n_2)}  \nn \\
& + \, \h_{(\,\m_1\m_2|} K_{21} \cdot  \underline{\cJ}{}_{\,2}{}^a{}_{\n_1\n_2,\,|\,\m_3)} + \, \h_{(\,\n_1|(\,\m_1} K_{22} \cdot  \underline{\cJ}{}_{\,2}{}^a{}_{\m_2\m_3)\,,\,|\,\n_2)} + \, \h_{\,\n_1\n_2} K_{23} \cdot \underline{\cJ}{}_{\,2}{}^b{}_{\m_1\m_2\m_3} \nn \\
& - \, \g_{(\,\n_1|(\,\m_1} K_{24} \cdot  \underline{\cJ}{}_{\,2}{}^a{}_{\m_2\m_3)\,,\,|\,\n_2)} + \, \h_{(\,\m_1\m_2}\g_{\,\m_3)} K_{31} \cdot  \underline{\cJ}{}_{\,3\,(\,\n_1,\,\n_2)} + \, \h_{(\,\m_1\m_2|}\g_{\,(\,\n_1|} K_{32} \cdot \underline{\cJ}{}_{\,3\,|\,\m_3)\,,\,|\,\n_2)} \nn \\
& + \, \h_{(\,\n_1|(\,\m_1\!}\g_{\,\m_2|} K_{33} \cdot  \underline{\cJ}{}_{\,3\,|\,\m_3)\,,\,|\,\n_2)} + \, \h_{(\,\m_1|(\,\n_1\!}\g_{\,\n_2)} K_{34} \cdot \underline{\cJ}{}_{\,3\,|\,\m_2,\,\m_3)} + \, \h_{\,\n_1\n_2}\g_{\,(\,\m_1|} K_{35} \cdot \underline{\cJ}{}_{\,3\,|\,\m_2\,,\,\m_3)} \nn \\
& + \, \h_{(\,\m_1\m_2}\h_{\,\m_3)(\,\n_1|} K_{41} \cdot  \underline{\cJ}{}_{\,4\,|\,\n_2)} + \, \h_{\,\n_1\n_2}\h_{(\,\m_1\m_2|} K_{42}\cdot  \underline{\cJ}{}_{\,4\,|\,\m_3)} + \, \h_{(\,\m_1|(\,\n_1}\h_{\,\n_2)|\,\m_2|} K_{43} \cdot \underline{\cJ}{}_{\,4\,|\,\m_3)} \nn \\
& + \h_{(\,\m_1\m_2}\g_{\,\m_3)(\,\n_1|} K_{44} \cdot  \underline{\cJ}{}_{\,4\,|\,\n_2)} - \, \h_{(\,\m_1|(\,\n_1}\g_{\,\n_2)|\,\m_2|} K_{45} \cdot  \underline{\cJ}{}_{\,4\,|\,\m_3)} + \, \h_{(\,\m_1\m_2}\h_{\,\m_3)(\,\n_1}\g_{\,\n_2)} K_{51} \cdot \underline{\cJ}{}_{\,5} \nn \\
& + \h_{\,\n_1\n_2}\h_{(\,\m_1\m_2}\g_{\,\m_3)} K_{52} \cdot  \underline{\cJ}{}_{\,5} \, + \, \h_{(\,\m_1|(\,\n_1}\h_{\,\n_2)|\,\m_2}\g_{\,\m_3)} K_{53} \cdot  \underline{\cJ}{}_{\,5} \, .\label{propagator}
\end{align}
The resulting propagator is rather complicated, and contains as many as 106 distinct contributions.
Two types of vectors, denoted by $\underline{\cJ}{}_{\,i}$ and $K_{\,ij}$, actually build this expression, and their scalar products are denoted with a standard symbol, ``$\,\cdot\,$''. These vectors, however, have three, four, five or seven distinct components, depending on the case. The vectors of the first type collect essentially all possible constructs built from the current $\cJ$,
{\allowdisplaybreaks
\begin{align}
& \underline{\cJ}{}_{\,1}{}^a{}_{\m_1\m_2,\,\n_1\n_2} \! = \g^{\,\l} \big(\, \cJ_{\l\m_1\m_2,\,\n_1\n_2} \comma \cJ_{\l\,(\,\m_1|\,(\,\n_1,\,\n_2)\,|\,\m_2)} \comma \cJ_{\l\n_1\n_2,\,\m_1\m_2} \comma \cJ_{\m_1\m_2\,(\,\n_1,\,\n_2\,)\,\l} \comma \cJ_{\n_1\n_2\,(\,\m_1,\,\m_2\,)\,\l} \,\big) \, , \nn \\[2pt]
& \underline{\cJ}{}_{\,1}{}^b{}_{\m_1\m_2\m_3,\,\n} \! = \g^{\,\l} \big(\, \cJ_{\l\,(\,\m_1\m_2,\,\m_3)\,\n} \comma \cJ_{\l\,\n\,(\,\m_1,\,\m_2\m_3\,)} \comma \cJ_{\m_1\m_2\m_3,\,\n \l} \comma \cJ_{\n\,(\,\m_1\m_2,\,\m_3)\,\l} \,\big) \, , \nn \\[2pt]
& \underline{\cJ}{}_{\,2}{}^a{}_{\m_1\m_2,\,\n} \! = \big(\, \cJ^{\,\pe_\m}{}_{\!\n\,,\,\m_1\m_2} \comma \cJ^{\,\pe_\m}{}_{\!(\,\m_1,\,\m_2)\,\n} \comma \cJ^{\,\hpe}{}_{\!\m_1\m_2,\,\n} \comma \cJ^{\,\hpe}{}_{\!\n\,(\,\m_1,\,\m_2)} \comma \cJ^{\,\pe_\n}{}_{\!\m_1\m_2\n} \comma \g^{\,\l\r} \cJ_{\l\m_1\m_2,\,\n\r} \comma \nn \\
& \phantom{\underline{\cJ}{}_{\,2}{}^a{}_{\!\m_1\m_2,\,\n} \! = \big(\,} \g^{\,\l\r} \cJ_{\l\,\n\,(\,\m_1,\,\m_2\,)\,\r} \,\big) \, , \nn \\[2pt]
& \underline{\cJ}{}_{\,2}{}^b{}_{\m_1\m_2\m_3} \! = \big(\, \cJ^{\,\pe_\m}{}_{\!(\,\m_1,\,\m_2\m_3)} \comma \cJ^{\,\hpe}{}_{\!(\,\m_1\m_2,\,\m_3)} \comma \cJ^{\,\pe_\n}{}_{\!\m_1\m_2\m_3} \comma \g^{\,\l\r} \cJ_{\l\,(\,\m_1\m_2,\,\m_3)\,\r} \,\big) \, , \nn \\
& \underline{\cJ}{}_{\,3}{}_{\,\m\,,\,\n} \! = \g^{\,\l} \big(\, \cJ^{\,\pe_\m}{}_{\!\l\,,\,\m\n} \comma \12\, \cJ^{\,\pe_\m}{}_{\!(\,\m\,,\,\n\,)\,\l} \comma \12\, \cJ^{\,\pe_\m}{}_{\![\,\m\,,\,\n\,]\,\l} \comma \12\, \cJ^{\,\hpe}{}_{\!\l\,(\,\m\,,\,\n\,)} \comma \12\, \cJ^{\,\hpe}{}_{\!\l\,[\,\m\,,\,\n\,]} \comma \cJ^{\,\pe_\n}{}_{\!\l\m\n} \comma \cJ^{\,\hpe}{}_{\!\m\n\,,\,\l} \,\big) \, , \nn \\
& \underline{\cJ}{}_{\,4}{}_{\,\m} \! = \big(\, \cJ^{\,\pe_\m\hpe}{}_{\!\m} \comma \cJ^{\,\pe_\m\pe_\n}{}_{\!\m} \comma \cJ^{\,\hpe\,\hpe}{}_{\m} \comma \g^{\,\l\r} \cJ^{\,\pe_\m}{}_{\!\m} \comma \g^{\,\l\r} \cJ^{\,\hpe}{}_{\m} \,\big) \, , \nn \\[2pt]
& \underline{\cJ}{}_{\,5} \! = \g \cdot \big(\, \cJ^{\,\pe_\m\hpe} \comma \cJ^{\,\pe_\m\pe_\n} \comma \cJ^{\,\hpe\,\hpe} \,\big) \, ,
\end{align}}
while their partners $K_{\,ij}$ are vectors of coefficients that are defined below, and whose list closes the present section.

Let us begin by displaying the terms that in eq.~\eqref{propagator} are accompanied by a single $\g$-matrix:
\begin{align}
& K_{11} \, = \, \frac{\left(\, -D^4\!-4D^3\!+20D^2\!+48D\!-64 \comma -\,4\, \r_{-2}\,\r_4 \comma -\,64 \comma 2\, \r_{2}\, (D^2\!+2D\!-16)\comma 16\, \r_2 \,\right)}{\r_{-4}\,\r_{-2}\,\r_2\,\r_4\,\r_6} \, , \nn \\
& K_{12} \, = \, \frac{\left(\, 2\,(D^2\! + 2D\! - 16) \comma 16 \comma - \, \r_2\, ( D^2\! + 2D\! - 20 ) \comma - \, 4 \, \r_2 \,\right)}{\r_{-4}\,\r_{-2}\,\r_4\,\r_6} \, .
\end{align}
Notice that $K_{11}$ and $K_{12}$ have poles in $D=2,4$, precisely where the general considerations of Section \ref{sec:reduction2f} apply for a reducible $(3,2)$ field. The same poles are also present in most of the other coefficients, that we collect in a few groups. The first group thus comprises the terms that in eq.~\eqref{propagator} are accompanied by two $\g$-matrices
\begin{align}
& K_{21} \, = \, \frac{1}{\r_{-4}\,\r_{-2}\,\r_0\,\r_2\,\r_4\,\r_6} \left(\, -\, \r_{-2}\, ( D^4\!+6D^3\!-12D^2\!-96D\!-96) \comma -\,2\,D\,(D^2\!-2D\!-16) \comma \right. \nn \\
& \left. 16\,(D^2\!-3D\!-6) \comma 2\,( D^4\!+D^3\!-22D^2\!-8D\!+48 ) \comma -\,4\, \r_2 \, ( D^2\!-12 ) \comma - \, 16 \, \r_6 \comma -\, 2 \, \r_6 \, ( D^2\!-8) \,\right) , \nn \\
& K_{22} \, = \, \frac{1}{\r_{-4}\,\r_{-2}\,\r_0\,\r_2\,\r_4\,\r_6} \left(\, 8\, ( D^2\!-3D\!-6 ) \comma D^4\!+D^3\!-22D^2\,-8D\,+48 \comma -D^5\!-3D^4\!+23D^3 \right. \nn \\
& \left. +\,42D^2\!-68D\!-72 \comma -\,4 ( D^3\!-D^2\!-11D\!-6 ) \comma \r_2( D^3\!+D^2\!-16D\!-12 ) \comma \r_6 ( D^2-12 ) \comma 4\, \r_6   \,\right) , \nn  \\
& K_{23} \, = \, \frac{\left(\, -\,4\, (D^2\!-12) \comma 2\, ( D^3\!+D^2\!-16D\!-12 ) \comma -D^4\!-4D^3\!+22D^2\!+64D\!-72 \comma 2\,\r_{-2}\,\r_6 \,\right)}{\r_{-4}\,\r_{-2}\,\r_0\,\r_4\,\r_6} \, , \nn \\
& K_{24} \, = \, \frac{\left(\, 8 \comma D^2\!-8 \comma 12-D^2 \comma -\,4 \comma -\, \r_{-2}\,\r_2 \comma -\,\r_1\,( D^2\!-12 ) \comma -\,4\,\r_1 \,\right)}{\r_{-4}\,\r_{-2}\,\r_0\,\r_2\,\r_4} \, ,
\end{align}
and in a similar fashion the second group comprises the terms that are accompanied by three $\g$-matrices
\begin{align}
& K_{31} \, = \, \frac{1}{\r_{-4}\,\r_{-2}\,\r_0\,\r_2\,\r_4\,\r_6} \left(\, D\,( D^3\!+4D^2\!-16D\!-56 ) \comma -\,4\,\r_2\, (D^2\!-12) \comma 0 \comma \right. \nn \\
& \left. -\,4\,D\,( D^2\!-2D\!-16 ) \comma 0 \comma 4\, \r_{-2}\,\r_0 \comma 8\, \r_{-6}\,\r_2 \,\right) , \nn \\
& K_{32} \, = \, \frac{1}{\r_{-4}\,\r_{-2}\,\r_0\,\r_2\,\r_4\,\r_6} \left( -\,2\, \r_2\, (D^2\!-12) \comma \r_2^2\, (D^2\!-12) \comma \r_{-4}\,\r_0\,\r_2\,\r_6 \comma -\, 4 \,\r_{-2}\,\r_0\,\r_2 \comma \right. \nn \\
& \left. -\,4\,\r_{-4}\,\r_0\,\r_6 \comma 4\, \r_{-6}\,\r_2 \comma -\, 2 \, \r_{-6}\,\r_2^2 \,\right) , \nn \\
& K_{33} \, = \, \frac{1}{\r_{-4}\,\r_{-2}\,\r_2\,\r_4\,\r_6} \left(\, -D^2\!+2D\!+16 \comma -\,2\, \r_{-2}\,\r_2 \comma -\, 2 \, \r_{-4}\,\r_6 \comma D^3\!+2D^2\!-8D\!-32 \comma \right. \nn \\
& \left. \r_{-4}\,\r_0\,\r_6 \comma -D^2\!+2D\!+16 \comma -\,2\,\r_{-2}\,\r_2 \,\right) , \nn \\
& K_{34} \, = \, \frac{\left(\, 4\, \r_{-6} \comma -\,2\,\r_{-6}\,\r_2  \comma 0 \comma -\, 4 \, \r_{-2} \,\r_0 \comma 0 \comma -\, 2\, ( D^2\!-12 ) \comma \r_2\, ( D^2\!-12 ) \,\right)}{\r_{-4}\,\r_{-2}\,\r_0\,\r_4\,\r_6} \, , \nn \\
& K_{35} \, = \, \frac{1}{\r_{-4}\,\r_{-2}\,\r_0\,\r_2\,\r_4\,\r_6} \left(\, 4\,\r_{-2}\,\r_0 \comma 8\,\r_{-6}\,\r_2 \comma 0 \comma -\,4\,D\,( D^2\!-2D\!-16 ) \comma 0 \comma \right. \nn \\
& \left. D\,( D^3\!+4D^2\!-16D\!-56 ) \comma -\,4\,\r_2\, ( D^2\!-12 ) \,\right) .
\end{align}
Finally, the last two groups comprise the terms that are accompanied by four $\g$-matrices,
\begin{align}
& K_{41} \, = \, \frac{\left(\, D^3\!+D^2\!-17D\!-18 \comma -D^2\!+3D\!+6 \comma -\,2\,( D^2\!-3D\!-6 ) \comma \r_6 \comma 2\, \r_6 \,\right)}{\r_{-4}\,\r_0\,\r_2\,\r_4\,\r_6} \, , \nn \\
& K_{42} \, = \, \frac{1}{\r_{-4}\,\r_{-2}\,\r_{-1}\,\r_0\,\r_2\,\r_4\,\r_6} \left(\, -\,2\, \r_{-2}\,\r_{-1}\, (D^2\!-3D\!-6) \comma D^5\!+3D^4\!-16D^3\!-62D^2\! \right. \nn \\
& \left. +\,24D\!+120 \comma -\,2\, ( 3D^4\!-2D^3\!-46D^2\!+8D\!+72 ) \comma -\,2\,\r_6\, ( D^2\!-D\!-10 ) \comma 4\, \r_{-6}\,\r_6 \right) , \nn \\
& K_{43} \, = \, \frac{1}{\r_{-4}\,\r_{-2}\,\r_{-1}\,\r_0\,\r_2\,\r_4\,\r_6} \left(\, -\,2\,\r_{-2}\,\r_{-1}\, ( D^2\!-3D\!-6 ) \comma -\,3D^4\!+2D^3\!+46D^2\!-8D\!-72 \comma \right. \nn \\
& \left. D^5\!-14D^3\!-16D^2\!+16D\!+48 \comma 2\,\r_{-6}\,\r_6 \comma -\,2\,\r_6\, ( D^2\!-2D\!-4 ) \right) , \nn \\
& K_{44} \, = \, \frac{\left(\, -\,\r_{-2}\,\r_{-1} \comma D^2\!-D\!-10 \comma -\,2\,\r_{-6} \comma \r_1\,( D^2\!-D\!-10 ) \comma -\, 2\, \r_{-6}\,\r_1 \,\right)}{\r_{-4}\,\r_{-2}\,\r_{-1}\,\r_0\,\r_2\,\r_4} \, , \nn \\
& K_{45} \, = \, \frac{\left(\, -\, \r_{-2}\,\r_{-1} \comma -\, \r_{-6} \comma D^2\!-2D\!-4 \comma -\, \r_{-6}\,\r_{1} \comma \r_1\,( D^2\!-2D\!-4 )\,\right)}{\r_{-4}\,\r_{-2}\,\r_{-1}\,\r_0\,\r_2\,\r_4} \, ,
\end{align}
or by five $\g$-matrices,
\begin{align}
& K_{51} \, = \, \frac{\left(\, -\, \r_2 \comma 2 \comma 4 \,\right)}{\r_{-4}\,\r_2\,\r_4\,\r_6} \, , \nn \\
& K_{52} \, = \, \frac{\left(\, 4\, \r_{-2}\,\r_{-1} \comma -D^3\!-3D^2\!+10D\!+64 \comma 2\, ( 3D^2\!-6D\!-32 ) \,\right)}{\r_{-4}\,\r_{-2}\,\r_{-1}\,\r_2\,\r_4\,\r_6} \, , \nn \\
& K_{53} \, = \, \frac{\left(\, 4\,\r_{-2}\,\r_{-1} \comma 3D^2\!-6D\!-32 \comma -D^3\!+4D\!+32 \,\right)}{\r_{-4}\,\r_{-2}\,\r_{-1}\,\r_2\,\r_4\,\r_6} \, .
\end{align}
The reader could verify, as we also did, that the right-hand side of eq.~\eqref{propagator} defines an expression whose $\g$-traces in each of its two index sets vanish once the $D$-dependent factors that arise in this computation (but, of course, not those appearing in the original $K_{\,ij}$ vector coefficients) are shifted letting $D \to D-2$.

\vskip 24pt


\scs{General fermionic fields}\label{sec:generalf}


In this section we can finally derive Lagrangians and field
equations for general unconstrained fermionic fields with
arbitrary numbers of index families. We open our discussion by showing
in detail how to complete the work of Labastida
\cite{labferm} with general constrained Lagrangians yielding
the fermionic field equations that he introduced in the eighties. To
this end, we follow rather closely the approach developed in
\cite{bose_mixed} for Bose fields and based on the Bianchi identities, that as
we shall see is particularly convenient for the extension to the unconstrained case. However, we
also display an
alternative derivation, based on the request that the Rarita-Schwinger-like kinetic
operators be self-adjoint, that proceeds along the lines of the
original bosonic work of Labastida \cite{labastida}, before closing
the section with a discussion of the unconstrained Lagrangians,
their field equations and their reduction to the form of
\cite{labferm}.

\vskip 24pt


\scss{The constrained Lagrangians}\label{sec:lagrangianf}


The $\g$-traces of the Bianchi identities
are a key tool to derive the Lagrangians for mixed-symmetry
fermions with arbitrary numbers of index families. In the
constrained setting they take the form
\be \label{bianchi_cf}
\pr_{\,i}\,\cS \, - \, \12 \dsl \ssl_{i} \, - \, \12 \ \pr^{\,j}\,T_{ij}\,\cS \, - \, \frac{1}{6} \ \pr^{\,j}\,\g_{\,ij}\,\cS \, = \, 0 \, ,
\ee
where we have enforced the Labastida constraints
\be \label{constrpsi}
T_{(\,ij}\! \psisl_{k\,)} \, = \, 0
\ee
on the right-hand side of eq.~\eqref{bianchi_cf}. As we already saw
in Section \ref{sec:lagrangian2f}, the $\g$-trace of
eq.~\eqref{bianchi_cf},
\be
\pr_{\,[\,i} \ssl_{j\,]} \, - \, \frac{2}{3} \dsl \ \g_{\,ij} \, \cS \, + \, \frac{1}{6} \
\pr^{\,k} \, \g_{\,ijk} \, \cS \, - \, \frac{1}{6} \ \pr^{\,k} \left(\, 3 \, T_{ik} \, \g_j \, +
\, T_{j\,[\,i} \, \g_{\,k\,]} \,\right) \cS \, = \, 0\, ,
\ee
admits two independent projections. In analogy with the bosonic
case, the symmetric one yields the gradient of
\be \label{constrs}
T_{(\,ij}\! \ssl_{k\,)} \, = \, 0 \, ,
\ee
an identity implied by the constraint \eqref{constrpsi}, as one can
verify by a direct computation. The antisymmetric projection,
however, is more interesting, and reads
\be
\pr_{\,[\,i} \ssl_{j\,]} \, - \, \frac{2}{3} \dsl \, \g_{\,ij} \, \cS \, + \, \frac{1}{6} \ \pr^{\,k} \, \g_{\,ijk} \, \cS \, - \, \frac{1}{3} \ \pr^{\,k} \, T_{k\,[\,i} \, \g_{\,j\,]} \, \cS \, = \, 0  \, ,
\ee
so that it relates in a non-trivial fashion different structures.

This suggests that the Lagrangians should rest on a special
sub-chain of suitably projected consequences of the Bianchi
identities, in analogy with the behavior of Bose fields described in
\cite{bose_mixed}. In order to sort it out, it is convenient to
first identify the non trivial ($\g$-)traces of the kinetic tensor $\cS$,
in order to build the most general Ansatz. In
Appendix \ref{app:idsf} we shall prove that, for any given number of
traces and antisymmetric $\g$-traces of the kinetic tensor $\cS$,
\emph{all Young projections in family indices with more than two
columns vanish in the constrained theory} on account of
eq.~\eqref{constrs}. This condition identifies the wide class of
terms that can enter fermionic Lagrangians which, as a result, are
far more involved than their bosonic counterparts. For $N$ families
the Lagrangians are in fact bound to rest on ${\cal O}(N^2)$
distinct Young projections, with a first column of length $l_1\leq
N$ and a second column of length $l_2 \leq l_1$, that in general is
actually shorter, since antisymmetric $\g$-traces $\g_{\,i_1
\ldots\, i_n}$ of $\cS$ can accompany the ordinary $T_{ij}$ traces
that already enter the bosonic Lagrangians. Interestingly,
\emph{any} given combination of these operators admits \emph{a
single two-column projection}, and it is then convenient to adopt
the short-hand notation
\bea \label{shortnots}
&&\cS^{\,\prime}{}_{ij} \, = \, T_{ij}\,\cS \, , \nonumber \\ && \ssl^{\,\prime}{}_{ij\,;\,k} \, = \, \frac{1}{3} \, \left(\, 2\,T_{ij}\,\g_{\,k} \, - \, T_{k\,(\,i}\,\g_{\,j\,)} \,\right) \cS \, , \nonumber \\
&& \hspace{0.9cm} \vdots \nonumber \\[5pt]
&& (\,\g^{\,[\,q\,]}\,\cS^{\,[\,p\,]}\,)_{\ i_1 j_1,\,\ldots\,,\,i_p
j_p \, ;\, k_1 \ldots\, k_q} \, = \ Y_{\{\,2^p,\,1^q\}} \ T_{i_1
j_1} \ldots \, T_{i_p j_p} \, \g_{\,k_1 \ldots\, k_q} \, \cS \,
.
\eea
The reader should appreciate that here colons separate again groups
of symmetric indices, while a semicolon precedes the left-over group
of antisymmetric ones.

In conclusion, the Lagrangian for generic $N$-family fermionic gauge
fields $\psi$ is of the form
\be
\cL \, = \, \frac{1}{2} \ \bra\, \bar{\psi} \,\comma\! \sum_{p\,,\,q\,=\,0}^{N} k_{\,p\,,\,q} \ \h^{\,p} \, \g^{\,q}\, (\,\g^{\,[\,q\,]}\,\cS^{\,[\,p\,]}\,) \,\ket \, + \, \textrm{h.c.}\, ,
\label{lagferconstr}
\ee
where
\be
\h^{\,p} \, \g^{\,q}\, (\,\g^{\,[\,q\,]}\,\cS^{\,[\,p\,]}\,) \, \equiv \, \h^{i_1 j_1} \ldots \, \h^{i_p j_p} \, \g^{\,k_1 \ldots\, k_q} \ (\,\g^{\,[\,q\,]}\,\cS^{\,[\,p\,]}\,)_{\, i_1 j_1\,,\,\ldots\,,\,i_p j_p \, ;\, k_1 \ldots\, k_q} \, .
\ee
As we shall see shortly, the $k_{\,p\,,\,q}$ coefficients
in eq.~\eqref{lagferconstr} are uniquely determined by the condition
that $\cL$ be gauge invariant, up to the convenient choice $k_{\,0,0} \, = \, 1$.

In order to proceed, we can begin by noticing that the identities
collected in Appendix \ref{app:fermi} make it possible to recast the
gauge variation of eq.~\eqref{lagferconstr}, up to total
derivatives, in the form
\be \label{gaugefgen}
\begin{split}
\d \, \cL \, = \, - \sum_{p\,,\,q\,=\,0}^{N} \, & \frac{1}{2^{\,p+1}} \ \bra \, T_{i_1j_1} \ldots T_{i_pj_p} \, \bar{\e}_{\,l} \, \g_{\,k_1 \ldots\, k_q} \,\comma\, k_{\,p\,,\,q} \ \pr_{\,l} \, (\,\g^{\,[\,q\,]}\,\cS^{\,[\,p\,]}\,)_{\, i_1 j_1,\,\ldots\,,\,i_p j_p \, ;\, k_1 \ldots\, k_q} \\[-5pt]
& + \,(\,q+1\,)\,k_{\,p\,,\,q+1} \dsl \, (\,\g^{\,[\,q+1\,]}\,\cS^{\,[\,p\,]}\,)_{\, i_1 j_1,\,\ldots\,,\,i_p j_p \, ;\, k_1 \ldots\, k_q \, l}  \\[5pt]
& + \,(\,p+1\,)\,k_{\,p+1,\,q} \ \pr^{\,m}\, (\,\g^{\,[\,q\,]}\,\cS^{\,[\,p+1\,]}\,)_{\, i_1 j_1,\,\ldots\,,\,i_p j_p\,,\,lm \, ;\, k_1 \ldots\, k_q}  \\[5pt]
& + \,(\,q+1\,)\,(\,q+2\,)\,k_{\,p\,,\,q+2} \ \pr^{\,m}\, (\,\g^{\,[\,q+2\,]}\,\cS^{\,[\,p\,]}\,)_{\, i_1 j_1,\,\ldots\,,\,i_p j_p \, ;\, k_1 \ldots\, k_q\,lm} \, \ket \, + \, \textrm{h.c.} \, .
\end{split}
\ee
Notice that all $\g^{\,[\,q\,]}\,\cS^{\,[\,p\,]}$ terms that are
right entries of the scalar product are two-column projected by
assumption. This statement applies to all free indices carried by
the last three terms, while the first is slightly
more complicated, since due to the $l$ index carried by the divergence
it also allows projections with three boxes
in the first row. However, the left entry of the scalar product
makes this type of projections proportional to the constraints on the
gauge parameters
\be \label{constrgaugef}
\g_{\,(\,i}\,\e_{\,j\,)} \, = \, 0\, ,
\ee
as expected, on account of the identity
\be \label{idconstrf}
\begin{split}
& \bra Y_{\{3,2^{p-1},1^q\}} \, T_{i_1j_1} \ldots T_{i_pj_p} \, \bar{\e}_{\,l} \, \g_{\,k_1 \ldots\, k_q} \comma \pr_{\,l} \, (\,\g^{\,[\,q\,]}\,\cS^{\,[\,p\,]}\,)_{\, i_1 j_1,\,\ldots\,,\,i_p j_p \, ;\, k_1 \ldots\, k_q} \ket \\[5pt]
= \ & \bra \, (\,p+1\,) \ T_{i_2j_2} \ldots T_{i_pj_p}\, T_{(\,i_1j_1} \, \bar{\e}_{\,l\,)} \, \g_{\,k_1 \ldots\, k_q} \, - \, T_{i_2j_2} \ldots T_{i_pj_p} \, T_{\,[\,k_1\,|\,(\,l} \, \bar{\e}_{\,i_1} \, \g_{\,j_1\,)}\, \g_{\,|\,k_2 \ldots\, k_q\,]} \comma \\
& \frac{p}{3\, (\,p+1\,)(\,p+q+2\,)} \ \pr_{\,(\,l\,|} \,
(\,\g^{\,[\,q\,]}\,\cS^{\,[\,p\,]}\,)_{\,|\, i_1
j_1\,),\,\ldots\,,\,i_p j_p \, ;\, k_1 \ldots\, k_q} \ket\, ,
\end{split}
\ee
that we shall prove in Appendix \ref{app:idsf}. In conclusion, up to the
Labastida constraints one
can effectively restrict the attention to two-column projections of
the gauge variation \eqref{gaugefgen}.

As we saw in the two-family case, fermionic fields bring about
a novel type of complication.
Indeed, the $\g$-traces of the Bianchi identities do not suffice to
set to zero the individual terms of eq.~\eqref{gaugefgen}, but
these, on the other hand, are not fully independent and can be
partly combined using the constraints \eqref{constrgaugef} on the
gauge parameters. Even in the general case, the possible Young projections
for the family indices are the guiding principle when trying to combine
different terms in eq.~\eqref{gaugefgen}. And indeed all terms of the type
\be \label{tpgammaq}
T_{i_1j_1} \ldots\, T_{i_pj_p} \, \bar{\e}_{\,l} \, \g_{\,k_1 \ldots\, k_q} \ee
that are left entries of the scalar products admit two distinct
Young projections with two columns, since
\be
\{\,2^{\,p},\,1^q\,\} \, \otimes \, \{1\} \, = \,  \{\,2^{\,p},\,1^{q+1}\} \, \oplus \, \{\,2^{\,p+1},\,1^{q-1}\} \, \oplus \, \{3,\,2^{\,p-1},\,1^q\} \, .
\ee
Leaving aside the last term, that is clearly related to the constraints, the first two projections can thus arise from
\emph{pairs} of neighboring contributions of the form
\eqref{tpgammaq}, with values of $p$ differing by one unit and with values of $q$
differing by two units.

As in the two-family case, the key step is to relate these two types
of terms, that are only apparently different, making use of the
constraints \eqref{constrgaugef} and of proper relabelings. To this
end, it proves particularly convenient to manipulate the
combinations $\bar{\e}_{\,l}\, \g_{\,k_1 \ldots\, k_q}$ according to
\be\label{q-terms}
\bar{\e}_{\,l} \, \g_{\,k_1 \ldots\, k_q} \, = \, \frac{1}{q+1} \ \bar{\e}_{\,[\,l} \, \g_{\,k_1 \ldots\, k_q\,]} \, + \, \frac{1}{q+1}\, \sum_{n\,=\,1}^{q} \, (-1)^{(n+1)} \ \bar{\e}_{\,(\,l} \, \g_{\,k_n\,)\,k_1 \ldots\, k_{r\neq n} \ldots\, k_q} \, .
\ee
The terms entering the sum can then be rewritten more conveniently as
\be \label{littlewood}
\bar{\e}_{\,(\,l} \, \g_{\,k_1\,) \, k_2 \ldots\, k_q} \, = \, \bar{\e}_{\,(\,l} \, \g_{\,k_1\,)} \, \g_{\, k_2 \ldots\, k_q} \, - \, T_{[\,k_2\,|\,(\,l} \, \bar{\e}_{\,k_1\,)} \, \g_{\,|\,k_3 \ldots\, k_q\,]} \, ,
\ee
where the first contribution on the right-hand side clearly vanishes
in the constrained setting, on account of eq.~\eqref{constrgaugef}.
Enforcing the constraints thus leads to
\be
\sum_{n\,=\,1}^{q} \, (-1)^{(n+1)} \ \bar{\e}_{\,(\,l} \, \g_{\,k_n\,)\,k_1 \ldots\, k_{r\neq n} \ldots\, k_q} \, \to \ T_{l\,[\,k_1} \, \bar{\e}_{\,k_2} \, \g_{\,k_3 \ldots\, k_q\,]} \, ,
\ee
since the sum with oscillating signs induces a complete antisymmetry in the $k_r$ indices, that as such forbids the simultaneous presence of two of them on the trace $T$. As a result, in the constrained theory eq.~\eqref{q-terms} can be finally cast in the form
\be \label{relstruct}
T_{i_1j_1} \ldots T_{i_pj_p} \, \bar{\e}_{\,l} \, \g_{\,k_1 \ldots\, k_q} \, = \, \frac{1}{q+1} \ T_{i_1j_1} \ldots\, T_{i_pj_p} \left(\ \bar{\e}_{\,[\,l} \, \g_{\,k_1 \ldots\, k_q\,]} \, + \, T_{\,l\,[\,k_1} \bar{\e}_{\,k_2} \, \g_{\,k_{3} \ldots\, k_q\,]} \
\right) \, ,
\ee
where the first term is annihilated by the $\{2^{p+1},1^{q-1}\}$
projection while the second is similarly annihilated by the
$\{2^{p},1^{q+1}\}$ projection, and in conclusion one is led to
\be
Y_{\{\,2^p,\,1^{q+1}\}}\, T_{i_1j_1} \ldots T_{i_pj_p} \, \bar{\e}_{\,l} \, \g_{\,k_1 \ldots\, k_q} \, = \, \frac{1}{q+1} \ Y_{\{\,2^p,\,1^{q+1}\}} \, T_{i_1j_1} \ldots T_{i_pj_p} \, \bar{\e}_{\,[\,l} \, \g_{\,k_1 \ldots\, k_q\,]}
\ee
and
\be \label{projfgauge}
Y_{\{\,2^{p+1},\,1^{q-1}\}}\, T_{i_1j_1}
\ldots T_{i_pj_p} \, \bar{\e}_{\,l} \, \g_{\,k_1 \ldots\, k_q} \, =
\, \frac{1}{q+1} \ Y_{\{\,2^{p+1},\,1^{q-1}\}} \, T_{i_1j_1} \ldots
T_{i_pj_p} \, T_{\,l\,[\,k_1} \bar{\e}_{\,k_2} \, \g_{\,k_{3}
\ldots\, k_q\,]}\, .
\ee

While the first of these relations is rather trivial, the second is quite
interesting, since it connects a relatively complicated projection
of $p$ traces and a single $q$-fold antisymmetric $\g$-trace to a
simpler one involving ${p+1}$ traces and a $(q-2)$-fold
antisymmetric $\g$-trace with the maximum possible number of
antisymmetrizations, generalizing in this way the relation \eqref{gammatot} that we used to build two-family Lagrangians. In this fashion, the current choice extends the $\k=1$ construction of Section \ref{sec:lagrangian2f}. Let us stress that one can also proceed in the opposite direction, decreasing the value of $p$. The relevant identity,
that can be proved using the techniques of Appendix \ref{app:idsf} and generalizes eq.~\eqref{ttogamma} to the $N$-family setting, is then
\be
Y_{\{\,2^p,1^{q+1}\}} \, T_{i_1j_1} \ldots T_{i_pj_p} \,
\bar{\e}_{\,l}
\, \g_{\,k_1 \ldots\, k_q}  = \,
\frac{(-1)^{\,q+1}}{p\,(\,q+1\,)} \, Y_{\{\,2^p,1^{q+1}\}} \sum_{n\,=\,1}^{p} \,
\left(\,\prod_{r\,\ne\,n}^p \, T_{i_rj_r}\, \right) \, \bar{\e}_{\,(\,i_n} \, \g_{\,j_n\,)\,k_1 \,
\ldots\, k_q \, l} \, ,
\ee
but it is clearly simpler to work with terms whose projections
involve the maximum number of antisymmetrizations compatible with
their tensorial character.

As we anticipated, making use of eq.~\eqref{idconstrf} one can eliminate all three column projections from the gauge variation \eqref{gaugefgen}, while the remaining two-column projected terms read
\begin{align}
\d \, \cL \, & = \, - \sum_{p\,,\,q\,=\,0}^{N} \, \frac{1}{2^{\,p+1}} \ \bra \, Y_{\{2^p,1^{q+1}\}}\, T_{i_1j_1} \ldots T_{i_pj_p} \, \bar{\e}_{\,l} \, \g_{\,k_1 \ldots\, k_q} \,\comma\, k_{\,p\,,\,q} \ \pr_{\,l} \, (\,\g^{\,[\,q\,]}\,\cS^{\,[\,p\,]}\,)_{\, i_1 j_1,\,\ldots\,,\,i_p j_p \, ;\, k_1 \ldots\, k_q} \nn \\
& + \,(\,q+1\,)\,k_{\,p\,,\,q+1} \dsl \, (\,\g^{\,[\,q+1\,]}\,\cS^{\,[\,p\,]}\,)_{\, i_1 j_1,\,\ldots\,,\,i_p j_p \, ;\, k_1 \ldots\, k_q \, l} \nn \\[5pt]
& + \,(\,p+1\,)\,k_{\,p+1,\,q} \ \pr^{\,m}\, (\,\g^{\,[\,q\,]}\,\cS^{\,[\,p+1\,]}\,)_{\, i_1 j_1,\,\ldots\,,\,i_p j_p\,,\,lm \, ;\, k_1 \ldots\, k_q} \nn \\[5pt]
& + \,(\,q+1\,)\,(\,q+2\,)\,k_{\,p\,,\,q+2} \ \pr^{\,m}\, (\,\g^{\,[\,q+2\,]}\,\cS^{\,[\,p\,]}\,)_{\, i_1 j_1,\,\ldots\,,\,i_p j_p \, ;\, k_1 \ldots\, k_q\,lm} \, \ket \nn \\
& - \sum_{p\,,\,q\,=\,0}^{N} \, \frac{1}{2^{\,p+1}} \ \bra \, Y_{\{2^{p+1},1^{q-1}\}}\, T_{i_1j_1} \ldots T_{i_pj_p} \, \bar{\e}_{\,l} \, \g_{\,k_1 \ldots\, k_q} \,\comma\, k_{\,p\,,\,q} \ \pr_{\,l} \, (\,\g^{\,[\,q\,]}\,\cS^{\,[\,p\,]}\,)_{\, i_1 j_1,\,\ldots\,,\,i_p j_p \, ;\, k_1 \ldots\, k_q} \nn \\
& + \, (\,p+1\,)\, k_{\,p+1\,,\,q} \ \pr^{\,m}\, (\,\g^{\,[\,q\,]}\,\cS^{\,[\,p+1\,]}\,)_{\, i_1 j_1,\,\ldots\,,\,i_p j_p\,,\,lm \, ;\, k_1 \ldots\, k_q} \, \ket \, + \, \textrm{h.c.}\, . \label{gaugefgen2}
\end{align}
Notice that the first term in eq.~\eqref{gaugefgen} gave contributions to both groups of projected terms above, while the second term gave a single contribution. The analysis of the gradient terms was slightly more subtle, but the end result was that one of them, the third term in eq.~\eqref{gaugefgen}, contributed to both. Indeed,
\begin{align}
& \{2^{\,p},\,1^{q+1}\} \, \otimes \, \{1\} \, = \,  \{\,2^{\,p},\,1^{q+2}\} \, \oplus \, \{\,2^{\,p+1},\,1^{q}\} \, \oplus \, \{3,\,2^{\,p-1},\,1^{q+1}\} \, , \nn \\
& \{2^{\,p+1},\,1^{q-1}\} \, \otimes \, \{1\} \, = \,  \{\,2^{\,p+1},\,1^{q}\} \, \oplus \, \{\,2^{\,p+2},\,1^{q-2}\} \, \oplus \, \{3,\,2^{\,p},\,1^{q-1}\} \, ,
\end{align}
so that both types of projections are compatible with the $\{2^{p+1},1^q\}$ projection carried by $(\,\g^{\,[\,q\,]}\,\cS^{\,[\,p+1\,]}\,)$. One can now manipulate the second group of terms in eq.~\eqref{gaugefgen2} using the key relation \eqref{projfgauge}, and
after a proper relabeling eq.~\eqref{gaugefgen2} can be finally turned into
\begin{align}
\d \, \cL \, & = \, - \sum_{p\,,\,q\,=\,0}^{N} \, \frac{1}{2^{\,p+1}} \ \bra \, Y_{\{\,2^p,\,1^{q+1}\}}\, T_{i_1j_1} \ldots T_{i_pj_p} \, \bar{\e}_{\,l} \, \g_{\,k_1 \ldots\, k_q} \,\comma \nn \\
& \phantom{+} \, k_{\,p\,,\,q} \ \pr_{\,l} \, (\,\g^{\,[\,q\,]}\,\cS^{\,[\,p\,]}\,)_{\, i_1 j_1,\,\ldots\,,\,i_p j_p \, ;\, k_1 \ldots\, k_q} \nn \\[2pt]
& + \, \frac{(\,q+1\,)\,(\,q+2\,)}{p\,(\,q+3\,)} \, k_{\,p-1,\,q+2}  \ \sum_{n\,=\,1}^{p}\, \pr_{\,(\,i_n\,|}\, (\,\g^{\,[\,q+2\,]}\,\cS^{\,[\,p-1\,]}\,)_{\,\ldots\,,\,i_{r\,\ne\,n} j_{r\,\ne\,n} \,,\,\ldots\, ;\,|\,j_n\,)\,l\, k_1 \ldots\, k_q} \nn \\[2pt]
& + \,(\,q+1\,)\,k_{\,p\,,\,q+1} \dsl \, (\,\g^{\,[\,q+1\,]}\,\cS^{\,[\,p\,]}\,)_{\, i_1 j_1,\,\ldots\,,\,i_p j_p \, ;\, k_1 \ldots\, k_q \, l} \nn \\[5pt]
& + \,(\,p+1\,)\,k_{\,p+1,\,q} \ \pr^{\,m}\, (\,\g^{\,[\,q\,]}\,\cS^{\,[\,p+1\,]}\,)_{\, i_1 j_1,\,\ldots\,,\,i_p j_p\,,\,lm \, ;\, k_1 \ldots\, k_q} \nn \\[3pt]
& + \, \frac{(\,q+1\,)\,(\,q+2\,)}{q+3} \, k_{\,p\,,\,q+2} \, \sum_{n\,=\,1}^{p} \, \pr^{\,m}\, (\,\g^{\,[\,q+2\,]}\,\cS^{\,[\,p\,]}\,)_{\,\ldots\,,\,i_{r\,\ne\,n} j_{r\,\ne\,n},\,\ldots\,,\,m\,(\,i_n ;\, j_n\,)\, l \, k_1 \ldots\, k_q} \nn \\[2pt]
& + \,(\,q+1\,)\,(\,q+2\,)\,k_{\,p\,,\,q+2} \ \pr^{\,m}\, (\,\g^{\,[\,q+2\,]}\,\cS^{\,[\,p\,]}\,)_{\, i_1 j_1,\,\ldots\,,\,i_p j_p \, ;\, k_1 \ldots\, k_q\,lm} \ket \, + \, \textrm{h.c.}\, ,
\end{align}
where we enforced the symmetry under interchanges of the $(i_n,j_n)$ pairs brought about by the product of identical $T$ tensors present on the right side of the scalar product. Furthermore, the contributions in the last two terms are proportional, since
\be
(\,\g^{\,[\,q+2\,]}\,\cS^{\,[\,p\,]}\,)_{\, i_1 j_1,\,\ldots\,,\,i_p j_p \, ;\, k_1 \ldots\, k_q\,lm} \, = \, (\,\g^{\,[\,q+2\,]}\,\cS^{\,[\,p\,]}\,)_{\, i_1 j_1,\,\ldots\,,\,i_{p-1} j_{p-1} \,,\,m\,(\,i_p\, ;\,j_p\,)\, l \, k_1 \ldots\, k_q} \, ,
\ee
because a symmetrization in $(i_p,j_p,m)$ vanishes on account of the two-column projection. Therefore, eq.~\eqref{gaugefgen} finally becomes
\be \label{finalgaugef}
\begin{split}
\d \, \cL \, &  = \, - \sum_{p\,,\,q\,=\,0}^{N} \, \frac{1}{2^{\,p+1}} \ \bra \, Y_{\{\,2^p,\,1^{q+1}\}}\, T_{i_1j_1} \ldots T_{i_pj_p} \, \bar{\e}_{\,l} \, \g_{\,k_1 \ldots\, k_q} \,\comma \\
& \phantom{+} \, k_{\,p\,,\,q} \ \pr_{\,l} \, (\,\g^{\,[\,q\,]}\,\cS^{\,[\,p\,]}\,)_{\, i_1 j_1,\,\ldots\,,\,i_p j_p \, ;\, k_1 \ldots\, k_q} \\[2pt]
& + \, \frac{(\,q+1\,)\,(\,q+2\,)}{p\,(\,q+3\,)} \, k_{\,p-1,\,q+2} \ \sum_{n\,=\,1}^{p}\, \pr_{\,(\,i_n\,|}\, (\,\g^{\,[\,q+2\,]}\,\cS^{\,[\,p-1\,]}\,)_{\,\ldots\,,\,i_{r\,\ne\,n} j_{r\,\ne\,n}\,,\,\ldots \, ;\,|\,j_n\,)\,l \, k_1 \ldots\, k_q}  \\[2pt]
& + \,(\,q+1\,)\,k_{\,p\,,\,q+1} \dsl \, (\,\g^{\,[\,q+1\,]}\,\cS^{\,[\,p\,]}\,)_{\, i_1 j_1,\,\ldots\,,\,i_p j_p \, ;\, k_1 \ldots\, k_q \, l}  \\[5pt]
& + \,(\,p+1\,)\,k_{\,p+1,\,q} \ \pr^{\,m}\, (\,\g^{\,[\,q\,]}\,\cS^{\,[\,p+1\,]}\,)_{\, i_1 j_1,\,\ldots\,,\,i_p j_p\,,\,lm \, ;\, k_1 \ldots\, k_q}  \\[5pt]
& + \, \frac{(\,q+1\,)\,(\,q+2\,)\,(\,q+p+3\,)}{q+3} \, k_{\,p\,,\,q+2}  \, \pr^{\,m} \, (\,\g^{\,[\,q+2\,]}\cS^{\,[\,p\,]}\,)_{\, i_1 j_1,\,\ldots\,,\,i_p j_p \, ;\, k_1 \ldots\, k_q\,lm}  \ket + \textrm{h.c.}\, .
\end{split}
\ee

In the fermionic case, the $\g$-traces of the Bianchi identities
take the rather involved form presented in Appendix
\ref{app:idsf}, but it is possible to extract from them the far simpler
chain of $\{\,2^{\,p},1^{q+1}\}$-projected identities
\be \label{finalbianchif}
\begin{split}
& (\,p+q+2\,) \ Y_{\{\,2^p,\,1^{q+1}\}} \, \pr_{\,l} \, (\,\g^{\,[\,q\,]}\,\cS^{\,[\,p\,]}\,)_{\, i_1 j_1,\,\ldots\,,\,i_p j_p \, ;\, k_1 \ldots\, k_q} \\
+ \ & \frac{1}{q+3} \ Y_{\{\,2^p,\,1^{q+1}\}} \, \sum_{n\,=\,1}^{p}\, \pr_{\,(\,i_n\,|}\, (\,\g^{\,[\,q+2\,]}\,\cS^{\,[\,p-1\,]}\,)_{\,\ldots\,,\,i_{r\,\ne\,n} j_{r\,\ne\,n} \,,\,\ldots\, ;\,|\,j_n\,)\,l \, k_1 \ldots\, k_q}  \\[2pt]
+ \ & (-1)^{\,q+1} \dsl \, (\,\g^{\,[\,q+1\,]}\,\cS^{\,[\,p\,]}\,)_{\, i_1 j_1,\,\ldots\,,\,i_p j_p \, ;\, k_1 \ldots\, k_q \, l}  \, - \, \pr^{\,m}\, (\,\g^{\,[\,q\,]}\,\cS^{\,[\,p+1\,]}\,)_{\, i_1 j_1,\,\ldots\,,\,i_p j_p\,,\,lm \, ;\, k_1 \ldots\, k_q}  \\[5pt]
- \ & \frac{1}{q+3} \ \pr^{\,m}\,
(\,\g^{\,[\,q+2\,]}\,\cS^{\,[\,p\,]}\,)_{\, i_1 j_1,\,\ldots\,,\,i_p
j_p \, ;\, k_1 \ldots\, k_q\,lm} \, = \, 0\, .
\end{split}
\ee
Comparing eqs.~\eqref{finalgaugef} and \eqref{finalbianchif} one can
now recognize that, in order to obtain a gauge invariant Lagrangian, the
coefficients should satisfy the two recursion relations
\bea
k_{\,p+1,\,q} & = & - \, \frac{1}{(\,p+1\,)\,(\,p+q+2\,)} \, k_{\,p\,,\,q} \, , \nonumber \\
k_{\,p\,,\,q+1} & = & \frac{(-1)^{\,q+1}}{(\,q+1\,)\,(\,p+q+2\,)} \, k_{\,p\,,\,q} \, ,
\eea
whose solution is
\be \label{solgenf}
k_{\,p\,,\,q} \, =  \,
\frac{(-1)^{\,p\,+\,\frac{q\,(q+1)}{2}}}{p\,!\,q\,!\,(\,p+q+1\,)\,!} \
\ee
if $k_{\,0,0}=1$, a result that is also fully consistent with the other terms in eq.~\eqref{finalbianchif}. Notice that this expression includes as special cases the coefficients first obtained by
Labastida for Bose fields: in fact, letting $q=0$ in eq.~\eqref{solgenf} recovers
the result presented in \cite{bose_mixed}, that differs from
the original one of \cite{labastida} only due to a slight change of
conventions and to the elimination of a typo in the relative signs.

The coefficients given in eq.~\eqref{solgenf} determine completely the constrained Lagrangians \eqref{lagferconstr} for mixed-symmetry Fermi fields. Together with the corresponding unconstrained Lagrangians that we shall soon come across in eq.~\eqref{laggenfunc}, these are the main results of this work.

Interestingly, eq.~\eqref{solgenf} can be also obtained demanding that the
kinetic operator be self-adjoint, along the lines of what was done
by Labastida for Bose fields in \cite{labastida}. The idea is
to build a Rarita-Schwinger-like tensor such that
\be \label{selfadjoint}
\bra \bar{\psi} \comma \cE \ket \, + \, \textrm{h.c.} \, = \, 2 \, \bra \bar{\psi} \comma \cE
\ket \, ,
\ee
up to constraints and total derivatives, starting from a general
Ansatz of the type
\be \label{einsteingen}
\cE \, = \, \sum_{p\,,\,q\,=\,0}^N \, k_{\,p\,,\,q} \ \h^{i_1j_1} \ldots\, \h^{i_pj_p} \,
\g^{\,k_1 \ldots\, k_q} \ T_{i_1j_1} \ldots\, T_{i_pj_p} \, \g_{\,k_1 \ldots\, k_q} \,
\cS \, .
\ee
Whereas this sum apparently differs from the one contained in
eq.~\eqref{lagferconstr}, in the constrained case they actually
coincide, since each term gives rise to a \emph{unique} two-column
projection. Hence, making use of the result for the general
$\g$-trace of the $\cS$ tensor given in eq.~\eqref{sgentrace}, up to
total derivatives and constraints one is led to the condition
\be
\begin{split}
& \bra \bar{\psi} \comma \cE \ket \, + \, \textrm{h.c.} \, = \, 2 \,
\bra \bar{\psi} \comma \cE \ket \, + \, \sum_{p\,,\,q\,=\,0}^N \,
\frac{i}{2^{\,p+1}} \, \bra\, Y_{\{2^p,1^q\}} \, T_{i_1j_1} \ldots\,
T_{i_pj_p} \, \bar{\psi} \, \g_{\,k_1 \ldots\, k_q} \comma \\
& - \,
\left[\, (\,p+q+1\,)\, k_{\,p\,,\,q} \, - \,
\frac{(-1)^{\,q}}{q} \, k_{\,p\,,\,q-1} \,\right] \, Y_{\{2^p,1^q\}}
\,\g_{\,[\,k_1 \ldots\, k_{q-1}\,} \pr_{\,k_q\,]} \prod_{r\,=\,1}^p
\, T_{i_rj_r} \, \psi \\
& + \, \left[\, k_{\,p\,,\,q} \, - \,
(-1)^{\,q} \, \frac{q+1}{p}
\, k_{\,p-1\,,\,q+1}\,\right] \, Y_{\{2^p,1^q\}} \, \sum_{n\,=\,1}^p
\, \g_{\,k_1 \ldots\, k_q \, (\,i_n} \, \pr_{\,j_n\,)}
\prod_{r\,\ne\,n}^p \, T_{i_rj_r} \, \psi \\
& + \, \left[\,
k_{\,p\,,\,q} \, + \, (-1)^{\,q} \, \frac{p+1}{q}
\, k_{\,p+1\,,\,q-1} \,\right] \, \pr^{\,l} \, Y_{\{2^{p+1},1^{q-1}\}}
\,  \g_{\,[\,k_1 \ldots\, k_{q-1}\,} T_{k_q\,]\,l} \prod_{r\,=\,1}^p
\, T_{i_rj_r} \, \psi \\
& + \, \bigg[\, k_{\,p\,,\,q} \, + \,
(-1)^{\,q} \, (\,q+1\,)\,(\,p+q+2\,) \, k_{\,p\,,\,q+1} \,\bigg] \,
\pr^{\,l} \, Y_{\{2^p,1^{q+1}\}} \, \g_{k_1 \ldots\, k_q\, l}
\prod_{r\,=\,1}^p \, T_{i_rj_r} \, \psi \ket \, ,
\label{selfadjferm}
\end{split}
\ee
so that the remainder vanishes precisely if the coefficients
$k_{\,p\,,\,q}$ are those of eq.~\eqref{solgenf}. To reiterate, the
coefficients for the general fermionic Lagrangians can be derived in
two distinct ways: enforcing the gauge symmetry via the Bianchi identities
along the lines of what was done for the symmetric case in
\cite{fsold2,fms}, or alternatively requiring that the kinetic operator
be self adjoint, as was done for Bose fields in \cite{labastida}.
We shall see again shortly that the first method appears more natural in
the unconstrained case.

\vskip 24pt


\scss{The unconstrained Lagrangians}\label{sec:lagrangianfunc}


In order to build the unconstrained Lagrangians, barring some
subtleties that we shall elaborate upon, one can follow the same
steps that led to their constrained counterparts. The new
ingredients are two kinds of terms that were previously
discarded, those annihilated by the Labastida constraints on the
gauge parameter \eqref{constrgaugef} and those generated by the
classical anomaly \eqref{constrpsi} that emerges on the right-hand
side of the Bianchi identities \eqref{bianchi_cf} for unconstrained
fields. In the unconstrained setting, however, the first class of
terms is proportional to the gauge transformation of the
$\x_{\,ij}(\Psi)$ introduced in Section
\ref{sec:lagrangian2f}, and can thus be balanced by suitable
compensator terms while, as we saw already in the two-family case,
the Bianchi identities become
\be \label{bianchi_W2} \mathscr{B}_i \, : \quad \pr_{\,i}\, \cW \, - \, \12 \dsl \ \cWsl_i \, - \,
\12 \ \pr^{\,j} \, T_{ij} \, \cW \, - \, \frac{1}{6} \ \pr^{\,j} \,
\g_{\,ij} \, \cW \, = \, \12 \ \pr^{\,j} \pr^{\,k} \, \cZ_{\,ijk} \, , \ee
with $\cZ_{\,ijk}$ the gauge-invariant constraint tensors introduced
in eq.~\eqref{z}. Clearly, the left-hand sides of the successive
$\g$-traces of eq.~\eqref{bianchi_W2} take the same form as the
$\g$-traces \eqref{gtrace_bianchi} of the constrained Bianchi
identities while, as for two-family fields, the right-hand sides
build combinations of the $\cZ_{\,ijk}$ tensors to be eventually
compensated by the gauge transformations of some Lagrange
multipliers, that here we shall denote by $\l_{\,ijk}$.

As for the Bose fields discussed in \cite{bose_mixed}, the natural
starting point to build the unconstrained Lagrangians is the
counterpart of eq.~\eqref{lagferconstr}, namely a trial Lagrangian
containing only the two-column projected combinations of $\g$-traces of the kinetic
tensor $\cW$ of eq.~\eqref{W}, that are not proportional to the
constraint tensors $\cZ_{\,ijk}$,
\be
\cL_0 \, = \, \frac{1}{2} \ \bra\, \bar{\psi} \,\comma\! \sum_{p,\,q\,=\,0}^{N}
k_{\,p\,,\,q} \ \h^{\,p} \, \g^{\,q}\, (\,\g^{\,[\,q\,]}\,\cW^{\,[\,p\,]}\,) \,\ket \, + \, \textrm{h.c.}\, ,
\label{lagferunconstr}
\ee
where the $k_{\,p\,,\,q}$ are given in eq.~\eqref{solgenf}.
Starting from eq.~\eqref{lagferconstr}, that has the same form as
eq.~\eqref{lagferunconstr}, in the constrained case we had to impose
the Labastida constraints \eqref{constrgaugef} on the gauge
parameters to annihilate all three-column projections in the gauge
variation \eqref{gaugefgen}. In the unconstrained setting, this
remainder can instead be compensated introducing a new term,
\be \label{comp1}
\begin{split}
\cL_1 &= \, - \, \frac{1}{4} \ \bra\, \bar{\x}_{\,ij} \ ,  \sum_{p\,,\,q\,=\,0}^N \, \h^{i_1j_1} \ldots \h^{i_pj_p}\, \g^{\,k_1 \ldots\, k_q} \, \times \\
& \times \bigg\{\, \frac{(\,p+1\,)\,k_{\,p+1\,,\,q-1}}{p+q+2} \ \pr_{\,(\,k_1\,|} (\,\g^{\,[\,q-1\,]}\,\cW^{\,[\,p+1\,]}\,)_{\,|\, ij\,),\,i_1j_1\,,\,\ldots\,,\,i_pj_p\,;\, k_2 \ldots\, k_q} \\[3pt]
& \phantom{\times}\, + \, \frac{2\,p\,(\,p-1\,)\,q\, k_{\,p\,,\,q+1}}{(\,p+1\,)\,(\,p+q+3\,)} \ \pr_{\,(\,i_1\,|}
(\g^{\,[\,q+1\,]}\,\cW^{\,[\,p\,]})_{\,|\, ij\,),\,i_2j
_2\,,\,\ldots\,,\,i_{p} j_{p}\,;\,j_1\, k_1 \ldots\, k_q} \,\bigg\}\,\ket \, + \, \textrm{h.c.} \ ,
\end{split}
\ee
whose form is determined by eq.~\eqref{idconstrf}.

The total gauge variation then takes the same form as in the
constrained case, barring the replacement of $\cS$ with $\cW$, and
can not be directly manipulated using the $\g$-traces of the
Bianchi identities. Rather, its terms should be rearranged and
joined in pairs using the unconstrained version of
eq.~\eqref{projfgauge}, that reads
\be \label{relstruct_unc}
\begin{split}
Y_{\{2^{p+1},1^{q-1}\}} \, T_{i_1j_1} \ldots T_{i_pj_p} \, \bar{\e}_{\,l} \, \g_{\,k_1 \ldots\, k_q} \, & = \, \frac{1}{q+1} \ Y_{\{2^{p+1},1^{q-1}\}} \, T_{i_1j_1} \ldots T_{i_pj_p} \, T_{\,l\,[\,k_1} \bar{\e}_{\,k_2\,} \g_{\,k_{3} \ldots\, k_q\,]} \\
& - \, \frac{2}{q+1} \ Y_{\{2^{p+1},1^{q-1}\}} \, \, T_{i_1j_1} \ldots T_{i_pj_p} \, \d \, \bar{\x}_{\,l\,[\,k_1\,} \g_{\,k_{2} \ldots\, k_q\,]}\ ,
\end{split}
\ee
and gives rise to a new contribution proportional to the gauge
variation of the $\x_{\,ij}$. This, in its turn, can be compensated adding to the Lagrangian
\be \label{comp2}
\begin{split}
& \cL_2 = \, - \, \frac{1}{4} \ \bra\, \bar{\x}_{\,ij} \ ,  \sum_{p\,,\,q\,=\,0}^N \, \h^{i_1j_1} \ldots \h^{i_pj_p}\, \g^{\,k_1 \ldots\, k_q} \, \times \\
& \times \bigg\{\, \frac{2\,(\,q+1\,)\,k_{\,p\,,\,q+1}}{q+2} \ Y_{\{2^{p+1},1^q\}} \, \pr_{\,(\,i\,|} (\,\g^{\,[\,q+1\,]}\,\cW^{\,[\,p\,]}\,)_{\,i_1j_1\,,\,\ldots\,,\,i_pj_p\,;\,|\,j\,)\, k_1 \ldots\, k_q} \\[3pt]
& \phantom{\times}\, - \, \frac{2\,(\,p+1\,)\,(\,q+1\,)\,k_{\,p+1\,,\,q+1}}{q+2} \ \pr^{\,m}\,
(\g^{\,[\,q+1\,]}\,\cW^{\,[\,p+1\,]})_{\,i_1j
_1\,,\,\ldots\,,\,i_{p} j_{p}\,,\,ij\,;\,m\, k_1 \ldots\, k_q} \,\bigg\} \ket \, + \, \textrm{h.c.} \ ,
\end{split}
\ee
which leads to the final gauge variation
\begin{align}
\d \left(\,\cL_0+\cL_1+\cL_2\,\right) &= \, - \sum_{\,p\,,\,q\,=\,0}^{N} \, \frac{1}{2^{\,p+1}} \, \frac{k_{\,p\,,\,q}}{(\,q+3\,)\,(\,p+q+2\,)} \ \bra \, T_{i_1j_1} \ldots T_{i_pj_p} \, \bar{\e}_{\,l} \, \g_{\,k_1 \ldots\, k_q} \,\comma \nn \\[2pt]
& \phantom{+} \, (\,q+3\,)\,(\,p+q+2\,) \ Y_{\{2^p,1^{q+1}\}} \, \pr_{\,l} \, (\,\g^{\,[\,q\,]}\,\cW^{\,[\,p\,]}\,)_{\, i_1 j_1,\,\ldots\,,\,i_p j_p \, ;\, k_1 \ldots\, k_q} \nn \\[2pt]
& + \, Y_{\{2^p,1^{q+1}\}} \, \sum_{n\,=\,1}^{p}\, \pr_{\,(\,i_n\,|}\, (\,\g^{\,[\,q+2\,]}\,\cW^{\,[\,p-1\,]}\,)_{\,\ldots\,,\,i_{r\,\ne\,n} j_{r\,\ne\,n}\,,\,\ldots \, ;\,|\,j_n\,)\,l \, k_1 \ldots\, k_q} \nn \\[2pt]
& + \, (-1)^{\,q+1}\, (\,q+3\,) \, \dsl \, (\,\g^{\,[\,q+1\,]}\,\cW^{\,[\,p\,]}\,)_{\, i_1 j_1,\,\ldots\,,\,i_p j_p \, ;\, k_1 \ldots\, k_q\, l} \nn \\[6pt]
& - \, (\,q+3\,)\ \pr^{\,m} \, (\,\g^{\,[\,q\,]}\,\cW^{\,[\,p+1\,]}\,)_{\, i_1 j_1,\,\ldots\,,\,i_p j_p \,,\, lm \, ;\, k_1 \ldots\, k_q} \nn \\[6pt]
& - \, \pr^{\,m} \, (\,\g^{\,[\,q+2\,]}\,\cW^{\,[\,p\,]}\,)_{\, i_1 j_1,\,\ldots\,,\,i_p j_p \, ;\, k_1 \ldots\, k_q\, lm} \, \ket \, + \, \textrm{h.c.} \ , \label{unc_finalvar}
\end{align}
which is the counterpart of eq.~\eqref{finalgaugef}, since as we already stated the $k_{\,p\,,\,q}$ are those of eq.~\eqref{solgenf}.

The terms that appear in eq.~\eqref{unc_finalvar} can be turned into
others involving the constraint tensors $\cZ_{\,ijk}$ by
suitable consequences of the Bianchi identities. As we already saw
for two-family fields, at this stage a new subtlety emerges with
respect to the constrained setting. For a gradient term
with the structure of the last contribution in
eq.~\eqref{unc_finalvar}, a $\{2^p,1^{q+1}\}$ projection in the
free indices does not suffice to enforce a two-column projection
extended to the additional index $m$. Hence, in order to reproduce the terms that enter the final gauge variation, one is to consider suitable linear combinations of the projected $\g$-traces of the Bianchi identities. The end result, that we shall prove in Appendix \ref{app:idsf}, is
\begin{align}
& \frac{p+q+3}{3}\, \bigg\{\, (\,q+3\,)\,(\,p+q+2\,) \ Y_{\{2^p,1^{q+1}\}} \, \pr_{\,l} \, (\,\g^{\,[\,q\,]}\,\cW^{\,[\,p\,]}\,)_{\, i_1 j_1,\,\ldots\,,\,i_p j_p \, ;\, k_1 \ldots\, k_q} \nn \\
& \phantom{\frac{p+q+3}{3}} + \, Y_{\{2^p,1^{q+1}\}} \, \sum_{n\,=\,1}^{p}\, \pr_{\,(\,i_n\,|}\, (\,\g^{\,[\,q+2\,]}\,\cW^{\,[\,p-1\,]}\,)_{\,\ldots\,,\,i_{r\,\ne\,n} j_{r\,\ne\,n}\,,\,\ldots \, ;\,|\,j_n\,)\,l \, k_1 \ldots\, k_q} \nn \\
& \phantom{\frac{p+q+3}{3}} + \, (-1)^{\,q+1}\, (\,q+3\,)  \dsl \ (\,\g^{\,[\,q+1\,]}\,\cW^{\,[\,p\,]}\,)_{\, i_1 j_1,\,\ldots\,,\,i_p j_p \, ;\, k_1 \ldots\, k_q\, l} \nn \\[2pt]
& \phantom{\frac{p+q+3}{3}} - \, (\,q+3\,)\ \pr^{\,m} \, (\,\g^{\,[\,q\,]}\,\cW^{\,[\,p+1\,]}\,)_{\, i_1 j_1,\,\ldots\,,\,i_p j_p \,,\, lm \, ;\, k_1 \ldots\, k_q} \nn \\[2pt]
& \phantom{\frac{p+q+3}{3}} - \, \pr^{\,m} \, (\,\g^{\,[\,q+2\,]}\,\cW^{\,[\,p\,]}\,)_{\, i_1 j_1,\,\ldots\,,\,i_p j_p \, ;\, k_1 \ldots\, k_q\, lm} \,\bigg\} = \nn \\
& \phantom{\frac{p+q+3}{3}}  = \, Y_{\{2^p,1^{q+1}\}} \bigg\{\, (\,p+2\,)\,(\,q+3\,)\ T_{i_1j_1} \!\ldots T_{i_pj_p} \, \g_{\,k_1 \ldots\, k_q} \pr^{\,m}\pr^{\,n} \cZ_{\,lmn} \nn \\ & \phantom{\frac{p+q+3}{3}} - \, \sum_{n\,=\,1}^p \, \prod_{r\,\ne\,n}^p \, T_{i_rj_r} \, \g_{\,k_1
\ldots\, k_q\,l\,(\,i_n\,|}\, \pr^{\,m}\pr^{\,n} \cZ_{\,|\,j_n\,) \,mn} \,\bigg\}\ . \label{combianchif}
\end{align}
In conclusion, the gauge variation \eqref{unc_finalvar} can be cast in the form
\be
\begin{split}
& \d \left(\,\cL_0+\cL_1+\cL_2\,\right) = \, - \sum_{\,p\,,\,q\,=\,0}^{N} \, \frac{1}{2^{\,p+2}} \, \frac{3\ k_{\,p\,,\,q}}{(\,q+3\,)\,(\,p+q+2\,)\,(\,p+q+3\,)}\, \times \\[2pt]
& \times \, \bra \, Y_{\{2^p,1^{q+1}\}} \, T_{i_1j_1} \ldots T_{i_pj_p} \, \bar{\e}_{\,l} \, \g_{\,k_1 \ldots\, k_q} \,\comma\, (\,p+2\,)\,(\,q+3\,)\ T_{i_1j_1} \!\ldots T_{i_pj_p} \, \g_{\,k_1 \ldots\, k_q} \pr^{\,m}\pr^{\,n}\, \cZ_{\,lmn} \\
& \phantom{\times \, \bra \, Y_{\{2^p,1^{q+1}\}} \, T_{i_1j_1} \ldots T_{i_pj_p} \, \bar{\e}}
- \, \sum_{n\,=\,1}^p \, \prod_{r\,\ne\,n}^p \, T_{i_rj_r} \, \g_{\,k_1
\ldots\, k_q\,l\,(\,i_n\,|}\, \pr^{\,m}\pr^{\,n}\, \cZ_{\,|\,j_n\,)\,mn} \, \ket \, + \, \textrm{h.c.} \ ,
\end{split}
\ee
and can be compensated by the Lagrange multiplier terms
\be
\cL_3 \, = \, \frac{1}{4} \, \bra \bar{\l}_{\,ijk}\comma \cZ_{\,ijk} \ket \, + \, \textrm{h.c.}\ ,
\ee
where the $\l_{\,ijk}$ transform as
\be \label{lagrangegen}
\begin{split}
& \d \l_{\,ijk} \, = \, \sum_{p\,,\,q\,=\,0}^N \, \h^{m_1n_1} \ldots\, \h^{m_pn_p} \, \g^{\,l_1 \ldots\, l_q} \,\times \\
& \times\, \pr_{\,(\,i\,}\pr_{\,j\,|} \, \bigg\{\, \frac{(\,p+2\,)\ k_{\,p\,,\,q}}{(\,p+q+2\,)\,(\,p+q+3\,)} \ Y_{\{2^p,1^{q+1}\}} \, T_{m_1n_1} \ldots T_{m_pn_p} \, \g_{\,l_1 \ldots\, l_q} \, \e_{\,|\,k\,)}\\[5pt]
& + \, \frac{(-1)^{\,q+1}\,(\,p+1\,)\ k_{\,p+1\,,\,q-2}}{q\,(\,q+1\,)\,(\,p+q+1\,)\,(\,p+q+2\,)} \ Y_{\{2^{p+1},1^{q-1}\}} \, T_{m_1n_1} \ldots T_{m_pn_p} \,T_{|\,k\,)\,[\,l_1} \g_{\,l_2 \ldots\, l_{q-1}\,} \e_{\,l_q\,]} \,\bigg\} \, .
\end{split}
\ee
For brevity, this expression is not presented here in its normal
ordered form, but the reader can verify that for two families it
reduces to eq.~\eqref{lagmultk1}.

Summarizing, the final result for the \emph{unconstrained} fermionic
Lagrangians of fields carrying an arbitrary number of
families is
\be \label{laggenfunc}
\begin{split}
\cL \, & = \, \12 \, \bra \bar{\psi} \comma \!\sum_{p\,,\,q\,=\,0}^{N} \, \frac{(-1)^{\,p\,+\,\frac{q\,(q+1)}{2}}}{p\,!\,q\,!\,(\,p+q+1\,)\,!} \ \h^{\,p} \, \g^{\,q}\, (\,\g^{\,[\,q\,]}\,\cW^{\,[\,p\,]}\,) \ket \, + \, \frac{1}{4} \, \bra \bar{\x}_{\,ij} \comma \Xi_{\,ij} \ket \\
& + \, \frac{1}{4} \, \bra \bar{\l}_{\,ijk}\comma \cZ_{\,ijk} \ket \, + \, \textrm{h.c.}\ ,
\end{split}
\ee
with
\begin{align}
\Xi_{\,ij} \, &= \, - \, \sum_{p\,,\,q\,=\,0}^N \, \h^{i_1j_1} \ldots\, \h^{i_pj_p} \, \g^{\, k_1 \ldots\, k_q}\, \times \nn \\
\times \, \bigg\{& \, \frac{(-1)^{\,(p+1)\,+\,\frac{q\,(q-1)}{2}}}{p\,!\,(\,q-1\,)\,!\,(\,p+q+2\,)\,!}\ \pr_{\,(\,k_1\,|} (\,\g^{\,[\,q-1\,]}\,\cW^{\,[\,p+1\,]}\,)_{\,|\, ij\,),\,i_1j_1\,,\,\ldots\,,\,i_pj_p\,;\, k_2 \ldots\, k_q} \nn \\[3pt]
+ & \, \frac{2\,(\,p-1\,)\,p\,q\,(-1)^{\,p\,+\,\frac{(q+1)(q+2)}{2}}}{(\,p+1\,)\,!\,(\,q+1\,)\,!\,(\,p+q+3\,)\,!}\ \pr_{\,(\,i_1\,|}
(\g^{\,[\,q+1\,]}\,\cW^{\,[\,p\,]})_{\,|\, ij\,),\,i_2j
_2\,,\,\ldots\,,\,i_{p} j_{p}\,;\,j_1\, k_1 \ldots\, k_q} \nn \\[3pt]
+ & \, \frac{2\,(\,q+1\,)\,(-1)^{\,p\,+\,\frac{(q+1)(q+2)}{2}}}{p\,!\,(\,q+2\,)\,!\,(\,p+q+2\,)\,!}\ Y_{\{2^{p+1},1^q\}} \, \pr_{\,(\,i\,|} (\,\g^{\,[\,q+1\,]}\,\cW^{\,[\,p\,]}\,)_{\,i_1j_1\,,\,\ldots\,,\,i_pj_p\,;\,|\,j\,)\, k_1 \ldots\, k_q} \nn \\[3pt]
- & \,
\frac{2\,(\,q+1\,)\,(-1)^{\,(p+1)\,+\,\frac{(q+1)(q+2)}{2}}}{p\,!\,(\,q+2\,)\,!\,(\,p+q+3\,)\,!}\
\pr^{\,m}\, (\g^{\,[\,q+1\,]}\,\cW^{\,[\,p+1\,]})_{\,i_1j
_1\,,\,\ldots\,,\,i_{p} j_{p}\,,\,ij\,;\,m\, k_1 \ldots\, k_q}\,
\bigg\}\ ,
\end{align}
and where the gauge transformations of the Lagrange multipliers are
given in eq.~\eqref{lagrangegen}.

As in the two-family case, the projected Bianchi identities
\eqref{combianchif} and the compensator terms share a portion of their
Young projections, and can thus be combined. The end result is the
option, analyzed in detail for two-family fields, of modifying at
the same time compensator terms and gauge transformations of the
multipliers without altering the form of the Rarita-Schwinger-like tensor and the types of Young projections appearing in the Lagrangians. All this has no analogue in the bosonic case.

Furthermore, as we already saw in Section \ref{sec:lagrangian2f},
the unconstrained Lagrangians \eqref{laggenfunc} can be also
presented in an alternative form that proves convenient when
deriving the field equations. To this end, one can make a first step
in the direction of the unconstrained theory considering
Lagrangians for fields $\psi$ \emph{not} subject to the Labastida
constraints \eqref{constrpsi} but \emph{only} invariant under
constrained gauge transformations. These can be obtained adding to
the constrained Lagrangians a set of multipliers enforcing
on-shell the Labastida constraints, and read
\be \label{lagmult}
\begin{split}
\cL_C & = \, \12 \, \bra \bar{\psi} \comma \!\sum_{p\,,\,q\,=\,0}^{N} \, \frac{(-1)^{\,p\,+\,\frac{q\,(q+1)}{2}}}{p\,!\,q\,!\,(\,p+q+1\,)\,!} \ \h^{\,p} \, \g^{\,q}\, (\,\g^{\,[\,q\,]}\,\cS^{\,[\,p\,]}\,) \ket \, + \, \frac{i}{12} \, \bra \bar{\z}_{\,ijk}\comma T_{(\,ij}\! \psisl_{k\,)} \ket + \, \textrm{h.c.}\, .
\end{split}
\ee
In this way the two classes of Lagrangians \eqref{lagferconstr} and
\eqref{lagmult} give rise to two sets of equations of motion that
are completely equivalent while, in order to guarantee the gauge
invariance of \eqref{lagmult}, the $\z_{\,ijk}$ Lagrange multipliers
are to transform as the $\l_{\,ijk}$ did in eq.~\eqref{lagrangegen}.
Therefore, the unconstrained Lagrangians of eq.~\eqref{laggenfunc}
can be recovered from eq.~\eqref{lagmult} via the Stueckelberg-like
shifts
\begin{align}
& \psi \, \rightarrow \, \psi \, - \, \pr^{\,i}\, \Psi_{\,i} \, , \nn \\
& \z_{\,ijk} \, \rightarrow \, \l_{\,ijk} \, - \, \Delta_{\,ijk}(\Psi) \, , \label{shift}
\end{align}
where the $\Psi_{\,i}$ are the compensator fields that we had
introduced in Section \ref{sec:lagrangian2f} and the $\D_{\,ijk}(\Psi)$ that
appear in the shift
of the Lagrange multipliers $\z_{\,ijk}$ can be formally obtained
replacing $\e_{\,i}$ with $\Psi_{\,i}$ in the gauge variation of the
Lagrange multipliers $\l_{\,ijk}$, so that the $\z_{\,ijk}$ are
mapped into a gauge invariant combination by eq.~\eqref{shift}. A detailed discussion
of the subtleties entailed by this step can be found in Section \ref{sec:motion2f}.

Both derivations of the unconstrained Lagrangians \eqref{laggenfunc}
rest on a suitable class of successive traces of the Bianchi
identities but, as was already stressed for Bose fields in
\cite{bose_mixed}, there is further freedom to redefine
simultaneously the Rarita-Schwinger-like tensor and the $\l_{\,ijk}$
or $\z_{\,ijk}$ Lagrange multipliers. For instance, the Lagrangian
for an unconstrained field $\psi$ that only allows constrained gauge
transformations can be also presented in the form
\be \label{lagmod}
\begin{split}
\widetilde{\cL}_C \, & = \, \12 \, \bra \bar{\psi} \comma \!\sum_{p\,,\,q\,=\,0}^{N} \, \frac{(-1)^{\,p\,+\,\frac{q\,(q+1)}{2}}}{p\,!\,q\,!\,(\,p+q+1\,)\,!} \ \h^{\,p} \, \g^{\,q}\, \widehat{\cS}^{\,[\,p\,,\,q\,]} \ket \, + \, \frac{i}{12} \, \bra \bar{\cY}_{\,ijk}\comma T_{(\,ij}\! \psisl_{k\,)} \ket \, + \, \textrm{h.c.}\ ,
\end{split}
\ee
where the $\cY_{\,ijk}$ are \emph{gauge invariant} Lagrange
multipliers and the $\widehat{\cS}^{\,[\,p\,,\,q\,]}$ are two-column
projected $\g$-traces of $\cS$ deprived of the $\{3,2^{p-1},1^{q}\}$
component originally present in the last term of
eq.~\eqref{sgentrace}~\footnote{Using the techniques of Appendix
\ref{app:idsf}, the reader can verify that the other gradient term
is already two-column projected, thanks to the projection induced in
its free indices.},
\be
\begin{split}
&\widehat{\cS}^{\,[\,p\,,\,q\,]}{}_{\,i_1j_1,\,\ldots\,,\,i_pj_p \,;\,k_1 \ldots\, k_q} \equiv \, i\, (-1)^{\,q} \, Y_{\{2^p,1^q\}}\, \bigg\{\, (\,q+1\,) \dsl \ T_{i_1j_1} \ldots\, T_{i_pj_p}\, \g_{\,k_1 \ldots\, k_q} \, \psi \\
& - \, (\,p+q+1\,)\, \pr_{\,[\,k_1\,|}\, T_{i_1j_1} \ldots\, T_{i_pj_p}\, \g_{\,|\,k_2 \ldots\, k_q\,]}\, \psi \, - \, \sum_{n\,=\,1}^p \, \pr_{\,(\,i_n\,|}\, \prod_{r\,\neq\,n}^p \, T_{i_rj_r}\, \g_{\,|\,j_n\,)\,k_1 \ldots\, k_q}\, \psi \\
& + \, \pr^{\,m}\, T_{i_1j_1} \ldots\, T_{i_pj_p}\, T_{m\,[\,k_1\,} \g_{\,k_2 \ldots\, k_q\,]}\, \psi \, - \, \pr^{\,m}\, Y_{\{2^p,1^{q+1}\}}\, T_{i_1j_1} \ldots\, T_{i_pj_p} \, \g_{\,m\,k_1 \ldots\, k_q}\, \psi \,\bigg\} \, .
\end{split}
\ee
In complete analogy with the bosonic case of \cite{bose_mixed}, the $\widehat{\cS}^{\,[\,p\,,\,q\,]}$ satisfy
two-column projected ``anomaly free" Bianchi identities that have the form of
eq.~\eqref{finalbianchif} and, as a result, do not contain any terms related to the constraints.
Looking back at the derivation of the Lagrangians presented in the previous
pages, the reader can easily recognize that the Lagrangians \eqref{lagmod} are gauge
invariant, and indeed the Rarita-Schwinger-like tensor in eq.~\eqref{lagmod}
is manifestly self-adjoint. One can then recover Lagrangians for fully unconstrained
gauge fields performing the Stueckelberg-like shift
\be
\psi \, \rightarrow \, \psi \, - \, \pr^{\,i}\, \Psi_{\,i} \, .
\ee
The result, however, differs from eq.~\eqref{laggenfunc}
by a field redefinition of the Lagrange multipliers $\l_{\,ijk}$, while the constrained
gauge invariance of \eqref{lagmod} guarantees that the $\Psi_{\,i}$ compensators still
enter only via the combinations $\xi_{\,ij}(\Psi)$.

Let us notice, to conclude, that the terms involving the Lagrange
multipliers have precisely the type of structure identified in the
two-family case. The arguments presented there still apply, so that
when working with the independent
$\Psi_i$ compensators all forms of the $N$-family Lagrangians are
invariant provided the Lagrange multipliers transform
like
\be
\delta \l_{\,ijk} \, = \, \eta^{\,lm}\, M_{\,ijk,\,lm} \, +\, \g^{\,lm}\, N_{\,ijk;\,lm} \, , \label{lambdalsymgen}
\ee
where $M_{\,ijk,\,lm}$ and $N_{\,ijk;\,lm}$ are $\{4,1\}$ projected in
their family indices and are mutually related according
to
\be
N_{\,ij\,(\,k\,;\,l\,)\,m} \, + \, N_{\,kl\,(\,i\,;\,j\,)\,m} \, = \, - \, \12 \left(\, M_{\,ijm,\,kl} + \, M_{\,klm,\,ij} \,\right) \, .  \label{MNrelgen}
\ee

\vskip 24pt


\scss{The field equations}\label{sec:motionf}


As was the case for generic Bose fields \cite{bose_mixed}, and as we saw for two-family
Fermi fields in Section \ref{sec:motion2f}, recovering the Lagrangian \eqref{laggenfunc}
from eq.~\eqref{lagmult} via Stueckelberg-like shifts proves particularly convenient when
deriving the field equations. In fact, retracing the derivation of the constrained
Lagrangian via the request of self adjointness and keeping track of the constraint terms,
one can conveniently obtain the equations of motion following from eq.~\eqref{lagmult}:
\begin{align}
& E_{\, \bar{\psi}}\, : \ \sum_{p\,,\,q\,=\,0}^N \, k_{\,p\,,\,q} \ \h^{i_1 j_1} \ldots \, \h^{i_p j_p} \, \g^{\,k_1 \ldots\, k_q} \ (\,\g^{\,[\,q\,]}\,\cS^{\,[\,p\,]}\,)_{\, i_1 j_1\,,\,\ldots\,,\,i_p j_p \, ;\, k_1 \ldots\, k_q} \nn \\
& \phantom{E_{\, \bar{\psi}}\, :}\ - \, \frac{i}{2} \! \sum_{p\,,\,q\,=\,0}^N (-1)^{\,q}\,k_{\,p\,,\,q} \ \h^{i_1j_1} \!\ldots \h^{i_pj_p}\, \g^{\,k_1 \ldots\, k_q} \, Y_{\{2^p,1^q\}} \, \pr^{\,m} \, Y_{\{3,2^{p-1},1^q\}} \, T_{i_1j_1} \!\ldots T_{i_pj_p} \g_{\,m\,k_1 \ldots\, k_q}\, \psi \label{eqgenpsi1} \nn \\
& \phantom{E_{\, \bar{\psi}}\, :}\ - \, \12 \ \h^{ij}\,\g^{\,k}\, \cY_{\,ijk} \, = \, 0 \, ,\\[5pt]
& E_{\, \bar{\z}}\, : \ \frac{i}{12}\ T_{(\,ij}\! \psisl_{k\,)} \, = \, 0 \, , \label{eqgenz}
\end{align}
where the $\psi$ terms contained in the $\cY_{\,ijk}$ originate from
the hermitian conjugates of those present in the second line of
eq.~\eqref{eqgenpsi1}. After some algebra, one can recast the
$\cY_{\,ijk}$ in the form
\be \label{Ygen}
\begin{split}
\cY_{\,ijk} \equiv \, i\ \z_{\,ijk} \,& + \,i \sum_{p\,,\,q\,=\,0}^N \, \frac{(-1)^{\,p\,+\,\frac{q\,(q+1)}{2}\,+\,1}}{p\,!\,q\,!\,(\,p+q+3\,)\,!}\ \h^{m_1n_1} \ldots\, \h^{m_pn_p}\, \g^{\,l_1 \ldots\, l_q} \, \times \\
& \times \, \pr_{\,(\,i\,|}\, Y_{\{2^{p+1},1^{q}\}}\, T_{|\,jk\,)}\,
T_{m_1n_1} \!\ldots T_{m_pn_p}\, \g_{\,l_1 \ldots\, l_q}\, \psi \, ,
\end{split}
\ee
and the arguments summarized in Appendix \ref{app:idsf} then show
that, due to the $\{3,2^{p-1},1^q\}$ projection, the whole second
line in eq.~\eqref{eqgenpsi1} vanishes on-shell on account of
eq.~\eqref{eqgenz}. As a consequence, the equation of motion obtained
from the constrained Lagrangian \eqref{lagferconstr} only contains the
first line of eq.~\eqref{eqgenpsi1} while, performing the
Stueckelberg-like shift \eqref{shift} as we did in Section
\ref{sec:motion2f}, one can recover the equations of motion for the
Lagrangian \eqref{laggenfunc}, that can be finally cast in the form
\begin{align}
& E_{\, \bar{\psi}} \, : \ \sum_{p\,,\,q\,=\,0}^N \, \frac{(-1)^{\,p\,+\,\frac{q\,(q+1)}{2}}}{p\,!\,q\,!\,(\,p+q+1\,)\,!} \ \h^{i_1 j_1} \ldots \, \h^{i_p j_p} \, \g^{\,k_1 \ldots\, k_q} \ (\,\g^{\,[\,q\,]}\,\cW^{\,[\,p\,]}\,)_{\, i_1 j_1\,,\,\ldots\,,\,i_p j_p \, ;\, k_1 \ldots\, k_q} \nn \\
& \phantom{E_{\, \bar{\psi}}\, : \ } - \, \12 \ \h^{ij}\,\g^{\,k}\, \cY_{\,ijk} \, = \, 0 \ , \label{eqgenpsi} \\[5pt]
& E_{\,\bar{\Psi}} \, : \ \pr_{\,l} \, E_{\,\bar{\psi}} \ - \, \sum_{\,p\,,\,q\,=\,0}^{N} \, \frac{3\, (\,q+1\,)\,(\,q+2\,)\, (-1)^{\,p\,+\,\frac{q\,(q+1)}{2}}}{p\,!\,(\,q+3\,)\,!\,(\,p+q+3\,)\,!}\ \h^{i_1j_1} \!\ldots \h^{i_pj_p}\, \g^{\,k_1 \ldots\, k_q}\, Y_{\{2^p,1^{q+1}\}}\, \times \nn \\[2pt]
& \phantom{E_{\,\bar{\Psi}} \, : \ \pr_{\,l} \, E_{\,\bar{\psi}} \ } \times \, \Big\{ \, (\,p+2\,)\,(\,q+3\,)\ T_{i_1j_1} \!\ldots T_{i_pj_p} \, \g_{\,k_1 \ldots\, k_q} \pr^{\,m}\pr^{\,n}\, (E_{\, \bar{\l}})_{\,lmn} \nn \\
& \phantom{E_{\,\bar{\Psi}} \, : \ \pr_{\,l} \, E_{\,\bar{\psi}} \ } - \, \sum_{n\,=\,1}^p \, \prod_{r\,\ne\,n}^p \, T_{i_rj_r} \, \g_{\,k_1
\ldots\, k_q\,l\,(\,i_n\,|}\, \pr^{\,m}\pr^{\,n}\, (E_{\, \bar{\l}})_{\,|\,j_n\,)\,mn} \, \Big\} \, = \, 0 \, , \\[5pt]
& E_{\, \bar{\l}} \, : \ \frac{1}{4} \ \cZ_{\,ijk} \, = \, 0 \, ,
\end{align}
where we have already enforced the last of these, $E_{\, \bar{\l}}$,
in the field equation for $\bar{\psi}$. As usual, the equations of motion for the
compensator fields $\bar{\Psi}_{\,i}$ simply guarantee the conservation
of external currents coupled to the field $\psi$,
while the $\cY_{\,ijk}$ tensors can be built performing the
Stueckelberg shift \eqref{shift} in eq.~\eqref{Ygen}, barring the
subtleties discussed in detail in Section \ref{sec:motion2f}.

\vskip 24pt


\scss{Comments on the on-shell reduction to $\cS=0$}\label{sec:reductionf}


The next step would be the reduction of the various choices for the field equations presented in the previous section to
the Labastida form $\cS=0$. For $N$-family fields, the operators on which the procedure rests
are $gl(N)$ counterparts of those exhibited in Section \ref{sec:reduction2f}, and the experience
developed with two-family fields suggests that, in low enough dimensions, similar
types of phenomena related to Weyl-like symmetries should emerge.

In the following we shall propose a set of conditions identifying the fields with this type of behavior, working directly at the level of the field equations, but we shall content ourselves with solving them explicitly in Section \ref{sec:weyl} for simple but instructive classes of examples. On the other hand, in analogy with the bosonic case discussed in \cite{bose_mixed}, one can argue rather simply that
in the $D \to \infty$ limit $N$-family gauge fields exhibit a universal behavior, with current exchanges that are free of special poles,
that can be captured ignoring altogether the intricacies brought about by the $S^{\,i}{}_j$ operators. In this
limit the reduction proceeds along the lines of the symmetric case, and the field equations can be directly turned into $\cS = 0$. Moreover, there are all reasons to expect
that, even for higher values of $N$, the special behavior is confined to dimensions $D \leq 2N + 1$,
where the corresponding models should be at most dual to other representations characterized by lower values of $N$, although we do not have a complete argument to this effect. The wider class of solutions that will be exhibited for $N$-family fields in Section \ref{sec:irr_example} vindicate these expectations.

Before coming to these points, however, we would like to dwell briefly upon the very meaning of the Labastida equation $\cS=0$, showing that it propagates in general representation of the little group $O(D-2)$, and thus the expected numbers of degrees of freedom. Combining this step with the results of the reduction procedure one can thus conclude that the Lagrangian systems that we are proposing describe indeed the correct free dynamics of mixed-symmetry Fermi fields.

\scsss{The degrees of freedom described by $\cS=0$} \label{sec:dofs}

In order to proceed, it is convenient to resort momentarily to an explicit oscillator realization. This is attained introducing a set of commuting vectors $u^{\,i\,\m}$ to cast a generic multi-symmetric
gauge field $\psi$ in the form
\be \label{psiosc}
\psi \, \equiv \, \frac{1}{s_{1}\,! \,\ldots\, s_{N}\,!} \ u^{\,1\,\m_1} \ldots\, u^{\,1\,\m_{s_1}} \, u^{\,2\,\n_1} \ldots\, u^{\,2\,\n_{s_2}} \ldots \ \psi_{\,\m_1 \ldots\, \m_{s_1},\,\n_1 \ldots\, \n_{s_2}\,,\, \ldots} \, .
\ee
As a result, divergences, gradients and $\g$'s with family indices take the form
\begin{alignat}{2}
& \pr_{\,i} \, \equiv \, \pr^{\,\m} \, \frac{\pr}{\pr \, u^{\,i\,\m}} \, , \qquad\qquad & & \pr^{\,i} \, \equiv \, \pr_{\,\m} \, u^{\,i\,\m} \, , \nn \\
& \g_{\,i} \, \equiv \, \g^{\,\m} \, \frac{\pr}{\pr \, u^{\,i\,\m}} \, , \qquad\qquad & & \g^{\,i} \, \equiv \, \g_{\,\m} \, u^{\,i\,\m} \, . \label{op_comm}
\end{alignat}

In studying the plane-wave solutions of the Labastida equation, that in momentum space takes the form
\be\label{momlab}
\psl \ \psi \, - \, (p \cdot u^i) \, \g_{\, i} \, \psi \, = \, 0 \ ,
\ee
it is convenient to decompose $\psi$ according to
\be\label{psidecomp}
\psi \, = \, (p \cdot u^i)\, \chi_{\,i} \, + \, \widehat{\psi} \, ,
\ee
sorting out its portion $\widehat{\psi}$ that is independent of $(p \cdot u^i)$. Substituting in \eqref{momlab} then gives
\be \label{eqpsub}
\psl \ \widehat{\psi} \, - \, (p \cdot u^i) \, \g_{\, i} \, \widehat{\psi} \,  - \ \frac{1}{2} \ (p \cdot u^i) \, (p \cdot u^j)\, \g_{\,(\,i} \, \chi_{\,j\,)} \, = \,
 0 \ .
\ee
Now the three terms must vanish independently, since they carry different powers of $(p \cdot u^i)$, so that \eqref{eqpsub} is equivalent to the three conditions
\begin{align}
& \psl \ \widehat{\psi} \, = \, 0 \ ,  \label{sys1p} \\
& \g_{\, i} \, \widehat{\psi} \, = \, 0 \ , \label{sys2p} \\
& \g_{\,(\,i} \, \chi_{\,j\,)} \, = \, 0 \ . \label{sys3p}
\end{align}
First of all, one can notice that \eqref{sys3p} subjects $\chi_{\,i}$ to the Labastida constraints, so that the first contribution in eq.~\eqref{psidecomp} can be gauged
away altogether. One is then left with $\widehat{\psi}$, that is subject to the
Dirac equation \eqref{sys1p}. This implies the mass-shell condition $p^{\,2}=0$,
so that one can take a momentum $p$ with a single non-vanishing light-cone component, $p^{\,+}$. As a result, the lack of dependence of $\widehat{\psi}$ on $(p \cdot u^i)$ translates into the conditions
\be \label{no-}
\frac{\pr}{\pr \, u^{\, i\, -}}\ \, \widehat{\psi} \, = \, 0 \ ,
\ee
while the Dirac equation \eqref{sys1p} turns into the standard projection
\be\label{dirac}
\g_{\, +} \, \widehat{\psi} \, = \, 0 \ ,
\ee
that halves the on-shell components of $\widehat{\psi}$. Finally,
eq.~\eqref{sys2p} becomes
\be\label{eqgamma+}
\left(\, \g^{\,+} \ \frac{\pr}{\pr \, u^{\, i\, +}} \, + \, \g^{\,\underline{m}} \ \frac{\pr}{\pr \, u^{\, i\, \underline{m}}} \,\right) \, \widehat{\psi} \, = \, 0 \ ,
\ee
where we have used $\underline{m}$ to denote the spatial Lorentz index in order to
distinguish it from the family indices. Multiplying eq.~\eqref{eqgamma+} by
$\g_+$ and making use of eq.~\eqref{dirac} one is finally led to the conditions
\begin{align}
& \frac{\pr}{\pr \, u^{\, i\, +}} \ \widehat{\psi} \, = \, 0 \ ,\label{no+}\\
& \g^{\,\underline{m}} \ \frac{\pr}{\pr \, u^{\, i\, \underline{m}}} \ \widehat{\psi}\, = \, 0 \ . \label{nogamma}
\end{align}

All components of $\widehat{\psi}$ along the two light-cone directions thus are absent on account of eqs.~\eqref{no-} and \eqref{no+}, while eq.~\eqref{nogamma} forces
the remaining transverse components to be $\g$-traceless. In conclusion, the Labastida
equation propagates in general representations of the little group $O(D-2)$, as expected on account of its uniqueness. Let us stress that at this stage these are reducible representations, since we did not enforce any projection on $\psi$. In Section \ref{sec:irreducible} we shall explain how to project fields and gauge parameters in order to describe irreducible representations.

We can conclude this short section by displaying a similar argument, not included in \cite{bose_mixed}, that proves that the bosonic Labastida equation propagates
in general representations of the little group $O(D-2)$. In the same notation used above for Fermi fields, the starting point is now
the decomposition
\be
\vf \, =\, (p \cdot u^i) \, \zeta_{\,i} \, +\, \widehat{\vf} \ ,
\ee
with $\widehat{\vf}$ independent of $(p \cdot u^i)$. Substituting in the
momentum-space field equation
\be
p^{\,2}\, \vf \, - \, (p \cdot u^i) \, \pr_{\,i} \, \vf \, +\, \frac{1}{2} \ (p \cdot u^i) \,(p \cdot u^j) \, T_{ij} \, \vf \, = \, 0
\ee
one obtains
\be
p^{\,2}\, \widehat{\vf} \, - \, (p \cdot u^i) \, \pr_{\,i} \, \widehat{\vf} \, +\, \frac{1}{2} \ (p \cdot u^i) \,(p \cdot u^j) \, T_{ij} \, \widehat{\vf}  \, +\,  \frac{1}{6} \ (p \cdot u^i)\, (p \cdot u^j)\, (p \cdot u^k)\, T_{(\,ij}\, \zeta_{\,k\,)} \, = \, 0 \, ,
\ee
where again the terms with different powers of $(p \cdot u^i)$ must vanish independently. As a result one is led to the conditions
\begin{align}
& p^{\,2}\, \widehat{\vf} \, = \, 0 \ , \label{sys1pb} \\
& \pr_{\,i}\, \widehat{\vf} \, = \, 0 \  , \label{sys2pb} \\
& T_{ij} \, \widehat{\vf} \, = \, 0  \ , \label{sys3pb} \\
& T_{(\,ij}\, \zeta_{\,k\,)} \, = \, 0 \ . \label{sys4pb}
\end{align}

Eq.~\eqref{sys4pb} now implies that one can gauge away the $\zeta_{\, i}$, since they are subject to the
Labastida constraints, and eq.~\eqref{sys1pb} then leads to the mass-shell condition, so that one can let again $p=p^{\, +}$. $\widehat{\vf}$ is by construction independent of $(p \cdot u^i)$, which now translates into its independence from $u^{i\, -}$. When combined with eq.~\eqref{sys2pb}, this implies that $\widehat{\vf}$ has only transverse components, while finally eq.~\eqref{sys3pb} implies that all traces of $\widehat{\vf}$ vanish. In conclusion, one is led once more to representations of the little group $O(D-2)$.

Notice that neither the triple $\g$-trace constraints for Fermi fields,
nor the double trace constraints for Bose fields, were needed to come to this point
directly from the non-Lagrangian equations. This is in contrast with what we shall see in Section \ref{sec:irreducible}, where they will have a direct bearing on a different, \emph{off-shell} counting of degrees of freedom.

\scsss{Weyl-like symmetries} \label{sec:weyl}

In Section~\ref{sec:reduction2f} we classified the two-family fields whose equations of motion do not
reduce directly to the Labastida form $\cS=0$. We saw that this can happen only if the $\cS$ spinor-tensors satisfy special conditions, and in order to pinpoint their nature we looked for solutions
preserving both the Bianchi identities and the triple $\g$-trace constraints induced on $\cS$ by the corresponding
constraints on the $\psi$ fields.
Finally, in Section~\ref{sec:weyl2} we showed explicitly that in \emph{all} these cases new Weyl-like gauge transformations emerge, that can be used to complete the reduction to $\cS=0$ by suitable gauge-fixings.

This suggests another convenient starting point to classify the sporadic cases exhibiting this type of behavior. To begin with, in complete analogy with the discussion presented for two-family fields,
all forms of the equations of motion can be reduced to
\begin{align}
E_{\, \bar{\psi}} \, : \ & \sum_{p\,,\,q\,=\,0}^N \, \frac{(-1)^{\,p\,+\,\frac{q\,(q+1)}{2}}}{p\,!\,q\,!\,(\,p+q+1\,)\,!}
\ \h^{i_1 j_1} \ldots \, \h^{i_p j_p} \, \g^{\,k_1 \ldots\, k_q} \ (\,\g^{\,[\,q\,]}\,\cS^{\,[\,p\,]}\,)_{\, i_1 j_1\,,\,\ldots\,,\,i_p j_p \, ;\, k_1 \ldots\, k_q} \nn \\
& - \, \12 \ \h^{ij}\,\g^{\,k}\, \cY_{\,ijk} \, = \, 0 \, , \label{eqgenred} \\
E_{\,\bar{\z}} \, : \ & \frac{i}{12} \ T_{(\,ij} \psisl_{k\,)} \, = \, 0\quad \Longrightarrow  \quad T_{(\,ij} \ssl_{k\,)} \, = \, 0 \, , \label{eqmultredgen}
\end{align}
possibly after a partial gauge fixing eliminating the compensators. Even in the
\mbox{$N$-family} case eq.~\eqref{eqgenred} generally sets to zero the $\g$-traceless part of $\cS$. As a result, one can reduce the problem of classifying the fields whose equations of motion are not invertible to that of classifying the Weyl-like gauge transformations of $\psi$ that leave invariant the $\g$-traceless part of $\cS$. In fact, by construction these transformations leave invariant both the Bianchi identities and the constraints induced on $\cS$ provided they also preserve the Labastida constraints on $\psi$, so that the fields allowing them are good candidates to look for cases where the direct reduction of the field equations is not possible. In the next few pages, following this approach, we shall propose a set of conditions identifying these Weyl-like shifts, and we shall also argue that they exhaust all relevant
shift symmetries of the Bianchi identities. Finally, we shall argue that, as we saw explicitly for two-family fields, the resulting transformations leave invariant the equation of motion \eqref{eqgenred} when they are combined with proper shifts of the $\cY_{\,ijk}$, and a fortiori that the conditions leading to them classify the fields whose equations of motion do not reduce directly to the Labastida form.

The discussion in Section \ref{sec:reduction2f} suggests to consider generic Weyl-like gauge transformations of the form
\be\label{varPSI}
\d \, \psi \, = \, \h^{i_1j_1} \ldots\, \h^{i_pj_p}\, \g^{\,k_1 \ldots\, k_q}\,
\Theta_{\,i_1j_1, \,\ldots\, ,\,i_pj_p;\,k_1 \ldots\, k_q} \, .
\ee
They affect the kinetic spinor-tensor $\cS$ according to
\be \label{varS}
\begin{split}
& \d \, \cS \, = \, - \, i\, q \ \h^{i_1j_1} \ldots\, \h^{i_pj_p}\, \g^{\,k_1 \ldots\, k_{q-1}}\, \pr^{\,l}\, \bigg\{\,
(\,D-q-1\,)\, \Theta_{\,i_1j_1, \,\ldots\, ,\,i_pj_p;\,l\,k_1 \ldots\, k_{q-1}} \\
& + \, \sum_{n\,=\,1}^p \, \Theta_{\,\ldots\, ,\,i_{r\neq n}j_{r\neq n},\, \ldots\, ,\,l\,(\,i_n;\,j_n\,)\,k_1 \ldots\, k_{q-1}}
+ \, 2\, S^{\,m}{}_l \, \Theta_{\, i_1j_1,\,\ldots\,,\,i_pj_p;\, m\,k_1 \ldots\, k_{q-1}} \,\bigg\} \\
& - \, i \, \frac{p}{q+1} \ \h^{i_1j_1} \ldots\, \h^{i_{p-1}j_{p-1}} \, \g^{\,k_1 \ldots\, k_{q+1}}\, \pr^{\,l}\,
\Theta_{\,i_1j_1,\,\ldots\,,\,i_{p-1}j_{p-1},\,l\,[\,k_1;\,k_2 \ldots\, k_{q+1}]} \\
& + \, (-1)^{\,q} \, \h^{i_1j_1} \ldots\, \h^{i_pj_p}\, \g^{\,k_1 \ldots\, k_q}\, \cS\,\left(\,
\Theta_{\,i_1j_1, \,\ldots\, ,\,i_pj_p;\,k_1 \ldots\, k_q} \,\right) \, ,
\end{split}
\ee
while in view of the detailed analysis of Section \ref{sec:reduction2f} the relevant conditions should reduce eq.~\eqref{varS} to its last term. Moreover, we saw in eq.~\eqref{shift2} that, for two family fields, one ought to combine different types of parameters carrying identical numbers of family indices. In the $N$-family case, one is thus led to consider Weyl-like shifts of the form
\be \label{shiftpsigen}
\d \, \psi \, = \, \sum_{p\,=\,0}^{\left[\frac{q}{2}\right]} \, \h^{i_1j_1} \ldots\, \h^{i_{p}j_{p}}\,
\g^{\,k_1 \ldots\, k_{q-2p}}\, \Theta^{\,[\,p\,,\,q-2p\,]}{}_{\,i_1j_1, \,\ldots\, ,\,i_pj_p;\,k_1 \ldots\, k_{q-2p}} \, ,
\ee
involving $q$ family indices distributed in all possible ways between $\eta$'s and $\g$'s. These affect the kinetic spinor-tensor $\cS$ according to
\be \label{shiftSgen}
\d \, \cS \, = \, \sum_{p\,=\,0}^{\left[\frac{q}{2}\right]} \, \h^{i_1j_1} \ldots\, \h^{i_{p}j_{p}}\,
\g^{\,k_1 \ldots\, k_{q-2p}}\, \cS\, \left(\, \Theta^{\,[\,p\,,\,q-2p\,]}{}_{\,i_1j_1, \,\ldots\, ,\,i_pj_p;\,k_1 \ldots\, k_{q-2p}} \,\right)\, ,
\ee
provided the chains of $p$-dependent conditions
\be \label{chain}
\begin{split}
& \frac{p+1}{(\,q-2\,p-1\,)\,(\,q-2\,p\,)} \ \Theta^{\,[\,p+1 \comma q-2(p+1)\,]}{}_{\,i_1j_1,\,\ldots\,,\,i_pj_p,\,l\,[\,k_1\,;\,k_2 \ldots\, k_{q-2p-1}\,]} \\[4pt]
& + \, (\,D-q+2\,p-1\,)\ \Theta^{\,[\,p \comma q-2p\,]}{}_{\,i_1j_1,\,\ldots\,,\,i_pj_p; \,l\,k_1 \ldots\, k_{q-2p-1}} \\
& + \, \sum_{n\,=\,1}^p \, \Theta^{\,[\,p \comma q-2p\,]}{}_{\,\ldots\, ,\,i_{r\neq n}j_{r\neq n},\, \ldots\, ,\,l\,(\,i_n;\,j_n\,)\,k_1 \ldots\, k_{q-2p-1}} \\
& + \, 2\, S^{\,m}{}_l \, \Theta^{\,[\,p \comma q-2p\,]}{}_{\, i_1j_1,\,\ldots\,,\,i_pj_p;\, m\,k_1 \ldots\, k_{q-2p-1}} = \, 0
\end{split}
\ee
hold for the allowed values of $q$.
Actually, when $N$ index families are present $p$ cannot take in general all values between zero and the integer part of $q/2$, so that the relevant range for eq.~\eqref{chain} is
\be
\max \left(\, 0 \comma \bigg[\,\frac{q-N}{2}\,\bigg] \,\right) \leq p \leq \bigg[\,\frac{q-1}{2}\,\bigg] \, ,
\ee
simply because for a given $N$ there is an upper limit for the number of indices that can be carried by an antisymmetric product of $\g$-matrices. Similar considerations apply to the lower ends of the sums in eqs.~\eqref{shiftpsigen} and \eqref{shiftSgen}, while their upper limits are explicitly displayed. Moreover, although in principle one should take into account all values of
$q$ not exceeding the rank $\sum_{i=1}^N s_i$ of the spinor-tensor $\psi$, the two-family example suggests to confine
the attention to $q \leq 2N$, since the (on-shell) Labastida constraints \eqref{constrpsi} annihilate all combinations of more than $2N$ $\g$-traces acting on $\psi$.

An admissible shift symmetry of $\psi$ must also preserve the Labastida constraints \eqref{constrpsi}, that hold off-shell in the constrained theories and are enforced on-shell in the unconstrained theories by the equations of motion \eqref{eqmultredgen} for the Lagrange multipliers. Our experience with two-family fields gives us confidence that the conditions \eqref{chain}, together with the Labastida-like constraints
\be \label{constrtheta}
T_{(\,lm}\, \g_{\,n\,)}\, \Theta^{\,[\,p\,,\,q-2p\,]}{}_{\,i_1j_1, \,\ldots\, ,\,i_pj_p;\,k_1 \ldots\, k_{q-2p}} \, = \, 0 \, ,
\ee
suffice to preserve the Labastida constraints on $\psi$ in eq.~\eqref{eqmultredgen}, although we have not
carried out in full detail the type of analysis leading to eqs.~\eqref{trace1}, \eqref{trace2} and \eqref{trace3} for the general shifts of eq.~\eqref{shiftpsigen}.
Once the Labastida constraints on $\psi$ are satisfied, similar constraints on $\cS$ hold by construction, and moreover the shifts \eqref{shiftSgen} are bound to preserve the Bianchi identities, simply because they originate from $\psi$ shifts. Furthermore, the $\cS\,(\,\Theta\,)$ satisfy the same set of conditions \eqref{chain} as the $\Theta$ parameters, since the $S^{\,i}{}_j$ operators commute with $\cS$.

On the other hand, one can sharpen the correspondence with the analysis of Section \ref{sec:reduction2f}  considering directly the shifts
\be
\d \, \cS \, = \, \h^{i_1j_1} \ldots\, \h^{i_pj_p}\, \g^{\,k_1 \ldots\, k_q}\,
\O_{\,i_1j_1, \,\ldots\, ,\,i_pj_p;\,k_1 \ldots\, k_q}
\ee
of the Labastida spinor-tensors $\cS$, where for simplicity we consider a single term of the sum corresponding to eq.~\eqref{shiftSgen}. They lead to the variations
\be \label{shiftBgen}
\begin{split}
& \d \, \mathscr{B}_l \, = \, - \, \frac{q}{2} \dsl \ \h^{i_1j_1} \ldots\, \h^{i_pj_p}\, \g^{\,k_1 \ldots\, k_{q-1}} \,
\bigg\{\, (\,D-q-1\,)\, \O_{\,i_1j_1, \,\ldots\, ,\,i_pj_p;\,l\,k_1 \ldots\, k_{q-1}} \\
& + \, \sum_{n\,=\,1}^p \, \O_{\,\ldots\, ,\,i_{r\neq n}j_{r\neq n},\, \ldots\, ,\,l\,(\,i_n;\,j_n\,)\,k_1 \ldots\, k_{q-1}}
+ \, 2\, S^{\,m}{}_l \, \O_{\, i_1j_1,\,\ldots\,,\,i_pj_p;\, m\,k_1 \ldots\, k_{q-1}} \,\bigg\} \\
& - \, \12 \, \frac{p}{q+1} \dsl \ \h^{i_1j_1} \ldots\, \h^{i_{p-1}j_{p-1}}\, \g^{\,k_1 \ldots\, k_{q+1}}
\O_{\,i_1j_1,\,\ldots\,,\,i_{p-1}j_{p-1},\,l\,[\,k_1;\,k_2 \ldots\, k_{q+1}]} \, + \, \ldots \\
& + \, \h^{i_1j_1} \ldots\, \h^{i_pj_p}\, \g^{\,k_1 \ldots\, k_q}\,
\mathscr{B}_l \,(\, \O_{\,i_1j_1, \,\ldots\, ,\,i_pj_p;\,k_1 \ldots\, k_q} \,)
\end{split}
\ee
of the Bianchi identities, where the omitted terms contain overall gradients. Notice
that the $\dsl$ terms have the same structure as the terms within brackets in eq.~\eqref{varS}, and
are thus eliminated by the conditions \eqref{chain}, that are hence plausibly
not only sufficient but also necessary to preserve the Bianchi identities. All gradient
terms, that we have not displayed for brevity in \eqref{shiftBgen}, must in fact disappear if eq.~\eqref{chain} holds, since
they cannot cancel against the last term, that vanishes if $\O \sim \cS\,(\Theta)$ with a $\Theta$-parameter satisfying eq.~\eqref{constrtheta}, so that \emph{a fortiori} the $\O$-parameters are also to satisfy the Bianchi identities. In conclusion, if as we expect the conditions \eqref{chain} preserve the Labastida constraints when they are combined with eq.~\eqref{constrtheta}, the transformations \eqref{shiftSgen} share all the key features of the Ansatz
that in Section~\ref{sec:reduction2f} made it possible to classify the sporadic two-family cases
with pathological reductions. This should not surprise the reader: looking for shift symmetries that preserve the Bianchi identities as we did in Section \ref{sec:reduction2f} is equivalent to looking for shift symmetries of the field equations that can be realized directly in terms of $\psi$.

The next issue is to understand whether the shifts of $\cS$ that we have identified
lead to symmetries of the field equation \eqref{eqgenred}, possibly when combined with
proper shifts of the $\cY_{\,ijk}$ spinor-tensors.
Since the structure of the Rarita-Schwinger-like tensor is fully determined by the
Bianchi identities and their $\g$-traces, the promotion of the symmetries of the Bianchi
identities to symmetries of the equation of motion for $\bar{\psi}$ is really expected. For instance, one
can neatly recover it for the simplest Weyl-like transformation of the Labastida tensor,
\be \label{shift1gen}
\d \, \cS \, = \, \g^{\,i}\, \O_{\,i} \, ,
\ee
since our previous analysis selects $\O_{\,i}$ parameters satisfying
\be \label{shiftcond1gen}
(\,D-2\,)\, \O_{\,i} + \, 2\, S^{\,j}{}_{\,i}\, \O_{\,j} \, = \, 0 \, ,
\ee
while eq.~\eqref{shift1gen} induces the variation
\begin{align}
& \d \, \cE \, = \sum_{p\,,\,q\,=\,0}^N  k_{\,p \comma q} \ \h^{i_1j_1} \!\ldots \h^{i_pj_p}\, \g^{\,k_1 \ldots\, k_q}\, [\,Y_{\{2^p,1^q\}}\, T_{i_1j_1} \!\ldots T_{i_pj_p}\, \g_{\,k_1 \ldots\, k_q} \comma \g^{\,l} \,]_{\,(-1)^{\,q+1}} \, \O_{\,l} \nn \\
& + \sum_{p\,,\,q\,=\,0}^N \frac{(-1)^{\,q}}{q+1}\, k_{\,p \comma q} \ \h^{i_1j_1} \!\ldots \h^{i_pj_p}\, \g^{\,k_1 \ldots\, k_{q+1}}\, Y_{\{2^p,1^{q+1}\}}\, T_{i_1j_1} \!\ldots T_{i_pj_p}\, \g_{\,[\,k_1 \ldots\, k_q} \, \O_{\,k_{q+1}\,]} \nn \\
& + \sum_{p\,,\,q\,=\,0}^N \frac{(-1)^{\,q}\,q}{p+1}\, k_{\,p \comma q} \ \h^{i_1j_1} \!\ldots \h^{i_{p+1}j_{p+1}}\, \g^{\,k_1 \ldots\, k_{q-1}}\, Y_{\{2^{p+1},1^{q-1}\}}\, \sum_{n\,=\,1}^{p+1}\, \prod_{r\,\neq\,n}^p \, T_{i_rj_r}\, \g_{\,k_1 \ldots\, k_{q-1}\,(\,i_n}\, \O_{\,j_n\,)} \nn \\
& + \sum_{p\,,\,q\,=\,0}^N \! (-1)^{\,q} \, k_{\,p \comma q} \, \h^{i_1j_1} \!\ldots
\h^{i_pj_p}\, \g^{\,k_1 \ldots\, k_q}\, \g^{\,l}\, Y_{\{3,2^{p-1},1^q\}} \left(\,
Y_{\{2^p,1^q\}}\, T_{i_1j_1} \!\ldots T_{i_pj_p}\, \g_{\,k_1 \ldots\, k_q} \,\right)
\O_{\,l} \label{varEgen}
\end{align}
of the Rarita-Schwinger-like tensor $\cE$ of eq.~\eqref{einsteingen}. In the variation of
eq.~\eqref{eqgenred}, the last line in eq.~\eqref{varEgen} can be compensated by
$\O$-dependent shifts of the $\cY_{\,ijk}$ tensors, on account of its three-column
projection. On the other hand, the first three lines should vanish directly, and they
indeed do provided they combine to rebuild eq.~\eqref{shiftcond1gen}. Starting from the rules
collected in Appendix~\ref{app:MIX}, one can indeed prove that
\begin{align}
& [\, T_{i_1j_1} \!\ldots T_{i_pj_p} \, \g_{\,k_1 \ldots\, k_q} \comma \g^{\,l} \,]_{\,(-1)^{q+1}} \, \O_{\,l}\, = \, (-1)^{\,q} \, \sum_{n\,=\,1}^p
\, \prod_{r\,\neq\,n}^p \, T_{i_rj_r} \, T_{(\,i_n\,|\,[\,k_1\,} \g_{\,k_2 \ldots\, k_q\,]}\, \O_{\,|\,j_n\,)} \label{commg1} \\
& + \, T_{i_1j_1} \!\ldots T_{i_pj_p}\, \g_{\,[\,k_1 \ldots\, k_{q-1}\,|}\! \left\{\, (\,D+q-1\,)\, \O_{\,|\,k_q\,]}\! + \, 2\, S^{\,l}{}_{|\,k_q\,]}\, \O_{\,l} \,\right\} + \sum_{n\,=\,1}^p
\prod_{r\,\neq\,n}^p T_{i_rj_r} \, \g_{\,k_1 \ldots\, k_q\,(\,i_n} \O_{\,j_n\,)} \, , \nn
\end{align}
so that, restricting the attention to two-column Young projections, one can finally show
that
\be \label{commgproj}
\begin{split}
& [\, Y_{\{2^p,1^q\}} \, T_{i_1j_1} \ldots\, T_{i_pj_p} \, \g_{\,k_1 \ldots\, k_q} \comma \g^{\,l} \,]_{\,(-1)^{q+1}} \, \O_{\,l} \, = \, Y_{\{2^p,1^q\}} \, \sum_{n\,=\,1}^p
\, \prod_{r\,\neq\,n}^p \, T_{i_rj_r} \, \g_{\,k_1 \ldots\, k_q\,(\,i_n}\, \O_{\,j_n\,)} \\[2pt]
& + \,Y_{\{2^p,1^q\}} \, T_{i_1j_1} \ldots\, T_{i_pj_p}\, \g_{\,[\,k_1 \ldots\, k_{q-1}\,|} \left\{\, (\,D+p+q-1\,)\, \O_{\,|\,k_q\,]}\, + \, 2\, S^{\,l}{}_{\,|\,k_q\,]}\, \O_{\,l} \,\right\} \, .
\end{split}
\ee
Substituting this identity in eq.~\eqref{varEgen} and combining the terms containing the
same invariant tensors finally gives
\be
\begin{split} \label{varEred}
\d \, \cE \, & = \, \sum_{p\,,\,q\,=\,0}^N \, k_{\,p\,,\,q} \ \h^{i_1j_1} \ldots\, \h^{i_pj_p}\, \g^{\,k_1 \ldots\, k_q}\, \times \\
& \times \, Y_{\{2^p,1^q\}}\, T_{i_1j_1} \ldots\, T_{i_pj_p}\, \g_{\,[\,k_1 \ldots\, k_{q-1}\,|} \left\{\, (\,D-2\,)\, \O_{\,|\,k_q\,]}\, + \, 2\, S^{\,l}{}_{\,|\,k_q\,]}\, \O_{\,l} \,\right\} \, + \, \ldots \, ,
\end{split}
\ee
where we have omitted the last line of eq.~\eqref{varEgen}, since we have already stressed
that proper shifts of the $\cY_{\,ijk}$ tensors can cancel it. This expression extends
the two-family result of eq.~\eqref{shiftE1} and shows clearly that the field equation
\eqref{eqgenred} is left invariant by the transformation \eqref{shift1gen} with
parameters subject to the conditions \eqref{shiftcond1gen}. It is reasonable to expect
that similar results hold for the generic Weyl-like shifts \eqref{shiftSgen} whose
parameters are subject to the conditions \eqref{chain}, but we do not have a complete
proof of this fact, although concrete indications to this effect can be found in Appendix \ref{app:redN}.

Let us conclude by stressing that the emergence of Weyl-like symmetries could be also investigated
directly at the level of the action, and a brief digression on this matter is perhaps
instructive. First of all, let us stress that one should also take into account the
degenerate cases in which the tensor $\cS$ (or, more generally, one of its $\g$-traces
$T_{i_1j_1} \ldots\, T_{i_pj_p}\, \g_{\,k_1 \ldots\, k_q}\, \cS$) vanishes identically.
These are not captured by our previous analysis that was just meant to identify the
conditions under which the equation $\cE = 0$ \emph{does not} reduce to $\cS = 0$.
Actually, we already came across an example of this sort in Section
\ref{sec:reduction2f}, an irreducible $\{1,1\}$ field in two dimensions. And indeed, as
we shall explain in some detail in Appendix~\ref{app:redN}, $\cS \equiv 0$ for all fully
antisymmetric spinor-tensors carrying $N$ space-time indices in $D = N$. Barring these
special cases, we can now discuss how to restate our previous results from this
viewpoint. For brevity, we shall however content ourselves with rephrasing in these terms
the simplest Weyl-like symmetry involving a single $\g^{\,i}$, already analyzed for
two-family fields in Section \ref{sec:reduction2f} and for $N$-family fields here. Let us
therefore consider a transformation of the field $\psi$ of the form
\be
\d \, \psi \, = \, \g^{\, i} \, \Theta_{\, i} \, .
\ee
The corresponding variation of the constrained Lagrangian \eqref{lagferconstr}, to which we restrict our attention for simplicity, reads then
\be \label{Lavarg}
\d \, \cL \, = \, - \, 2 \ \bra \bar{\Theta}_{\, i} \, , \, \g_{\,i} \, \cE \ket \, ,
\ee
thus stressing again that the emergence of Weyl-like symmetries is actually related to conditions on $\g$-traces of $\cE$. Following steps similar to those leading to eq.~\eqref{varEred}, one can show that
\be \label{giE}
\begin{split}
\g_{\,l}\, \cE \, & = \, \left[\, (\,D  -  2\,) \, \d^{\, m}{}_{\, l} \, + \, 2 \, S^{\, m}{}_{\, l}\,\right]\, \times \\
& \times \sum_{p\,,\,q\,=\,0}^N \, (\,q+1\,)\, k_{\,p\,,\,q+1} \ \h^{i_1j_1} \ldots\, \h^{i_pj_p} \, \g^{\,k_1 \ldots\, k_q} \, Y_{\{2^p,1^{q+1}\}}\, T_{i_1j_1} \ldots\, T_{i_pj_p}\, \g_{\,m\,k_1 \ldots\, k_q} \, \cS \, ,
\end{split}
\ee
since the three-column projected terms are annihilated by the Labastida-like constraints \eqref{constrs} on $\cS$. All in all, the invariance of $\cL$ therefore translates into the conditions \eqref{shiftcond1gen}, that we already met for two-family fields in eq.~\eqref{shiftcond1} and that can be recovered in this fashion moving the $S^{\, i}{}_{\, j}$-dependent operator to the left in the scalar product.

\vskip 24pt


\scs{Irreducible fields of mixed symmetry}\label{sec:irreducible}


The gauge fields $\psi$ considered in the previous sections were only symmetric under interchanges of pairs of vector indices belonging to the same set, but did not possess any symmetry relating different sets. In other words, they were \emph{reducible} $gl(D)$
spinor-tensors. Fields of this type naturally emerge in String Theory, where they are
associated to products of bosonic oscillators, together with more general reducible
$gl(D)$ tensors, some sets of whose indices are actually antisymmetrized rather than
symmetrized. We shall explain how to adapt our construction to this other class of fields in Section \ref{sec:multiforms}, but in the following we would like to concentrate briefly on \emph{irreducible} $gl(D)$ spinor-tensors, still with manifestly symmetric index sets. Fields of this type can be associated to Young diagrams, and are more
akin to the familiar low-spin examples. In contrast with the more familiar (Majorana-)Weyl projections, these irreducibility conditions only affect in a non-trivial fashion the vector indices carried by the $\psi$ fields, so that the changes induced in the present case are actually along the lines of what we already described in \cite{bose_mixed} for Bose fields. For completeness, however, we now briefly adapt the irreducibility constraints to the present fermionic setting and display their solutions. We also present an instructive counting argument for the propagating degrees of freedom
for irreducible two-family spinor-tensors.

Irreducibility requires that a vanishing result obtain if one tries to symmetrize all vector indices of $\psi$ belonging to a given family with any additional index of the lower lines of its Young diagram.  In terms of the $S^{\,i}{}_j$ operators introduced in Appendix \ref{app:MIX}, this condition reads
\be \label{condirr}
S^{\,i}{}_j \, \psi \, = \, 0 \, , \quad \forall \ \, i < j \, , \quad 2 \leq j \leq N \, ,
\ee
where $N$ is the number of families. The Fang-Fronsdal-Labastida kinetic operator
\eqref{fermikin} \emph{commutes with the $S^{\,i}{}_j$ operators}, as was the case for
its bosonic counterpart \cite{bose_mixed}, and in a similar fashion the
Rarita-Schwinger-like tensors of eq.~\eqref{einsteingen} also commute with the
$S^{\,i}{}_j$ operators, on account of the relations
\be
[\,S^{\,m}{}_n \comma \h^{i_1j_1} \ldots\, \h^{i_pj_p}\, \g^{\,k_1 \ldots\, k_q}\, T_{i_1j_1} \ldots\, T_{i_pj_p}\, \g_{\,k_1 \ldots\, k_q} \,] \, = \, 0 \, ,
\ee
or more simply because they are ``family'' scalars.
As a consequence, in the constrained theory the Lagrangians
can be presented in \emph{exactly} the same form \eqref{lagferconstr} and
\eqref{solgenf} for both reducible and irreducible gauge fields. For
irreducible fields, however, not all $\g$-traces of $\psi$ remain independent, and therefore it is also possible to reduce the number of terms appearing in them, although the redundant description is anyway correct and has the further advantage of being universal. An illustration of this fact may be found in Section 4 of \cite{bose_mixed}.

At any rate, in general irreducible $\psi$ fields involve fewer
independent gauge parameters and constraints with respect to their
reducible counterparts, and it is interesting to characterize
them. The gauge transformations
\eqref{gaugef} take the same form as their bosonic analogues, and
therefore the considerations of \cite{bose_mixed} apply verbatim. In
particular, the requirement that the constraints \eqref{condirr} be preserved
by gauge transformations leads to
\be \label{parirr1}
\pr^{\,k}\, \left(\, S^{\,i}{}_j \, \e_{\,k} \, + \, \d_{\,k}{}^{\,i} \, \e_{\,j} \,\right) \, = \, 0 \, , \quad \forall \ \, i < j \, , \quad 2 \leq j \leq N \, ,
\ee
and thus, up to gauge-for-gauge transformations, to
\be \label{parirr}
S^{\,i}{}_j \, \e_{\,k} \, + \, \d_{\,k}{}^{\,i} \, \e_{\,j} \, = \, 0 \, , \quad \forall \ \, i < j \, , \quad 2 \leq j \leq N \, .
\ee
As for Bose fields, the $\e_{\,i}$ that solve this system can be built from the available Young-projected $\widetilde{\e}_{\,i}$, that exist only when the corresponding Young diagrams, obtained stripping one box from the original diagram of $\psi$, are admissible. With this proviso, one can solve eqs.~\eqref{parirr}, obtaining
\be \label{parirrsol}
\e_{\,N-k} \, = \, \widetilde{\e}_{\,N-k} \, + \, \sum_{l\,=\,1}^k \ \sum_{i_1\,>\,\ldots\,>\,i_l\,\geq\,0}^{k-1} \, \left[\, l_{\,k} \,\right]_{\,i_1 \ldots\, i_l}\, S^{\,N-i_1}{}_{N-k}\, S^{\,N-i_2}{}_{N-i_1}\, \ldots\, S^{\,N-i_l}{}_{N-i_{l-1}} \, \widetilde{\e}_{\,N-i_l}\, ,
\ee
where
\be \label{parirrdetail}
\left[\, l_{\,k} \,\right]_{\,i_1 \ldots\, i_l} \, = \, \frac{1}{s_{N-i_l}-s_{N-k}+i_{l}-k}\, \prod_{h\,=\,1}^l \,  \frac{1}{s_{N-i_l}-s_{N-i_h}+\,i_l-i_h} \, .
\ee
For each of the original gauge parameters $\epsilon_{\,i}$, the expansion \eqref{parirrsol} in powers of the $S^{\,i}{}_j$ operators thus starts with the corresponding irreducible $\widetilde{\epsilon}_{\,i}$, if this exists, but in higher orders it also receives contributions from other irreducible $\widetilde{\e}_{\, N-k}$ parameters corresponding to lower values of $k$. Let us stress again that the irreducible parameters are in one-to-one correspondence with the admissible Young diagrams obtained stripping \emph{one} box from the original diagram of the spinor-tensor $\psi$. As a result, they are in general fewer than the number of families. For instance, a set of $k$ lines of identical length gives rise to $k$ distinct gauge parameters in the reducible case but to a single gauge parameter in the irreducible case. This was first pointed out in \cite{firstlaba,labferm}, and here we are thus providing an explicit derivation of the result.

Eqs.~\eqref{parirr} induce an identical set of conditions on the $\Psi_{\,i}$
compensators,
\be \label{compirr}
S^{\,m}{}_n \, \Psi_{\,i} \, + \, \d^{\,m}{}_{\,i} \, \Psi_{\,n} \, = \, 0 \, , \quad \forall \ \, m < n \, , \quad 2 \leq n \leq N \, ,
\ee
and computing a $\g$-trace of eq.~\eqref{compirr} then leads to the
relations satisfied by the composite compensators $\x_{\,ij}(\Psi)$,
\be \label{xijirr}
S^{\,m}{}_n \, \x_{\,ij}(\Psi) \, + \, \d^{\,m}{}_{\,(\,i}\, \x_{\,j\,)\,n}(\Psi) \, = \, 0 \, , \quad \forall \ \, m < n \, , \quad 2 \leq n \leq N \, .
\ee
In a similar fashion, applying $S^{\,m}{}_n$ to the symmetrized triple $\g$-trace of the
gauge field $\psi$, one can identify the conditions satisfied
by the constraint tensors $\cZ_{\,ijk}$, and consequently by the Lagrange
multipliers $\lambda_{\,ijk}$,
\be \label{lijkirr}
S^{\,m}{}_n \, \l_{\,ijk} \, + \, \d^{\,m}{}_{\,(\,i}\, \l_{\,jk\,)\,n} \, = \, 0 \, , \quad \forall \ \, m < n \, , \quad 2 \leq n \leq N \, .
\ee
In analogy with the result described for the gauge parameters, one can see that eqs.~\eqref{lijkirr} associate an independent irreducible Lagrange multiplier to each
admissible Young diagram obtained stripping \emph{three} boxes from that of the
original spinor-tensor gauge field. Actually, the $S^{\,i}{}_j$ operators satisfy Leibnitz's rule, and as a result explicit solutions of eqs.~\eqref{xijirr} and \eqref{lijkirr} can be obtained tensoring the solutions \eqref{parirrsol} of eq.~\eqref{parirr}.

This concludes our cursory look at the restrictions on the field content in the irreducible case. As we already stated, similar conditions can be used to relate the irreducible Young projections of the various $\g$-traces appearing in the Rarita-Schwinger-like tensors, in complete analogy with what was illustrated in \cite{bose_mixed} for some types of Bose fields.

An important, related issue, has to do with the propagating modes of these systems. While the argument presented in Section \ref{sec:dofs} can be adapted to this irreducible case
by simply enforcing on $\psi$ the irreducibility conditions \eqref{condirr}, possibly together with the available (Majorana-)Weyl conditions, it is perhaps instructive to
count in detail the propagating modes, taking into account all Labastida constraints
on the independent components of the field $\psi$ and of the parameters selected by
the procedure that we have just illustrated. We shall content ourselves with the
case of generic irreducible two-family gauge fields~\footnote{The fields could be also subject to (Majorana-)Weyl projections, but these do not play a direct role in our arguments. For brevity, we assume that $s_1> s_2+1>2$ in order to leave out special degenerate cases, and that $D>5$ so that the corresponding $\{s_1,s_2\}$ representations of $o(D-2)$ exist.} $\psi_{\,\mu_1 \ldots\, \mu_{s_1} ,\,\nu_1 \ldots\, \nu_{s_2}}$,
since the combinators behind this argument becomes readily clumsy as the number of
index families increases.

Following the procedure proposed in \cite{siegelcount}, the number of physical degrees of freedom propagated by an $\{s_1,s_2\}$ Young projected two-family gauge field $\psi$ is
\be
N_{\,\textrm{d.o.f.}} \, = \, \frac{1}{2}\, \left(\, N_{\,\psi} \, -3\, N_{\,\tilde{\e}_{\,i}} \, + \, 5\,  N_{\,\tilde{\e}_{\,[\,ij\,]}} \,\right) \, ,
\ee
where $N_{\,\psi}$ , $N_{\,\tilde{\e}_{\,i}}$ and $N_{\,\tilde{\e}_{\,[\,ij\,]}}$ denote the numbers of independent components of the constrained irreducible gauge field $\psi$, of the two constrained irreducible gauge parameters $\widetilde{\e}_{\,i}$ and of the single constrained irreducible gauge-for-gauge parameters $\widetilde{\e}_{\,[\,ij\,]}$. In order to identify these contributions, let us begin by decomposing the spinor-tensors of interest in $o(D)$ representations. This procedure differs slightly from the one used in the examples of \cite{labastida}, since it focusses directly on quantities that survive the Labastida constraints \eqref{constrpsi} and \eqref{constrgaugef}, bypassing somehow the potential difficulties introduced by the linear dependence of the constraints.

The decompositions of $gl(D)$ unconstrained two-family gauge fields $\psi$ and gauge parameters $\widetilde{\e}_{\,i}$ in terms of $o(D)$ representations would read
\be
\begin{split}\label{decomppsi}
\psi : \: \{s_1,s_2\}_{gl} = \ & \{s_1,s_2\}_o \oplus\, \{s_1-1,s_2\}_o \oplus\, \{s_1,s_2-1\}_o \oplus\, \{s_1-2,s_2\}_o  \oplus\, \{s_1,s_2-2\}_o \\
& \oplus \, 2\, \{s_1-1,s_2-1\}_o \oplus\, 2\, \{s_1-2,s_2-1\}_o \oplus\, 2\, \{s_1-1,s_2-2\}_o \\
& \oplus\, 3\, \{s_1-2,s_2-2\}_o \oplus\, \ldots \ ,
\end{split}
\ee
where the $gl$ and $o$ subscripts are meant to stress that on the left-hand side we refer to $gl(D)$ representations, that as such include all ($\g$-)traces, while on the right-hand side we refer to $o(D)$ representations, where all ($\g$-)traces have been removed. In a similar fashion
\begin{align}
\widetilde{\e}_{\,1} : \: \{s_1-1,s_2\}_{gl} = \ & \{s_1-1,s_2\}_o \oplus\, \{s_1-2,s_2\}_o \oplus\, \{s_1-1,s_2-1\}_o \oplus\, \{s_1-1,s_2-2\}_o \nn \\
& \oplus\, 2\, \{s_1-2,s_2-1\}_o \oplus\, \ldots \ , \nn \\[5pt]
\widetilde{\e}_{\,2} : \: \{s_1,s_2-1\}_{gl} = \ & \{s_1,s_2-1\}_o \oplus\, \{s_1-1,s_2-1\}_o \oplus\, \{s_1,s_2-2\}_o \oplus\, \{s_1-2,s_2-1\}_o \nn \\
& \oplus\, 2\, \{s_1-1,s_2-2\}_o \oplus\, \ldots \ , \label{decompei}
\end{align}
where we have refrained from displaying terms that are manifestly related to the Labastida constraints. These include all contributions to
$\psi$ obtained removing at least three boxes from the same row and all contributions to
the $\widetilde{\e}_{\,i}$ obtained removing at least two boxes from the same row.
For two-family fields it is then relatively simple to sort out in eqs.~\eqref{decomppsi} and \eqref{decompei} the $o(D)$ components that survive the Labastida constraints. Denoting by $[\,s_1,s_2\,]$ the dimension of the $o(D)$ representation with $s_1$ boxes in its first row and $s_2$ boxes in its second row, one thus obtains
\be
\begin{split}
N_{\,\psi} \, & = \, [\,s_1,s_2\,] \, + \, [\,s_1-1,s_2\,] \, + \, [\,s_1,s_2-1\,] \, + \, [\,s_1-2,s_2\,] \, + \, 2\, [\,s_1-1,s_2-1\,] \\
& + \, [s_1,s_2-2] \, + \, [\,s_1-2,s_2-1\,] \, + \, [\,s_1-1,s_2-2\,] \, + \, [\,s_1-2,s_2-2\,]
\end{split}
\ee
and
\be
N_{\,\tilde{\e}_{\,i}} \, = \, [\,s_1-1,s_2\,] \, + \, [\,s_1,s_2-1\,] \, + \, [\,s_1-1,s_2-1\,] \, + \, [\,s_1-2,s_2-1\,] \, + \, [\,s_1-1,s_2-2\,] \, .
\ee
Furthermore, the same constraints imply that the only available $\widetilde{\e}_{\,[\,ij\,]}$ is $\g$-traceless, so that
\be
N_{\,\tilde{\e}_{\,[\,ij\,]}} \, = \, [\,s_1-1,s_2-1\,] \, .
\ee
Taking into account all these contributions and making use of eq.~\eqref{dim1}, one is finally led to
\be\label{fermidofcount}
N_{\,\textrm{d.o.f.}} \, = \, 2^{\,\left[\frac{D-2}{2}\right]}\, \frac{s_1-s_2+1}{(\,s_1+1\,)\,!\,s_2\,!} \, \frac{(\,D+s_1-5\,)\,!}{(\,D-4\,)\,!} \, \frac{(\,D+s_2-6\,)\,!}{(\,D-6\,)\,!} \, (\,D+s_1+s_2-4\,) \, ,
\ee
which is precisely the dimension of the $\{s_1,s_2\}$ irreducible representation of $o(D-2)$, up to a possible overall factor reflecting a (Majorana-)Weyl projection that we have not considered for brevity. Similar results can be presented for Bose fields, and in principle one could also adapt these arguments to fields with larger numbers of index families, taking into account larger numbers of leftover $o(D)$ components and the presence of further gauge-for-gauge parameters.

\vskip 24pt


\scss{An application: some Lagrangians with Weyl-like symmetries}\label{sec:irr_example}


The search for Weyl-invariant Lagrangians can be performed considering directly irreducible fields and, as we shall see shortly, the relations displayed in the previous section greatly simplify the analysis of the conditions identifying those symmetries. We thus conclude this section by displaying some non-trivial solutions of eqs.~\eqref{chain}. To begin with, we can show how to solve \emph{all} these conditions for the special case of single-column spinor-tensors carrying $s$ vector indices. In Section~\ref{sec:topology} we
presented concise forms for their Lagrangians, that vanish manifestly for $s \leq D \leq 2\,s$, and we noticed that the formal expressions for their field equations in terms of $\cS$ should possess Weyl-like symmetries. We can now recover this result solving the conditions \eqref{chain} for spinor-tensor forms. First of all, in this case the $\h^{ij}$ tensors do not enter Lagrangians and field equations, so that the
chains of conditions \eqref{chain} reduce to the far simpler independent equations
\be \label{eqantisymm}
(\,D-q-1\,)\ \Theta^{\,[\,q\,]}{}_{\,k_1 \ldots\, k_{q-1}l} + 2 \, S^{\,m}{}_l \, \Theta^{\,[\,q\,]}{}_{\,k_1 \ldots\, k_{q-1}m} = \, 0 \, ,
\ee
whose solutions identify Weyl-like shifts of the form
\be \d \, \psi \, = \, \g^{\,k_1 \ldots\, k_q}\, \Theta^{\,[\,q\,]}{}_{\,k_1 \ldots\,
k_q} \, . \ee
Let us also stress that no other conditions are to be imposed on these parameters, because all spinor-tensor forms are \emph{unconstrained}.
Furthermore, for this type of fields $S^{\,m}{}_l\, \psi = 0$ for all $l\neq m$,
since it is clearly impossible to symmetrize their space-time indices. As a result
\be
0 \, = \, S^{\,m}{}_l \, \d\, \psi \, = \, \g^{\,k_1 \ldots\, k_q}\, \left(\, S^{\,m}{}_l\, \Theta^{\,[\,q\,]}{}_{\,k_1 \ldots\, k_q} + \, (-1)^{\,q+1}\, \d^{\,m}{}_{[\,k_1} \Theta^{\,[\,q\,]}{}_{\,k_2 \ldots\, k_q\,]\,l} \,\right) \, , \qquad \, l \neq m \, ,
\ee
so that eq.~\eqref{eqantisymm} becomes
\be \label{eqantisymmbis}
(\,D-2\,N\!+q-1\,)\ \Theta^{\,[\,q\,]}{}_{\,k_1 \ldots\, k_{q-1}l} \, = \, 0 \, .
\ee
Fully antisymmetric fields can admit a single independent $q$-th $\g$-trace,
that as we now see is left undetermined in $D = 2\,N\!-q+1$,
since it can be shifted by non-vanishing $\Theta^{\,[\,q\,]}$ parameters. For one-column
fields the number $N$ of families appearing in eq.~\eqref{eqantisymmbis} is equivalent to
the total number $s$ of space-time indices carried by the field $\psi$.
Therefore, as expected, we are recovering the existence of
Weyl-like symmetries for these fields for $s+1 \leq D \leq 2\,s$. Notice, finally,
that the Lagrangians
\eqref{lag_anti} also vanish manifestly for $D = s$, but in these cases the
$\cS$ spinor-tensors vanish as well, as one can see computing explicitly their only
available component. This explains why these special space-time dimensions do not emerge
from the previous discussion. Further comments on this case can be found in Appendix \ref{app:redN}.

We can now supplement the classification of Section \ref{sec:reduction2f} of the Weyl-like symmetries emerging in the two-family case by displaying some interesting classes of irreducible $N$-family fields that allow Weyl-like shifts of the form
\be\label{Weyl1G}
\d \, \psi \, = \, \g^{\,i} \, \Theta_{\,i} \, .
\ee
In this case the irreducibility of $\psi$ translates into the conditions
\be \label{irredO}
\d^{\,i}{}_k \, \Theta_{\, j} \, +\, S^{\,i}{}_j \, \Theta_{\, k} \, = \, 0 \qquad (\, i < j \,)
\ee
on the $\Theta_{\,i}$, to be combined with the conditions allowing \eqref{Weyl1G} to define a Weyl symmetry, that were given in eq.~\eqref{shiftcond1gen} and that we can rewrite in the convenient form \footnote{Here and in the rest of this section no summations are left implicit.}
\be \label{shiftcond1gen2}
(\,D+2\, s_i -4\,)\, \Theta_{\,i} + \, 2\, \sum_{j\,<\,i} S^{\,j}{}_{\,i}\, \Theta_{\,j} \,+\, 2\, \sum_{j\,>\,i} S^{\,j}{}_{\,i}\, \Theta_{\,j} \, = \, 0 \qquad (\,i=1\,,\, \ldots \,, N\,)\ ,
\ee
where $s_i$ denotes the number of Lorentz labels in the $i$-th family for the spinor-tensor $\psi$. Eqs.~\eqref{irredO} finally reduce \eqref{shiftcond1gen2} to
\be \label{shiftcond1gen3}
\left[\,D+2\, s_i -2\,(\,i+1\,)\,\right]\, \Theta_{\,i}\, - \, 2\, \sum_{j\,>\,i} S^{\,j}{}_{\,i}\, S^{\,i}{}_{\,j}\, \Theta_{\,i} \, = \, 0 \qquad (\,i=1\,,\, \ldots \,, N\,)\ .
\ee

The sum is absent in the last of these equations, that is simply
\be
\left[\,D+2\, s_N -2\,(\,N+1\,)\,\right]\, \Theta_{\,N} \, = \, 0 \, ,
\ee
and selects the condition
\be\label{DsN}
D+2\, s_N \, = \, 2\,(\,N+1\,)
\ee
to recover a Weyl-like symmetry with a non-vanishing $\Th_{\,N}$ parameter. As a result, one can conclude right away that no solutions of this kind exist in odd dimensions, or if $D > 2\,N$, consistently with the tables of Section \ref{sec:reduction2f} and with the considerations made at the beginning of Section \ref{sec:reductionf}. Inverting the diagonal equations in \eqref{irredO} would give
\be \label{irrinv}
\Theta_{\,i} \, = \, - \, \frac{1}{s_i-s_j+1}\ S^{\,j}{}_{\,i}\, \Theta_{\,j} \qquad (\, i < j \,) \ ,
\ee
up to possible homogeneous solutions. At any rate, if $\Theta_{\,N}$ is non-trivial, all the $\Theta_{\,i}$ for $i<N$ are also non trivial, since in our conventions $s_i \geq s_j$.

Turning to the other equations with $i<N$, one can notice that the irreducibility conditions \eqref{irredO} also imply that the operators $S^{\,j}{}_{\,i}\, S^{\,i}{}_{\,j}$ act \emph{diagonally} on $\Theta_{\,i}$. In fact, $S^{\,j}{}_{\,i}$ and $S^{\,i}{}_{\,j}$ are but the $L_-^{(ij)}$ and $L_+^{(ij)}$ operators for the $su(2)$ subalgebra of $gl(N)$ connecting the two rows $i$ and $j$, for which
\be
L_3^{(ij)} \, = \, \frac{1}{2} \, \left(\, S^{\,i}{}_{\,i} \, - \, S^{\,j}{}_{\,j} \,\right)
\ee
while eq.~\eqref{irredO} implies that
\be
L_+^{(ij)}\, \Theta_{\,j} = 0 \ , \qquad \Theta_{\,j} \, = \, - \, L_+^{(ij)}\, \Theta_{\,i} \, .
\ee
As a result, $\Theta_{\,i}$ and $\Theta_{\,j}$ have the same ``total angular momentum'' quantum number
\be
\ell_{\,ij}\,(\, \Theta_{\,i} \,) \, = \, \ell_{\,ij}\,(\, \Theta_{\,j} \,) \, = \, \frac{s_i - s_j+1}{2} \, ,
\ee
that is also the ``magnetic'' quantum number of $\Theta_{\,j}$, that lies at the tip of the $su(2)$ chain and lacks one space-time index of the $j$-th row when compared to the gauge field $\psi$. At the same time, the ``magnetic'' quantum number of $\Theta_{\,i}$, that lacks an index in the $i$-th row compared to $\psi$, is
\be
m_{\,ij}\,(\, \Theta_{\,i} \,) \, = \, \frac{s_i - s_j-1}{2} \, ,
\ee
so that finally
\be
S^{\,j}{}_{\,i}\, S^{\,i}{}_{\,j}\, \Theta_{\,i} \, = \, (\,s_i - s_j+1\,)\, \Theta_{\,i} \, .
\ee
In conclusion, the remaining equations \eqref{shiftcond1gen3} with $i<N$ reduce to the chain of conditions
\be \label{conds1gamma}
\left[\, (\,N-i-1\,) \, s_i - \sum_{j\,=\,i+1}^{N-1} s_j \,\right] \, \Theta_{\,i} \, = \, 0\, \qquad (\,i=1,\,\ldots\,, N-2\,) \, ,
\ee
while the $(N-1)$-th condition is also identically satisfied on account of eq.~\eqref{DsN}. Notice that, as we anticipated, on account of the irreducibility all the $\Theta_{\,i}$ must be non-vanishing to define a non-trivial solution when eq.~\eqref{DsN} holds. One thus identifies the $\{s,\ldots,s,s_N\}$ irreducible fields with $s$ arbitrary and $s_{N}\leq N$ in the space-time dimensions determined by eq.~\eqref{DsN}. We shall content ourselves with this general class of solutions, although others with vanishing $\Theta_{\,N}$ exist in low-enough dimensions.

The solutions that we identified are also to preserve the Labastida constraints \eqref{constrpsi}, and in eq.~\eqref{tripleOi} we have seen that, whenever eqs.~\eqref{shiftcond1gen2} hold, for the Weyl-like shifts that we are considering
this is guaranteed by the additional conditions
\be \label{triplered}
T_{(\,ij}\, \g_{\,k\,)}\, \Theta_{\,l} \, = \, 0 \, .
\ee
On the other hand, the conditions \eqref{irredO} imply that all $\Theta_{\, l}$
parameters have the same $gl(D)$ structure as the irreducible $\Theta_{N}$, that in this
sense represents the most convenient choice for the single independent quantity left over
in eq.~\eqref{shiftcond1gen2}. Therefore, the transformations of the $\{s,\ldots,s,1 \}$
fields in $D = 2\,N$ rest on this $\{s,\ldots,s,0 \}$ conformal Weyl parameter, that in this
space-time dimension admits a $\g$-traceless component that as such cannot be affected by
the triple $\g$-trace constraints \eqref{triplered}. This is consistent with the fact
that for two-family fields we did find explicitly that $\{s,1\}$ fields admit Weyl-like
symmetries of the type \eqref{Weyl1G} compatible with the Labastida constraints. In this fashion, we have shown that Weyl-like symmetries emerge in a class of examples that are richer that fully antisymmetric spinor-tensors, and already display the key features of generic mixed-symmetry fields. On the
other hand, for higher values of $s_N$ a standard theorem already quoted in Section
\ref{sec:topology} (for which we refer the reader to the first reference in \cite{branching}, \textsection $10$-$6$) implies that a $\g$-traceless component does not exist for $\{s,\ldots,s,s_N-1\}$-projected parameters in $D = 2\,(\,N-s_N+1\,)$, so that the previous considerations protecting this type of shift are no
more available. While we do not have a complete argument, our explicit analysis of
two-family fields makes us expect that the triple $\g$-trace conditions eliminate at least some of these solutions \eqref{DsN} in low enough space-time dimensions. Let us close this section by stressing that even these new classes of solutions emerge in space-time dimensions $D \leq 2\,N + 1$, in agreement with the considerations made at the beginning of Section \ref{sec:reductionf}. Moreover, the standard theorem just recalled shows that \emph{all} these fields do not propagate any local degrees of freedom. We do not
know whether all these $\{s,\ldots,s,s_N\}$ models have vanishing actions, but in Section \ref{sec:topology} we have already seen that the simplest member of this class, a $\{2,1\}$ field in two dimensions, does not, so that it is reasonable to expect that they all behave in the same fashion.

\vskip 24pt


\scs{Multi-form spinor-tensors}\label{sec:multiforms}


In the previous sections we have described the properties of multi-symmetric gauge fields
of the type $\psi_{\,\m_1 \ldots \m_{s_1},\,\n_1 \ldots\, \n_{s_2}\,,\, \ldots\,}$, that in
general propagate the degrees of freedom of reducible representations of the Poincar\'e
group. As we have seen, this ``redundant'' reducible description is particularly
convenient to manifest the structure of the algebraic constraints on fields and gauge parameters, but it is also tailored for the comparison with massive string
spectra, where fields of this type are the natural counterparts of products of bosonic
oscillators. Superstring spectra, however, also involve fermionic oscillators, and it is
thus interesting to describe how the previous results can be adapted to the case of
multi-form fermionic gauge fields, for which the indices belonging to a given family are fully
antisymmetric. As we shall see shortly, the algebraic properties of the basic operators
are altered to some extent, but the structure of the whole construction remains
pretty much unaffected in this variant of the formalism.

The first ingredient for moving from multi-symmetric fields
to multi-forms is a proper definition of the operators that we used
in the preceding sections. As we have seen in Section \ref{sec:dofs}, one
can describe multi-symmetric spinor-tensors in an explicit oscillator
realization contracting their vector indices with commuting vectors
$u^{\,i\,\m}$, as in eq.~\eqref{psiosc}.
One can then adapt the basic operations of eq.~\eqref{op_comm}
to the case of multi-form gauge fields replacing the
\emph{commuting} $u^{\,i\,\m}$ with
\emph{anticommuting} $\theta^{\,i\,\m}$ vectors, so that
\begin{alignat}{2}
& \pr_{\,i} \, \equiv \, \pr^{\,\m} \, \frac{\pr}{\pr \, \theta^{\,i\,\m}} \, , \qquad\qquad & & \pr^{\,i} \, \equiv \, \pr_{\,\m} \, \theta^{\,i\,\m} \, , \nn \\
& \g_{\,i} \, \equiv \, \g^{\,\m} \, \frac{\pr}{\pr \, \theta^{\,i\,\m}} \, , \qquad\qquad & & \g^{\,i} \, \equiv \, - \, \g_{\,\m} \, \theta^{\,i\,\m} \, , \label{op_anticomm}
\end{alignat}
where for convenience we have also flipped the sign in the
definition of the $\g^{\,i}$~\footnote{With the definitions of
eq.~\eqref{op_anticomm}, the terms in the Lagrangians that already
show up for spin-$3/2$ fields retain the usual coefficients.}. Moreover, it is convenient to adopt the conventions
\be
\{\, \g_{\,\m} \comma \theta^{\,i\,\n} \,\} \, = \, 0 \, , \qquad\qquad \{\, \g_{\,\m} \comma \frac{\pr}{\pr \, \theta^{\,i\,\n}} \,\} \, = \, 0 \, .
\ee
The key ingredients of our construction are the algebraic properties of
the operators that we have introduced, and with the definitions \eqref{op_anticomm}
divergences and gradients become anticommuting objects satisfying
\be
\{\, \pr_{\,i} \comma \pr^{\,j} \,\} \, = \, \Box \, \d_{\,i}{}^{\,j} \, ,
\ee
while the $\g$'s now satisfy the algebra
\begin{align}
& [\, \g_{\,i} \comma \g_{\,j} \,] \, = \, 2 \, T_{ij} \, , \nn \\
& [\, \g^{\,i} \comma \g^{\,j} \,] \, = \, - \, 4 \, \h^{\,ij} \, ,
\end{align}
that defines the \emph{antisymmetric} $T_{ij}$ and $\h^{\,ij}$
operators of the present setting.

Furthermore, multi-form gauge fields $\psi$ have the gauge
transformations
\be
\d \, \psi \, = \, \pr^{\,i} \, \e_{\,i} \, ,
\ee
that maintain the form of eq.~\eqref{gaugef} but now
admit the gauge-for-gauge transformations
\be
\d \, \e_{\,i} \, = \, \pr^{\,j}\, \e_{\,(\,ij\,)} \, , \quad \d\, \e_{\,(\,ij\,)} \, = \, \pr^{\,k}\, \e_{\,i\,(\,jk\,)}\, , \quad \ldots
\ee
with parameters that are \emph{symmetric} under the interchange of
their family indices. This property reflects the well-known fact
that gauge-for-gauge transformations are already present for
one-family fields, that correspond to spinorial forms in the
antisymmetric basis. Taking into account that
\be
\{\, \pr_{\,i} \comma \dsl \, \}\, = \, 0
\ee
and following almost verbatim the reasoning that led Labastida
\cite{labferm} to the $\cS$ tensor of eq.~\eqref{fermikin}, one can
now identify the basic kinetic tensor of the theory,
\be \label{sform}
\cS \, = \, i \left(\, \dsl \, \psi \, + \, \pr^{\,i} \! \psisl_i \,\right) \, ,
\ee
as the gauge invariant completion of $\not \! \pr \, \psi$.
As in the multi-symmetric setting, the gauge invariance of $\cS$ can
be attained at the price of suitable constraints on the gauge
parameters,
\be \g_{\,[\,i}\,\e_{\,j\,]} \, = \, 0 \, , \label{constrganti} \ee
since
\be
\d \, \cS \, = \, \frac{i}{2} \ \pr^{\,i} \pr^{\,j} \, \g_{\,[\,i}\,\e_{\,j\,]} \,
.
\ee
In analogy with the presentation of Section \ref{sec:lagrangian2f},
unconstrained equations of motion obtain performing the
Stueckelberg-like shift
\be
\psi \ \to \ \psi \,-\, \partial^{\,i}\, \Psi_{\,i}  \quad \textrm{with} \quad \d \, \Psi_{\,i} \, = \, \e_{\,i} \, ,
\ee
that builds the unconstrained gauge invariant kinetic tensor
\be
\cW \, = \, \cS \, - \, i \ \pr^{\,i} \pr^{\,j} \, \x_{\,ij}\,(\Psi) \, .
\ee
As in the multi-symmetric case, only the composite compensator
fields
\be
\xi_{\,ij}\,(\Psi) =
\frac{1}{2} \ \gamma_{\,[\,i} \,
\Psi_{j\,]} \, , \label{xianti}
\ee
that are now \emph{antisymmetric}, appear in $\cW$, due to the
constrained gauge invariance of $\cS$.

The other key ingredients of the construction developed in the
previous sections were the Bianchi identities satisfied by $\cS$,
that now become
\be \label{bianchiform} \pr_{\,i}\,\cS \, - \, \12 \dsl \,
\g_{\,i}\,\cS \, - \, \12 \ \pr^{\,j}\,T_{ij}\,\cS \, - \,
\frac{1}{6} \ \pr^{\,j}\,\g_{\,ij}\,\cS \, = \, - \, \frac{i}{6} \
\pr^{\,i}\pr^{\,j} \, T_{[\,ij}\,\g_{\,k\,]}\, \psi \, , \ee
where one can recognize the first two members of the family of
\emph{symmetric} operators
\be \g_{\,i_1 \ldots \, i_n} \, \equiv \, \frac{1}{n!} \
\g_{\,(\,i_1} \g_{\phantom{)}\!i_2} \ldots \, \g_{\,i_n \, )} \, . \ee
The Bianchi identities \eqref{bianchiform} identify the analogues of
the Labastida constraints \eqref{labacf} for multi-form fields,
\be
T_{[\,ij}\,\g_{\,k\,]}\, \psi \, = \, 0 \, , \label{constranti}
\ee
that can be enforced off-shell, from the very beginning, or equivalently
only on-shell, via Lagrange multipliers $\l_{\,ijk}$, that are now \emph{totally
antisymmetric} in their family indices. For brevity, in the following we shall
confine ourselves to showing how the details that led to the constrained Lagrangians are
modified in the present multi-form setting, since the unconstrained Lagrangians behave similarly.

First of all, one can verify by a direct computation that for
multi-forms the constraints \eqref{constranti} induce the relations
\be
T_{[\,ij}\,\g_{\,k\,]}\, \cS \, = \, 0 \, ,
\ee
so that the only non vanishing combinations of $\g$-traces of $\cS$
are now \emph{two-row} projected,
\be
(\,\g^{\,[\,q\,]}\,\cS^{\,[\,p\,]}\,)_{\, k_1 \ldots\, k_q;\,i_1 j_1;\,\ldots\,;\,i_p j_p} \, \equiv \, Y_{\{p+q \comma p\}} \, T_{i_1j_1} \ldots\, T_{i_pj_p} \, \g_{\,k_1 \ldots\, k_q} \, \cS \, .
\ee
This property can be easily proved adapting the arguments of
Appendix \ref{app:idsf}. Working in the \emph{antisymmetric basis}
for Young tableaux is particularly convenient in this context, while
the relevant consequences of the Bianchi identities are now their
maximally symmetric \emph{two-row} projected $\g$-traces. Starting from
\begin{align}
& [\, T_{i_1j_1} \ldots\, T_{i_pj_p} \comma \pr^{\,l} \,] \, = \, - \, \sum_{n\,=\,1}^p \, \pr_{\,[\,i_n}\, \d_{\,j_n\,]}{}^{\,l} \, \prod_{r\,\neq\,n}^p \, T_{i_rj_r} \, , \\
& [\, \g_{\,k_1 \ldots\, k_q} \comma \dsl \ ] \, = \, - \, 2 \ \pr_{\,(\,k_1} \g_{\,k_2 \ldots\, k_q\,)} \, , \\[7pt]
& [\, \g_{\,k_1 \ldots\, k_q} \comma \pr^{\,l} \,] \, = \ \dsl \ \g_{\,(\,k_1 \ldots\, k_{q-1}\,} \d_{\,k_q\,)}{}^{\,l} \, - \, \pr_{\,(\,k_1} \g_{\,k_2 \ldots\, k_{q-1}\,} \d_{\,k_q\,)}{}^{\,l} \, ,
\end{align}
and considering the composition rules for the $\g_{\,k_1 \ldots\, k_q}$ operators,
\begin{align}
& \g_{\,k_1 \ldots\, k_q} \, \g_{\,l} \, = \, \g_{\,k_1 \ldots\, k_q \, l} \, + \, \g_{\,(\,k_1 \ldots\, k_{q-1}} \, T_{k_q\,)\,l} \, , \\[3pt]
& \g_{\,k_1 \ldots\, k_q} \, \g_{\,lm} \, = \, \g_{\,k_1 \ldots\, k_q\, lm} \, + \left(\, \g_{\,(\,k_1 \ldots\, k_{q-1}\,|\,l} \, T_{\,|\,k_q\,)\,m} \, + \, \g_{\,(\,k_1 \ldots\, k_{q-1}\,|\,m} \, T_{\,|\,k_q\,)\,l} \,\right) \nonumber \\
& \phantom{\g_{\,k_1 \ldots\, k_q} \, \g_{\,lm} \,} - \, \g_{\,(\,k_1 \ldots\, k_{q-2}} \, T_{k_{q-1}\,|\,l}\, T_{\,|\,k_q\,)\,m} \, ,
\end{align}
one can then compute the $\{\,p+q+1 \comma p\,\}$ projection of the
generic $T_{i_1j_1} \ldots\, T_{i_pj_p} \, \g_{\,k_1 \ldots\, k_q}$
$\g$-trace of the Bianchi identities \eqref{bianchiform}:
\be
\begin{split}
& (\,p+q+2\,) \ Y_{\{p+q+1 \comma p\}} \, \pr_{\,l} \, (\,\g^{\,[\,q\,]}\,\cS^{\,[\,p\,]}\,)_{\, k_1 \ldots\, k_q;\, i_1 j_1;\,\ldots\,;\,i_p j_p} \\
+ \ & \frac{1}{q+3} \ Y_{\{p+q+1 \comma p\}} \, \sum_{n\,=\,1}^{p}\, \pr_{\,[\,i_n\,|}\, (\,\g^{\,[\,q+2\,]}\,\cS^{\,[\,p-1\,]}\,)_{\,|\,j_n\,]\, k_1 \ldots\, k_q \, l\,;\,\ldots\,;\,i_{r\,\ne\,n} j_{r\,\ne\,n} \,;\,\ldots}  \\[2pt]
- \ & \!\!\dsl \, (\,\g^{\,[\,q+1\,]}\,\cS^{\,[\,p\,]}\,)_{\, k_1 \ldots\, k_q \, l\,;\, i_1 j_1;\,\ldots\,;\,i_p j_p} \, - \,  \pr^{\,m}\, (\,\g^{\,[\,q\,]}\,\cS^{\,[\,p+1\,]}\,)_{\, k_1 \ldots\, k_q;\, i_1 j_1;\,\ldots\,;\,i_p j_p\,;\,lm}  \\[5pt]
- \ & \frac{1}{q+3} \ \pr^{\,m}\,
(\,\g^{\,[\,q+2\,]}\,\cS^{\,[\,p\,]}\,)_{\, k_1 \ldots\, k_q\,lm\,;\, i_1 j_1;\,\ldots\,;\,i_p
j_p} \, = \, 0\, .
\end{split}
\ee
As was the case for the Bianchi identities \eqref{bianchiform}, these results maintain
the same form as their analogues presented in eq.~\eqref{finalbianchif}, barring some sign
differences and the replacement of symmetric sets of family indices with
antisymmetric ones and vice versa.

In a similar fashion, the counterpart of the Ansatz \eqref{lagferconstr} is
\be \label{lagform}
\cL \, = \, \frac{1}{2} \ \bra\, \bar{\psi} \,\comma\! \sum_{p\,,\,q\,=\,0}^{N} \widehat{k}_{\,p\,,\,q} \ \h^{i_1j_1} \ldots\, \h^{i_pj_p} \, \g^{\,k_1 \ldots\, k_q} \, (\,\g^{\,[\,q\,]}\,\cS^{\,[\,p\,]}\,)_{\, k_1 \ldots\, k_q;\, i_1 j_1;\,\ldots\,;\,i_p j_p} \,\ket \, + \, \textrm{h.c.}\, ,
\ee
where we would like to stress again that only \emph{two-row} projected
quantities appear here, to be contrasted with the two-column
projected ones entering eq.~\eqref{lagferconstr}. Finally, using
\begin{align}
& [\, \pr_{\,l} \comma \h^{i_1j_1} \ldots\, \h^{i_pj_p} \,] \, = \, \12 \, \sum_{n\,=\,1}^p \, \d_{\,l}{}^{\,[\,i_n}\, \pr^{\,j_n\,]} \, \prod_{r\,\neq\,n}^p \, \h^{\,i_rj_r} \, , \\
& [\, \pr_{\,l} \comma \g^{\,k_1 \ldots\, k_q} \,] \, = \, \d_{\,l}{}^{\,(\,k_1\,} \g^{\,k_2 \ldots\, k_q\,)} \dsl \ - \, \d_{\,l}{}^{\,(\,k_1\,} \g^{\,k_2 \ldots\, k_{q-1}\,} \pr^{\,k_q\,)} \, ,
\end{align}
and the relation
\be Y_{\{\,2^{p+1},\,1^{q-1}\}}\, T_{i_1j_1} \ldots T_{i_pj_p} \,
\bar{\e}_{\,l} \, \g_{\,k_1 \ldots\, k_q} \, = \, \frac{1}{q+1} \
Y_{\{\,2^{p+1},\,1^{q-1}\}} \, T_{i_1j_1} \ldots T_{i_pj_p} \,
T_{\,l\,(\,k_1} \bar{\e}_{\,k_2} \, \g_{\,k_{3} \ldots\, k_q\,)}\, ,
\ee
that holds in the constrained theory, one can recognize that, up to
the convenient choice of an overall normalization $\widehat{k}_{\,0 \comma 0}
= 1$, the constrained Lagrangians \eqref{lagform} are gauge
invariant only if
\be
\widehat{k}_{\,p\,,\,q} \, =  \,
\frac{(-1)^{\,p\,+\,q}}{p\,!\,q\,!\,(\,p+q+1\,)\,!}
\, .
\ee
Aside from a slight difference in the oscillating
signs, that are simpler for multi-forms, the coefficients are thus
the same as in eq.~\eqref{solgenf}.

It should be clear that, in this presentation of the formalism,
the need for constraints is determined solely by the number
of columns of the Young tableaux associated to generic
representations of the Poincar\'e group. In fact, the antisymmetry
in the family indices of eqs.~\eqref{constrganti} and \eqref{constranti}
shows that all single-column fields, like the gravitino, are not subject to any constraints, while two-column fields are only subject to constraints on the gauge
parameters. This is the case, for instance, for a spin $5/2$ field.
Furthermore, in this setting the Fang-Fronsdal-Labastida operator
\eqref{sform} can be recast in the form
\be
\cS \, = \, i \ \g_{\,k}\,  \pr^{\,k} \, \psi \, ,
\ee
and for one-family spinorial forms it is manifestly related to the
corresponding curvature via a single $\g$-trace.

\vskip 24pt


\scs{Elimination of higher derivatives}\label{sec:lowerder}


Different types of Lagrangian descriptions for unconstrained
irreducible higher-spin fields were explored in the last years.
Still, until recently, they all displayed some unusual features with
respect to their lower-spin
counterparts: as recalled in the Introduction, they
typically involve many auxiliary fields, in numbers that increase
linearly with the spin
\cite{bpt}, or alternatively they contain non-local terms
\cite{fsoldnl,mixednl,fms,dario07}, or finally local descriptions
with minimal (and fixed) sets of auxiliary fields contain some
higher-derivative terms. This is certainly the case for the
constructions of \cite{fsold2,fms,bose_mixed}, but it is
equally true for the unconstrained formulations proposed in the
previous sections for mixed-symmetry fermions, as can be seen for
instance in eqs.~\eqref{W} and
\eqref{laggenfunc}.

A more conventional solution to the same problem was first obtained
in \cite{buchnew} for symmetric bosons and fermions. There it was
shown how to complement with suitable constraints the Lagrangians
given in \cite{fsoldl1} for the ``triplets'' of String Field Theory
\cite{sftheory}, so as to avoid the propagation of the lower spin
components contained in the symmetric gauge
potentials. The end result of this analysis is an elegant off-shell
extension of the constraint relating triplets and
compensators equations that was first discussed in
\cite{fsoldl1}. A detailed analysis of the triplet systems in the
``frame-like'' formalism of \cite{lopvas} was also recently presented in \cite{sorokvas}.

An alternative construction, that is more in the spirit of the minimal Lagrangians of
\cite{fsold2,fms}, was proposed in \cite{dario07} for symmetric Bose fields, and was then
extended to mixed-symmetry Bose fields in \cite{bose_mixed}. In this section we would
like to further extend those results to Fermi fields of mixed symmetry. The logic behind
the construction will mirror the procedure that we followed to build the minimal local
Lagrangians: as a first step, we shall identify a candidate kinetic tensor that is gauge
invariant under unconstrained gauge transformations, but is now devoid of higher
derivatives. Several choices are indeed possible, and for each of them the second step
will be determined by the corresponding form of the Bianchi identities, and will be aimed
at building the gauge invariant completion of a trial action suggested by the
analogy with the constrained case. One is then to make sure that the resulting equations of motion propagate only the desired degrees of freedom, taking a closer look at the candidate kinetic tensors, in particular in relation with the
structure of their Bianchi identities, in order to identify the proper choices. While the
result of \cite{dario07, bose_mixed} was more involved than its minimal counterparts
involving higher derivatives, the clearcut logic behind the construction makes it
possible to extend the program to Fermi fields in a systematic way.

We shall describe the procedure and the results in the next two
sections. First, in Section \ref{lowersymmfermi} we discuss in
some detail the case of symmetric Fermi fields, for which the
general construction is presented here for the first time. Having
introduced all basic ideas and problems in a relatively simple
setting, in Section \ref{lowermixfermi} we then
turn to the general solution for Fermi fields of mixed
symmetry.

\vskip 24pt

\scss{Symmetric fermions} \label{lowersymmfermi}
Let us begin by analyzing the behavior of the Fang-Fronsdal
theory when the constraints are not enforced, with the aim of building
eventually gauge invariant kinetic
tensors that are free from higher derivatives. Indeed, in order to
compensate the variation of the Fang-Fronsdal tensor~\footnote{Consistently with the choice made in \cite{bose_mixed}, here we
largely abide to the conventions followed in the rest of the present
paper, even though for one-family fields the notation of
\cite{fsoldnl,fsoldl1, fsold2, fms, dario07} would allow further simplifications,
in particular for what concerns products of mutually commuting
tensors, obtained at the minor price of defining some new operators
like the double gradient in \eqref{varFF}. For the sake of brevity, however, we use ``primes''
to denote traces and we still resort to that notation in eq.~\eqref{gtrproj}.}
\be \label{varFF}
\d \, \cS \, = \, -  \, i \, \pr \, \pr \not \! \e
\ee
without introducing higher derivatives, the simplest options would
be provided by the following alternative definitions for the compensator
fields:
\begin{align}
& \s_{\, (1)} \, : & &\d \, \s_{\, (1)} \, = \, \pr  \esl && \, \ra \, & &
\cW_{\, \s_{\, (1)}} \, = \, \cS \, + \, i \, \pr \, \s_{\, (1)} \, ,& \nn \\
& \s_{\, (2)} \, : & &\d \,\s_{\, (2)} \, = \, \pr\, \pr  \esl && \, \ra \, & &
\cW_{\, \s_{\, (2)}} \, = \, \cS \, + \, i \, \s_{\, (2)} \, . &  \label{compensfer}
\end{align}
Actually, in contrast with the choice made in \cite{fsoldl1,fsold2,fms}, corresponding to the
one-family case of eq.~\eqref{fermicompens}, the fields in \eqref{compensfer} are \emph{not}
pure gauge contributions. On the other hand, this observation does not represent a true
obstruction, since in principle one could add a Lagrange multiplier term
relating these fields to gradients of the usual compensators in order to reduce them to
the status of ``effective'' Stueckelberg-like fields. A closer look at the corresponding
Bianchi identities
\begin{align}
& \s_{\, (1)} : \ \prd \cW_{\, \s_{\, (1)}} - \, \frac{1}{2} \, \pr \, \cW^{\, \pe}_{\, \s_{\, (1)}} - \,
\frac{1}{2} \dsll \ {\cWsl}_{\, \s_{\, (1)}}  = \, \fr{i}{2} \, (\, \pr \, \pr \,
\psisl^{\, \pe} \, + \, \Box \, \s_{\, (1)} \, - \, \!\dsll \, \pr \, {\not \! \s_{\, (1)}} \,
- \, \pr \, \pr \, \s_{\, (1)}^{\, \pe} \,) \, ,  \nn \\
& \s_{\, (2)}: \ \prd \cW_{\, \s_{\, (2)}} - \, \frac{1}{2} \, \pr \, \cW^{\, \pe}_{\, \s_{\, (2)}} - \,
\frac{1}{2} \!\dsll \ {\cWsl}_{\, \s_{\, (2)}} = \, \fr{i}{2} \, (\, \pr \, \pr \,
\psisl^{\, \pe} \, +  \, 2\, \prd \s_{\, (2)} \, - \, \dsll \, {\not \! \s_{\, (2)}} \, -
\, \pr \, \s_{\, (2)}^{\, \pe}\, ) \, ,  \label{bianchiferm}
\end{align}
can clearly exhibit the difficulties potentially entailed by these simple choices.
Indeed, if one were to pursue one of the two options in
\eqref{compensfer}, starting from trial Lagrangians built from
either $\cW_{\, \s_{\, (1)}}$ or $\cW_{\, \s_{\, (2)}}$ the gauge
variations would give rise to the remainders in
eq.~\eqref{bianchiferm}. In order to compensate them, proceeding
along the lines that led to \eqref{laggenfunc}, one would be
tempted to introduce Lagrange multipliers $\l_{\, (1)}$ and $\l_{\,
(2)}$, but here they would transform
\emph{proportionally to the bare gauge parameter} $\e$, since the
double gradient acting on ${\not {\! \!\psi}}^{\, \pe}$ in these
cases cannot be factored out of the other terms. A different
manifestation of these difficulties is that, in contrast with what
happened in the minimal theory of
\cite{fsold2, fms}, the multipliers would appear in the equations of motion
for $\bar{\psi}$ only via their double divergence, so that they would
not be expressible algebraically, on shell, in terms of the gauge
potential itself, which would have guaranteed against the danger
that they carry additional degrees of freedom.

This approach works nicely for symmetric bosons, since in that case one of the
multipliers transforms proportionally to the divergence of the gauge parameter. This
makes it possible to replace it with a combination of the gauge potential and the
ordinary compensator that possesses the same kind of gauge transformation, as shown in
\cite{dario07}. On the other hand, this alternative is clearly not viable in the
present case, since no single combination of the fields $\psi$ and $\s_{\, (i)}$ can be defined
that transforms like the \emph{bare} gauge parameter $\e$. In the same spirit of the
solution that we gave in \cite{bose_mixed} for mixed-symmetry bosons, we are thus led to
look for more general definitions for the gauge invariant kinetic tensors, where a certain
amount of additional technical complications can still be acceptable provided the
corresponding Bianchi identities are devoid of the previous difficulties. To this end,
mimicking what we did in \cite{bose_mixed} for bosons, let us  reconsider the gauge
transformation of $\psi$, separating in the gauge parameter $\e$ its $\g$-traceless part
$\e^{\, (t)}$ from the rest,
letting
\be
\begin{split}
&\d \, \psi \, = \, \pr \, \e \, , \\
& \e \, = \, \e^{\, (t)} \, + \, \g \, \e^{\, (g)} \, ,
\end{split}
\ee
so that
\be
\esl \, = \, \g \, \cdot (\g \, \e^{\, (g)}) \, .
\ee
The solution for $\e^{\, (g)}$ is then
\be \label{gtrproj}
\e^{\, (g)} \, = \,  \sum_{n=1}^{[\fr{s}{2}]}\, \rho_{n}(D, \, s)\,
\h^{\, n-1}\, \left\{\esl^{\ [n-1]}\, + \, \fr{1}{2 \, n} \, \g \, \e^{\, [n]} \right\}\, ,
\ee
with
\be
\r_{n}(D, \, s)\, = \, (- \, 1)^{\, n + 1} \, \prod_{k = 1}^n \, \fr{1}{D \, + \, 2 \, (s \, - \, k \, - \, 1)} \, .
\ee
Here $s$ is the tensorial rank of $\psi$ and $\e^{\, [n]}$ denotes the $n$-th
trace of $\e$, while $\h^{\, n}$ denotes a combination of products of $n$ Minkowski
metric tensors defined with unit overall normalization and with the minimal number of
terms needed in order to obtain a totally symmetric expression. The corresponding
decomposition of the unconstrained variation of $\psi$,
\be
\d \, \psi \, = \, \pr \, \e^{\, (t)} \, + \, \g \, \pr \, \e^{\, (g)}\, ,
\ee
allows one to express the gauge variation of $\cS$ in terms of $\e^{\,
(g)}$ alone,
\be
\d \, \cS \, = \, - \, i \, (\,D \, + \, 2 \, s \, - \, 4\,) \, \pr \, \pr \,
\e^{\, (g)} \, + \, i \, \g \, \pr \, \pr \, {\not \! \e^{\, (g)}} \, .
\ee
Introducing a compensator $\s_{\, (g)}$ such that
\be
\d \, \s_{\, (g)} \, = \, \pr \, \e^{\, (g)} \, ,
\ee
one can now define a new unconstrained tensor, $\cW_{\,
\s_{\, (g)} }$, as
\be \label{kinferm}
\cW_{\, \s_{\, (g)} } \, = \, \cS  \, +  \, i \, (\,D \, + \, 2 \, s \, - \, 4\,) \, \pr \,
\s_{\, (g)} \, + \, \g \, \cS \, (\s) \, ,
\ee
where $\cS \, (\s)$ is the Fang-Fronsdal tensor for $\s_{\, (g)}$,
\be
\cS \, (\s) \, = \, i \, (\, \dsll \, \s_{\, (g)} \, - \, \pr \,  {\not \! \s_{\, (g)}} \,) \, .
\ee
Once more, the somewhat involved structure of $\cW_{\, \s_{\, (g)} }$
can well be justified if it avoids the complications that emerged
for the simpler choices of eq.~\eqref{compensfer}. In fact, in
particular thanks to the presence of the tensor $\cS \, (\s)$ in
$\cW_{\,
\s_{\, (g)} }$, the Bianchi identity satisfied by the latter
tensor reduces to
\be
\prd \cW_{\, \s_{\, (g)} }\, - \, \frac{1}{2} \ \pr \, \cW^{\, \pe}_{\, \s_{\, (g)} } \, - \,
\frac{1}{2} \dsll \, \, {\cWsl}_{\, \s_{\, (g)}}  \, = \, \fr{i}{2} \, \pr \, \pr\,  \left\{\,
\psisl^{\, \pe} \, + \, \g \, {\not \! {\s_{\, (g)}}^{\, \pe}} \, - \, (\,D \, + \, 2 \, s \, - \, 4\,) \, \s_{\, (g)}^{\, \pe} \,\right\}\, ,
\ee
so that its structure makes it manifest that the Lagrange multiplier to be
introduced does not carry any additional degrees of freedom.

For this reason, one can now consider the variation of a new trial
Lagrangian built out of $\cW_{\, \s_{\, (g)} }$ and eventually one can complete the
construction in the usual way. However, in order to better
understand the rationale behind the kinetic tensor
\eqref{kinferm}, it is useful to trace its origin to the substitution
\be \label{trueshift}
\psi \, \ra \, \psi \, - \, \g \, \s_{\, (g)} \, ,
\ee
to be performed in the Fang-Fronsdal tensor $\cS$.

Eq.~\eqref{trueshift} should be compared to the
Stueckelberg shift \eqref{stueckf}, that in the symmetric case would simply read
\be \label{2stueckf}
\psi \ \to \ \psi \,-\, \partial \, \Psi \, ,
\ee
and would give rise to the higher-derivative tensor \eqref{W}.
In this case, one defines an \emph{identically} gauge
invariant combination of fields that, as such, would make \emph{any}
tensor built out of it gauge invariant as well. The price to pay, however, is
to introduce a compensator field whose dimension is different from
that of $\psi$. This choice is then ultimately responsible for the
appearance of higher derivatives after the substitution
\eqref{2stueckf} is performed in a Lagrangian that initially involved only $\psi$
and one-derivative terms. On the other hand,
eq.~\eqref{trueshift} defines a combination of fields displaying
\emph{the same gauge transformation as the Fang-Fronsdal field}.
As a result, that substitution would guarantee the \emph{unconstrained} gauge invariance of any
expression that is gauge invariant, to begin with, under
\emph{constrained} gauge transformations.
This property is all one really needs. Moreover, it
is manifest from \eqref{trueshift} that the compensator $\s_{\,
(g)}$ has the same dimension as $\psi$, so that all kinetic
operators in the Fang-Fronsdal Lagrangian maintain by construction
the same number of derivatives even after the substitution
\eqref{trueshift} is implemented.

Thus, enforcing eq.~(\ref{trueshift}) in the constrained Lagrangian and adding a Lagrange
multiplier for the triple $\gamma$-trace of the field $\psi$ results in a new Lagrangian
with only one-derivative terms that is invariant under \emph{unconstrained} gauge
transformations. Adding to it a further constraint to guarantee the equivalence between
$\s_{\, (g)}$ and a suitable linear combination of $\gamma$-traces of the ``old''
compensator $\xi$ finally leads to
\be \label{fermlagr}
\begin{split}
\cL & \,  =  \, \12 \, \bra \bar{\psi} \, + \, \bar{\s}_{\, (g)} \, \g \comma
\cW_{\, \s_{\, (g)}}\, - \, \frac{1}{2} \, \g \ {\cWsl}_{\, \s_{\, (g)} } \,
- \, \12 \, \h \, \cW^{\, \pe}_{\, \s_{\, (g)} }  \ket \\
& + \, \bra \bar{\l} \comma \g \cdot (\, \psi \, - \, \g \, \s_{\, (g)} \, )^{\, \pe} \ket \,
+ \, \bra \bar{\c} \comma \s \, - \, \pr \, \P \, \xi \ket \, + \, h.c. \, ,
\end{split}
\ee
that does not contain any higher-derivative contributions. The projector $\P$ entering
this expression, whose explicit form can be obtained from eq.~\eqref{gtrproj}, is such that
$\P \, \d \, \xi \, = \, \e^{\, (g)}$, while the full set of gauge transformations is
\be
\begin{split}
&\d \, \psi \, = \, \pr \, \e^{\, (t)} \, + \, \g \, \pr \, \e^{\, (g)} \, , \\
&\d \, \s_{\, (g)} \, = \, \pr \, \e^{\, (g)} \, , \\
&\d \, \xi \, = \, \esl \, = \, \g \cdot (\g \,\e^{\, (g)})\,  ,\\
&\d \, \l \, = \, \prd \prd \e^{\, (t)} \, , \\
&\d \, \c \, = \, 0 \, .
\end{split}
\ee
The consistency of the system can be easily verified considering the
equations of motion obtained after removing $\xi$ by a partial gauge
fixing. In this case the equation for $\bar{\psi}$ reduces to
\be E_{\, \bar{\psi}} \, : \ \cS \, - \, \12 \, \g \, {\not \! \cS} \, - \, \12 \, \h \,
\cS^{\, \pe } \, - \, \fr{i}{2} \, \h \, \g \, \left(\, \l \, - \, \12 \, \prd
\psi^{\, \pe}\, \right)\, = \, 0 \, , \ee
that can be shown to imply $\l =  \12 \, \prd \psi^{\, \pe}$ and
$\cS = 0$, as usual, while the equations for the multipliers $\bar{\c}$
and $\bar{\l}$, in their turn, set to zero $\s_{\, (g)}$ and
$\not{\! \! \psi}^{\, \pe}$. Finally, the equation for $\bar{\s}_{\,
(g)}$ determines the multiplier $\c$ as a function of $\cS$, and
also guarantees, together with the equation for $\xi$, the
consistency conditions on the divergence of $E_{\,
\bar{\psi}}$, which is to vanish as a consequence of the
unconstrained gauge invariance of  \eqref{fermlagr}.

\vskip 24pt

\scss{Mixed-symmetry fermions}\label{lowermixfermi}

The extension of these results to mixed-symmetry fermions proceeds
along the same lines, and we can therefore confine ourselves to
illustrating the main steps that lead to the final result.
To begin with, in the gauge variation \eqref{gaugef} of the field
$\psi$,
\be
\d \, \psi \, = \, \pr^{\,i} \, \e_{\,i} \, ,
\ee
one should isolate the contribution to the gauge parameters which is
to vanish in the constrained theory. To this end, let us consider
the decomposition
\be \label{gammadec}
\e_{\,i} \, = \, \e^{\,(t)}{}_{\,i} \, + \, \g^{\, k} \, \e^{\,(g)}{}_{\, i\, ; \, k} \, ,
\ee
where
\be
\begin{split}
& \g_{\,(\, i}\,\e^{\,(t)}{}_{\,j\,)} \, \equiv \, 0 \, , \\
& \g_{\,(\,i}\, \g^{\, k} \, \e^{\,(g)}{}_{\, j\,)\, ; \, k}  \, \equiv \,
\g_{\,(\,i}\,\e_{\,j\,)} \, .
\end{split}
\ee
In this fashion, the parameters $\e^{\,(g)}{}_{\, i\, ; \, k}$ embody
precisely the gauge freedom that is absent in the Labastida
formulation, and that we would like to allow here without
introducing new fields whose dimensions differ from that of $\psi$.
In particular, the variation of the Labastida tensor
\eqref{fermikin} under the decomposition \eqref{gammadec} reads
\be
\d \, \cS \, = \, - \, \frac{i}{2} \ \pr^{\, i} \pr^{\, j} \, \left(\, D \, \d^{\, k}{}_{\, (\,i\,|} \,
+ \, 2 \, S^{\, k}{}_{\, (\,i\,|}\, \right)\, \e^{\,(g)}{}_{\, |\,j\,)\, ; \, k} \, + \, \fr{i}{2} \
\g^{\, k} \, \pr^{\, i} \pr^{\, j} \, \g_{\,(\,i}\, \e^{\,(g)}{}_{\, j\,)\, ; \, k} \, .
\ee
Guided by the solution of the problem for symmetric fermions presented in the previous
section and from the general solution proposed for bosons in \cite{bose_mixed}, we are
thus led to introduce new compensator fields $\s_{\, k}$ such that
\be
\d \, \s_{\, k} \, = \, \pr^{\, l} \, \e^{\,(g)}{}_{\, l\, ; \, k} \, .
\ee
This makes it possible to define the fully gauge invariant kinetic
tensor
\be \label{lowerkin}
\cW_{\, \s} \, = \, \cS \, + \, i \, \left[\,(\, D \, - \, 2 \, ) \, \pr^{\, j} \, + \,
2 \, \pr^{\, i} \, S^{\, j}{}_{\, i}\,\right]\, \s_{\, j} \, + \, \g^{\, k} \, \cS_{\, k} \, (\s) \, ,
\ee
where
\be
\cS_{\, k} \, (\s) \, = \, i \, (\, \dsll \, \s_{\, k} \, - \, \pr^{\, j} \, \g_{\, j} \, \s_{\, k} \, )
\ee
is the Labastida tensor for $\s_{\, k}$, and where, let us reiterate,
no higher derivatives appear, precisely because the compensators
$\s_{\, k}$ have the same physical dimensions as $\psi$.

The Bianchi identity for $\cW_{\, \s}$,
\be \label{bianchi_Ws}
\pr_{\,i}\, \cW_{\, \s} \, - \, \12 \dsl \, \g_{\, i} \, \cW_{\, \s} \, - \,
\12 \ \pr^{\,j} \, T_{ij} \, \cW_{\, \s} \, - \, \frac{1}{6} \ \pr^{\,j} \,
\g_{\,ij} \, \cW_{\, \s} \, = \, \frac{i}{6} \ \pr^{\,j} \pr^{\,k} \, T_{(\,ij}\, \gamma_{\, k\,
)}\, ( \, \psi \, - \, \g^{\, l} \, \s_{\, l} \,) \, ,
\ee
takes a particularly nice form, so that the introduction of Lagrange multipliers is not
expected to be problematic, and actually provides a clue to a more complete understanding
of the structure of $\cW_{\, \s}$. Indeed, the compensators $\s_{\, k}$ allow the
redefinition
\be \label{fermishift}
\psi \, \ra \, \psi \, - \, \g^{\, k} \, \s_{\, k}
\ee
of the field $\psi$, that has the virtue of identifying a combination of fields
possessing the same gauge transformation as the \emph{constrained} Labastida field under
variations involving \emph{unconstrained} gauge parameters. As we already observed,
this is to be contrasted with the Stueckelberg shift \eqref{stueckf}. It is then possible
to verify that, after the substitution \eqref{fermishift}, the Labastida tensor
\eqref{fermikin} takes the form of eq.~\eqref{lowerkin}.

One can thus follow a relatively straightforward route to
a Lagrangian based on the tensor \eqref{lowerkin},
starting from the Lagrangian defined by
eq.~\eqref{lagmult}, that we display again here for the sake of
clarity,
\be
\cL \, = \, \frac{1}{2} \ \bra\, \bar{\psi} \,\comma\! \sum_{p\,,\,q\,=\,0}^{N} k_{\,p\,,\,q} \ \h^{\,p} \, \g^{\,q}\, (\,\g^{\,[\,q\,]}\,\cS^{\,[\,p\,]}\,) \,\ket \,
+ \, \frac{i}{12} \, \bra \bar{\l}_{\,ijk}\comma \, T_{(\,ij}\, \gamma_{\, k\,
)}\, \psi  \, \ket + \, \textrm{h.c.}\, \, ,
\label{2lagferconstr}
\ee
with
\be 
k_{\,p\,,\,q} \, =  \,
\frac{(-1)^{\,p\,+\,\frac{q\,(q+1)}{2}}}{p\,!\,q\,!\,(\,p+q+1\,)\,!} \ , \ee
where the constraints on the symmetrized triple $\gamma$-trace of $\psi$ are enforced by
Lagrange multipliers. Since \eqref{2lagferconstr} is gauge invariant under
\emph{constrained} gauge transformations, by construction it will become gauge invariant
under unconstrained ones if one performs everywhere the substitution \eqref{fermishift},
that leads to
\be
\begin{split}
\cL \, &= \, \frac{1}{2} \ \bra\, \bar{\psi} \,+ \, \bar{\s}_{\, k}\, \g^{\, k}
\comma\! \sum_{p\,,\,q\,=\,0}^{N} k_{\,p\,,\,q} \ \h^{\,p} \, \g^{\,q}\, (\,\g^{\,[\,q\,]}\,
\cW_{\, \s}^{\,[\,p\,]}\,) \,\ket \\
& + \, \frac{i}{12} \, \bra \bar{\l}_{\,ijk}\comma T_{(\,ij}\, \gamma_{\, k\,
)}\, ( \, \psi \, - \, \g^{\, l} \, \s_{\, l} \,) \ket + \, \textrm{h.c.}\,  .
\label{3lagferconstr}
\end{split}
\ee
Finally, in order to guarantee that the new compensators $\s_{\, k}$ be effectively pure
gauge, eq.~\eqref{3lagferconstr} should be supplemented with further
constraints, meant to enforce a linear relation between the $\s_{\, k}$ and the
compensators $\xi_{\,ij}\,(\Psi)$ defined in \eqref{xipsi}, so that the complete
Lagrangian takes the form
\be \label{basiclagfer}
\begin{split}
\cL \, & = \, \frac{1}{2} \ \bra \bar{\psi} \,+ \, \bar{\s}_{\, k}\, \g^{\, k}
\, \comma\! \sum_{p\,,\,q\,=\,0}^{N} k_{\,p\,,\,q} \ \h^{\,p} \, \g^{\,q}\, (\,\g^{\,[\,q\,]}\,
\cW_{\, \s}^{\,[\,p\,]}\,) \,\ket \,  \\
& + \, \frac{i}{12} \, \bra \bar{\l}_{\,ijk}\comma T_{(\,ij}\, \gamma_{\, k\,
)}\, ( \, \psi \, - \, \g^{\, l} \, \s_{\, l} \,) \ket \,
+ \, \bra \bar{\chi}_{\, k} \comma \s_{\, k} \, - \, \pr^{\, l} \, \P^{\, i j }{}_{\, k l } \, \xi_{\,ij} \ket \, + \,
\textrm{h.c.}\,  ,
\end{split}
\ee
where $\P^{\, i j }{}_{\, k l }$ is the projector allowing to express the solution of
eq.~\eqref{gammadec} as
\be \label{fermiproj}
\e^{\,(g)}{}_{\, k\, ; \, l} \, = \,  \P^{\, i j }{}_{\, k l} \, \g_{\,(\,i}\,\e_{\,j\,)} \, ,
\ee
and where, summarizing, the transformations leaving
eq.~\eqref{basiclagfer} invariant are
\be
\begin{split}
&\d \, \psi \, = \, \pr^{\, i} \, \left(\,\e^{\, (t)}{}_{\, i} \, + \, \g^{\, k} \, \e^{\, (g)}{}_{\, i\, ; \, k}\,\right) \, , \\
&\d \, \s_{\, k} \, = \, \pr^{\, l} \, \e^{\,(g)}{}_{\, l\, ; \, k} \, , \\
&\d \, \xi_{\, ij} \, = \, \g_{\,(\,i}\,\e_{\,j\,)} \, ,\\
&\d \, \chi_{\, k} \, = \, 0 \, ,
\end{split}
\ee
with $\d \, \l_{\,ijk}$ given in \eqref{lagrangegen} after replacing $\e$ with $\e^{\,
(t)}$. Our derivation of the Lagrangian \eqref{basiclagfer} makes it also manifest that
the equations of motion reduce generically to $\cS=0$, since the fields $\s_{\, k}$ and
$\xi_{\, ij}$ can be gauged away simultaneously, while the Lagrange multipliers
$\l_{\,ijk}$ and $\c_{\, k}$ can be expressed in terms of the physical field $\psi$ by the
equations for $\bar{\psi}$ and $\bar{\s}_{\, k}$, respectively.

We would like to conclude this section with some general comments on
the unconstrained Lagrangians without higher-derivative terms that
we proposed here and in \cite{dario07, bose_mixed}. Since the
compensators $\s_{\, k}$ (as well as their bosonic counterparts $\th_{\,
ij}$ of \cite{dario07, bose_mixed}) add
\emph{exactly} the amount of gauge freedom that was frozen in the
Fronsdal-Labastida formulation, we believe that the field content of
these Lagrangians is the minimal possible one, if one requires
that the theory be unconstrained and contain no higher-derivative
operators. In addition, in the symmetric case there is still a
one-to-one correspondence between the field contents of bosonic and fermionic theories, as was
the case for the minimal Lagrangians with higher derivatives. Strictly speaking, this
correspondence is not maintained with mixed symmetry, simply because some of the auxiliary fields bear different numbers of family indices, but this construction of single derivative Lagrangians for Fermi fields mirrors nonetheless the corresponding one for Bose fields of \cite{bose_mixed}, which could prove useful when dealing with higher-spin supermultiplets.
On the other hand, as for the bosonic fields discussed in \cite{bose_mixed},
the gauge transformations for the $\s_{\, k}$ compensators are typically rather involved, and we do not know them in closed form in general, beyond the symmetric case that was described in the previous section. This is clearly a disadvantage with respect
to the minimal theory with higher derivatives, that in this sense
appears technically simpler, not only in view of its reduced
field content. Finally, we would like to stress again that, although they are somewhat
unusual, higher derivatives accompany in our minimal Lagrangians
fields that can be gauged away algebraically.
In this sense, consistently with all the tests that one can perform
at the present time \cite{fms}, they are not expected to represent a
real source of difficulties, even in future generalizations of our
setting to the non-linear level.


\scs{Conclusions}\label{sec:conclusions}


In this paper we have constructed Lagrangian theories
for free mixed-symmetry gauge fermions in a Minkowski background.  These types of
fields are inevitable in all dimensions $D>5$, and indeed higher-spin fields of mixed
symmetry are key ingredients of all massive string spectra. Whereas their dynamics is
still poorly understood, it is difficult to escape the feeling that these
excitations, together with their bosonic counterparts,
should be held ultimately responsible for many of the most spectacular
properties of String Theory. And, in a wider sense, that a closer look at their dynamics
has a real potential to shed some light on the very meaning of String Theory and on its
possible generalizations.

We based our construction on the important results obtained by Labastida in
\cite{labferm} (see also \cite{siegel,mixed2,BMV,mixednl,extraframe,skvortsov,pernicolas} for other relevant works on
mixed-symmetry fields), where he did not arrive at a Lagrangian formulation, but
identified nonetheless the constraints on gauge fields and gauge parameters that let the
field equations propagate the correct degrees of freedom, as we show in Section \ref{sec:dofs}. Since our main interest lies
in unconstrained formulations, our basic kinetic spinor-tensors are gauge invariant completions
of \eqref{fermikin}, that was introduced by Labastida in \cite{labferm}. In their
simplest form, they are obtained introducing compensator fields $\xi_{\,ij}$ that transform
proportionally to the constraints \eqref{labacfermg}, but in analogy with the bosonic
case of \cite{bose_mixed} the very fact that the constraints are not independent forces
one to relate the $\xi_{\,ij}$ to more fundamental objects, here denoted by $\Psi_{\,i}$, as in \eqref{xipsi}. We could thus generalize the basic dynamical setting of Labastida, defined
by eqs.~\eqref{fermikin}, \eqref{gaugef} and \eqref{labacfermg},
\be
\begin{split}
&\cS \, = \, i \left(\, \dsl \, \psi \, - \, \pr^{\,i} \! \psisl_i \,\right) \,  = 0 \, , \\
&\d \, \psi \, = \, \pr^{\, i} \, \e_{\, i} \, , \\
&\g_{\,(\,i}\,\e_{\,j\,)} \, = \, 0 \, ,
\end{split}
\ee
to its unconstrained counterpart, summarized by eqs.~\eqref{W},
\eqref{gaugef}, \eqref{xipsi} and \eqref{deltapsi},
\be
\begin{split}
&\cW \, = \, \cS \, + \, i \, \pr^{\, i} \, \pr^{\, j} \, \xi_{\, ij} \, = \, 0 \, , \\
&\d \, \psi \, = \, \pr^{\, i} \, \e_{\, i} \, , \\
& \xi_{\,ij}\,(\Psi) \, = \, \frac{1}{2} \ \gamma_{\,(\,i} \, \Psi_{j\,)}\, , \\
& \d \, \Psi_{\, i} \, = \, \e_{\, i} \, ,
\end{split}
\ee
that describes the same physical polarizations.

As in the bosonic construction of
\cite{bose_mixed}, our method to build the Lagrangians
was largely driven by the Bianchi identities, but for Fermi
fields we had to face a number of novel complications.
These can be traced, one way or another, to the fact
that fermionic constraints and compensator terms are \emph{not} well distinct a
priori, so that disentangling the truly independent contributions to the Lagrangians is
not straightforward. This led, for two-family fields, to the \emph{one-parameter}
family of Lagrangians \eqref{lag_fer}, equivalent to one another up to
redefinitions of the Lagrange multipliers and whose derivation is spelled out in detail
in Section \ref{sec:lagrangian2f}. The general $N$-family Lagrangians illustrated in Sections \ref{sec:lagrangianf} and \ref{sec:lagrangianfunc} were also constructed exploiting the Bianchi identities \eqref{bianchi_cf}, or their unconstrained counterparts \eqref{bianchi_W2}. However, in this case we refrained from displaying the full one-parameter family, and confined our attention to a specific form of
the Lagrangians that has the virtue of relative simplicity. Moreover, we also presented a derivation that stresses the self-adjoint nature of the resulting Rarita-Schwinger-like tensors. Eqs.~\eqref{lagferconstr} and \eqref{solgenf} are the complete Lagrangians in the constrained setting, and together with eq.~\eqref{laggenfunc}, that provides the corresponding unconstrained extension, represent the main results of the present work.

We also discussed the reduction of the field equations to the Labastida form $\cS=0$,
with particular attention to the sporadic low-dimensional cases where the naive procedure
encounters some problems and, at the same time, new Weyl-like symmetries emerge. Just as we did for
bosons in \cite{bose_mixed}, for two-family Fermi fields we thus arrived at a
classification of these special cases, while for general $N$-family fields we contented
ourselves with a discussion of the general setup, together with the explicit analysis of
some classes of examples.

We have analyzed the properties of spinor-tensors of mixed symmetry mostly with reference
to a specific type of redundant presentation, via \emph{multi-symmetric
reducible} fields of the type $\psi_{\,\m_1 \ldots \, \m_{s_1} , \, \n_1
\ldots \, \n_{s_2} \, , \, \ldots}$ , that is capable of encompassing all possibilities.
However, for the sake of completeness we have also described how the theory
adapts itself to the case of \emph{irreducible} spinor-tensors. In addition, in order to allow a
better comparison with String Theory, we have explained how the formalism can be
modified in order to deal with \emph{multi-antisymmetric reducible} spinor-tensors.

While the presence of higher-derivative terms related to the compensators in the
unconstrained theory is somehow the price to pay for the simplest possible
setting, we also showed in Section \ref{sec:lowerder} how to properly modify the form of
$\cW$ so as to recover a formulation that is a bit more complicated but has the virtue of
involving only one-derivative terms. This can be attained with additional auxiliary fields, whose numbers depend only on the number of index families carried by the original gauge fields but not on their ranks, following \cite{dario07, bose_mixed}, albeit with some technical complications that emerge when one tries to define explicitly the corresponding gauge parameters.

The present work substantially completes the construction of ``minimal'' Lagrangian
theories for massless, unconstrained higher-spin fields in a flat background. This
program was developed in a series of previous papers exploiting the ``metric-like''
formalism, in an attempt to keep as close as possible to the conventional presentation for
lower-spin theories, while also exhibiting the geometrical interpretation of the free
dynamics, that is foreign to the original, constrained formulation of (Fang-)Fronsdal and
Labastida. In particular, the works \cite{fsoldnl, fms, dario07} were devoted to
characterizing the role played by linearized higher-spin curvatures in the dynamics of free massless or massive higher-spin fields. On the other hand, in
\cite{fsold2, fms, dario07, bose_mixed}, and finally in the present paper, we have
presented a complete local Lagrangian description for symmetric higher-spin fields in
flat and (A)dS backgrounds, and for fields of any symmetry type in flat
backgrounds, making use of a minimal amount of additional auxiliary fields that make up
for the direct use of higher-spin curvatures. This avoids all non-localities, but also
bypasses, at the same time, the need for algebraic constraints of any sort on gauge fields
and gauge parameters. In this local setting we mostly focused our attention on massless
fields, but in a flat space time the treatment can be directly
extended to the massive case, for instance via the harmonics of a Kaluza-Klein circle
reduction. The description of (partially) massive higher-spin fields \cite{partlymass} on (A)dS backgrounds
is less straightforward. It was discussed, in particular, in the second of \cite{fms} for
symmetric tensors, while a number of features of the mixed-symmetry case were recently
discussed in \cite{pernicolas}, extending the unfolded equations in Minkowski space time presented in \cite{skvortsov}. These papers also shed more light on the flat limit of the (A)dS description, supporting the BMV conjecture of \cite{BMV}.

Other interesting aspects of the free theory can still
benefit from further analysis. These include the quantization of the Lagrangians we are
proposing, the construction of supersymmetric higher-spin Lagrangians, possibly even resting on higher-spin analogues of the ordinary superalgebras, the analysis of
(partially) massive, mixed symmetry representations on (A)dS backgrounds, and a
closer look at the higher-spin geometry for the mixed symmetry, fermionic case. It is
clear, however, that the real issue is now to try and reach a comparable understanding of
the interactions, although progress in this respect is expected to be far more difficult
and slower. Indeed, the radical difficulties that are met in any naive attempts to go
beyond the linear regime for higher-spin fields stimulated along the years a number of
different approaches to the problem (see \cite{otherold, othernew} for an inevitably
incomplete list, including however some recent contributions). The approach initially proposed by Fradkin and Vasiliev
and then extensively pursued by Vasiliev himself and by his collaborators
\cite{vasold, vasnew} (see also \cite{sezsun} for relevant contributions along the same
lines) has proved, unquestionably, to be most successful to date, since it has led to non-linear equations for systems of infinitely many massless
fully symmetric higher-spin fields. Still, a better understanding along these more conventional lines can definitely spur a better grasp of the systematics of these constructions. All this, in one way or another,
supports the definite hope that the ``metric form'' of the theory of free higher-spin
fields attained by now can provide some useful insights both into the nature of
higher-spin interactions and, in a wider sense perhaps, into the very nature of String Theory.

\vskip 24pt


\section*{Acknowledgments}


We are very grateful to X. Bekaert and N. Boulanger for extended discussions
at the beginning
of this project. We are also grateful to the APC-Paris VII, to the \'Ecole
Polyt\'echnique, to the Chalmers University of Technology, to the
Scuola Normale Superiore di Pisa and to the ``Galileo
Galilei Institute'' (GGI) of Florence
for the kind hospitality extended to one or more of us at various
stages while this work was in
progress. Finally, we would like to thank the referee for urging us to extend Section 4
and to add Section 3.4.1. The present research was supported in part by APC-Paris
VII, by Scuola Normale Superiore, by INFN, by the CNRS, in
particular through the P2I program, by the MIUR-PRIN contract
2007-5ATT78, by the EU contracts MRTN-CT-2004-503369 and
MRTN-CT-2004-512194, by the NATO grant PST.CLG.978785,
and by the ERC Advanced Investigator Grant no. 226455
``Supersymmetry, Quantum Gravity and Gauge Fields'' (SUPERFIELDS).

\newpage

\begin{appendix}


\scs{Notation and conventions}\label{app:MIX}


In this paper we use the ``mostly plus'' convention for the
space-time signature and resort to a compact notation, eliminating
all space-time indices from tensor relations. The fields of interest
are here multi-symmetric spinor-tensors $\psi_{\,\mu_1 \ldots\,
\mu_{s_1},\,\nu_1 \ldots\, \nu_{s_2}\,,\,\ldots\,}$, whose sets of
symmetric indices are referred to as ``families''. In Section
\ref{sec:multiforms} we also describe how to adapt our results to
the case of multi-form spinor-tensors, where the indices belonging to a
given set are antisymmetrized, but in all the appendices, for
definiteness, we refer explicitly to the case of multi-symmetric
spinor-tensors, as in most of the present paper.

The fields $\psi_{\,\mu_1 \ldots\, \mu_{s_1},\,\nu_1 \ldots\,
\nu_{s_2}\,,\,\ldots\,}$, in the following simply denoted by $\psi$,
are thus fully symmetric under interchanges of pairs of indices
belonging to the same set, but have no prescribed symmetry relating
different sets, and are thus reducible $gl(D)$ spinor-tensors. As a
result, they are perhaps less familiar than Young projected
spinor-tensors, but are most convenient for the present discussion
and play a natural role in String Theory. For instance, in the five
superstring models these reducible spinor-tensors accompany
products of bosonic oscillators of the type $\alpha_{\,-1}^{\,\mu_1
\phantom{\mu_{s_1}}} \hspace{-10pt} \ldots\,
\alpha_{\,-1}^{\,\mu_{s_1}} \, \alpha_{\,-2}^{\,\nu_1
\phantom{\mu_{s_1}}} \hspace{-10pt} \ldots\,
\alpha_{\,-2}^{\,\nu_{s_2}} \ldots\,$, that are only symmetric under
interchanges of pairs of identical oscillators, while the reducible
multi-forms described in Section \ref{sec:multiforms} accompany
products of fermionic oscillators. In general, the massive
superstring spectra contain spinor-tensors with some symmetric and
some antisymmetric index sets, whose theory can be deduced from the
results of this paper. Recovering conventional field
theories from the present formulation requires suitable projections,
and this procedure is discussed in Section \ref{sec:irreducible}.

Aside from these basic fields, Lagrangians and field equations
involve $\g$-traces, traces, gradients and divergences of $\psi$, as
well as Minkowski metric tensors related to one (or two) of the
previous index sets and $\g$-matrices. As a result, ``family
indices" are often needed in order to specify the sets to which some
tensor indices belong. As in \cite{bose_mixed}, these family indices
are here denoted by small-case Latin letters, and the Einstein
convention for summing over pairs of them is used throughout. It
actually proves helpful to be slightly more precise: \emph{upper}
family indices are thus reserved for operators, like a gradient,
which \emph{add} space-time indices, while \emph{lower} family
indices are used for operators, like a divergence, which
\emph{remove} them. For instance gradients, divergences and traces
of a field $\psi$ are denoted concisely by $\pr^{\, i} \, \psi$,
$\pr_{\, i}\, \psi$ and $T_{ij} \, \psi$. This shorthand notation
suffices to identify the detailed meaning of these symbols, so that
\begin{eqnarray} \label{operations}
\pr^{\, i} \, \psi & \equiv & \pr_{\,(\,\m^i_1|} \, \psi_{\,\ldots \,,\, | \, \m^i_2 \, \ldots \, \m^i_{s_i+1} ) \,,\, \ldots} \ , \nonumber \\
\pr_{\, i}\, \psi & \equiv & \pr^{\,\l} \, \psi_{\,\ldots \,,\, \l \, \m^i_1 \, \ldots \, \m^i_{s_i-1} \,,\, \ldots} \ , \nonumber \\
T_{ij} \, \psi & \equiv & \psi^{\phantom{\,\ldots \,,\,} \l}{}_{\hspace{-19pt} \ldots \,,\,\phantom{\l} \ \m^i_1 \, \ldots \, \m^i_{s_i-1} \,,\, \ldots \,,\, \l \,
\m^j_1 \ldots \, \m^j_{s_j-1} \,,\, \ldots} \ ,
\end{eqnarray}
where, as in \cite{fsoldnl,fsoldl1,fsold2,fms,dario07,bose_mixed}, we work
with symmetrizations that are \emph{not} of unit strength, but
involve nonetheless the minimum possible number of terms, and we use
round brackets to denote them. Thus, for instance, the product
$T_{(\,ij}\, T_{kl\,)}$ here stands for
$T_{ij}\,T_{kl}+T_{ik}\,T_{jl}+T_{il}\,T_{jk}$. Moreover, we use
square brackets to denote antisymmetrizations.

In the mixed-symmetry case, one must also introduce mixed metric tensors
\be \label{metric} \h^{\,ij} \, \psi \, \equiv \, \12 \, \sum_{n\,=\,1}^{s_i+1}  \, \h_{\,\m^i_n \, (\,\m^j_1\,|}\, \psi_{\,\ldots \,,\, \m^i_{1} \ldots \, \m^i_{r\neq n} \ldots \, \m^i_{s_i+1} \,,\, \ldots
\,,\, |\,\m^j_2 \ldots \m^j_{s_j+1}) \,,\, \ldots} \ . \ee
Notice that the $\h^{ij}$ are
rescaled for $i \neq j$, in order that the diagonal terms retain the
conventional normalization. Aside from these basic operations, that
were already introduced in \cite{bose_mixed}, for the Fermi fields
treated in this paper one must also define some additional
operations involving $\g$-traces,
\begin{eqnarray} \label{operations_ferm}
\g^{\, i} \, \psi & \equiv & \g_{\,(\,\m^i_1|} \, \psi_{\,\ldots \,,\, |\, \m^i_2 \, \ldots \, \m^i_{s_i+1} ) \,,\, \ldots} \ , \nonumber \\
\g_{\, i}\, \psi & \equiv & \g^{\, \l} \, \psi_{\,\ldots \,,\, \l \, \m^i_1 \, \ldots \, \m^i_{s_i-1} ,\, \ldots} \ , \nonumber \\
\dsl \,\, \psi & \equiv & \g_{\,\l} \, \pr^{\,\l} \, \psi_{\,\m^1_1 \, \ldots \, \m^1_{s_1} ,\, \ldots} \ .
\end{eqnarray}
The space-time $\g$-matrices introduced here satisfy the usual
Clifford algebra
\be \label{cliffordspace}
\{\,\g^{\,\m} \comma \g^{\,\n} \,\} \, = \, 2 \, \h^{\,\m\n} \, ,
\ee
and moreover
\be
(\g_{\,0})^{\, \dagger} \, = \, - \, \g_{\,0} \, , \qquad (\g_{\,\m\,\neq\,0})^{\, \dagger} \, = \, \g_{\,\m\,\neq\,0} \, .
\ee
In the present paper we also use the antisymmetric
basis for ``family'' $\g$-matrices, introducing
\be \label{anti}
\g_{\,i_1 \ldots \, i_n} \, \equiv \, \frac{1}{n!} \, \g_{\,[\,i_1} \, \g_{\,i_2} \ldots \, \g_{\,i_n \, ]} \, ,
\ee
and corresponding objects with raised indices. Furthermore, different groups of \emph{symmetrized} indices borne by a given quantity are separated by colons, while a semicolon signals the beginning of a group of \emph{antisymmetrized} indices.

In order to further simplify the combinatorics, as in
\cite{bose_mixed} it proves convenient to introduce the scalar
product
\be \label{scalar} \bra \bar{\psi} \comma \c \ket \, \equiv \, \frac{1}{s_1! \ldots s_n!} \ \bar{\psi}_{\,\m^1_1 \ldots \, \m^1_{s_1} \, , \, \ldots \, , \,
\m^n_1 \ldots \, \m^n_{s_n}} \, \c^{\,\m^1_1 \ldots \, \m^1_{s_1} \, , \, \ldots \, , \, \, \m^n_1 \ldots \, \m^n_{s_n}} \, \equiv \,
\frac{1}{s_1! \, \ldots s_n!} \ \bar{\psi} \, \c \, . \ee
Inside the brackets it is then possible to integrate by parts and to
turn $\h$'s into traces without introducing any $s_i$-dependent
combinatoric factors, since
\be \label{deriv}
\bra \bar{\psi} \comma \pr^{\,i} \, \c \ket \, \equiv \,  \frac{1}{s_1! \ldots s_n!} \ \bar{\psi} \ \pr^{\,i} \, \c \, = \, - \, \frac{s_i}{s_1! \ldots s_n!} \
(\,\pr_{\,i}\, \bar{\psi}\,) \, \c \, \equiv \, - \, \bra \pr_{\,i}\, \bar{\psi} \comma \c \ket \, , \ee
and, for instance, if $i \neq j$,
\be
\bra \bar{\psi} \comma \h^{\,ij} \, \c \ket \, \equiv \, \frac{1}{s_1! \ldots s_n!} \ \bar{\psi} \ \h^{\,ij} \, \c \, = \, \12 \, \frac{s_i \, s_j}{s_1! \ldots
s_n!} \ (\,T_{ij} \, \bar{\psi}\,) \, \c \, \equiv \, \12 \, \bra T_{ij} \, \bar{\psi}
\comma \c \ket \, , \label{byparts}
\ee
where the reader should notice the somewhat unusual factor
$\frac{1}{2}$, originating from our choice of normalization
in eq.~(\ref{metric}). In a similar fashion, for
$\g$-matrices
\be
\bra \bar{\psi} \comma \g^{\,i} \, \c \ket \, \equiv \, \frac{1}{s_1! \ldots s_n!} \ \bar{\psi} \ \g^{\,i} \, \c \, = \, \frac{s_i}{s_1! \ldots s_n!} \ (\,\bar{\psi}\,\g_{\,i}\,) \, \c \, \equiv \, \bra \bar{\psi}\, \g_{\,i} \comma \c \ket \, .
\ee
However, in the main body of this paper we are ignoring, for
simplicity, the overall factors $\prod_{i=1}^N s_i!$, that should
accompany the Lagrangians of multi-symmetric spinor-tensors to grant
them the conventional normalization.

The scalar products prove quite convenient to derive Lagrangians and
field equations for mixed-symmetry fields, but they would be as
convenient for symmetric tensors. In this notation the Lagrangian
for symmetric, or one-family, spinor-tensors, would simply read
\be {\cal L} \, = \, \frac{1}{2} \, \bra \bar{\psi} \comma \cW \,-\,
\12 \, \g \, \cWsl \, - \, \12 \, \h \, \cW^{\,\pe} \ket \, + \,
\frac{1}{4} \, \bra \bar{\x} \comma \pr\, \cdot \,  \cWsl \ket \, -
\, \frac{1}{8} \, \bra \xibsl \comma \pr\, \cdot\, \cW^{\,\pe} \ket
\, + \, \frac{1}{4} \, \bra \bar{\l} \comma \cZ \ket \, +
\,\textrm{h.c.} \ ,\ee
to be compared with the corresponding expression of \cite{fms},
\be
\begin{split}
{\cal L} \, & = \, \frac{1}{2} \  \bar{\psi} \left( \cW - \frac{1}{2} \, \g \, \cWsl \, - \, \frac{1}{2} \, \h \, \cW^{\; \prime} \right) \, - \, \frac{3}{4} \left( s \atop 3 \right)
\, \bar{\xisl} \ \partial \cdot \cW^{\; \prime} \\
& + \, \frac{1}{2} \left( s \atop 2 \right) \, \bar{\xi} \ \pr \, \cdot \, \cWsl \, + \, \frac{3}{2} \left( s \atop 4 \right) \, \bar{\l} \, \cZ \, + \, \textrm{h.c.} \, ,
\end{split}
\ee
that contains explicit combinatoric factors.

In order to take proper advantage of this compact notation, it is
convenient to collect a number of identities that are used
recurrently in this type of analysis. As was already pointed out in
\cite{bose_mixed}, these have a more complicated structure
than their symmetric counterparts of \cite{fms}, since they involve
a genuinely new type of operation. This turns a tensor with $s_i$
indices in the $i$-th group into others with $s_i+1$ indices in the
$i$-th group and $s_j-1$ indices in the $j$-th group, according to
\be \label{flip} S^{\,i}{}_{j} \, \psi \, \equiv \, \psi_{\ldots \,,\, ( \m^i_1 \ldots\, \m^i_{s_i} | \,,\, \ldots \,,\, |\, \m^i_{s_i+1} ) \, \m^j_1
\ldots\, \m^j_{s_j-1} \,,\, \ldots}  \ . \ee
The new rules that are needed follow from the algebra of the various
operators, and can be also derived from a realization in terms of
bosonic oscillators, along the lines of bosonic String Field Theory
and of \cite{firstlaba,labastida,labferm}. The subalgebra
that does not contain the operators built using $\g$-matrices reads
\begin{align}
&[ \, \pr_{\,i} \comma \pr^{\,j} \, ] \ = \ \Box \, \d_{\,i}{}^{\,j} \, , \\[0pt]
&[\, T_{ij} \comma \pr^{\,k} \,] \ = \ \pr_{\,(\,i}\,\d_{\,j\,)}{}^k \, , \\[0pt]
&[\, \pr_{\,k} \comma \h^{\,ij} \,] \ = \ \12 \ \d_{\,k}{}^{\,(\,i} \, \pr^{\,i\,)} \, , \\[0pt]
&[\, T_{ij} \comma \h^{\,kl} \,] \ = \ \frac{D}{2} \ \d_{\,i}{}^{\,(\,k\,}\,\d_{\,j}{}^{\,l\,)} \, + \,
\12 \left( \, \d_{\,i}{}^{\,(\,k}S^{\,l\,)}{}_j \, + \, \d_{\,j}{}^{\,(\,k}S^{\,l\,)}{}_i \, \right) , \\[0pt]
&[\, S^{\,i}{}_j \comma \h^{\,kl} \,] \ = \ \d_{\,j}{}^{\,(\,k} \, \h^{\,l\,)\,i} \, , \\[0pt]
&[\, T_{ij} \comma S^{\,k}{}_l \,] \ = \ T_{l\,(\,i}\, \d_{\,j\,)}{}^{\,k} \, , \\[0pt]
&[\, S^{\,i}{}_j \comma \pr^{\,k} \,] \ = \ \pr^{\,i} \, \d_{\,j}{}^{\,k}  \, ,  \\[0pt]
&[\, \pr_{\,k} \comma S^{\,i}{}_j \,] \ = \ \d_{\,k}{}^{\,i} \, \pr_{\,j}   \, , \\[0pt]
&[\, S^{\,i}{}_j \comma S^{\,k}{}_l \,] \ = \ \d_{\,j}{}^{\,k}S^{\,i}{}_l \, - \, \d_{\,l}{}^{\,i}S^{\,k}{}_j \, . \label{SS}
\end{align}
Notice that the commutators of the $S^{\,i}{}_{j}$ operators, the key novelty of the mixed-symmetry case, build the $gl(N)$ Lie algebra if $N$ index families are
present.

The usual anti-commutation rules \eqref{cliffordspace} of the
Clifford algebra induce the following relations for the $\g$'s
carrying family indices,
\begin{alignat}{2}
& \{\, \g_{\,i} \comma \g_{\,j} \,\} & & = \, 2 \, T_{ij} \, , \\
& \{\, \g^{\,i} \comma \g^{\,j} \,\} & & = \, 4 \, \h^{\,ij} \, ,
\end{alignat}
where the additional factor in the second of these relations
originates from our choice of normalization for $\eta^{ij}$ in
eq.~(\ref{metric}). The reader should notice that the structure of
the Clifford algebra is only apparently transferred to the $\g$'s
with family indices, that are actually more complicated because the
off-diagonal $T_{ij}$ and $\h^{ij}$ do not vanish in general. The
difference is already manifest in the anticommutator
\be
\{\, \g_{\,i} \comma \g^{\,j} \,\} \, = \, D \, \d_{\,i}{}^{\,j} \, + \, 2 \, S^{\,j}{}_{\,i} \ ,  \label{anticomgamma}
\ee
that also contains the $S^{\,i}{}_{j}$ operators. The other relevant basic (anti)commutation relations are
\begin{align}
&[\, \g_{\,i} \comma \pr^{\,j} \,] \ = \ \, \dsl \ \d_{\,i}{}^{\,j} \, , \\[0pt]
&[\, \pr_{\,i} \comma \g^{\,j} \,] \ = \ \, \dsl \ \d_{\,i}{}^{\,j} \, , \\[0pt]
&[\, \g_{\,i} \comma \h^{\,jk} \,] \ = \ \12 \ \d_{\,i}{}^{\,(\,j}\, \g^{\,k\,)} \, ,  \\[0pt]
&[\, T_{ij} \comma \g^{\,k} \,] \ = \ \g_{\,(\,i}\, \d_{\,j\,)}{}^{\,k} \, ,  \\[0pt]
&[\, S^{\,i}{}_{j} \comma \g^{\,k} \,] \ = \ \g^{\,i} \, \d_{\,j}{}^{\,k} \, ,  \\[0pt]
&[\, \g_{\,k} \comma S^{\,i}{}_j \,] \ = \ \d_{\,k}{}^{\,i} \,\g_{\,j} \, ,  \\[0pt]
&\{ \, \g_{\,i} \, , \dsl \, \} \ = \ 2 \, \pr_{\,i} \, , \\[0pt]
&\{ \dsl \,,\, \g^{\,i} \, \} \ = \ 2 \, \pr^{\,i} \, ,
\end{align}
and the previous relations give rise to the basic rules
collected in Appendix \ref{app:fermi}.

Finally, in this paper we make extensive use of a number of standard
tools related to the symmetric group. These include, in particular,
the Young projectors $Y$, that are used repeatedly to separate
irreducible components in family-index space. In the text the
various components are often specified by ordered lists of the
lengths of the rows for the associated diagrams, so that, for
instance, the $\{3,2\}$ component corresponds to
\be
\begin{picture}(0,25)(20,0)
\multiframe(0,10.5)(10.5,0){1}(10,10){}
\multiframe(10.5,10.5)(10.5,0){1}(10,10){}
\multiframe(21,10.5)(10.5,0){1}(10,10){}
\multiframe(0,0)(10,0){1}(10,10){}
\multiframe(10.5,0)(10,0){1}(10,10){}
\end{picture}
\ee
In general the Young projectors $Y$ can be built combining the
contributions of different Young tableaux, that can be identified
associating integer labels to the tensor indices to be projected and
allowing within the given graph all their arrangements such that
these integers grow from left to right and from top to bottom. In
some cases, however, this simple procedure can fail to
produce an orthogonal decomposition, which can still be attained by
a further Graham-Schmidt orthogonalization. This difficulty is not
present if, for any pair of tableaux, there is at least a couple of
indices belonging to a row of the first that lie in the same column
of the second, and vice versa. Let us stress that this
problem is never to be faced in our constructions, as a result of
the particular symmetry properties of our basic objects. Outside
Section \ref{sec:multiforms}, in this paper we adopt the conventions
of \cite{bose_mixed}, and Young tableaux are then defined in the
\emph{symmetric} basis, so that the projector corresponding to a
tableau containing $n$ boxes takes the form
\be
Y_\tau \, = \, \frac{\lambda(\tau)}{n\,!} \ S\, A \, , \label{young_proj_gen}
\ee
where S and A are the corresponding products of ``row symmetrizers''
and ``column antisymmetrizers''. On the other hand, in Section
\ref{sec:multiforms} it is more convenient to work in the
\emph{antisymmetric} basis, where the roles of S and A are
interchanged. Moreover, $\lambda(\tau)$ is the dimension of the
associated representation of the symmetric group, that can be
computed, for instance, counting the standard ways of filling the
boxes of the corresponding diagram with the numbers $1,2,\ldots,n$
in increasing order from left to right and from top to bottom. We
often denote the ratio that appears in eq.~\eqref{young_proj_gen} as
\be \label{hooklenght}
\frac{\lambda(\tau)}{n\,!} \, = \, \frac{1}{h(\tau)} \, ,
\ee
where $h(\t)$ is the ``hook length'' of the corresponding tableau.
Similar techniques are used extensively in Section
\ref{sec:irreducible} to identify irreducible $gl(D)$ and Lorentz
tensors.

The hook length $h$ also enters the formulae for the dimensions
of $gl(D)$ and $o(D)$ (spinorial) representations, that might prove useful to the reader. Thus, the number of components of an irreducible $gl(D)$ \emph{spinor-tensor} with
vector indices that are $\{s_1, \ldots ,s_N\}$ projected is
\be
\dim_{\,gl(D)}\, [\{s_1,\ldots,s_N\}] \, = \, \frac{2^{\,[\frac{D}{2}]}}{h} \, \prod_{i\,=\,1}^N \, \frac{(\,D+s_i-i\,)\,!}{(\,D-i\,)\,!} \, ,
\ee
while for the corresponding $o(D)$ \emph{spinor-tensor} this number is
\be \label{dim1}
\dim_{\,S}\, [\{s_1,\ldots,s_N\}] \, = \, \frac{2^{\,[\frac{D}{2}]}}{h} \, \prod_{i\,=\,1}^N \, \frac{(\,D+s_i-N-i\,)\,!}{(\,D-2\,i\,)\,!} \, \prod_{j\,=\,i+1}^N \, (\,D+s_i+s_j-i-j+1\,) \, ,
\ee
to be compared with the number of components for the $o(D)$ \emph{tensor} with
the same Young projection in its vector indices:
\be \label{dim2}
\dim_{\,T}\, [\{s_1,\ldots,s_N\}] \, = \, \frac{1}{h} \, \prod_{i\,=\,1}^N \, \frac{(\,D+s_i-N-i-1\,)\,!}{(\,D-2\,i\,)\,!} \, \prod_{j\,=\,i}^N \, (\,D+s_i+s_j-i-j\,) \, .
\ee
These last relations also appear in \cite{labastida}, but for a typo in the last products of eqs.~\eqref{dim1} and \eqref{dim2}.

We refrain from adding further details, since these and other related standard facts are discussed extensively in the literature, and in particular in \cite{branching}.

\vskip 24pt


\scs{Useful identities for fermions}\label{app:fermi}


Using repeatedly the (anti)commutation rules of Appendix
\ref{app:MIX} yields the useful identities
\begin{align}
& [\,\pr_{\,l} \, , \, \h^{i_1j_1} \ldots\, \h^{i_pj_p} \,] \, = \, \12 \, \sum_{n\,=\,1}^p \, \prod_{r\,\neq\,n}^p \, \h^{i_rj_r} \, \d_{\,l}{}^{\,(\,i_n}\, \pr^{\,j_n\,)} \, , \\[2pt]
& [\,T_{i_1j_1} \ldots\, T_{i_pj_p} \, , \, \pr^{\,l}\,] \, = \, \sum_{n\,=\,1}^p \, \prod_{r\,\neq\,n}^p \, T_{i_rj_r} \, \pr_{\,(\,i_n}\, \d_{\,j_n\,)}{}^{\,l} \, ,
\end{align}
that were already presented in \cite{bose_mixed}, and in a similar
fashion
\begin{align}
& [\,\pr_{\,l} \, , \, \g^{\,k_1 \ldots\, k_q} \,] \, = \, (-1)^{\,q+1} \ \d_{\,l}{}^{\,[\,k_1}\, \g^{\,k_2 \ldots\, k_q\,]} \dsl \, + \, \d_{\,l}{}^{\,[\,k_1\,} \pr^{\,k_2}\, \g^{\,k_3 \ldots\, k_q\,]} \, , \\[5pt]
& [\,\g_{\,k_1 \ldots\, k_q} \, , \, \pr^{\,l}\,] \, = \, (-1)^{\,q+1} \dsl \ \g_{\,[\,k_1 \ldots\, k_{q-1} \,} \d_{\,k_q\,]}{}^{\,l} \, + \, \g_{\,[\,k_1 \ldots\, k_{q-2}\,} \pr_{\,k_{q-1}\,} \d_{\,k_q\,]}{}^{\,l} \, , \\[5pt]
& [\,\g_{\,k_1 \ldots\, k_{q}} \, , \dsl \ ]_{\,(-1)^{\,q+1}} = \, 2 \ \g_{\,[\,k_1 \ldots\, k_{q-1}\,} \pr_{\,k_q\,]} \, , \\[5pt]
& [\, \dsl \, , \, \g^{\,k_1 \ldots\, k_q} \,]_{\,(-1)^{\,q+1}} = \, 2 \ \pr^{\,[\,k_1}\, \g^{\,k_2 \ldots\, k_q\,]} \, ,
\end{align}
where the symbols $[\, \comma \,]_{\,(-1)^{\,q+1}}$ can
denote both commutators and anticommutators, depending on the sign
of $(-1)^{\,q+1}$. They are widely used in Section \ref{sec:2fermi}
and in Section \ref{sec:generalf} in the construction of the
unconstrained Lagrangians. For the reduction of the equation of
motion to the Labastida form it is instead convenient to introduce
the new operators
\be \label{newgamma}
\g^{\,k_1 \ldots\, k_q}{}_{\,l} \, = \, \g^{\,k_1 \ldots\, k_q}\, \g_{\,l} \, - \, \g^{\,[\,k_1 \ldots\, k_{q-1}\,} S^{\,k_q\,]}{}_{\,l} \, ,
\ee
that appear in the identities
\begin{align}
& [\,\g_{\,l} \, , \, \g^{\,k_1 \ldots\, k_q} \,]_{\,(-1)^{\,q+1}} = \, (\,D-q+1\,)\ \d_{\,l}{}^{\,[\,k_1}\, \g^{\,k_2 \ldots\, k_q\,]} \, + \, 2\, (-1)^{\,q+1}\, \g^{\,[\,k_1 \ldots\, k_{q-1}\,} S^{\,k_q\,]}{}_{\,l} \, , \\[5pt]
& [\,\g_{\,k_1 \ldots\, k_{q}} \, , \, \g^{\,l} \,]_{\,(-1)^{\,q+1}} = \, (\,D-q+1\,)\ \g_{\,[\,k_1 \ldots\, k_{q-1}\,} \d_{\,k_q\,]}{}^{\,l} \, + \, 2 \, (-1)^{q+1}\, S^{\,l}{}_{\,[\,k_1\,} \g_{\,k_2 \ldots\, k_q\,]}  \, , \\[5pt]
& [\,T_{\,lm} \, , \, \g^{\,k_1 \ldots\, k_q} \,] \, = \, (-1)^{\,q+1} \, \d_{\,(\,l\,|}{}^{\,[\,k_1}\, \g^{\,k_2 \ldots\, k_q\,]}{}_{\,|\,m\,)} \, , \\[5pt]
& [\,\g_{\,lm} \, , \, \g^{\,k_1 \ldots\, k_q} \,] \, = \, - \, (\,D-q+1\,)\,(\,D-q+2\,)\ \d_{\,l}{}^{\,[\,k_1}\, \d_{\,m}{}^{\,k_2}\, \g^{\,k_3 \ldots\, k_q\,]} \nn \\[2pt]
& \phantom{[\,\g_{\,lm} \, , \, \g^{\,k_1 \ldots\, k_q} \,] \,} + \, (-1)^{\,q+1} \, (\,D-q+1\,)\ \d_{\,[\,l\,|}{}^{\,[\,k_1}\, \g^{\,k_2 \ldots\, k_{q-1}\,} S^{\,k_q\,]}{}_{\,|\,m\,]} \nn \\[2pt]
& \phantom{[\,\g_{\,lm} \, , \, \g^{\,k_1 \ldots\, k_q} \,] \,} + \, (-1)^{\,q} \, (\,D-q+2\,)\ \d_{\,[\,l\,|}{}^{\,[\,k_1}\, \g^{\,k_2 \ldots\, k_q\,]}{}_{\,|\,m\,]} + \, 2 \ \g^{\,[\,k_1 \ldots\, k_{q-1}\,|}{}_{\,[\,l}\, S^{\,|\,k_q\,]}{}_{\,m\,]}  \, , \\[5pt]
& [\,S^{\,l}{}_{m} \, , \, \g^{\,k_1 \ldots\, k_q} \,] \, = \, (-1)^{\,q+1} \, \d_{\,m}{}^{\,[\,k_1}\, \g^{\,k_2 \ldots\, k_q\,]\,l} \, , \\[5pt]
& [\,\g_{\,k_1 \ldots\, k_q} \, , \, S^{\,l}{}_{m} \,] \, = \, (-1)^{\,q+1} \, \g_{\,m\,[\,k_1 \ldots\, k_{q-1}\,} \d_{\,k_q\,]}{}^{\,l} \, .
\end{align}
These are used together with the corresponding commutators involving products of metric tensors:
\begin{align}
& [\,\g_{\,l} \, , \, \h^{i_1j_1} \ldots\, \h^{i_pj_p} \,] \, = \, \12 \, \sum_{n\,=\,1}^p \, \prod_{r\,\neq\,n}^p \, \h^{i_rj_r} \, \d_{\,l}{}^{\,(\,i_n}\, \g^{\,j_n\,)} \, , \\[2pt]
& [\,T_{\,lm} \, , \, \h^{i_1j_1} \ldots\, \h^{i_pj_p} \,] \, = \, \12 \, \sum_{n\,=\,1}^p \, \Big\{ \ D \, \prod_{r\,\neq\,n}^p \h^{i_rj_r}\, \d_{\,l}{}^{\,(\,i_n}\,  \d_{\,m}{}^{\,j_n\,)} \\
& \phantom{[\,T_{\,lm} \, , \, \h^{i_1j_1} \ldots\, \h^{i_pj_p} \,] \,} + \, \sum_{m \, < \, n} \, \prod_{r\,\neq\,m,n}^p \!\!\h^{i_rj_r} \d_{\,(\,l\,|}{}^{(\,i_m\,} \h^{\,j_m\,)\,(\,i_n\,} \d^{\,j_n\,)}{}_{\,|\,m\,)} + \prod_{r\,\neq\,n}^p \h^{i_rj_r} \, \d_{\,(\,l\,|}{}^{(\,i_n\,} S^{\,j_n\,)}{}_{|\,m\,)} \, \Big\} \nn \, , \\[2pt]
& [\,\g_{\,lm} \, , \, \h^{i_1j_1} \ldots\, \h^{i_pj_p} \,] \, = \, \frac{1}{4} \, \sum_{n\,=\,1}^p \, \Big\{\, \sum_{m \, < \, n} \, \prod_{r\,\neq\,m,n}^p \!\!\h^{i_rj_r} \, \d_{\,[\,l\,|}{}^{(\,i_m\,} \g^{\,j_m\,)\,(\,i_n\,} \d^{\,j_n\,)}{}_{\,|\,m\,]} \nn \\
& \phantom{[\,\g_{\,lm} \, , \, \h^{i_1j_1} \ldots\, \h^{i_pj_p} \,] \,} + \, 2\, \prod_{r\,\neq\,n}^p \h^{i_rj_r} \, \d_{\,[\,l\,|}{}^{(\,i_n\,} \g^{\,j_n\,)}{}_{|\,m\,]} \, \Big\} \, .
\end{align}
Finally, we are making a wide use of the following composition rules
for family $\g$-matrices:
\begin{align}
& \g_{\,k_1 \ldots\, k_q} \, \g_{\,l} \, = \, \g_{\,k_1 \ldots\, k_q \, l} \, + \, \g_{\,[\,k_1 \ldots\, k_{q-1}} \, T_{k_q\,]\,l} \, , \label{compgamma1} \\[3pt]
& \g_{\,k_1 \ldots\, k_q} \, \g_{\,lm} \, = \, \g_{\,k_1 \ldots\, k_q\, lm} \, + \left(\, \g_{\,[\,k_1 \ldots\, k_{q-1}\,|\,m} \, T_{\,|\,k_q\,]\,l} \, - \, \g_{\,[\,k_1 \ldots\, k_{q-1}\,|\,l} \, T_{\,|\,k_q\,]\,m} \,\right) \nonumber \\
& \phantom{\g_{\,k_1 \ldots\, k_q} \, \g_{\,lm} \,} + \, \g_{\,[\,k_1 \ldots\, k_{q-2}} \, T_{k_{q-1}\,|\,m}\, T_{\,|\,k_q\,]\,l} \, . \label{compgamma2}
\end{align}
Similar rules also apply when one raises indices, but the
normalization of $\h^{\,ij}$ introduced in eq.~\eqref{metric}
requires that this operation be supplemented by the substitution
\be
T_{ij} \, \to \, 2 \, \h^{ij} \, .
\ee
In order to derive the equations of motion, or in order to impose
the condition \eqref{selfadjferm} of self-adjointness of the
constrained Lagrangians, it is also convenient to keep in mind the
further, standard relation
\be
\g_{\,0} \, (\g_{\,k_1 \ldots\, k_q})^{\,\dagger}\, \g_{\,0} \, = \, (-1)^{\frac{(q-1)(q-2)}{2}} \, \g_{\,k_1 \ldots\, k_q} \, .
\ee

The previous identities also suffice to compute generic $\g$-traces
of the kinetic tensors $\cS$ and $\cW$ and of the Bianchi
identities. In particular, starting from the Fang-Fronsdal-Labastida
tensor
\be \cS \, = \, i \left(\, \dsl \, \psi \, - \, \pr^{\,i} \!
\psisl_i \,\right)\, , \ee
one can obtain
\be \label{sgentrace}
\begin{split}
& \prod_{r\,=\,1}^p \, T_{i_rj_r} \, \g_{\,k_1 \ldots\, k_q} \, \cS \, = \, (-1)^{\,q} \, i \, (\,q+1\,) \dsl \ \g_{k_1 \ldots\, k_q} \prod_{r\,=\,1}^p \, T_{i_rj_r} \, \psi  \\
& + \,  i\,(\,q+1\,) \, \g_{\,[\,k_1 \ldots\, k_{q-1}\,} \pr_{\,k_q\,]} \prod_{r\,=\,1}^p \, T_{i_rj_r} \, \psi \, - \,i\, \sum_{n\,=\,1}^p \, \g_{\,[\,k_1 \ldots\, k_{q-1}\,}  T_{k_q\,]\,(\,i_n} \, \pr_{\,j_n\,)} \prod_{r\,\ne\,n}^p \, T_{i_rj_r} \, \psi \\
& - \, i \, \sum_{n\,=\,1}^p \, \g_{\,k_1 \ldots\, k_q \, (\,i_n} \, \pr_{\,j_n\,)} \prod_{r\,\ne\,n}^p \, T_{i_rj_r} \, \psi \, - \, i\ \pr^{\,l} \, \g_{\,[\,k_1 \ldots\, k_{q-1}\,} T_{k_q\,]\,l} \prod_{r\,=\,1}^p \, T_{i_rj_r} \, \psi \\
&  - \, i\ \pr^{\,l} \, \g_{k_1 \ldots\, k_q l} \prod_{r\,=\,1}^p \, T_{i_rj_r} \, \psi \, ,
\end{split}
\ee
and this result is at the root of the derivation, in
eq.~\eqref{selfadjferm}, of the constrained Lagrangians via the
identification of a constrained self-adjoint operator. It is also
needed in a direct derivation of the equations of motion, and in
this context the relevant $\g$-traces for a two-family gauge field
read
\begin{align}
& \ssl_i \, = \, i \left(\, - \, 2 \dsl \, \psisl_i \, + \, 2 \, \pr_{\,i} \, \psi \, - \, \pr^{\,j} \, T_{ij} \, \psi \, - \, \pr^{\,j} \, \g_{\,ij} \, \psi \,\right) \, ,  \\[10pt]
& T_{ij} \, \cS \, = \, i \left(\, \dsl \ T_{ij} \, \psi \, - \, \pr_{\,(\,i} \, \!\psisl_{j\,)} \, - \, \pr^{\,k} \, T_{ij} \psisl_k  \,\right) \, ,  \\[10pt]
& \g_{\,ij} \, \cS \, = \, i \left(\, 3 \!\dsl \ \g_{\,ij} \, \psi \, + \, 3 \, \g_{\,[\,i}\,\pr_{\,j\,]} \, \psi \, + \, \pr^{\,k} \, T_{k\,[\,i} \, \g_{\,j\,]} \, \psi \, - \, \pr^{\,k} \, \g_{\,ijk} \, \psi \,\right) \, , \\[10pt]
& T_{ij} \ssl_k \, = \, i \, \Big(\, \left(\, 2 \, T_{ij} \, \pr_{\,k} \, - \, T_{k\,(\,i} \,\pr_{\,j\,)} \,\right) \psi \, - \, \g_{\,k\,(\,i} \, \pr_{\,j\,)} \, \psi \, - \, 2 \dsl \ T_{ij} \, \g_{\,k} \, \psi \nn \\
& \phantom{T_{ij} \ssl_k \, = \, i \, \Big(\,} - \, \pr^{\,l} \, T_{ij}\,T_{kl} \, \psi \, - \, \pr^{\,l} \, T_{ij} \, \g_{\,kl} \, \psi \,\Big) \, , \\[10pt]
& T_{ij} \, T_{kl} \, \cS \, = \, i \left( \dsl \ T_{ij}\,T_{kl} \, \psi
\, - \, T_{ij} \, \pr_{\,(\,k} \psisl_{\,l\,)} \, - \, T_{kl} \,
\pr_{\,(\,i} \psisl_{\,j\,)} \, - \, \pr^{\,m} \, T_{ij}\,T_{kl}
\psisl_m \,\right) \, .
\end{align}

Other identities that are useful for a direct construction of the field
equations in the two-family case are:
{\allowdisplaybreaks
\begin{align}
& \cWsl_i \, = \, \ssl_i \, + \, 2 \, i \, \pr^{\,j} \!\!\dsl \ \x_{ij} \, + \, i \, \pr^{\,j}\pr^{\,k} \, \g_{\,i} \, \xi_{jk} \, ,  \\[10pt]
& T_{ij} \, \cW \, = \, T_{ij} \, \cS \, + \, 2 \, i \, \Box \, \xi_{ij} \, + \, 2 \, i \, \pr^{\,k}\,\pr_{\,(\,i}\,\x_{j\,)\,k} \, + \, i \, \pr^{\,k}\pr^{\,l} \, T_{ij}\,\x_{kl}  \, ,  \\[10pt]
& \g_{\,ij} \, \cW \, = \, \g_{\,ij} \, \cS \, - \, 2 \, i \, \pr^{\,k} \!\! \dsl \ \g_{\,[\,i} \, \x_{j\,]\,k} \, + 2 \, i \, \pr^{\,k}\, \pr_{\,[\,i}\,\x_{j\,]\,k} \, + \, i \, \pr^{\,k}\pr^{\,l} \, \g_{\,ij} \, \x_{kl} \, , \\[10pt]
& T_{ij} \, \g_{\,k} \, \cW \, = \,  T_{ij} \, \g_{\,k} \, \cS \, + \, 2 \, i \, \Box \, \g_{\,k}\,\x_{ij} \, + \, 2 \, i \!\! \dsl \ \pr_{\,(\,i}\,\x_{\,j\,)\,k} \, + \, 2 \, i \, \pr^{\,l}\, \g_{\,k} \, \pr_{\,(\,i} \, \x_{j\,)\,l} \nn \\[2pt]
& \phantom{T_{ij} \, \g_{\,k} \, \cW \,} + \, 2 \, i \, \pr^{\,l} \! \dsl \ T_{ij}\,\x_{kl} \, + \, i \, \pr^{\,l}\pr^{\,m} \, T_{ij}\,\g_{\,k} \, \x_{\,lm}  \, , \\[10pt]
& T_{ij} \, T_{kl} \, \cW \, = \, T_{ij}\,T_{kl} \, \cS \, + \, 2 \, i \, \Box \, \left(\, T_{ij}\,\x_{kl} \, + \, T_{kl}\,\x_{ij} \,\right) \, + \, 2 \, i \, \left(\, \pr_{\,i\,}\pr_{\,(\,k}\,\x_{\,l\,)\,j} \, + \, \pr_{\,j\,}\pr_{\,(\,k}\,\x_{\,l\,)\,i} \,\right) \nn \\[2pt]
& \phantom{T_{ij} \, T_{kl} \, \cW \,} + \, 2 \, i \, \pr^{\,m} \left(\,
T_{ij}\,\pr_{\,(\,k}\,\x_{\,l\,)\,m} \, + \,
T_{kl}\,\pr_{\,(\,i}\,\x_{\,j\,)\,m} \,\right) \, + \, i \,
\pr^{\,m}\pr^{\,n} \, T_{ij}\,T_{kl} \, \x_{mn} \, .
\end{align}}

\vskip 24pt

\scs{Proof of some results used in Section \ref{sec:generalf}}
\label{app:idsf}

The construction of the Lagrangians for general Fermi fields of mixed symmetry
presented in Sections \ref{sec:lagrangianf} and \ref{sec:lagrangianfunc} rests upon a number of statements that were
referred to without presenting their proofs. In this Appendix we would like to provide
the missing details:
\begin{itemize}
\item the proof that, for any given number of traces
and antisymmetric $\g$-traces of the kinetic tensor $\cS$, all Young
projections in family indices with more than two columns
vanish in the constrained theory;
\item the proof of eq.~\eqref{idconstrf}, that links terms in the gauge variation of the Lagrangians involving projections with more than two columns to the Labastida constraints on the gauge parameters;
\item the generic traces and $\g$-traces of the Bianchi
identities and some of their projections that play an important role
in our constructions;
\item the actual link between eq.~\eqref{combianchif} and a combination of
$\g$-traces of the Bianchi identities involving only two-column
projected $\g$-traces of the unconstrained kinetic tensor $\cW$.
\end{itemize}
These four points will be discussed in detail in the following four subsections.

\subsection*{\sc Young projections and constraints}

In the symmetric basis that we adopt in most of this paper for the
Young tableaux, all projections with more than two columns of a
generic combination of traces and antisymmetrized $\g$-traces can be
expressed as a sum of terms where at least three family indices are
symmetrized. We can now show that they are all directly related
to the Labastida constraints \eqref{constrpsi} if the $\g$-traces
are applied to $\psi$, or to the constraints \eqref{constrs} if the
$\g$-traces are applied to $\cS$.

To begin with, one can hide the traces that do not carry any of the
symmetrized indices, because they do not play a role in the
following analysis. Then, out of the possible placements of
these indices the combination
\be
T_{i\,(\,a} \, T_{\,bc\,)} \, \g_{\,k_1 \ldots\, k_q}
\ee
is directly related to the constraints. This is so since the cycle
can be directly extended to contain four indices, due to the
presence of two identical tensors, while the traces commute with
$\g_{\,k_1 \ldots\, k_q}$. There is another possibility that involves only
traces,
\be
T_{i\,(\,a\,|}\, T_{j\,|\,b}\, T_{\,c\,)\,l}\, \g_{\,k_1 \ldots\, k_q}\, .
\ee
It can be treated as in \cite{bose_mixed}, rewriting it in the form
\be
T_{i\,(\,a}\, T_{\,bc}\, T_{\,j\,)\,l}\, \g_{\,k_1 \ldots\, k_q} - \, T_{ij}\, T_{(\,ab}\, T_{\,c\,)\,l}\, \g_{\,k_1 \ldots\, k_q} \, - \, T_{i\,(\,a}\, T_{\,bc\,)}\, T_{jl}\, \g_{\,k_1 \ldots\, k_q} \, ,
\ee
so that it vanishes manifestly in the constrained theory, since all these terms
can be related to the constraints after extending the symmetrizations as pertains to
products of identical $T$ tensors. On the other hand, if one of the symmetrized indices is carried by a
(multiple) $\g$-trace, one can resort to eq.~\eqref{compgamma1}, obtaining
\begin{align}
& T_{(\,ab}\, \g_{\,c\,)\,k_1 \ldots\, k_{q-1}} = \, (-1)^{\,q+1} \, \g_{\,k_1 \ldots\, k_{q-1}}\, \g_{\,(\,a}\, T_{\,bc\,)} \, + \, (-1)^{\,q} \, \g_{\,[\,k_1 \ldots\, k_{q-2}}\, T_{\,k_{q-1}\,]\,(\,a}\, T_{\,bc\,)} \, , \nn \\[2pt]
& T_{i\,(\,a\,|}\, T_{\,j\,|\,b}\, \g_{\,c\,)\,k_1 \ldots\, k_{q-1}} = \, (-1)^{\,q+1}\, \g_{\,k_1 \ldots\, k_{q-1}}\, \g_{\,(\,a}\, T_{\,b\,|\,i}\, T_{\,|\,c\,)\,j} \nn \\
& \phantom{T_{i\,(\,a\,|}\, T_{\,j\,|\,b}\, \g_{\,c\,)\,k_1 \ldots\, k_{q-1}}} + \, (-1)^{\,q} \, \g_{\,[\,k_1 \ldots\, k_{q-2}}\, T_{\,k_{q-1}\,]\,(\,a}\, T_{\,b\,|\,i}\, T_{\,|\,c\,)\,j} \, . \label{relconstr}
\end{align}
All these terms therefore vanish either directly or after enlarging
the cycle as above.

Alternatively, as in the bosonic case discussed in \cite{bose_mixed} one can also
derive these results noticing that the expression
\be \label{expr1}
T_{i_1j_1} \ldots\, T_{i_{p-1}j_{p-1}} T_{(\,ab}\, \g_{\,c\,)\,k_1 \ldots\, k_{q-1}} \, ,
\ee
admits all Young projections allowed by
\be \label{expr2}
T_{i_1j_1} \ldots\, T_{i_{p-1}j_{p-1}} T_{ab}\, \g_{\,c\,k_1 \ldots\, k_{q-1}} \, ,
\ee
aside from the two-column one. Furthermore, acting on each
irreducible component of \eqref{expr1} the permutation group can
generate the entire corresponding irreducible subspaces, and in
particular the irreducible components of \eqref{expr2} that it
contains. This suffices to show that all Young projections
of the generic expression \eqref{expr2} with more than two columns are
actually related to the Labastida constraints. In fact, the previous
observation shows that they can be expressed as linear combinations
of \eqref{expr1} with similar quantities obtained permuting indices, and
\eqref{expr1} is related to the Labastida constraints via the first
of eqs.~\eqref{relconstr}.

\subsection*{\sc Proof of eq.~\eqref{idconstrf}}

The presence of the scalar product makes it possible to extract the
$\{3,2^{p-1}\}$ component of
\be \label{iniproof}
\bra T_{i_1j_1} \ldots T_{i_pj_p} \, \bar{\e}_{\,l} \, \g_{\,k_1 \ldots\, k_q} \comma \pr_{\,l} \, (\,\g^{\,[\,q\,]}\,\cS^{\,[\,p\,]}\,)_{\, i_1 j_1,\,\ldots\,,\,i_p j_p \, ;\, k_1 \,\ldots\, k_q} \ket
\ee
acting on the left entry with the projector associated to the \emph{single} Young tableau
\be \label{proof_tableau1}
\textrm{
\begin{picture}(30,100)(0,0)
\multiframe(0,85)(15,0){1}(15,15){{\footnotesize $i_1$}}
\multiframe(15.5,85)(15,0){1}(15,15){{\footnotesize $j_1$}}
\multiframe(31,85)(15,0){1}(15,15){{\footnotesize $l$}}
\multiframe(0,65)(15,0){1}(15,19.5){\vspace{7pt}$\vdots$}
\multiframe(15.5,65)(15,0){1}(15,19.5){\vspace{7pt}$\vdots$}
\multiframe(0,50)(15,0){1}(15,14.8){{\footnotesize $i_p$}}
\multiframe(15.5,50)(15,0){1}(15,14.8){{\footnotesize $j_p$}}
\multiframe(0,35)(15,0){1}(15,14.8){{\footnotesize $k_1$}}
\multiframe(0,15)(15,0){1}(15,19.5){\vspace{7pt}$\vdots$}
\multiframe(0,0)(15,0){1}(15,14.8){{\footnotesize $k_q$}}
\end{picture}
}
\ee
In fact, one can decompose $Y_{\{3,2^{p-1}\}}$ into a sum of Young
tableaux where only the projector associated to
\eqref{proof_tableau1} is not annihilated when contracted against the
right entry of the scalar product, where
$(\g^{\,[\,q\,]}\,\cS^{\,[\,p\,]})$ carries the projection
\be \label{proof_tableau2}
\textrm{
\begin{picture}(30,100)(0,0)
\multiframe(0,85)(15,0){1}(15,15){{\footnotesize $i_1$}}
\multiframe(15.5,85)(15,0){1}(15,15){{\footnotesize $j_1$}}
\multiframe(0,65)(15,0){1}(15,19.5){\vspace{7pt}$\vdots$}
\multiframe(15.5,65)(15,0){1}(15,19.5){\vspace{7pt}$\vdots$}
\multiframe(0,50)(15,0){1}(15,14.8){{\footnotesize $i_p$}}
\multiframe(15.5,50)(15,0){1}(15,14.8){{\footnotesize $j_p$}}
\multiframe(0,35)(15,0){1}(15,14.8){{\footnotesize $k_1$}}
\multiframe(0,15)(15,0){1}(15,19.5){\vspace{7pt}$\vdots$}
\multiframe(0,0)(15,0){1}(15,14.8){{\footnotesize $k_q$}}
\end{picture}
}\ee
since all other tableaux would result in symmetrizations beyond a given a line.

Furthermore, one can recognize that the operations needed to build the projection
associated to the tableau \eqref{proof_tableau1} differ from those
implied by the tableau \eqref{proof_tableau2} only in the symmetrization
of the three indices $(i_1,j_1,l)$. More precisely, denoting the
tableau \eqref{proof_tableau1} by $\t_1$ and the tableau
\eqref{proof_tableau2} by $\t_2$ and using the notation of
eq.~\eqref{young_proj_gen}, when acting on the left entry of eq.~\eqref{iniproof}
\be
Y_{\t_2} \, = \, \frac{2}{h(\t_2)} \, {\widetilde S}_{\,\t_2} \, A_{\,\t_2} \, ,
\ee
where the product ${\widetilde S}_{\,\t_2}$ of ``row symmetrizers'' does not
include the operator $S_{(i_1,j_1)}$, because this symmetrization is
already induced by the others. Acting with $Y_{\t_1}$ gives instead
\be
Y_{\t_1} \, Y_{\t_2} \, = \, \frac{1}{h(\tau_1)} \,
S_{\,(i_1,\,j_1,\,l)} \, {\widetilde S}_{\,\t_2} \, A_{\,\t_2} \, Y_{\,\t_2} \, =
\, \frac{h(\tau_2)}{2\,h(\tau_1)} \,
S_{\,(i_1,\,j_1,\,l)} \, (\,Y_{\t_2}\,)^{\,2} \ ,
\ee
so that, taking into account the ratio between the hook lengths, one ends up with
\be
\begin{split}
& \bra Y_{\{3,2^{p-1},1^q\}} \, T_{i_1j_1} \ldots T_{i_pj_p} \, \bar{\e}_{\,l} \, \g_{\,k_1 \ldots\, k_q} \comma \pr_{\,l} \, (\,\g^{\,[\,q\,]}\,\cS^{\,[\,p\,]}\,)_{\, i_1 j_1,\,\ldots\,,\,i_p j_p \, ;\, k_1 \,\ldots\, k_q} \ket \\
& = \frac{p\,(\,p+q+1\,)}{(p+1)(p+q+2)} \, \bra Y_{\t_2} \, T_{i_1j_1} \!\ldots T_{i_pj_p} \, \bar{\e}_{\,l} \, \g_{\,k_1 \ldots\, k_q} \comma \pr_{\,(\,l\,|} \, (\,\g^{\,[\,q\,]}\cS^{\,[\,p\,]}\,)_{\,|\, i_1 j_1\,)\,,\,\ldots\,,\,i_p j_p \, ;\, k_1 \,\ldots\, k_q} \ket \, .
\end{split}
\ee
In a similar discussion presented in Appendix C of
\cite{bose_mixed}, these steps sufficed to recover the Labastida
constraints on the gauge parameters, due to the presence of only
identical $T$ tensors. Here one needs to refine the arguments,
because the family indices $(i_1,j_1)$ that are symmetrized with $l$
can be carried both by a trace $T$ and by an antisymmetrized $\g$,
as can be easily seen absorbing the symmetrizations induced by
$\t_2$ in the right entry of the scalar product, so that
\be \label{scalar_antisymm}
\begin{split}
& \bra Y_{\{3,2^{p-1},1^q\}} \, T_{i_1j_1} \ldots T_{i_pj_p} \, \bar{\e}_{\,l} \, \g_{\,k_1 \ldots\, k_q} \comma \pr_{\,l} \, (\,\g^{\,[\,q\,]}\,\cS^{\,[\,p\,]}\,)_{\, i_1 j_1,\,\ldots\,,\,i_p j_p \, ;\, k_1 \,\ldots\, k_q} \ket \\
& = \, \bra \frac{2^{\,p}\,(\,q+1\,)\,!}{(\,p+q+1\,)\,!} \ T_{[\,i_1\,|\,j_1} \ldots\, T_{|\,i_p\,|\,j_p}\, \bar{\e}_{\,l}\, \g_{\,|\,k_1 \ldots\, k_q\,]} \comma \\
& \phantom{=\, \bra} \frac{p\,(\,p+q+1\,)}{(\,p+1\,)(\,p+q+2\,)} \ \pr_{\,(\,l\,|} \, (\,\g^{\,[\,q\,]}\,\cS^{\,[\,p\,]}\,)_{\,|\, i_1 j_1\,)\,,\,\ldots\,,\,i_p j_p \, ;\, k_1 \,\ldots\, k_q} \, \ket \, .
\end{split}
\ee

In order to proceed, it is convenient to separate the terms in the
antisymmetrization that differ in the position of $(i_1,j_1)$,
\be \label{expanti}
\begin{split}
& T_{[\,i_1\,|\,j_1} \ldots T_{|\,i_p\,|\,j_p}\, \bar{\e}_{\,l}\, \g_{\,|\,k_1 \ldots\, k_q\,]} \, = \, T_{i_1j_1}\,T_{[\,i_2\,|\,j_2} \ldots\, T_{|\,i_p\,|\,j_p} \, \bar{\e}_{\,l}\, \g_{\,|\,k_1 \ldots\, k_q\,]} \\
& - \, \sum_{n\,=\,2}^p\, (-1)^{\,n\,p}\ T_{i_1j_n}\,T_{j_1\,[\,i_2}\, T_{i_3\,|\,j_{n+1}} \ldots T_{|\,i_p\,|\,j_{n-1}} \, \bar{\e}_{\,l}\, \g_{\,|\,k_1 \ldots\, k_q\,]} \\
& - \, T_{[\,i_2\,|\,j_2} \ldots\, T_{|\,i_p\,|\,j_p}\, T_{|\,k_1\,|\,j_1} \, \bar{\e}_{\,l}\, \g_{\,i_1\,|\,k_1 \ldots\, k_q\,]} \, .
\end{split}
\ee
Forcing again the symmetrization in $(i_1,j_1,l)$, in the scalar
product the second group of terms in eq.~\eqref{expanti} becomes proportional
to the first term, because
\be
\begin{split}
& \sum_{n\,=\,2}^p\, (-1)^{\,n\,p}\ T_{[\,i_2\,|\,j_n} \ldots\, T_{|\,i_{p-1}\,|\,j_{n-2}} \ldots T_{|\,i_p\,|\,(\,i_1} \, T_{j_1\,|\,j_{n-1}} \bar{\e}_{\,|\,l\,)}\, \g_{\,|\,k_1 \ldots\, k_q\,]} \\
= \, & \sum_{n\,=\,2}^p\, (-1)^{\,n\,p}\ T_{[\,i_2\,|\,j_n} \ldots\, T_{|\,i_{p-1}\,|\,j_{n-2}} \ldots T_{|\,i_p\,|\,(\,i_1} \, T_{j_1l} \, \bar{\e}_{\,j_{n-1}\,)}\, \g_{\,|\,k_1 \ldots\, k_q\,]} \\
- \, & \sum_{n\,=\,2}^p\, (-1)^{\,n\,p}\ T_{(\,i_1j_1}\,T_{l\,)\,[\,i_2}\,T_{i_3\,|\,j_{n+1}} \ldots\, T_{|\,i_p\,|\,j_{n-1}}\, \bar{\e}_{\,j_n}\, \g_{\,|\,k_1 \ldots\, k_q\,]} \\
- \, & (\,p-1\,) \, T_{[\,i_2\,|\,j_n} \ldots\,
T_{|\,i_p\,|\,j_{n-1}}\, T_{(\,i_1j_1\,} \bar{\e}_{\,l\,)}\,
\g_{\,|\,k_1 \ldots\, k_q\,]}\, ,
\end{split}
\ee
and all antisymmetrizations over four indices (including those
induced on a couple of traces by a symmetrization over three
indices) are annihilated by the right entry of the scalar product
\eqref{scalar_antisymm}, so that only the last term survives.

In conclusion, the left entry of eq.~\eqref{scalar_antisymm} gives
rise to two terms when one forces the symmetrization in
$(i_1,j_1,l)$,
\be
\begin{split}
T_{[\,i_1\,|\,j_1} \ldots\, T_{|\,i_p\,|\,j_p}\, \bar{\e}_{\,l}\, \g_{\,|\,k_1 \ldots\, k_q\,]} \, \longrightarrow \, \frac{1}{6}\, \Big\{\, & (\,p+1\,)\, T_{[\,i_2\,|\,j_n} \ldots\, T_{|\,i_p\,|\,j_{n-1}}\, T_{(\,i_1j_1\,} \bar{\e}_{\,l\,)}\, \g_{\,|\,k_1 \ldots\, k_q\,]} \\
& - \, T_{[\,i_2\,|\,j_n} \ldots\, T_{|\,i_p\,|\,j_{n-1}}\, T_{|\,k_1\,|\,(\,i_1} \bar{\e}_{\,j_1}\, \g_{\,l\,)\,|\,k_1 \ldots\, k_q\,]} \,\Big\} \, ,
\end{split}
\ee
and imposing the remaining symmetrizations one can recover in each
of them the projection associated to the Young tableau
\be \label{proof_tableau3}
\textrm{
\begin{picture}(30,100)(0,0)
\multiframe(0,85)(15,0){1}(15,15){{\footnotesize $i_2$}}
\multiframe(15.5,85)(15,0){1}(15,15){{\footnotesize $j_2$}}
\multiframe(0,65)(15,0){1}(15,19.5){\vspace{7pt}$\vdots$}
\multiframe(15.5,65)(15,0){1}(15,19.5){\vspace{7pt}$\vdots$}
\multiframe(0,50)(15,0){1}(15,14.8){{\footnotesize $i_p$}}
\multiframe(15.5,50)(15,0){1}(15,14.8){{\footnotesize $j_p$}}
\multiframe(0,35)(15,0){1}(15,14.7){{\footnotesize $k_1$}}
\multiframe(0,15)(15,0){1}(15,19.5){\vspace{7pt}$\vdots$}
\multiframe(0,0)(15,0){1}(15,14.8){{\footnotesize $k_q$}}
\end{picture}
}\ee
However, this projection is automatically enforced by the right
entry of the scalar product, so that
\be \label{laststep}
\begin{split}
& \bra Y_{\{3,2^{p-1},1^q\}} \, T_{i_1j_1} \ldots T_{i_pj_p} \, \bar{\e}_{\,l} \, \g_{\,k_1 \ldots\, k_q} \comma \pr_{\,l} \, (\,\g^{\,[\,q\,]}\,\cS^{\,[\,p\,]}\,)_{\, i_1 j_1,\,\ldots\,,\,i_p j_p \, ;\, k_1 \,\ldots\, k_q} \ket \\
& = \, \frac{p\,(\,p+q+1\,)}{(\,p+1\,)(\,p+q+2\,)} \ \bra\, \frac{1}{3\,(\,p+q+1\,)} \, Y_{\t_3} \, \Big\{\, (\,p+1\,)\, T_{[\,i_1\,|\,j_1} \ldots\, T_{|\,i_p\,|\,j_p}\, \bar{\e}_{\,l}\, \g_{\,|\,k_1 \ldots\, k_q\,]} \\
& - \, T_{[\,i_2\,|\,j_n} \ldots\, T_{|\,i_p\,|\,j_{n-1}}\, T_{|\,k_1\,|\,(\,i_1} \bar{\e}_{\,j_1}\, \g_{\,l\,)\,|\,k_1 \ldots\, k_q\,]} \,\Big\} \comma \pr_{\,(\,l\,|} \, (\,\g^{\,[\,q\,]}\,\cS^{\,[\,p\,]}\,)_{\,|\, i_1 j_1\,)\,,\,\ldots\,,\,i_p j_p \, ;\, k_1 \,\ldots\, k_q} \ket  \\
& = \, \bra \, (\,p+1\,) \ T_{i_2j_2} \ldots T_{i_pj_p}\, T_{(\,i_1j_1} \, \bar{\e}_{\,l\,)} \, \g_{\,k_1 \ldots\, k_q} \, - \, T_{i_2j_2} \ldots T_{i_pj_p} \, T_{\,[\,k_1\,|\,(\,i_1} \, \bar{\e}_{\,j_1} \, \g_{\,l\,)\,|\,k_2 \ldots\, k_q\,]} \comma \\
& \frac{p}{3\, (\,p+1\,)(\,p+q+2\,)} \ \pr_{\,(\,l\,|} \,
(\,\g^{\,[\,q\,]}\,\cS^{\,[\,p\,]}\,)_{\,|\, i_1
j_1\,),\,\ldots\,,\,i_p j_p \, ;\, k_1 \,\ldots\, k_q} \ket\, ,
\end{split}
\ee
where $\t_3$ denotes the tableau of eq.~\eqref{proof_tableau3} and
the last two lines lead to the result in
eq.~\eqref{idconstrf}, using eq.~\eqref{compgamma1}.

\subsection*{\sc Bianchi identities}

The generic $\g$-trace of the constrained Bianchi identities
\be \mathscr{B}_l \, : \ \pr_{\,l}\,\cS \, - \, \12 \dsl \
\g_{\,l}\,\cS \, - \, \12 \ \pr^{\,m}\,T_{lm}\,\cS \, - \,
\frac{1}{6} \ \pr^{\,m}\,\g_{\,lm}\,\cS \, = \, 0 \ee
reads
\begin{align}
& (-1)^{\,q} \prod_{r\,=\,1}^{p} T_{i_r j_r} \, \g_{\,k_1 \ldots\, k_q} \, \mathscr{B}_l \, : \ \frac{1}{6} \, \sum_{n\,=\,1}^{p} \, \pr_{\,(\,i_n\,}\g_{\,j_n)\,k_1 \ldots\, k_q\,l}\, \prod_{r\,\ne\,n}^{p} T_{i_r j_r} \, \cS \nn \\
& + \, \pr_{\,[\,k_1\,}\g_{\,k_2 \ldots\, k_q\,l\ ]} \prod_{r\,=\,1}^{p} T_{i_r j_r} \, \cS \, + \, \frac{q-1}{6} \ \pr_{\,[\,k_1\,}\g_{\,k_2 \ldots\, k_q\,]\,l} \prod_{r\,=\,1}^{p} T_{i_r j_r} \, \cS \nn \\
& - \, \frac{(-1)^{\,q}}{2} \sum_{n\,=\,1}^{p} \g_{\,k_1 \ldots\, k_q}\,\pr_{\,(\,i_n}\,T_{j_n\,)\,l}\! \prod_{r\,\ne\,n}^{p} T_{i_r j_r} \, \cS \, + \, \frac{1}{6} \, \sum_{n\,=\,1}^{p} \pr_{\,(\,i_n\,}\g_{\,j_n\,)\,[\,k_1 \ldots\, k_{q-1}\,} T_{\,k_q\,]\,l}\! \prod_{r\,\ne\,n}^{p} T_{i_r j_r} \, \cS \nn \\
& - \, \frac{1}{6} \, \sum_{n\,=\,1}^{p} \, \g_{\,l\,[\,k_1 \ldots\, k_{q-1}\,} T_{\,k_q\,]\,(\,i_n}\,\pr_{\,j_n\,)} \prod_{r\,\ne\,n}^{p} T_{i_r j_r} \, \cS \, + \, \frac{q+1}{6} \ \pr_{\,[\,k_1\,} \g_{\,k_2 \ldots\, k_{q-1}\,} T_{\,k_q\,]\,l} \, \prod_{r\,=\,1}^{p} T_{i_r j_r} \, \cS \nn \\
& + \frac{(-1)^{\,q}}{6} \, \sum_{n\,=\,1}^{p} \, \g_{\,[\,k_1 \ldots\, k_{q-2}\,} T_{\,k_{q-1}\,|\,l}\,T_{\,|\,k_q\,]\,(\,i_n}\,\pr_{\,j_n\,)} \prod_{r\,\ne\,n}^{p} T_{i_r j_r} \, \cS \nn \\
& - \, \dsl \left\{\, \frac{q+3}{6} \ \g_{\,k_1 \ldots\, k_q \, l} + \, \frac{q-1}{6} \ \g_{\,[\,k_1 \ldots\, k_{q-1}\,} T_{\,k_q\,]\,l} \,\right\}\! \prod_{r\,=\,1}^{p} T_{i_r j_r}\, \cS  - \, \frac{(-1)^{\,q}}{6} \ \pr^{\,m} \, \g_{\,k_1 \ldots k_q \,lm} \prod_{r\,=\,1}^{p} T_{i_r j_r}\, \cS \nn \\
& - \, \frac{(-1)^{\,q}}{2} \, \pr^{\,m} \, \g_{\,k_1 \ldots k_q} \, T_{lm} \prod_{r\,=\,1}^{p} T_{i_r j_r} \, \cS - \frac{1}{6} \, \pr^{\,m} \!\left(\, \g_{\,l\,[\,k_1 \ldots\, k_{q-1}\,} T_{\,k_q\,]\,m}  - \, \g_{\,m\,[\,k_1 \ldots\, k_{q-1}\,} T_{\,k_q\,]\,l} \,\right) \prod_{r\,=\,1}^{p} T_{i_r j_r} \, \cS \nn \\
& + \, \frac{(-1)^{\,q}}{6} \ \pr^{\,m} \, \g_{\,[\,k_1 \ldots\, k_{q-2}\,} T_{\,k_{q-1}\,|\,l}\,T_{\,|\,k_q\,]\,m} \prod_{r\,=\,1}^{p} T_{i_r j_r} \, \cS \, = \, 0 \, . \label{gtrace_bianchi}
\end{align}

In order to recover the result in eq.~\eqref{finalbianchif}
for the $\{2^p,1^{q+1}\}$ projection of this expression, we shall see shortly
that the terms containing the same number of traces
and an antisymmetrized $\g$-trace with the same number of indices become
proportional when they are $\{2^p,1^{q+1}\}$-projected. Furthermore,
some of the contributions in eq.~\eqref{gtrace_bianchi}
are annihilated by the Young projector, while the proportionality
factors can be fixed resorting to the procedures explained in
Appendix C of \cite{bose_mixed}, that we shall also briefly recall in the
present context.

First of all, it is convenient to recognize that the
$\{2^p,1^{q+1}\}$ component can be extracted acting with the
projector associated to the \emph{single} Young tableau
\be \label{proof_tableau4}
\textrm{
\begin{picture}(30,115)(0,0)
\multiframe(0,100)(15,0){1}(15,15){{\footnotesize $i_1$}}
\multiframe(15.5,100)(15,0){1}(15,15){{\footnotesize $j_1$}}
\multiframe(0,80)(15,0){1}(15,19.5){\vspace{7pt}$\vdots$}
\multiframe(15.5,80)(15,0){1}(15,19.5){\vspace{7pt}$\vdots$}
\multiframe(0,65)(15,0){1}(15,14.8){{\footnotesize $i_p$}}
\multiframe(15.5,65)(15,0){1}(15,14.8){{\footnotesize $j_p$}}
\multiframe(0,50)(15,0){1}(15,14.8){{\footnotesize $k_1$}}
\multiframe(0,30)(15,0){1}(15,19.5){\vspace{7pt}$\vdots$}
\multiframe(0,15)(15,0){1}(15,14.8){{\footnotesize $k_q$}}
\multiframe(0,0)(15,0){1}(15,14.8){{\footnotesize $l$}}
\end{picture}
}\ee
on account of the symmetry properties of the combination $T_{i_1j_1} \ldots \, T_{i_pj_p} \, \g_{\,k_1 \ldots\, k_q} \, \mathscr{B}_l$.
As a consequence, all symmetrizations in the generic set of three
indices $(i_m,j_m,k_n)$ vanish due to a standard property of Young
tableaux. One can now use this fact to relate various terms, so that
for instance
\be
Y_{\{2^p,1^{q+1}\}}\, \sum_{n\,=\,1}^p \, \pr_{\,(\,i_n}\, T_{j_n\,)\,l} \prod_{r\,\neq\,n}^p \, T_{i_rj_r} \g_{\,k_1 \ldots\, k_q} \, \cS \, = \, - \, p \ Y_{\{2^p,1^{q+1}\}}\, \partial_{\,l} \, T_{i_1j_1} \ldots\, T_{i_pj_p} \, \g_{\,k_1 \ldots\, k_q} \, \cS \, .
\ee
To conclude one must also consider that, in order to build the
projection associated to \eqref{proof_tableau4}, an
antisymmetrization over $l$ and all the $k_n$ indices is required.
The $\{2^p,1^{q+1}\}$ projections of expressions that become
proportional after this operation remain proportional with the
same overall factor even after the full projection is enforced, and this finally leads to
\be \label{firstprojbianchi}
\begin{split}
& Y_{\{2^p,1^{q+1}\}}\, T_{i_1j_1} \ldots\, T_{i_pj_p} \, \g_{\,k_1 \ldots\, k_q}\, \mathscr{B}_l \, : \\[4pt]
& \left[\, (\,q+1\,) \,+\, \frac{q\,(\,q-1\,)}{6} \,+\, \frac{p}{2} \,+\, \frac{p\,q}{6} \,\right] \, Y_{\{\,2^p,\,1^{q+1}\}} \, \pr_{\,l} \, T_{i_1j_1} \ldots\, T_{i_pj_p}\, \g_{\,k_1 \ldots\, k_q} \, \cS \\
& + \, \frac{1}{6} \ Y_{\{\,2^p,\,1^{q+1}\}} \, \sum_{n\,=\,1}^{p}\, \pr_{\,(\,i_n\,|}\, \prod_{r\,\neq\,n}^p \, T_{i_rj_r} \, \g_{\,|\,j_n\,)\,l\,k_1 \ldots\, k_q}\, \cS \\[1pt]
& + \, (-1)^{\,q+1} \, \frac{q+3}{6} \ Y_{\{\,2^p,\,1^{q+1}\}} \dsl \ T_{i_1j_1} \ldots\, T_{i_pj_p}\, \g_{\,k_1 \ldots\, k_q\, l} \, \cS  \\[4pt]
& - \, \frac{q+3}{6} \ Y_{\{\,2^p,\,1^{q+1}\}} \, \pr^{\,m}\, T_{i_1j_1} \ldots\, T_{i_pj_p}\, T_{lm}\, \g_{\,k_1 \ldots\, k_q} \, \cS  \\[4pt]
& - \, \frac{1}{6} \ Y_{\{\,2^p,\,1^{q+1}\}} \, \pr^{\,m}\, T_{i_1j_1} \ldots\, T_{i_pj_p}\, \g_{\,k_1 \ldots\, k_q\, lm} \, \cS \, = \, 0 \, .
\end{split}
\ee

Furthermore, in the two divergence terms only the two-column
projected combination of the traces and the antisymmetrized $\g$-traces can
contribute to the full two-column projection also involving the
divergence. In a similar fashion, a two-column projection in all
indices is enforced in the first gradient term. In fact, the
manifest symmetries for its lower family indices are
\be \label{manifestsym}
\{2^{\,p},1^{q+1}\} \, \otimes \, \{1\} \, = \, \{2^{\,p},1^{q+2}\} \, \oplus \, \{2^{\,p+1},1^q\} \, \oplus \, \{3,2^{\,p-1},1^{q+1}\} \, ,
\ee
but on the other hand this term results from the product of $(p+1)$ traces and a
$q$-fold $\g$-trace. As a result, the irreducible components with less than four columns that it
admits originate from the decomposition
\be
\{2^{\,p+1}\} \, \otimes \, \{1^q\} \, = \, \{2^{\,p+1},1^q\} \, \oplus \, \{3,2^{\,p},1^{q-1}\} \, \oplus \, \{3^2,2^{\,p-1},1^{q-2}\} \, \oplus \, \ldots \ ,
\ee
so that the corresponding term is actually $\{2^{p+1},1^q\}$
projected. This is no longer true for the last term of
eq.~\eqref{firstprojbianchi}, where the components displayed in
eq.~\eqref{manifestsym} and those carried by the product of $p$
traces and a $(q+2)$-fold $\g$,
\be
\{2^{\,p}\} \, \otimes \, \{1^{q+2}\} \, = \, \{2^{\,p},1^{q+2}\} \, \oplus \, \{3,2^{\,p-1},1^{q+1}\} \, \oplus \, \{3^2,2^{\,p-2},1^q\} \, \oplus \, \ldots \ ,
\ee
share two admissible components. However, in the constrained setting
one can elude this problem and simply select the $\{2^p,1^{q+2}\}$
component, because it is the only non-vanishing one, so that
eq.~\eqref{firstprojbianchi} directly leads to
eq.~\eqref{finalbianchif}, while this detailed analysis is
the starting point for the next subsection.

\subsection*{\sc Analysis of eq.~\eqref{combianchif}}

Most of the results of the previous subsection can be transferred to
the unconstrained setting by simply replacing $\cS$ with $\cW$.
Thus, the $\{2^p,1^{q+1}\}$ projection of the generic $\g$-trace of
the Bianchi identities \eqref{bianchi_W2},
\be
Y_{\{2^p,1^{q+1}\}}\, T_{i_1j_1} \ldots\, T_{i_pj_p} \, \g_{\,k_1 \ldots\, k_q} \, \mathscr{B}_l \, ,\ee
reads
\be \label{projunc}
\begin{split}
& \frac{(\,q+3\,)\,(\,p+q+2\,)}{6} \ Y_{\{\,2^p,\,1^{q+1}\}} \, \pr_{\,l} \, (\,\g^{\,[\,q\,]}\,\cW^{\,[\,p\,]}\,)_{\, i_1 j_1,\,\ldots\,,\,i_p j_p \, ;\, k_1 \ldots\, k_q} \\
+ \ & \frac{1}{6} \ Y_{\{\,2^p,\,1^{q+1}\}} \, \sum_{n\,=\,1}^{p}\, \pr_{\,(\,i_n\,|}\, (\,\g^{\,[\,q+2\,]}\,\cW^{\,[\,p-1\,]}\,)_{\,\ldots\,,\,i_{r\,\ne\,n} j_{r\,\ne\,n} \,,\,\ldots\, ;\,|\,j_n\,)\,l \, k_1 \ldots\, k_q}  \\[2pt]
+ \ & (-1)^{\,q+1} \, \frac{q+3}{6} \dsl \, (\,\g^{\,[\,q+1\,]}\,\cW^{\,[\,p\,]}\,)_{\, i_1 j_1,\,\ldots\,,\,i_p j_p \, ;\, k_1 \ldots\, k_q \, l}  \\[5pt]
- \ & \frac{q+3}{6} \ \pr^{\,m}\, (\,\g^{\,[\,q\,]}\,\cW^{\,[\,p+1\,]}\,)_{\, i_1 j_1,\,\ldots\,,\,i_p j_p\,,\,lm \, ;\, k_1 \ldots\, k_q}  \\[5pt]
- \ & \frac{1}{6} \ Y_{\{\,2^p,\,1^{q+1}\}} \, \pr^{\,m}\, T_{i_1j_1} \ldots\, T_{i_pj_p}\, \g_{\,k_1 \ldots\, k_q\, lm} \, \cW \\[5pt]
= \ & \12 \ Y_{\{\,2^p,\,1^{q+1}\}} \, T_{i_1j_1} \ldots\, T_{i_pj_p}\, \g_{\,k_1 \ldots\, k_q} \, \pr^{\,m}\pr^{\,n} \, \cZ_{\,lmn} \, .
\end{split}
\ee

As we just saw, the peculiar feature of this expression is that the
$\{2^p,1^{q+1}\}$ projection in its free indices does not suffice to
induce a two-column projection in all indices carried by the last
$\cW$ term. This is not a problem in the constrained setting, where
the other projections vanish on account of the constraints
\eqref{constrs}, but here the last $\cW$ term of eq.~\eqref{projunc}
can be only presented in this form, with the lower index $m$ not taking
part in the projection. However, an expression
containing only two-column projected contributions obtains
combining eq.~\eqref{projunc} with a different product of traces and
antisymmetric $\g$-traces of $\mathscr{B}_l$ carrying the same
projection in the family indices, as we can now see. In particular, a $\{2^p,1^{q+1}\}$
projection is also allowed for the product of $p-1$ traces and a
$(q+2)$-fold $\g$-trace acting on $\mathscr{B}_l$. In order to
maintain the symmetry under interchanges of the $(i_n,j_n)$ couples
manifest, it is then natural to consider
\be
Y_{\{2^p,1^{q+1}\}}\, \sum_{n\,=\,1}^p \, \prod_{r\,\ne\,n}^p \, T_{i_rj_r} \, \g_{\,k_1
\ldots\, k_q\,l\,(\,i_n} \mathscr{B}_{j_n\,)} \, ,
\ee
that using the techniques of the previous subsection can be cast in
the form
{\allowdisplaybreaks
\begin{align}
& p\,(\,q+1\,) \, \bigg\{\, \frac{(\,q+3\,)\,(\,p+q+2\,)}{6} \ Y_{\{\,2^p,\,1^{q+1}\}} \, \pr_{\,l} \, (\,\g^{\,[\,q\,]}\,\cW^{\,[\,p\,]}\,)_{\, i_1 j_1,\,\ldots\,,\,i_p j_p \, ;\, k_1 \ldots\, k_q} \nn \\
& \phantom{p\,(\,q+1\,)} + \, \frac{1}{6} \ Y_{\{\,2^p,\,1^{q+1}\}} \, \sum_{n\,=\,1}^{p}\, \pr_{\,(\,i_n\,|}\, (\,\g^{\,[\,q+2\,]}\,\cW^{\,[\,p-1\,]}\,)_{\,\ldots\,,\,i_{r\,\ne\,n} j_{r\,\ne\,n} \,,\,\ldots\, ;\,|\,j_n\,)\,l \, k_1 \ldots\, k_q} \nn \\[2pt]
& \phantom{p\,(\,q+1\,)} + \, (-1)^{\,q+1} \, \frac{q+3}{6} \dsl \, (\,\g^{\,[\,q+1\,]}\,\cW^{\,[\,p\,]}\,)_{\, i_1 j_1,\,\ldots\,,\,i_p j_p \, ;\, k_1 \ldots\, k_q \, l} \nn \\[5pt]
& \phantom{p\,(\,q+1\,)} - \, \frac{q+3}{6} \ \pr^{\,m}\, (\,\g^{\,[\,q\,]}\,\cW^{\,[\,p+1\,]}\,)_{\, i_1 j_1,\,\ldots\,,\,i_p j_p\,,\,lm \, ;\, k_1 \ldots\, k_q} \, \bigg\} \nn \\[5pt]
& - \, \frac{p\,(\,q+3\,)}{6}\ Y_{\{2^p,1^{q+1}\}}\, \pr^{\,m}\, T_{i_1j_1} \ldots\, T_{i_pj_p} \, \g_{\,k_1 \ldots\, k_q\, lm}\, \cW \nn \\
& + \, \frac{(-1)^{\,q}}{3}\ Y_{\{2^p,1^{q+1}\}}\, \pr^{\, m}\, \sum_{n\,=\,1}^p \, \prod_{r\,\neq\,n}^p \, T_{i_rj_r} \, T_{m\,(\,i_n}\, \g_{\,j_n\,)\,k_1 \ldots\, k_q\, l} \, \cW \nn \\
& = \, \12 \ Y_{\{2^p,1^{q+1}\}} \, \sum_{n\,=\,1}^p \, \prod_{r\,\ne\,n}^p \, T_{i_rj_r} \, \g_{\,k_1 \ldots\, k_q\,l\,(\,i_n\,|}\, \pr^{\,m}\pr^{\,n}\, \cZ_{\,|\,j_n\,)\,mn} \, . \label{projunc2}
\end{align}}
Hence, the combination
\be
Y_{\{2^p,1^{q+1}\}} \bigg\{\, (\,p+2\,)\,(\,q+3\,)\ T_{i_1j_1} \!\ldots T_{i_pj_p} \, \g_{\,k_1 \ldots\, k_q} \mathscr{B}_l \, - \, \sum_{n\,=\,1}^p \, \prod_{r\,\ne\,n}^p \, T_{i_rj_r} \, \g_{\,k_1
\ldots\, k_q\,l\,(\,i_n} \mathscr{B}_{j_n\,)} \,\bigg\}
\ee
in eq.~\eqref{combianchif} contains the following terms
with $p$ traces and a $(q+2)$-fold $\g$\ :
\be
\begin{split}
& - \, \frac{1}{3} \ Y_{\{2^p,1^{q+1}\}} \, \pr^{\,m} \bigg\{\, (\,q+3\,)\ T_{i_1j_1} \ldots\, T_{i_pj_p}\, \g_{\,k_1 \ldots\, k_q\, lm}\, \cW \\
& + \, (-1)^{\,q+1}\, \sum_{n\,=\,1}^p\, \prod_{r\,\neq\,n}^p\, T_{i_rj_r}\, T_{m\,(\,i_n}\, \g_{\,j_n\,)\,k_1 \ldots\, k_q\, l}\, \cW \,\bigg\} \, .
\end{split}
\ee

In order to verify that this expression is indeed $\{2^p,1^{q+2}\}$
projected, one can first recognize that the
$\{2^p,1^{q+1}\}$ projection can be enforced by the projector
associated to the single Young tableau displayed in
eq.~\eqref{proof_tableau4}. Then, using the notation of
eq.~\eqref{young_proj_gen} and applying the column antisymmetrizer
$A$ associated to \eqref{proof_tableau4} to the expression within
parentheses gives
\begin{align}
& A_{\{2^p,1^{q+1}\}}\, \pr^{\,m} \bigg\{\, (\,q+3\,)\ T_{i_1j_1} \!\ldots T_{i_pj_p}\, \g_{\,k_1 \ldots\, k_q\, lm}\, \cW + \, (-1)^{\,q+1}\! \sum_{n\,=\,1}^p \prod_{r\,\neq\,n}^p T_{i_rj_r}\, T_{m\,(\,i_n} \g_{\,j_n\,)\,k_1 \ldots\, k_q\, l}\, \cW \,\bigg\} \nn \\
& = \, p\,!\,(\,q+1\,)\,!\ \pr^{\,m} \bigg\{\, (\,q+3\,)\ T_{[\,i_1\,|\,j_1} \ldots\, T_{|\,i_p\,|\,j_p}\, \g_{\,|\,k_1 \ldots\, k_q\, l\,]\,m}\, \cW \nn \\
& + \, (-1)^{\,q+1} \, (\,q+2\,)\, \sum_{n\,=\,1}^p \, T_{[\,i_1\,|\,j_n} \ldots\, T_{|\,i_{p-1}\,|\,j_{n-2}}\, T_{m\,j_{n-1}}\, \g_{\,|\,i_p\,k_1 \ldots\, k_q\,l\,]}\, \cW \nn \\
& + \, (-1)^{\,q+1}\, \sum_{n\,=\,1}^p\, T_{[\,i_1\,|\,j_n} \ldots\, T_{|\,i_{p-1}\,|\,j_n-2}\,T_{m\,|\,i_p\,|}\, \g_{\,j_{n-1}\,|\,k_1 \ldots\, k_q\, l\,]}\, \cW \, \bigg\} \, . \label{anti0}
\end{align}
When acting with the row symmetrizer $S$, the last two terms
become proportional, since they only differ because of an
antisymmetrization over all the $i_n$ indices and a single $j_n$ index,
that is annihilated by $S$. In conclusion, the right-hand side of
eq.~\eqref{anti0} is equivalent to
\be \label{anti1}
(\,q+3\,)\,p\,!\,(\,q+1\,)\,!\ \pr^{\,m}\, T_{[\,i_1\,|\,j_1} \ldots\, T_{|\,i_p\,|\,j_p}\, \g_{\,|\,k_1 \ldots\, k_q\, lm\,]}\, \cW \, ,
\ee
and the remaining symmetrizations induced by $S$ build an expression
proportional to that associated to a Young tableau with the index
$m$ attached at the end of the longer column of
\eqref{proof_tableau4}. Due to the symmetry properties of the
underlying combination, this suffices to build the Young projector
$Y_{\{2^p,1^{q+2}\}}$. In order to fix the proportionality factor
and to reach the result displayed in eq.~\eqref{combianchif}, one
can then compare eq.~\eqref{anti1} with an expression obtained
acting with the column antisymmetrizer of the $\{2^p,1^{q+2}\}$
projector,
\be \label{anti2}
A_{\{2^p,1^{q+2}\}}\, T_{i_1j_1} \ldots\, T_{i_pj_p}\, \g_{\,k_1 \ldots\, k_q\, lm}\, \cW \, = \, p\,!\,(\,q+2\,)\,!\ T_{[\,i_1\,|\,j_1} \ldots\, T_{|\,i_p\,|\,j_p}\, \g_{\,|\,k_1 \ldots\, k_q\, lm\,]}\, \cW \, .
\ee
Taking into account the hook lengths of the corresponding
expressions,
\be
h(\,\{2^p,1^{q+1}\}\,) \, = \, \frac{p\,!\,(\,p+q+2\,)\,!}{q+2} \, , \qquad h(\,\{2^p,1^{q+2}\}\,) \, = \, \frac{p\,!\,(\,p+q+3\,)\,!}{q+3} \, ,
\ee
finally gives
\begin{align}
& Y_{\{2^p,1^{q+1}\}} \, \pr^{\,m} \bigg\{\, (\,q+3\,)\ T_{i_1j_1} \!\ldots T_{i_pj_p}\, \g_{\,k_1 \ldots\, k_q\, lm}\, \cW  + \, (-1)^{\,q+1}\! \sum_{n\,=\,1}^p \prod_{r\,\neq\,n}^p T_{i_rj_r}\, T_{m\,(\,i_n\,} \g_{\,j_n\,)\,k_1 \ldots\, k_q\, l}\, \cW \,\bigg\} \nn \\
& = \, (\,p+q+3\,)\ \pr^{\,m} \, Y_{\{2^p,1^{q+2}\}}\, T_{i_1j_1} \!\ldots T_{i_pj_p}\, \g_{\,k_1 \ldots\, k_q\, lm}\, \cW \, ,
\end{align}
that when added to the other contributions contained in
eqs.~\eqref{projunc} and \eqref{projunc2} proves the result
in eq.~\eqref{combianchif}.

\vskip 24pt

\scs{Some properties of the reduction for $N$-family fields} \label{app:redN}

This last appendix collects some results on the reduction to $\cS=0$ of the field
equations for \mbox{$N$-family} spinor-tensors that complement our discussion in
Section~\ref{sec:reductionf}.

Let us begin with some comments on the possibility that the tensor $\cS$ itself vanish
identically. To this end, let us resort again to an oscillator representation for the relevant
operators, as in Section \ref{sec:multiforms}, that allows to rewrite $\cS$ in the form
\be
\cS \, = \, i \ \pr^{\,\m}\, \g^{\,\n}\, \left(\, \h_{\,\m\n}\, - \, u^{\,i}_{\,\m}\, \frac{\pr}{\pr\, u^{\,i}_{\,\n}} \,\right) \, \psi \, ,
\ee
where
\be \psi \, \equiv \, \frac{1}{s_{1}\,! \,\ldots\, s_{N}\,!} \ u^{\,1}_{\,\m_1} \ldots\,
u^{\,1}_{\,\m_{s_1}} \, u^{\,2}_{\,\n_1} \ldots\, u^{\,2}_{\,\n_{s_2}} \ldots \
\psi^{\,\m_1 \ldots\, \m_{s_1},\,\n_1 \ldots\, \n_{s_2}\,,\, \ldots} \, . \ee
In general, the spinor-tensor $\cS$ vanishes identically provided the tensor
\be \label{zeroS}
\left(\, \h_{\,\m\n}\, - \, u^{\,i}_{\,\m}\, \frac{\pr}{\pr\, u^{\,i}_{\,\n}} \,\right) \, \psi
\ee
is $\g$-traceless. We can now see that the fully antisymmetric fields in $D=N$
mentioned in Section \ref{sec:reductionf} are the unique class for which the combination \eqref{zeroS} is identically zero. While it is relatively easy
to verify that in these cases the only available component of $\cS$ indeed vanishes, the
absence of other solutions can be appreciated in two steps. To begin with, setting the trace of
\eqref{zeroS} to zero yields
\be
\left(\, D \, - \, \sum_{i\,=\,1}^N \, s_i \,\right)\, \psi \, = \, 0 \, ,
\ee
which in its turn implies the inequality $D \geq N$. On the other hand, for $\m \neq \n$
one ends up with
\be \label{scalarzero}
u^{\,i}_{\,\m}\, \frac{\pr}{\pr\, u^{\,i}_{\,\n}} \, = \, 0 \, ,
\ee
that can be read as orthogonality conditions for the vectors $u^{\,i}_{\,\m}$ and
$\frac{\pr}{\pr\, u^{\,i}_{\,\n}}$. More in detail, in order for \eqref{scalarzero} to
hold at fixed $\m$, it would be necessary to have at least $(D-1)$ orthogonal directions in
the $N$-dimensional space to which those vectors belong, which can be possible if $D \leq
N$. One can therefore conclude that $D = N$, so that all $s_i$ are equal to 1.

Furthermore, in Section \ref{sec:reductionf} we proposed the set of conditions
\eqref{chain}, in order to select the shift transformations
\be \label{shiftSred}
\d \, \cS \, = \, \sum_{n\,=\,0}^{\left[\frac{m}{2}\right]} \, \h^{i_1j_1} \ldots\, \h^{i_{n}j_{n}}\,
\g^{\,k_1 \ldots\, k_{m-2n}}\, \Omega^{\,[\,n\,,\,m-2n\,]}{}_{\,i_1j_1, \,\ldots\, ,\,i_nj_n;\,k_1 \ldots\, k_{m-2n}}
\ee
that leave eq.~\eqref{eqgenred} invariant. Actually, we showed that eqs.~\eqref{chain}
build these transformations in terms of Weyl-like $\psi$ shifts with parameters $\Theta$
such that $\O \sim \cS\,(\Theta)$. As a result, they are expected to preserve the Bianchi
identities and the (on-shell) Labastida-like constraints induced on $\cS$, provided the
parameters also satisfy the triple $\g$-trace constraints \eqref{constrtheta}. A missing
step was an explicit proof that all transformations of this type, when combined with
proper shifts of the $\cY_{\,ijk}$, are symmetries of the field equations for
$\bar{\psi}$, a result that we verified in Section~\ref{sec:reductionf} only for the
simplest shifts with a single $\g^{\, i}$.

While we did not arrive at a complete proof, we can nonetheless display the cancelation mechanism required to this effect for an interesting class of terms. In Sections \ref{sec:reduction2f} and \ref{sec:reductionf} we repeatedly stressed that, when varying the $\cS$-dependent terms in eq.~\eqref{eqgenred}, all Young projections with more than two columns can be compensated by proper $\O$-dependent shifts of the $\cY_{\,ijk}$ tensors. The rest is of the form
\begin{align}
& \d \, E_{\,\bar{\psi}} \, = \, \sum_{p\,,\,q\,=\,0}^N \, k_{\,p\,,\,q} \ \h^{i_1j_1} \ldots\, \h^{i_pj_p}\, \g^{\,k_1 \ldots\, k_q}\, \times \label{shiftEgen} \\
& \times Y_{\{2^p,1^q\}} \sum_{n\,=\,0}^{[\frac{m}{2}]} \, \sum_{r_1 < \ldots < \,r_n}\, \prod_{s\,\neq\,r_l}^{p} \, T_{i_sj_s} \g_{\,[\,k_1 \ldots\, k_{q-m+2n}|}\, \theta^{\,[\,n\,,\,m-2n\,]}(\,\O\,){}_{\,i_{r_1}j_{r_1}, \,\ldots\, ,\,i_{r_n}j_{r_n},\, |\, k_{q-m+2n+1} \ldots\, k_{q}\,]} \, , \nn
\end{align}
so that the symmetry is present provided the combinations
$\theta^{\,[\,n\,,\,m-2n\,]}(\,\O\,)$ vanish whenever eqs.~\eqref{chain} hold. In the following, however, we shall only prove this result for the terms $\theta^{\,[\,0\,,\,m\,]}(\,\O\,)$, or $\theta^{\,[\,m\,]}$ for brevity, that are fully antisymmetric in their family indices and are thus relatively handy to control.
In displaying their structure, it is convenient to follow an iterative approach, building $\theta^{\,[\,m+1\,]}$ from $\theta^{\,[\,m\,]}$ via proper redefinitions of the $\O$ parameters with $m$ family indices that enter the latter quantity. In particular, for even values of $m$ one can let
\be \label{redef1}
\O^{\,[\, n \comma m-2n \,]}{}_{\,i_1j_1, \,\ldots\, ,\,i_nj_n;\,k_1 \ldots\, k_{m-2n}} \!\to \, \g^{\,l}\, \O^{\,[\, n \comma m-2n+1 \,]}{}_{\,i_1j_1, \,\ldots\, ,\,i_nj_n;\,k_1 \ldots\, k_{m-2n}\,l} \, ,  \quad\ 0 \leq n \leq \frac{m}{2} \, ,
\ee
while for odd values of $m$ one can let
\begin{alignat}{2}
& \O^{\,[\, n \comma m-2n \,]}{}_{\,i_1j_1, \,\ldots\, ,\,i_nj_n;\,k_1 \ldots\, k_{m-2n}} \!\to \, \g^{\,l}\, \O^{\,[\, n \comma m-2n+1 \,]}{}_{\,i_1j_1, \,\ldots\, ,\,i_nj_n;\,k_1 \ldots\, k_{m-2n}\,l} \, ,  \quad\! && 0 \leq n \leq \frac{m-3}{2} \, , \nn \\[3pt]
& \O^{\,[\, n \comma 1 \,]}{}_{\,i_1j_1, \,\ldots\, ,\,i_nj_n;\,k} \!\to \, \g^{\,l} \left\{\, \O^{\,[\, n \comma 2 \,]}{}_{\,i_1j_1, \,\ldots\, ,\,i_nj_n;\,kl} + \12\, \O^{\,[\, n \comma 0 \,]}{}_{\,i_1j_1, \,\ldots\, ,\,i_nj_n,\,kl} \,\right\} ,  \quad\! && n = \frac{m-1}{2} \, . \label{redef2}
\end{alignat}
These redefinitions turn a shift of $\cS$ with parameters carrying $m$ family indices into the next member of the set \eqref{shiftSred}, with $(m+1)$ family indices. One can therefore use them to determine how the field equations vary under shifts of $\cS$ with $(m+1)$ family indices, starting from the redefinition
\be
\theta^{\,[\,m\,]}{}_{\,k_1 \ldots\, k_{m}} \, \to \, \g^{\,l}\, \tilde{\theta}^{\,[\,m\,]}{}_{\,l\,;\,k_1 \ldots\, k_{m}}
\ee
that the procedure induces on $\theta^{\,[\,m\,]}$ and substituting it in eq.~\eqref{shiftEgen} in order to determine the structure of $\theta^{\,[\,m+1\,]}$. Confining our attention to the $n=0$ part in \eqref{shiftEgen} gives
\be \label{shiftEanti1}
\begin{split}
& \d \, E_{\,\bar{\psi}} \, = \, \sum_{p\,,\,q\,=\,0}^N \, \h^{i_1j_1} \ldots\, \h^{i_pj_p}\, \g^{\,k_1 \ldots\, k_q}\, Y_{\{2^p,1^q\}}\, \times \\
& \times \bigg\{\, k_{\,p\,,\,q} \, \left[\ T_{i_1j_1} \ldots\, T_{i_pj_p}\, \g_{\,[\,k_1 \ldots\, k_{q-m}\,|} \comma  \g^{\,l}\ \right]_{\,(-1)^{q-m+1}} \, \tilde{\theta}^{\,[\,m\,]}{}_{\,l\,;\,|\,k_{q-m+1} \ldots\, k_{q}\,]} \\
& +\, (-1)^{\,q+1}\, \frac{k_{\,p\,,\,q-1}}{q}\ T_{i_1j_1} \ldots\, T_{i_pj_p} \, \g_{\,[\,k_1 \ldots\, k_{q-m-1}}\, \tilde{\theta}^{\,[\,m\,]}{}_{\,k_{q-m};\, k_{q-m+1} \ldots\, k_{q}\,]} \\
& +\, (-1)^{\,q}\, \frac{q+1}{p}\, k_{\,p-1\,,\,q+1} \, \sum_{n\,=\,1}^p\, \prod_{r\,\neq\,n}^p \, T_{i_rj_r}\, \g_{\,[\,k_1 \ldots\, k_{q-m+1}\,|}\, \tilde{\theta}^{\,[\,m\,]}{}_{(\,i_n;\,j_n\,)\,|\,k_{q-m+2} \ldots\, k_q\,]} \\
& +\, (-1)^{\,q+1}\, \frac{q+1}{p}\, k_{\,p-1\,,\,q+1} \, \sum_{n\,=\,1}^p\, \prod_{r\,\neq\,n}^p \, T_{i_rj_r}\, \g_{\,[\,k_1 \ldots\, k_{q-m}\,|\,(\,i_n}\, \tilde{\theta}^{\,[\,m\,]}{}_{\,j_n\,)\,;\,|\,k_{q-m+1} \ldots\, k_q\,]} \bigg\} + \, \ldots \, ,
\end{split}
\ee
where we are tracking only terms that can contribute to $\theta^{\,[\,m+1\,]}$. Taking into account the (anti)commutators \eqref{commgproj} and the coefficients $k_{\,p\,,\,q}$ of eq.~\eqref{solgenf}, eq.~\eqref{shiftEanti1} finally becomes
\be \label{shiftEanti2}
\begin{split}
& \d \, E_{\,\bar{\psi}} \, = \, \sum_{p\,,\,q\,=\,0}^N k_{\,p\,,\,q}\ \h^{i_1j_1} \!\ldots \h^{i_pj_p}\, \g^{\,k_1 \ldots\, k_q}\, Y_{\{2^p,1^q\}}\, \times \\
& \times\, \bigg\{\, T_{i_1j_1} \!\ldots T_{i_pj_p} \, \g_{\,[\,k_1 \ldots\, k_{q-m-1}\,|}\, \cO\,(\,D-m-2\,)^{\,l}{}_{\,|\,k_{q-m}\,|}\, \tilde{\theta}^{\,[\,m\,]}{}_{\,l\,;\,|\,k_{q-m+1} \ldots\, k_q\,]} \\
& +\, \sum_{n\,=\,1}^p\, \prod_{r\,\neq\,n}^p \, T_{i_rj_r}\, \g_{\,[\,k_1 \ldots\, k_{q-m+1}\,|}\, \tilde{\theta}^{\,[\,m\,]}{}_{(\,i_n;\,j_n\,)\,|\,k_{q-m+2} \ldots\, k_q\,]} \,\bigg\} \, + \, \ldots \, ,
\end{split}
\ee
where we have introduced the operators
\be \label{O}
\cO\,(\,\l\,)^{\,l}{}_{\,m} \, = \, \l\ \d^{\,l}{}_{\,m} + \, 2 \, S^{\,l}{}_{\,m} \, .
\ee
Comparing this result with eq.~\eqref{shiftEgen}, one can thus recognize that $\theta^{\,[\,m+1\,]}$ is of the form
\be \label{theta_to_tilde}
\theta^{\,[\,m+1\,]}{}_{\,k_1 \ldots\, k_{m+1}} \, \equiv \, \cO\,(\,D-m-2\,)^{\,l}{}_{\,[\,k_1\,|}\, \tilde{\theta}^{\,[\,m\,]}{}_{\,l\,;\,|\,k_2 \ldots\, k_{m+1}\,]} \, ,
\ee
while other contributions in eq.~\eqref{shiftEanti2} are not relevant for the present discussion, but would enter the definition of other $\theta$'s.

The next issue is to exhibit the $\O$-dependence of the $\tilde{\theta}^{\,[\,m\,]}$ and $\theta^{\,[\,m\,]}$ spinor-tensors, and the $\cO\,(\,\l\,)$ operators introduced in eq.~\eqref{O} are particularly convenient to this end. In fact, let us begin by guessing the result,
\be \label{thetam}
\begin{split}
& \theta^{\,[\,m\,]}{}_{\,k_1 \ldots\, k_m} \, = \, \sum_{n\,=\,0}^{[\frac{m}{2}]}\ (-1)^{\,\frac{(m-n)(m-n-1)}{2}} \, \prod_{r\,=\,1}^{n} \, S^{\,p_r}{}_{|\,k_r\,|} \, \times \\
& \times \, \prod_{s\,=\,1}^{m-2n} \cO\,(\,D-m-s\,)^{\,q_s}{}_{|\,k_{n+s}\,|} \, \O^{\,[\,n\,,\,m-2n\,]}{}_{p_1\,|\,k_{m-n+1}\,|\,, \,\ldots\, ,\,p_n\,|\,k_m\,]\,;\,q_1 \ldots\, q_{m-2n}} \, ,
\end{split}
\ee
where the indices within vertical bars are fully antisymmetrized and the products where the maximum value of the dummy index lies below the lower end are to be regarded as equal to unity.

Our task is now to prove that eq.~\eqref{thetam} is indeed correct, performing the redefinitions \eqref{redef1} or \eqref{redef2} and commuting the additional $\g$-matrix past the other operators acting on the $\O$'s. This identifies $\tilde{\theta}^{\,[\,m\,]}$, and multiplying it by a further $\cO\,(\,D-m-2\,)$ as in eq.~\eqref{theta_to_tilde} should yield eq.~\eqref{thetam} with $m \to m+1$. Moving the $\g$-matrix to the left of all the operators in eq.~\eqref{thetam} transforms a generic term of the sum into
\be \label{comm_theta1}
\begin{split}
& (-1)^{\,\frac{(m-n)(m-n-1)}{2}} \left(\, \g^{\,l}\, \prod_{r\,=\,1}^{n}\, S^{\,p_r}{}_{|\,k_r\,|} + \, \Big[\, \prod_{r\,=\,1}^{n}\, S^{\,p_r}{}_{|\,k_r\,|} \comma \g^{\,l} \,\Big] \,\right) \times \\
& \times \prod_{s\,=\,1}^{m-2n} \cO\,(\,D-m-s-2\,)^{\,q_s}{}_{|\,k_{n+s}\,|} \, \O^{\,[\,n\,,\,m-2n+1\,]}{}_{p_1\,|\,k_{m-n+1}\,|\,, \,\ldots\, ,\,p_n\,|\,k_m\,]\,;\,q_1 \ldots\, q_{m-2n}\,l} \, ,
\end{split}
\ee
while for odd values of $m$ the last term of the sum becomes
\be \label{last_odd}
\begin{split}
& \left(\, \g^{\,l}\, \prod_{r\,=\,1}^{\frac{m-1}{2}}\, S^{\,p_r}{}_{|\,k_r\,|} + \, \Big[\, \prod_{r\,=\,1}^{\frac{m-1}{2}}\, S^{\,p_r}{}_{|\,k_r\,|} \comma \g^{\,l} \,\Big] \,\right) \, \times \\
& \times \, \bigg\{\, \cO\,(\,D-m-3\,)^{\,q}{}_{|\,k_{\frac{m+1}{2}}\,|} \,  \O^{\,[\,\frac{m-1}{2}\,,\,2\,]}{}_{p_1\,|\,k_{\frac{m+3}{2}}\,|\,, \,\ldots\, ,\,p_{\frac{m-1}{2}}\,|\,k_m\,]\,;\,q\,l} \\
& + \, \12 \ \cO\,(\,D-m+1\,)^{\,q}{}_{|\,k_{\frac{m+1}{2}}\,|} \, \O^{\,[\,\frac{m+1}{2}\,,\,0\,]}{}_{p_1\,|\,k_{\frac{m+3}{2}}\,|\,, \,\ldots\, ,\,p_{\frac{m-1}{2}}\,|\,k_m\,]\,,\,q\,l} \,\bigg\} \, .
\end{split}
\ee
The symmetries of the parameter $\O^{\,[\,n\,,\,m-2n+1\,]}$ then reduce the commutator in eq.~\eqref{comm_theta1} to
\be \label{commS1}
\Big[\, \prod_{r\,=\,1}^{n}\, S^{\,p_r}{}_{|\,k_r\,|} \comma \g^{\,l} \,\Big] \, \to \, \sum_{r\,=\,1}^n \, \g^{\,p_r} \, \d_{\,|\,k_r\,|}{}^{\,l} \, \prod_{s\,\neq\,r}^n \, S^{\,p_s}{}_{|\,k_s\,|} \, ,
\ee
since they force $S^{\,p_s}{}_{|\,k_s\,|}$ and $\g^{\,p_r}$ to commute.

One should now multiply $\tilde{\theta}^{\,[\,m\,]}$, that is obtained combining eqs.~\eqref{comm_theta1} and \eqref{last_odd} and stripping off an overall $\g$-matrix, by the additional $\cO\,(\,D-m-2\,)$ as in eq.~\eqref{theta_to_tilde}. One can conveniently begin by considering the contributions in \eqref{comm_theta1}, whose first line thus turns into
\begin{align}
& (-1)^{\,\frac{(m-n)(m-n-1)}{2}} \, \bigg\{\, (-1)^{\,n} \prod_{r\,=\,1}^{n}\, S^{\,p_r}{}_{|\,k_r\,|} \, \cO\,(\,D-m-2\,)^{\,l}{}_{\,|\,k_{n+1}\,|} \, + \, 2\, \Big[\, S^{\,l}{}_{[\,k_1\,|} \comma \prod_{r\,=\,1}^{n}\, S^{\,p_r}{}_{|\,k_{r+1}\,|} \,\Big] \nn \\
& + \, 2\, \sum_{r\,=\,1}^n \, S^{\,p_r}{}_{[\,k_1\,|} \, \d_{\,|\,k_{r+1}\,|}{}^{\,l} \, \prod_{s\,\neq\,r}^n \, S^{\,p_s}{}_{|\,k_s\,|} \, \bigg\} \, ,  \label{Spart}
\end{align}
after moving to the right the overall $\cO\,(\,D-m-2\,)$, again on account of the symmetries of the $\O$ parameters. In a similar fashion, the commutator in eq.~\eqref{Spart} reduces to
\be \label{commS2}
\Big[\, S^{\,l}{}_{[\,k_1\,|} \comma \prod_{r\,=\,1}^{n}\, S^{\,p_r}{}_{|\,k_{r+1}\,|} \,\Big] \, \to \, - \, \sum_{r\,=\,1}^n \, S^{\,p_r}{}_{[\,k_1\,|} \, \d_{\,|\,k_{r+1}\,|}{}^{\,l} \, \prod_{s\,\neq\,r}^n \, S^{\,p_s}{}_{|\,k_s\,|} \, ,
\ee
so that it cancels against the last term in eq.~\eqref{Spart}, that originates from eq.~\eqref{commS1}. Keeping track of an additional factor $(-1)^{\,m-2n}$ from the reordering of the antisymmetric indices of $\O^{\,[\,n\,,\,m-2n+1\,]}$ finally leads to eq.~\eqref{thetam} with $m \to m+1$, which vindicates our original guess, up to the terms in eq.~\eqref{last_odd}, that however behave in the same fashion since
\be
\cO\,(\,\l\,)^{\,l}{}_{\,[\,i\,|} \, \cO\,(\,\r\,)^{\,q}{}_{\,|\,j\,]} \, \O^{\,[\,n\,,\,0\,]}{}_{\,\ldots\,,\,l\,q} \, = \, 2\,(\,\r-\l-2\,) \, S^{\,l}{}_{\,[\,i\,|}\, \O^{\,[\,n\,,\,0\,]}{}_{\,\ldots\,,\,|\,j\,]\,l}\, ,
\ee
as one can prove with arguments similar to those of eq.~\eqref{commS}.

Having identified the structure of the $\theta^{\,[\,m\,]}$ spinor-tensors, we can finally show that they vanish when eqs.~\eqref{chain} hold. To this end, it is convenient to recast the sum \eqref{thetam} in the form
\begin{align}
& \theta^{\,[\,m\,]}{}_{\,k_1 \ldots\, k_m} = \sum_{n\,=\,0}^{[\frac{m}{2}]}\, (-1)^{\,\frac{(m+n-1)(m+n-2)}{2}} \left(\frac{m-2\,n}{m}\right) \prod_{r\,=\,1}^{n} \, S^{\,p_r}{}_{|\,k_r\,|} \!\! \prod_{s\,=\,1}^{m-2n-1} \!\!\cO\,(\,D-m-s-1\,)^{\,q_s}{}_{|\,k_{n+s}\,|} \nn \\
& \times \bigg\{\, \cO\,(\,D-m+n-1\,)^{\,l}{}_{\,|\,k_{m-n}\,|} \, \O^{\,[\,n\,,\,m-2n\,]}{}_{p_1\,|\,k_{m-n+1}\,|\,, \,\ldots\, ,\,p_n\,|\,k_m\,]\,;\,l\,q_1 \ldots\, q_{m-2n-1}} \label{thetam2}\\
& + \frac{(-1)^{\,n}\,(\,n+1\,)}{(\,m-2\,n\,)\,(\,m-2\,n-1\,)}\ \O^{\,[\,n+1\,,\,m-2(n+1)\,]}{}_{p_1\,|\,k_{m-n+1}\,|\,, \,\ldots\, ,\,p_n\,|\,k_{m-1},\,k_{m}\,]\,[\,q_1;\,q_2 \ldots\, q_{m-2n-1}\,]} \,\bigg\} . \nn
\end{align}
Before proving that this is indeed possible, let us verify that eq.~\eqref{thetam2}  suffices to reach our conclusion. In fact, the $\O$-parameters can only contribute to eqs.~\eqref{thetam} and \eqref{thetam2} via their unique two-column Young projections in the family indices, since expanding them as in eq.~\eqref{O} and making use of eqs.~\eqref{chain} one can effectively replace all the $S^{\,q_s}{}_{|k_{n+s}|}$ with various $\d$'s. As a result, all projections with more than two columns of the resulting terms vanish, simply because it is not possible to antisymmetrize them over $m-n$ indices. On the other hand, the terms within braces in \eqref{thetam2} vanish if eqs.~\eqref{chain} hold, since they are just the antisymmetrization in its $(j_r,k)$ indices of
\be \label{projected_chain}
\begin{split}
& \cO\,(\,D-m+n-1\,)^{\,l}{}_{\,k} \, \O^{\,[\,n\,,\,m-2n\,]}{}_{i_1j_1, \,\ldots\, ,\,i_nj_n\,;\,l\,q_1 \ldots\, q_{m-2n-1}} \\
& + \frac{(\,n+1\,)}{(\,m-2\,n\,)\,(\,m-2\,n-1\,)}\ \O^{\,[\,n+1\,,\,m-2(n+1)\,]}{}_{i_1j_1, \,\ldots\, ,\,i_nj_n,\,k\,[\,q_1;\,q_2 \ldots\, q_{m-2n-1}\,]} \, = \, 0 \, ,
\end{split}
\ee
that collects the contributions to eqs.~\eqref{chain} from two-column projected parameters. Let us stress that the $\cO$ operators make it possible to deal with these expressions rather efficiently, so that we expect a similar behavior for the generic $\theta^{\,[\,n\,,\,m-2n\,]}$ appearing in eq.~\eqref{shiftEgen}~\footnote{The proof that the bosonic field equations are invariant under the shift symmetries of the Bianchi identities, not presented explicitly in \cite{bose_mixed}, also rests on the emergence of products of $\cO$ operators, that are actually commuting when the conditions imposed by the Bianchi identities hold.}.

We can now close the present appendix, and hence the whole paper, proving the identities \eqref{thetam2}. First of all, the symmetries of the $\O$ parameters make it possible to replace any of the $S^{\,p_r}{}_{|k_r|}$ operators in eq.~\eqref{thetam} with $\frac{1}{2} \, \cO\,(\,\l\,)^{\,p_r}{}_{|k_r|}$, with an arbitrary argument $\l$. One can thus recast the individual contributions labeled by $n$ in eq.~\eqref{thetam} in the form
\begin{align}
& (-1)^{\frac{(m-n)(m-n-1)}{2}} \, \frac{n}{m\,(\,m-2\,n+1\,)} \, \prod_{r\,=\,1}^{n-1} \, S^{\,p_r}{}_{|\,k_r\,|} \prod_{s\,=\,1}^{m-2n+1} \cO\,(\,D-m-s-1\,)^{\,q_s}{}_{|\,k_{n+s-1}\,|}\, \times \nn \\
& \times \, \O^{\,[\,n\comma m-2n\,]}{}_{p_1\,|\,k_{m-n+1}\,|\,, \,\ldots\, ,\,p_{n-1}\,|\,k_{m-1},\,k_m\,]\,[\,q_1;\, q_2 \ldots\, q_{m-2n+1}\,]} \nn \\[5pt]
& - \, \frac{n\, (-1)^{\frac{(m-n)(m-n-1)}{2}}}{m\,(\,m-2\,n+1\,)} \, \prod_{r\,=\,1}^{n-1} \, S^{\,p_r}{}_{|\,k_r\,|} \, \bigg\{\, 2 \sum_{s\,=\,1}^{m-2n+1} \!(-1)^{\,s+1} \prod_{a\,=\,1}^{s-1} \cO\,(\,D-m-a-1\,)^{\,q_a}{}_{|\,k_{n+a-1}\,|} \, \times \nn \\
& \times \, S^{\,p_n}{}_{|\,k_{n+a}\,|} \prod_{b\,=\,s+1}^{m-2n+1} \cO\,(\,D-m-b-1\,)^{\,q_{b-1}}{}_{|\,k_{n+b-1}\,|} \ - \ \frac{m\,(\,m-2\,n+1\,)}{n}\ S^{\,p_n}{}_{|\,k_n\,|}\, \times \nn \\
& \times \, \prod_{s\,=\,1}^{m-2n} \, \cO\,(\,D-m-s\,)^{\,q_s}{}_{|\,k_{n+s}\,|} \,\bigg\} \ \O^{\,[\,n\,,\,m-2n\,]}{}_{p_1\,|\,k_{m-n+1}\,|\,, \,\ldots\, ,\,p_n\,|\,k_m\,]\,;\,q_1 \ldots\, q_{m-2n}} \, , \label{step1}
\end{align}
adding and subtracting the first term, that actually coincides with the second contribution within braces in \eqref{thetam2} after replacing $n$ with $n-1$. Moving all ``naked'' $S$ operators to the left of the $\cO$ operators in this expression using
\be
\cO\,(\,\l\,)^{\,q_s}{}_{[\,i\,|} \, S^{\,p_n}{}_{\,|\,j\,]} \, \to \, - \, S^{\,p_n}{}_{[\,i\,|} \, \cO\,(\,\l+2\,)^{\,q_s}{}_{|\,j\,]} \, ,
\ee
the sum within braces can now be turned into
\begin{align}
& S^{\,p_n}{}_{|\,k_n\,|} \, \bigg\{\, 2 \sum_{s\,=\,1}^{m-2n+1} \prod_{a\,=\,1}^{s-1} \cO\,(\,D-m-a+1\,)^{\,q_a}{}_{|\,k_{n+a}\,|} \prod_{b\,=\,s}^{m-2n} \cO\,(\,D-m-b-2\,)^{\,q_b}{}_{|\,k_{n+b}\,|} \label{step2} \\
& - \, \frac{m\,(\,m-2\,n+1\,)}{n}\, \prod_{s\,=\,1}^{m-2n} \cO\,(\,D-m-s\,)^{\,q_s}{}_{|\,k_{n+s}\,|} \,\bigg\} \, \O^{\,[\,n\,,\,m-2n\,]}{}_{p_1\,|\,k_{m-n+1}\,|\,, \,\ldots\, ,\,p_n\,|\,k_m\,]\,;\,q_1 \ldots\, q_{m-2n}} \, . \nn
\end{align}
One can finally observe that the $\cO$ operators in \eqref{thetam} commute,
\be \label{commO}
\begin{split}
& \left[\, \cO\,(\,\l\,)^{\,p}{}_{\,[\,i\,|} \comma \cO\,(\,\r\,)^{\,q}{}_{|\,j\,]} \,\right] \, \O^{\,[\,n\,,\,m-2n\,]}{}_{\ldots\,; \,\ldots\, p \,\ldots\, q \,\ldots} \\
& = \, 4 \, \left[\, S^{\,p}{}_{\,[\,i\,|} \comma S^{\,q}{}_{|\,j\,]} \,\right] \, \O^{\,[\,n\,,\,m-2n\,]}{}_{\ldots\,; \,\ldots\, p \,\ldots\, q \,\ldots} = \, 0 \, ,
\end{split}
\ee
since the antisymmetrization in $(i,j)$ tries to make the commutator symmetric in $(p,q)$. Furthermore, expanding the $\cO$ operators, the resulting terms where only products of $\d$'s appear suggest that eq.~\eqref{step2} can be recast in the form
\begin{align}
& (-1)^{\,\frac{(m+n-1)(m+n-2)}{2}} \, \left(\frac{m-2\,n}{m}\right) \, \prod_{r\,=\,1}^{n} \, S^{\,p_r}{}_{|\,k_r\,|} \prod_{s\,=\,1}^{m-2n-1} \cO\,(\,D-m-s-1\,)^{\,q_s}{}_{|\,k_{n+s}\,|} \nn \\
& \times \, \cO\,(\,D-m+n-1\,)^{\,l}{}_{\,|\,k_{m-n}\,|} \, \O^{\,[\,n\,,\,m-2n\,]}{}_{p_1\,|\,k_{m-n+1}\,|\,, \,\ldots\, ,\,p_n\,|\,k_m\,]\,;\,l\,q_1 \ldots\, q_{m-2n-1}} \, , \label{step3}
\end{align}
which completes the proof. The factors accompanying these terms are indeed
\begin{align}
& 2 \sum_{r\,=\,1}^{m-2n+1} \, \prod_{a\,=\,1}^{r-1} (\,D-m-a+1\,) \! \prod_{b\,=\,r}^{m-2n}  (\,D-m-b-2\,) \, - \, \frac{m\,(\,m-2\,n+1\,)}{n} \prod_{a\,=\,1}^{m-2n} (\,D-m-a\,) \nn \\
& = \, - \, \frac{(\,m-2\,n+1\,)\,(\,m-2\,n\,)}{n} \, \prod_{a\,=\,1}^{m-2n} (\,D-m-a-1\,) \, (\,D-m+n-1\,) \, , \label{prod_num}
\end{align}
and agree with the structure in eq.~\eqref{step3} taking into account that in the resulting product one is to move the $l$ index to the first position, which gives rise to an additional factor $(-1)^{m+1}$. While we have not fully proved eq.~\eqref{step3}, we have verified that terms of higher order in $S$ work out in a number of cases.
\end{appendix}

\newpage


\end{document}